\documentclass[a4paper, 12pt,  epsfig]{report}
\usepackage[T1,T2A]{fontenc}
\usepackage[utf8]{inputenc}
\usepackage[english, russian]{babel}
\usepackage{amsmath}
\usepackage{amsfonts, amssymb,  longtable}
\usepackage[dvips]{graphicx}
\usepackage{caption}

\newcommand{\eqdef }{\stackrel{ def}{=}}
\begin{document}
\title{\bf On the theory of a vector field with symmetric affinors.\\
 Real vector field in the framework of the standard methods.}
\author{Yu. A. Alebastrov \thanks{jalebastrov@gmail.com} \\
 141980 Dubna, Moscow region, RUSSIA }
\date{2.10.2015}
\maketitle
\renewcommand{\abstractname}{Abstract}
\begin{abstract}

\qquad  Attention is drawn to the mathematical equality of rights of symmetrical constituents derived affinor of a vector field
   in relation to its antisymmetric constituents.

In this regard, raises the question  not only of equitable accounting, but  and mainly question of the real existence of fields, represented by these constituents.

In particular, we conclude that the classical electromagnetic field at any point of space\,-\,time accompanied, in the General case,
independent  {\em physical} field, defined symmetrical derived affinor of  4-potential of classical electrodynamics.

 Discussed, within the framework of  the Bogolyubov and Shirkov axiomatic, a theory of real vector field, clearly and equitably
 taking into account the symmetric derived affinors this field and found a number  of important distinguishing features this model.

 Despite accounting explicitly gauge-noninvariant constituents,  the proposed theory has specialized
  gauge invariance, which provides, in particular, conservation of electric current.

 In this connection, the difficulties with the probabilistic interpretation, in the case of indefinite-metric version of the theory, are
 surmountable by standard methods.

 Finally, is proposed the physical content of the fields, defined of a symmetrical derived affinor  of 4-potential of classical electrodynamics.


\end{abstract}

\maketitle
\renewcommand{\abstractname}{Аннотация}
\begin{abstract}

Обращается внимание на математическое равноправие симметрических составляющих производного аффинора векторного поля по отношению к его антисимметрической составляющей.

 В связи с этим поднимаются вопросы не только равноправного учёта, но и, главным образом, равноправного и реального  существования полей представленных этими составляющими.

 Заключается, в частности, что классическое электромагнитное поле в любой точке пространства\,-\,времени сопровождается, в общем случае, самостоятельным  {\em физическим} полем, определяемым   симметрическим производным аффинором 4-потенциала  электродинамики.

Рассмотрена, в рамках аксиоматики Боголюбова и Ширкова, теория вещественного векторного поля, {\em явно и равноправно} учитывающая симметрические производные аффиноры этого поля и найден целый ряд важных отличительных особенностей
данной модели.

Несмотря на учёт явно калибровочно-неинвариантных составляющих, предложенная теория обладает специализированной
калибровочной инвариантностью, обеспечивающей, в частности, сохранение тока.

В связи с этим, трудности с вероятностной интерпретацией, при индефинитно\,-\,метрическом варианте теории, оказываются преодолимыми стандартными способами.

Наконец, предлагается физическое содержание полей, определяемых   симметрическим производным аффинором 4-потенциала  классической электродинамики.

Ключевые слова: векторное поле, симметрический производный аффинор, калибровочная инвариантность.

\end{abstract}

\part* {Часть I}
\addcontentsline{toc}{part}{Часть I}

\chapter {Классическое действительное векторное поле в рамках стандартных методов}

\section* {Введение}
\addcontentsline{toc}{section}{Введение}

\qquad Обращается внимание на математическое равноправие и самостоятельность симметрических составляющих  производного аффинора векторного поля \cite{Rash} по отношению к его антисимметрической составляющей.

В связи с этим, поднимаются вопросы не только равноправного учёта, но и, главным образом, равноправного и реального  существования полей представленных этими составляющими.

 Заключается, в частности, что классическое электромагнитное поле в любой точке пространства--времени сопровождается, в общем случае,  самостоятельным {\em физическим} полем, определяемым   симметрическим производным аффинором 4-потенциала классической  электродинамики.

Последнее составляет определяющую идеологию представленной теории векторного поля и ее главный отличительный признак.

Рассмотрена, в рамках аксиоматики \cite{BSh}, теория вещественного векторного поля, {\em явно и равноправно} учитывающая симметрические производные аффиноры этого поля.

Получены коммутационные соотношения для последних.

При этом, коммутационные соотношения для антисимметрической составляющей в точности соответствуют их стандартному виду.

Найдены так же  следующие основные отличительные особенности обсуждаемой теории.

1. В частности, имеем систему дифференциальных уравнений, описывающую полевую систему  определяемую
   симметрическим производным аффинором 4-потенциала классической электродинамики.

2. Общее решение уравнений движения рассматриваемого векторного поля удовлетворяет двухмассовому  уравнению четвертого порядка.

3. Коммутатор рассматриваемого поля допускает безмассовый предел и в этом пределе получаем перестановочные соотношения для компонент 4-потенциала электромагнитного поля.

4. Самая  сильная  особенность  функции  Паули\,--\,Йордана  на световом конусе,  $\delta$-функция, исключена из-под знака производной второго порядка коммутатора.

5. Теория  допускает  “причинное” определение  причинной функции векторного  поля, аналогичное
 определению причинной функции  скалярного  поля.

6.  Представляются  две возможные формулировки теории: в оснащённом гильбертовом пространстве состояний и в пространстве состояний с индефинитной метрикой.

7. Пропагатор свободного векторного поля:

а) не содержит под знаком производной второго порядка самых сильных особенностей, $\delta (\lambda)$ и $\lambda ^{-1}$, на световом конусе;

б) не содержит члена вида  $k_{m}k_{n}/m^{2}$, приводящего к суровой \cite{Bern} ультрафиолетовой проблеме;

в) допускает безмассовый предел, приводящий к фотонной функции Грина в общековариантной калибровке;

г) приводит к ренормируемой теории, вне зависимости от условия сохранения тока;

д) позволяет провести процедуру ренормализации последовательно, без дополнительной перестройки пропагатора на промежуточном этапе;

е) произвол калибровки фиксируется коэффициентами при членах лагранжиана, учитывающих симметрические аффиноры.

Несмотря на учёт явно калибровочно-неинвариантных составляющих, предложенная теория обладает специализированной калибровочной инвариантностью, обеспечивающей, в частности, сохранение тока.

В связи с этим, трудности с вероятностной интерпретацией, при индефинитно-метрическом варианте теории, оказываются преодолимыми стандартными способами.

Наконец, предлагается физическое содержание полей, определяемых симметрическим  производным аффинором
 4-потенциала классической электродинамики.

Подчеркивается, что все вышеперечисленные особенности  обсуждаемой модели являются
 не определяющими, а сопровождающими ее, что, в свою очередь, является важным  идеологическим отличием
  данной теории от теорий векторного поля, предлагаемых вплоть до настоящего  времени.

\section*{\S \,1  Аксиоматика теории }
\addcontentsline{toc}{section}{\S \,1  Аксиоматика теории}

\qquad Общеизвестно, что представление группы Лоренца, определенное на пространстве тензоров
 второго ранга, в общем случае, не является неприводимым.

При этом хорошо известно также, что произвольный тензор этого пространства однозначно представляется в релятивистски-инвариантном виде
$$
T(x) = T_{1}(x)+T_{2}(x)+T_{3}(x),             \eqno\ldots (1.1)
$$
где ковариантные компоненты (координаты) составляющих правой части определены соотношениями
$$
  T_{1mn}(x) = T_{[mn]}(x) = \frac{1}{2}\,(T_{mn}(x) - T_{nm}(x)),
$$
$$
 T_{2mn}(x) = T_{d(mn)}(x) = \frac{1}{2}\,(T_{mn}(x) + T_{nm}(x)) - \frac{1}{4}\,T_{k}^{k}(x)\,\textsl{g}_{mn},
$$
$$
 T_{3mn}(x) = T_{\ast (mn)}(x) =  \frac{1}{4}\,T_{k}^{k}(x)\,\textsl{g}_{mn}.
$$

Представление (1.1) приводит к соответствующему представлению ковариантных компонент
 произвольного тензора второго ранга
$$
T_{mn}(x) = T_{[mn]}(x)+T_{d(mn)}(x)+T_{\ast(mn)}(x).            \eqno\ldots (1.2)
$$

Таким образом, рассматриваемое тензорное пространство, благодаря существованию инвариантных, относительно  преобразований Лоренца, операций альтернирования, симметризации и свёртки пар тензорных индексов, представляется прямой суммой трех инвариантных подпространств, имеющих размерности 6, 9 и 1.

 Данная инвариантность данных подпространств, в частности, означает, что, при переходе от одной инерциальной
  системы  отсчета  к  другой  инерциальной  системе  отсчета, симметрия данного тензора из данного инвариантного подпространства сохраняется.

 То есть, путём выбора соответствующей инерциальной системы отсчета нельзя симметрические составляющие свести к антисимметрическим и наоборот.

 Именно это обстоятельство мы, в первую очередь,  имеем в виду,  когда говорим о самостоятельности или равноправии этих составляющих, либо о самостоятельности или равноправии полей представленных (описывающихся) этими составляющими.

 Как частный случай (1.1), производный аффинор $ \partial A(x)$ \cite{Rash} действительного векторного поля  $ A(x)$ имеет следующее локальное представление, обладающее теми же свойствами инвариантности, что и (1.1),
 $$
\partial A(x) = F(x) +  G(x) + H(x),              \eqno\ldots (1.3)
 $$
где ковариантные компоненты (координаты) производных аффиноров правой и левой частей  определены соотношениями
$$
\partial A_{mn}(x) = \partial_{m} A_{n}(x) ,
 $$
$$
F_{mn}(x) =\partial A _{[mn]}(x) = \frac{1}{2}\,(\partial_{m} A_{n}(x) - \partial_{n} A_{m}(x)),
$$
$$
 G_{mn}(x) = \partial A_{d(mn)}(x) = \frac{1}{2}\,(\partial_{m} A_{n}(x) + \partial_{n} A_{m}(x)) - \frac{1}{4}\,\partial_{k}A^{k}(x)\,\textsl{g}_{mn},
$$
$$
 H_{mn}(x) =  \partial A_{\ast(mn)}(x) =   \frac{1}{4}\,\partial_{k}A^{k}(x)\,\textsl{g}_{mn}.
$$

Тут и далее можно было бы не использовать термин  "производный  аффинор векторного поля"\,, а  говорить о производном тензоре векторного поля или о градиенте векторного поля, однако мы предпочитаем следовать терминологии \cite{Rash},  с учетом того, что координаты всякого двухвалентного тензора, в том числе и координаты рассматриваемого производного тензора, могут быть истолкованы как координаты некоторого аффинора \cite{Rash}.

Представление (1.3) приводит к соответствующему представлению ковариантных компонент
 производного аффинора $ \partial A(x)$
$$
\partial _{m}A_{n}(x) = F_{mn}(x) +  G_{mn}(x) + H_{mn}(x).              \eqno\ldots (1.4)
 $$

Таким образом, поля, определяемые симметрическими производными аффинорами, $ G(x)$  и $ H(x)$, выступают совершенно равноправно и самостоятельно по отношению к полю, определяемому антисимметрическим производным аффинором $F(x)$.

Но, как общеизвестно, поле, определяемое соответствующим антисимметрическим производным аффинором электродинамического потенциала классической электродинамики, -- это традиционное классическое электромагнитное поле.

Возникает совершенно  естественный вопрос о судьбе, как в математическом, так и в физическом отношениях, {\em инвариантного} полевого объекта, представленного  симметрическим производным аффинором классического электродинамического потенциала,  $ G(x)$, попытке ответа на который и посвящается  обсуждаемая   теория.

Вышеуказанное равноправие и самостоятельность сопоставляемых аффиноров, {\em с самого  начала} рассматриваются  как фундаментальное  {\em математическое} основание к тому, чтобы утверждать,  что, {\em наряду} с общеизвестным традиционным {\em классическим} электромагнитным полем, представляемым  антисимметрическим производным аффинором,  существует и равноправное самостоятельное {\em физическое} поле, определяемое симметрическим производным аффинором  классического  электродинамического 4-потенциала \cite{Al1}--\cite{Al3}.

При этом, в  силу того, что множество симметрических тензоров второго ранга, так же  как и множество антисимметрических тензоров того же ранга, являются линейными многообразиями \cite{LS}, то  любой симметрический тензор второго ранга не представляется  линейной комбинацией соответствующих антисимметрических тензоров того же ранга, вследствие чего,  поле, определяемое симметрическим производным аффинором электродинамического 4-потенциала $A(x)$, $ G(x)$,
не рассматривается нами в качестве {\em составляющей} классического электромагнитного поля, определяемого соответствующим антисимметрическим производным аффинором,  и,  в связи с этим, мы предпочитаем присвоить  ему и в дальнейшем
использовать, в рамках обсуждаемой теории, {\em самостоятельное} наименование, а именно,  поле, определяемое симметрическим производным аффинором  классического электродинамического 4-потенциала $A(x)$, $G(x)$, будем называть далее электроджейтонным полем.

Однако, известные нам до сих пор теории векторного поля, оперирующие с $F(x)$, полностью игнорируют $G(x)$  и лишь иногда привлекают $H(x)$ \cite{Schw}, но рассматривают введение последнего как вспомогательную процедуру  при проведении канонического квантовании (общеизвестный приём Ферми) или при получении желаемого пропагатора \cite{Coest}--\cite{GO}.

В силу вышеуказанного, мы поднимаем вопрос не только равноправного учёта $ G(x)$  и $ H(x)$  в общей теории
 векторного поля, но и, главным образом, вопрос равноправного и самостоятельного существования поля, определяемого
 симметрическим производным аффинором электродинамического 4-потенциала $A(x)$.

Таким образом, в силу локального представления (1.3), если в некоторой точке пространства-времени мы имеем поле,
определяемое антисимметрическим аффинором  (4-тензором электромагнитного поля) $F(x)$, то в этой же точке, в общем случае, существует и самостоятельное равноправное поле, определяемое симметрическим аффинором (4-тензором электроджейтонного поля) $ G(x)$,  обладающее, как увидим ниже,  так же и самостоятельным  {\em физическим} содержанием.

С другой стороны, $F(x)$,  $ G(x)$  и $ H(x)$  нельзя представлять совершенно оторванными друг от друга, так как они являются дифференциальными характеристиками одного и того же векторного поля $A(x)$, тем самым, классический электродинамический потенциал $A(x)$, например, является, в общем случае, носителем как традиционного классического электромагнитного поля, так и
классического электроджейтонного поля, определяемого 4-тензором электроджейтонного поля, $ G(x)$, в связи с чем, $A(x)$ рассматривается нами как потенциал электромагнитного и электроджейтонного полей (как электромагнитоджейтонный потенциал),
выступающий в качестве {\em единого} полевого объекта (единого "начала"), представляющего одновременно  {\em оба} рассматриваемых поля.

Последнее обстоятельство необходимо учитывать   в квантовой теории, когда электродинамический потенциал, совместно с соответствующей калибровочной инвариантностью, начинает играть определяющую (ключевую) роль, в связи с чем само электромагнитное поле, например, относят к классу векторных полей, в частности, к классу безмассовых абелевых векторных  калибровочных полей с соответствующей группой внутренней симметрии.

В данном контексте,  мы начнём с  рассмотрения вопросов, связанных с  явным и  равноправным учетом полей определяемых симметрическими производными аффинорами в теории классического вещественного векторного поля, используя при этом только рамки стандартного формализма аксиоматики  \cite{BSh}.

Тут и далее, в целях упрощения записей, используется один и тот же буквенный символ, $A(x)$, как для обозначения
 функций поля \cite{BSh}, представляющих электродинамический 4-потенциал, так и для для обозначения
полевых функций, представляющих "массивное"\, действительное векторное поле \cite{IZ}.

Метрика, общие обозначения и соглашение о суммировании по дважды повторяющимся индексам соответствуют  \cite{BSh}.

И так, мы исследуем модель теории действительного векторного поля $A(x)$, в которой осуществлен переход
 $$
\{ \partial A(x)\} \Rightarrow \cup\{F(x)\},\{ G(x)\}, \{H(x)\},              \eqno\ldots (1.5)
 $$
то есть, при описании свойств самого поля \cite{La}, будем исходить из локального лагранжиана, являющегося функцией
$$
L(x) = L(A(x), F(x), G(x), H(x)).                   \eqno\ldots (1.6)
$$

При этом, так как стандартное условие классической теории,
$$
\partial \!\cdot \!A(x) = 0,                    \eqno\ldots (1.7)
$$
восстанавливающее, в частности, положительную определенность энергии векторного поля, означает, что
$$
H(x) = \widetilde{0},                      \eqno\ldots (1.8)
$$
то есть, вновь возвращает нас к неучету одной из симметрических составляющих,  то, отдавая предпочтение математической полноте теории, мы не предполагаем выполнения (1.7) с самого начала.

Уместно подчеркнуть также, что уравнение (1.7), в электродинамическом случае, предпочтительнее рассматривать не просто как соответствующее вспомогательное условие, а как равноправное полевое уравнение, определяющее универсальную связь полевых переменных, дивергенции векторного потенциала и частной производной по времени от скалярного потенциала.

С другой стороны, в виду того, что вплоть до настоящего времени претендентом на физическое существование безоговорочно предполагается, в традиционной классической электродинамике, только поле, определяемое антисимметрическим производным аффинором электродинамического 4-потенциала $A(x)$, то это предположение одновременно является главным физическим основанием для введения требования соответствующей локальной калибровочной инвариантности.

 В нашем случае, в силу того, что мы отказались от данного предположения, то последнее фундаментальное требование вышеуказанной традиционной калибровочной инвариантности теряет вышеуказанное основание.

В свою очередь, как общеизвестно  \cite{Schw}, попытка сформулировать теорию электромагнитного поля только через антисимметрические комбинации первых частных производных 4-потенциала,  в рамках стандартного метода построения теории, сразу же наталкивается на осложнения при каноническом квантовании, обусловленные обращением, канонически сопряжённого с $A^{0}(x)$,  импульса $\pi^{0}(x)$  в 0.

При формулировке теории с учетом  $G(x) $  и $H(x)$, вышеуказанного обращения $\pi^{0}(x)$  в 0 не происходит именно потому, что эти составляющие производного аффинора являются симметрическими.

В связи с этим мы склонны заключить, что и данное осложнение говорит, в свою очередь, о непоследовательности предположения реального существования только поля, определяемого только антисимметрическим  производным аффинором электродинамического 4-потенциала.

\section*{\S \,2  Уравнения движения}
\addcontentsline{toc}{section}{\S \,2 Уравнения движения}

\qquad Самый общий вид локального лагранжиана (1.6) не квантованного действительного векторного поля, определяемого стандартными условиями \cite{BSh}, есть
$$
L(x) = - \frac{\alpha}{8}\,F_{mn}F^{mn}- \frac{\beta}{8}\,G_{mn}G^{mn}- \frac{\gamma}{8}H_{mn}H^{mn}+ \frac{m^{2}}{2}\,A_{m}A^{m},                            \eqno\ldots (2.1)
$$
где   $F_{mn}$, $G_{mn}$    и $H_{mn}$, в отличие от (1.3),  теперь  представлены традиционными соотношениями
$$
F_{mn}= F_{mn}(x) = \partial_{m} A_{n}(x) - \partial_{n} A_{m}(x),               \eqno\ldots (2.2)
$$
$$
G_{mn}=  G_{mn}(x)  = \partial_{m} A_{n}(x) + \partial_{n} A_{m}(x) - \frac{1}{2}\,\partial_{k}A^{k}(x)\,\textsl{g}_{mn},    \eqno\ldots (2.3)
$$
$$
H_{mn}= H_{mn}(x) = \frac{1}{2}\,\partial_{k}A^{k}(x)\,\textsl{g}_{mn}.            \eqno\ldots (2.4)
$$

То есть, в лагранжиане, описывающим классическую систему взаимодействующих вещественного векторного и спинорного полей  \cite{BSh},
 $$
 L  = L_{0}(A) + L_{0}(\psi) + L_{I},               \eqno\ldots (2.5)
 $$
лагранжиан $L_{0}(A) $  теперь представлен соотношением (2.1).

(2.1) демонстрирует, что стандартные лагранжианы свободного вещественного поля \cite{BSh} являются частными случаями лагранжиана (2.1), определяемыми частными значениями коэффициентов $\alpha, \beta$ и $\gamma$.

При этом важно подчеркнуть, что в каждом из этих частных случаев определяющим лагранжианом обычно считается \cite{BSh} лагранжиан представленный только $F$-частью, вследствие чего полностью исключается из рассмотрения поле, определяемое  $G$-частью производного аффинора векторного поля и, в частности, поле соответствующее  симметрическому
производному аффинору  4-потенциала электродинамики.

В соответствии с поставленной задачей, определяющим лагранжианом рассматриваемой модели является лагранжиан (2.1), {\em явно и равноправно} учитывающий как  $F$-, так и  $G$-составляющие соответствующего производного аффинора.

Выделим в $L$ стандартную $F$-часть, переписав (2.5) в виде $L = \displaystyle   \frac{\alpha}{2}\,L^{'}$, и перейдём к построению теории на основании лагранжиана $L^{'}$, в котором лагранжиан свободного векторного поля   теперь имеет  вид
$$
L^{'}(x) = L_{0}^{'}(A) = - \frac{1}{4}\,F_{mn}F^{mn}- \frac{a}{4}\,G_{mn}G^{mn}- \frac{b}{4}H_{mn}H^{mn}+ \frac{\mu^{2}}{2}\,A_{m}A^{m},                            \eqno\ldots (2.6)
$$
где
$$
a = \beta/\alpha, \quad b = \gamma/\alpha, \quad  \mu^{2} = 2 m^{2}/\alpha.
$$

Переход $L\rightarrow L^{'}$, как известно, оставляет неизменными уравнения движения, а восстановление исходных динамических переменных, таких как тензор энергии--импульса, момента и т. д., определяемых исходным лагранжианом (2.5), является тривиальным.

Уравнения Лагранжа\,--\,Эйлера, при использовании (2.6), приводят к следующим полевым уравнениям
$$
\partial_{m}F^{mn}(x) + a \,\partial_{m}G^{mn}(x) +b\, \partial_{m}H^{mn}(x) + \mu^{2}A^{n}(x) = 0.     \eqno\ldots (2.7)
$$

В частности, вместо соответствующих уравнений Максвелла, теперь имеем
$$
\partial_{m}F^{mn}(x) + a \,\partial_{m}G^{mn}(x) +b\, \partial_{m}H^{mn}(x)  = 0.     \eqno\ldots (2.8)
$$

Если не разделять симметрический производный аффинор  потенциала $A(x)$ на соответствующие инвариантные части, то полевое уравнение (2.8) принимает упрощенный частный вид
$$
\partial_{m}F^{mn}(x) + a \,\partial_{m}G^{mn}(x) = 0,     \eqno\ldots (2.9)
$$
где
$$
 G_{mn}(x) \eqdef  \partial_{m} A_{n}(x) + \partial_{n} A_{m}(x).         \eqno\ldots (2.10)
$$

В отношении интерпретации (2.9) важно подчеркнуть, что, во-первых, так как уравнение (2.9) можно переписать в двух эквивалентных формах
$$
\partial_{m}F^{mn}(x) +\frac{2a}{1+a} \,\partial^{n} \partial_{m}A^{m}(x) = 0,     \eqno\ldots (2.11)
$$
$$
\partial_{m}G^{mn}(x) - \frac{2}{1+a} \,\partial^{n}\partial_{m}A^{m}(x) = 0,     \eqno\ldots (2.12)
$$
демонстрирующих, в свою очередь, равноправие электромагнитного и электроджейтонного полей, то в (2.9) нельзя отдавать предпочтение ни $F$-, ни $G$-составляющим производного аффинора электродинамического потенциала $A(x)$, то есть, уравнение (2.9) не является ни соответствующим уравнением Максвелла, ни уравнением на $G$-поле.

Во-вторых, при стандартном условии классической теории, условии Лоренца, уравнения (2.9), (2.11) и (2.12) представляют собой одно и то же уравнение -- уравнение Даламбера на компоненты 4-потенциала $A(x)$, то есть, в отношении уравнений на $A(x)$, (2.9), (2.11) и (2.12) эквивалентны друг другу, в рассматриваемом случае.

Однако, с точки зрения уравнений на компоненты электромагнитного и электроджейтонного полей, (2.11) и (2.12) описывают самостоятельную динамику соответствующих полевых переменных.

А именно, если уравнение (2.11), в рассматриваемом случае, приводит к известным соотношениям для компонент электромагнитного поля
$$
\partial_{\alpha}F^{\alpha0}(x) = 0,           \eqno\ldots (2.13)
$$
$$
\partial_{0}F^{0\beta}(x) = - \partial_{\alpha}F^{\alpha \beta}(x),       \eqno\ldots (2.14)
$$
то уравнение (2.12), в данном случае, приводит к следующим соотношениям для компонент электроджейтонного поля
$$
\partial_{\alpha}G^{\alpha0}(x) = - \partial_{0}G^{00}(x),           \eqno\ldots (2.15)
$$
$$
\partial_{0}G^{0\beta}(x) = - \partial_{\alpha}G^{\alpha \beta}(x).           \eqno\ldots (2.16)
$$

Таким образом, эквивалентность уравнений (2.11) и (2.12), с точки зрения уравнения на 4-потенциал $A(x)$, не означает эквивалентность этих уравнений как уравнений на компоненты электромагнитного и электроджейтонного полей.

В частности, (2.15), в отличие от (2.13), демонстрирует, что напряжённость электрического поля, сопровождающего джейтонное поле, вне источника поля  имеет, в общем случае,  отличную от нуля расходимость.

С другой стороны, уравнения (2.14) и (2.16)  интерпретируются идентично, но при этом, если уравнение (2.14) говорит о том, что переменное электрическое поле порождает (сопровождает)  магнитное поле, представленное в этом уравнении  компонентами $F^{\alpha\beta}(x)$, то уравнение (2.16) демонстрирует то, что соответствующее переменное электрическое поле порождает (сопровождает) джейтонное поле, представленное в уравнении (2.16) компонентами $G^{\alpha\beta}(x)$.

При этом уместно отметить так же и принципиальную отличительную особенность  напряженностей данных электрических полей, заключающуюся в том, что если первая, как известно, представляет собой, в волновой зоне поля излучения ограниченных источников, поперечное поле, то вторая содержит, в общем случае,  и продольную составляющую.

Таким образом, переменное электроджейтонное поле, определяемое симметрическим производным аффинором электродинамического потенциала $A(x)$, в волновой зоне представляет собой, в частности, продольную электроджейтонную волну, в которой, однако, продольное электрическое поле связано не с общеизвестным магнитным полем, а с джейтонным полем, представленным симметрической составляющей 3-градиента векторного электродинамического потенциала $\vec{A}(x)$.

В общем случае псевдориманова пространства событий, (2.8) представляется
\begin{multline*}
  \frac{1}{\sqrt{-\textsl{g}(x)}}\,\partial_{m}(\sqrt{-\textsl{g}(x)}\, \tilde{F}^{mn}(x)) +\frac{a}{\sqrt{-\textsl{g}(x)}}\,\partial_{m}(\sqrt{-\textsl{g}(x)}\, \tilde{G}^{mn}(x))+\\ a\,\Gamma^{n}_{mk}(x)\tilde{G}^{mk}(x)+
\frac{b}{\sqrt{-\textsl{g}(x)}}\,\partial_{m}(\sqrt{-\textsl{g}(x)}\, \tilde{H}^{mn}(x))+\\
 \frac{b}{2\textsl{g}(x)}\,\partial_{m}
(\sqrt{-\textsl{g}(x)}\textsl{g}^{mn}(x))\partial_{k}(\sqrt{-\textsl{g}(x)}A^{k}(x))  = 0,  \qquad\ldots (2.17.\text {\em а})
\end{multline*}
где
$$
 \tilde{F}_{mn}(x)=  F_{mn}(x),         \eqno\ldots (2.17.\text {\em б})
$$
$$
 \tilde{G}_{mn}(x)= G_{mn}(x) - 2\,\Gamma^{k}_{mn}(x)A_{k}(x)- \frac{1}{2 \textsl{g}(x)}\,\partial_{k}\textsl{g}(x) A^{k}(x)\textsl{g}_{mn}(x),         \eqno\ldots (2.17.\text {\em в})
$$
$$
 \tilde{H}_{mn}(x)= H_{mn}(x) + \frac{1}{2 \textsl{g}(x)}\,\partial_{k}\textsl{g}(x) A^{k}(x)\textsl{g}_{mn}(x).       \eqno\ldots (2.17.\text {\em г})
$$

(2.17.{\em а})–(2.17.{\em г}) демонстрируют, одновременно, специфическую реакцию компонент {\em симметрических} производных аффиноров электродинамического потенциала  на пространство событий общей теории относительности и далеко не тривиальный их вклад в полевые уравнения.



Считая $a \neq -1 $   и  используя соотношения (2.2)--(2.4), уравнение (2.7) можно представить в виде
 $$
\Box A^{n}(x)- \frac{(a+b- 2)}{1+a}\,\partial^{n}\partial_{m}A^{m}(x)- \frac{\mu^{2}}{1+a}\, A^{n}(x) = 0.   \eqno\ldots (2.18)
 $$

Заметив, что уравнения движения (2.18) не приводит к автоматическому выполнению дополнительного условия (1.7) и, учтя, что мы отказались от его наложения извне, приходим к выводу: решения уравнений движения рассматриваемой теории, в общем случае, не принадлежат к решениям уравнений движения стандартной теории
$$
(\Box - m^{2}) A_{n}(x) = 0.                                         \eqno\ldots (2.19)
$$

При  этом, оказывается можно найти  дифференциальный  оператор $D\,(\Box, {m_{1}}^{2}, {m_{2}}^{2})$,  такой, что
$$
D\,(\Box, {m_{1}}^{2}, {m_{2}}^{2})A_{n}(x) = 0                 \eqno\ldots (2.20)
$$
для любого решения уравнения (2.18).

Действительно, действуя оператором  $\partial _{n}$  на  (2.18) и считая $3a+b \neq 0$, имеем
$$
\left(\Box - \frac{2 \mu^{2}}{3a+b}\right)\partial _{n}A^{n}(x)= 0.        \eqno\ldots (2.21)
$$

С другой стороны, действуя на  (2.18)  оператором  Д’Аламбера  и учитывая  (2.21), получаем
$$
\Box\Box A^{n}(x) - \frac{\mu^{2}\,(5a+b+2)}{(3a+b)(1+a)}\,\Box A^{n}(x) + \frac{2\mu^{4}}{(3a+b)(1+a)}A^{n}(x) =0  \eqno\ldots (2.22)
$$
и, после введения обозначений
$$
{m_{1}}^{2} \eqdef \frac{\mu^{2}}{1+a}\,,         \eqno\ldots (2.23)
$$
$$
{m_{2}}^{2} \eqdef \frac{2\mu^{2}}{3a+b}\,,           \eqno\ldots (2.24)
$$
переписываем (2.22) в компактном виде
$$
(\Box - {m_{1}}^{2}) (\Box - {m_{2}}^{2}) A^{n}(x) = 0.   \eqno\ldots (2.25)
$$

Таким образом, приходим к очередному существенному отличию от стандартной теории: любое решение уравнений движения рассматриваемой теории необходимо удовлетворяет (2.25) вместо (2.19).

В свою очередь, (2.25), на основании теоремы Боджио \cite{Bogg},  приводит к заключению, что общее решение уравнений движения (2.18) представляется
$$
A_{n}(x) = A_{n}^{(1)}(x) + A_{n}^{(2)}(x),        \eqno\ldots (2.26)
$$
где  $A_{n}^{(1)}(x)$  и  $A_{n}^{(2)}(x)$  -- независимые функции, подчиняющиеся уравнениям
$$
(\Box - {m_{1}}^{2})A_{n}^{(1)}(x)= 0,     \eqno\ldots (2.27)
$$
$$
(\Box - {m_{2}}^{2}) A_{n}^{(2)}(x) = 0.    \eqno\ldots (2.28)
$$

Математическая независимость этих функций, как увидим ниже, позволит нам провести процедуру квантования, отличную от процедур квантования в рамках теорий, аналогичных исследуемой в отношении использования полного векторного представления однородной группы Лоренца, и, как следствие этого, избежать тех или иных недостатков  \cite{Hall} таких теорий.

В заключение настоящего раздела отметим, что если любое решение уравнения (2.18) удовлетворяет (2.25), то обратное утверждение не имеет места, то есть, уравнения (2.25) являются более общими по отношению к (2.18).

Следовательно, возникает необходимость выделения из множества решений уравнения (2.25), такого подмножества, каждый элемент которого удовлетворяет (2.18).

Это будет проделано в разделе 4, после чего, одновременно, станет ясным и спиновое содержание векторных полей  $A^{(1)}(x) $   и   $A^{(2)}(x) $, определяющих класс общих решений уравнения (2.18).

\section*{\S \,3 Динамические переменные}
\addcontentsline{toc}{section}{\S \,3 Динамические переменные}

\qquad Канонический тензор энергии--импульса, согласно (2.6), принимает вид
$$
T'^{kl}_{can}= - F^{kl}\partial^{l}A_{i}- a\,G^{kl}\partial^{l}A_{i}- b\,H^{kl}\partial^{l}A_{i}- L'\textsl{g}^{kl}.  \eqno\ldots (3.1)
$$

Тогда, орбитальная и спиновая части тензора момента количества  движения представляются, соответственно, в виде
\begin{multline*}
 {{M'_{0}}_{lm}}^{k}= x_{l}F^{ki}\partial _{m}A_{i}-  x_{m}F^{ki}\partial _{l}A_{i}+a\,(x_{l}G^{ki}\partial _{m}A_{i}-  x_{m}G^{ki}\partial _{l}A_{i})+  \\
   b\,(x_{l}H^{ki}\partial _{m}A_{i}-  x_{m}H^{ki}\partial _{l}A_{i}) - (x_{l}\delta _{m}^{k}- x_{m}\delta _{l}^{k})L',  \qquad\ldots (3.2)
\end{multline*}
$$
{S'_{lm}}^{k}={F^{k}}_{l}A_{m} - {F^{k}}_{m}A_{l}+ a\,({G^{k}}_{l}A_{m} - {G^{k}}_{m}A_{l}) +b\,({H^{k}}_{l}A_{m} - {H^{k}}_{m}A_{l}).     \eqno\ldots (3.3)
$$

Соотношения (3.1)--(3.3), в свою очередь, демонстрируют равноправные и самостоятельные вклады в динамические переменные поля, определяемые рассматриваемыми симметрическими производными аффинорами.

Интегрируя пространственные плотности этих моментов, легко получить соответствующие выражения для компонент тензоров орбитального и спинового моментов.

После этого, выражения для компонент вектора спина
 $$
 S^{\alpha}= \varepsilon^{\alpha\beta\gamma} S_{\beta\gamma},  \eqno\ldots (3.4)
$$
где
$$
S_{\alpha\beta}= \int \!\!{S'_{\alpha\beta}}^{0}d\vec{x},
$$
через очевидные обозначения, принимают вид
$$
 S^{\alpha}=  S^{\alpha}(F) +  S^{\alpha}(G)+ S^{\alpha}(H).        \eqno\ldots (3.5)
$$

Однако, так как для пространственных компонент
$$
{H^{0}}_{l}A_{m} - {H^{0}}_{m}A_{l} = 0,        \eqno\ldots (3.6)
$$
то $S^{\alpha}(H)= 0 $   и поэтому (3.5) принимает вид
$$
 S^{\alpha}=  S^{\alpha}(F) +  S^{\alpha}(G).       \eqno\ldots (3.7)
$$

Дальнейшее рассмотрение спина  продолжим после перехода к импульсному представлению, а теперь выполним симметризацию тензора энергии--импульса.

Восстанавливая исходные динамические переменные, соответствующие определяющему  лагранжиану (2.1),  имеем
$$
T^{kl}_{can}=\frac{\alpha}{2}\,T'^{kl}_{can}=  - \frac{\alpha}{2}\, F^{ki}\partial^{l}A_{i}- \frac{\beta}{2}\,G^{ki}\partial^{l}A_{i}- \frac{\gamma}{2}\,H^{ki}\partial^{l}A_{i}- L\textsl{g}^{kl}.  \eqno\ldots (3.8)
$$

При этом интересно отметить, что в частном случае рассматриваемой теории, когда $\alpha = \beta =\gamma = 1$, соответствующему лагранжиану [2], канонический тензор энергии--импульса (3.8) является уже симметрическим,  то есть, является, одновременно, метрическим тензором энергии--импульса.

Следовательно, нам остаётся, в рассматриваемом частном случае, только  перейти в (3.8) полностью к $F $- и  $ G$-компонентам, то есть, к компонентам электромагнитного и электроджейтонного полей.

Таким образом, в рассматриваемом частном случае, получаем
$$
T^{kl}=  - \frac{1}{4}\, (F^{ki}{F^{l}}_{i}+G^{ki}{G^{l}}_{i}+ F^{ki}{G^{l}}_{i}+ G^{ki}{F^{l}}_{i})- L\textsl{g}^{kl},  \eqno\ldots (3.9)
$$
где $G_{mn} $ определена соотношением (2.10).

В свою очередь, (3.9) приводит к следующему выражению для плотности импульса
$$
T^{0\alpha}=  - \frac{1}{4}\, (F^{0\beta}{F^{\alpha}}_{\beta}+G^{0i}{G^{\alpha}}_{i}+ F^{0\beta}{G^{\alpha}}_{\beta}+ G^{0i}{F^{\alpha}}_{i}),  \eqno\ldots (3.10)
$$
в котором компоненты $F$- и  $G$-полей входят совершенно симметрично и равноправно, демонстрируя, что наряду со стандартным потоком энергии электромагнитного поля,  существуют как поток энергии обусловленный электроджейтонным полем, так и смешанные потоки энергии обусловленные компонентами как электромагнитного, так и электроджейтонного полей.

Метрический тензор энергии--импульса, $T'^{kl}$, в общем случае, можно получить, например, варьированием действия, предварительно записанного в криволинейных координатах \cite{La} или симметризуя канонический тензор энергии--импульса методом Белинфанте\,--\,Розенфельда  \cite{Bel}.

При этом, доказательства эквивалентности данных методов, в общем случае, мы не имеем \cite{Goed}, поэтому, по очевидным соображениям, отдаём предпочтение последнему и находим
$$
T'^{kl}= {T'^{kl}}_{can}+ \frac{1}{2}\,\partial_{j}\left(\textsl{g}^{lm}\textsl{g}^{kn}{S_{mn}}^{j}+ \textsl{g}^{jm}\textsl{g}^{ln}{S_{mn}}^{k}- \textsl{g}^{km}\textsl{g}^{jn}{S_{mn}}^{l}\right),       \eqno\ldots (3.11)
$$
где спиновая часть тензора момента количества  движения, ${S_{lm}}^{k}$,  определена  соотношением  (3.3).

Подставляя  (3.3) в (3.11) и используя уравнения движения (2.7), получаем
\begin{multline*}
  T'^{kl}= {F^{k}}_{i}F^{il}+ \mu^{2}A^{k}A^{l}+ \frac{a}{2}\,\left({G^{l}}_{i}F^{ik} +{G^{k}}_{i}F^{il}\right)+
a \left(\partial_{i}G^{ik}A^{l}+ \partial_{i}G^{il}A^{k}\right)+ \\
b \left(\partial_{i}H^{ik}A^{l}+ \partial_{i}H^{il}A^{k} \right)+
a\,\partial_{i}\left(G^{kl}A^{i} \right)- b\,\partial_{i}\left(H^{kl}A^{i} \right)- L'\textsl{g}^{kl}. \qquad \qquad\ldots (3.12)
\end{multline*}

В свою очередь, (3.12), с учётом  (2.6), приводит к следующим выражениям для плотности энергии и плотности импульса
\begin{multline*}
  T'^{00}= {F^{0}}_{\alpha}F^{\alpha0} +\frac{ \mu^{2}}{2}\,A_{m}A_{m}+ \frac{1}{4}\,F_{mn}F^{mn}+ \frac{a}{4}\, G_{mn}G^{mn}+  \frac{b}{4}\,H_{mn}H^{mn}+ \\
  2a\,\partial_{m}G^{m0}A^{0}
  +  2b\,\partial_{m}H^{m0}A^{0}+ a\,{G^{0}}_{\alpha}F^{\alpha0}- a\,\partial_{m}(G^{00}A^{m})- b\,\partial_{m}(H^{00}A^{m}), \ldots (3.14)
\end{multline*}
\begin{multline*}
  T'_{0\alpha}= - F_{\beta0}{F^{\beta}}_{\alpha}+\mu^{2}A_{0}A_{\alpha}+a\,(\partial^{m}G_{0m}A_{\alpha}+ \partial^{m}G_{\alpha m}A_{0})+b\,(\partial^{m}H_{0m}A_{\alpha}+\\ \partial^{m}H_{\alpha m}A_{0})+
 \frac{ a}{2}\,(G_{0m}{F^{m}}_{\alpha}+ G_{\alpha m}{F^{m}}_{0})- a\,\partial^{m}(G_{0\alpha}A_{m}).    \qquad \qquad\ldots (3.15)
\end{multline*}

\section*{\S \,4 Импульсное представление}
\addcontentsline{toc}{section}{\S \,4 Импульсное представление}

\qquad В виду того, что  $A_{m}(x)$   удовлетворяет уравнению (2.25), а не уравнению Клейна\,--\,Гордона (2.19), возникает необходимость замены уравнения четвёртого порядка, системой уравнений второго порядка, чтобы далее следовать обычной процедуре перехода к импульсному представлению согласно \cite{BSh}.

Теорема Боджио, позволившая выразить общее решение наших уравнений движения  (2.18)  в виде  (2.26), как раз и представляет собой, в силу (2.27) и (2.28), теорему существования конкретной реализации такой замены.

Таким образом, мы можем, без потери общности, продолжить построение теории стандартными методами, описывая нашу систему двумя независимыми функциями  $ A^{(1)}(x)$  и  $A^{(2)}(x)$.

Однако, возвращаясь к исходному уравнению (2.18), можно специализировать вид функций $ A^{(1)}(x)$  и  $A^{(2)}(x)$     и, тем самым, детализировать исследование свойств рассматриваемой теории.

Результат такой специализации можно сформулировать в виде следующей теоремы.

{\em Общее решение дифференциального уравнения (2.18), после ковариантного представления его в виде}   \cite{Feld, ArHS}
$$
A_{m}(x) =\; ^{\tau}\!\!A_{m}(x)+\; ^{\ell}\!\!A_{m}(x),        \eqno\ldots (4.1)
$$

{\em 1. содержит функции \, $^{\tau}\!\!A_{m}(x)$ и \, $^{\ell}\!\!A_{m}(x)$  как независимые решения уравнений (2.18) и}

{\em 2. эти функции удовлетворяют, соответственно, уравнениям (2.27) и (2.28).}

Поперечная и продольная части представления (4.1) определяются, соответственно,
$$
^{\tau}\!\!A_{m}(x) \eqdef {\tau_{m}}^{n}(x-y)\,A_{n}(y),      \eqno\ldots (4.2)
$$
$$
^{\ell}\!\!A_{m}(x) \eqdef  \partial_{m}^{x} \partial_{x}^{n} D_{0}^{c}(x-y) \,A_{n}(y),    \eqno\ldots (4.3)
$$
$$
{\tau_{m}}^{n}(x) \eqdef  {\textsl{g}_{m}}^{n}\,\delta (x) - \partial_{m}^{x} \partial_{x}^{n} D_{0}^{c}(x),    \eqno\ldots (4.4)
$$
где   $D_{0}^{c}(x)$ -- причинная функция скалярного поля при массе равной нулю и по повторяющемуся $"\!y"$ подразумевается интегрирование.

Независимость функций правой части (4.1), как увидим ниже, оказывается очень важным обстоятельством, как при проведении процедуры квантования, так и при определении вакуума и последующего определения причинной функции рассматриваемой теории.

Первой частью доказательства теоремы является получение уравнения (2.25) для любого решения уравнения движения (2.18) и последующее использование теоремы Боджио, для представления общего решения в виде (2.26), что было уже проделано ранее.

Вторая часть состоит в доказательстве соотношений
$$
 \left.
\begin{array}{c}
  A_{m}^{(1)}(x)=\; ^{\tau}\!\!A_{m}(x),\\
 A_{m}^{(2)}(x)=\; ^{\ell}\!\!A_{m}(x),
\end{array}
\right.         \eqno\ldots (4.5)
$$
из которых следует, как независимость поперечной и продольной частей $A_{m}(x)$, в силу независимости  $A_{m}^{(1)}(x)$  и  $A_{m}^{(2)}(x)$, так и представление общего решения уравнения (2.26) в более специализированном  для данной модели  виде (4.1).

Доказательство (4.5) начнем с нахождения уравнений для  $^{\tau}\!\!A_{m}(x)$ и $^{\ell}\!\!A_{m}(x)$.

Для этого заметим, что, используя соотношения
$$
\partial_{m}\,^{\ell}\!\!A_{n}(x) = \partial_{n}\,^{\ell}\!\!A_{m}(x),            \eqno\ldots (4.6)
$$
$$
\partial_{m}A^{m}(x) = \partial_{m}\,^{\ell}\!\!A^{m}(x),            \eqno\ldots (4.7)
$$
следующие из (4.3) и (4.4),  (2.18),  после представления его решения в виде (4.1), можно переписать
$$
\left(\Box - {m_{1}}^{2}\right) \,^{\tau}\!\!A^{n}(x)+ \frac{{m_{1}}^{2}}{{m_{2}}^{2}}\,\left(\Box - {m_{2}}^{2}\right) \,^{\ell}\!\!A^{n}(x)= 0,                      \eqno\ldots (4.8)
$$
где   ${m_{1}}^{2}$    и   ${m_{2}}^{2}$   определены, соответственно, согласно (2.23) и (2.24).

С другой стороны, (2.21) приводит к уравнению
$$
\left(\Box - {m_{2}}^{2}\right) \,^{\ell}\!\!A^{n}(x)= 0,                      \eqno\ldots (4.9)
$$
которое уже говорит о том, что $^{\ell}\!\!A_{m}(x)\in \{A_{m}^{(2)}(x)\}$.

В свою очередь,  из (4.8), согласно (4.9), имеем
$$
\left(\Box - {m_{1}}^{2}\right) \,^{\tau}\!\!A^{n}(x)= 0,                         \eqno\ldots (4.10)
$$
и, аналогично, заключаем, что $^{\tau}\!\!A_{m}(x)\in \{A_{m}^{(1)}(x)\}$.

А теперь, представляя функции  $A_{m}^{(1)}(x)$   и  $A_{m}^{(2)}(x)$  в виде (4.1) и заметив, что, в силу  (4.2)--(4.4),
$$
^{\tau}\!\!A_{m}^{(1)}(x)+\; ^{\tau}\!\!A_{m}^{(2)}(x) =\; ^{\tau}\!\!A_{m}(x),             \eqno\ldots (4.11)
$$
$$
^{\ell}\!\!A_{m}^{(1)}(x)+\; ^{\ell}\!\!A_{m}^{(2)}(x) =\; ^{\ell}\!\!A_{m}(x),             \eqno\ldots (4.12)
$$
находим
$$
\left(\Box - {m_{1}}^{2}\right)\left(^{\tau}\!\!A_{m}^{(1)}(x)+\; ^{\tau}\!\!A_{m}^{(2)}(x) \right)= 0,   \eqno\ldots (4.14)
$$
$$
\left(\Box - {m_{2}}^{2}\right)\left(^{\ell}\!\!A_{m}^{(1)}(x)+\; ^{\ell}\!\!A_{m}^{(2)}(x) \right)= 0,   \eqno\ldots (4.15)
$$
а эти уравнения, совместно с (2.27) и (2.28), требуют, в общем случае  ${m_{1}}^{2}\neq{m_{2}}^{2}$,  чтобы
$$
^{\tau}\!\!A_{m}^{(2)}(x) = \; ^{\ell}\!\!A_{m}^{(1)}(x).             \eqno\ldots (4.16)
$$

Но (4.16)  означает, в силу соотношения
$$
{\tau_{m}}^{n}(x-y)\,^{\ell}\!A_{n}(y)= 0,                                \eqno\ldots (4.17)
$$
что
$$
^{\ell}\!\!A_{m}^{(1)}(x)= 0,                                        \eqno\ldots (4.18)
$$
следовательно, и
$$
^{\tau}\!\!A_{m}^{(2)}(x)= 0.                                        \eqno\ldots (4.19)
$$

В итоге, (4.11) и (4.12), с учётом  (4.18) и (4.19), приводят к (4.5).

Приведённое доказательство демонстрирует, одновременно, что исходное уравнение движения (2.18) действует подобно дополнительному условию на решения уравнения (2.25), исключая из  $ A_{m}^{(1)}(x)$ часть, соответствующую спину  0, а из $ A_{m}^{(2)}(x)$,  часть, соответствующую спину 1.

Теперь, переходя к импульсному представлению, проводим, в силу (4.10), стандартное лоренц-инвариантное
 разбиение $ ^{\tau}\!A_{m}(x)$  на положительно- и отрицательно-частотные части
$$
^{\tau}\!A_{m}(x)= \frac{1}{(2\pi)^{3/2}}\int\!\frac{d\vec{k}}{\sqrt{2k^{0}}}\left( a_{m}^{+}(\vec{k})e^{ikx}+a_{m}^{-}(\vec{k})e^{-ikx}\right),          \eqno\ldots (4.20)
$$
где
$$
a_{m}^{\pm}(\vec{k})=\frac{ \theta(k^{0})\,^{\tau}\!A_{m}(\pm k)}{\sqrt{2k^{0}}},        \eqno\ldots (4.21)
$$
$$
k^{0}=\left(\vec{k}^{2}+ {m_{1}}^{2}\right)^{1/2}.                      \eqno\ldots (4.22)
$$

Как  следствие вещественности  рассматриваемого поля, имеем
$$
(a_{m}^{\pm}(\vec{k}))^{\ast}= a_{m}^{\mp}(\vec{k})               \eqno\ldots (4.23)
$$
и, как следствие  автоматической  поперечности $^{\tau}\!A_{m}(x)$, получаем
$$
k_{\alpha}a_{\alpha}^{\pm}(\vec{k})= k_{0}a_{0}^{\pm}(\vec{k}).        \eqno\ldots (4.24)
$$

Аналогично, на основании (4.9), выражаем продольную часть
$$
^{\ell}\!A_{m}(x)= \frac{1}{(2\pi)^{3/2}}\int\!\frac{d\vec{k}}{\sqrt{2k^{0}}}\left( b_{m}^{+}(\vec{k})e^{ikx}+b_{m}^{-}(\vec{k})e^{-ikx}\right),          \eqno\ldots (4.25)
$$
где
$$
b_{m}^{\pm}(\vec{k})=\frac{ \theta(k^{0})\,^{\ell}\!A_{m}(\pm k)}{\sqrt{2k^{0}}},        \eqno\ldots (4.26)
$$
$$
k^{0}= \left( \vec{k}^{2}+ {m_{2}}^{2}\right)^{1/2}.                     \eqno\ldots (4.27)
$$

Но, в силу соотношения (4.17), фурье-компоненты  продольного поля удовлетворяют уравнению
 $$
 ({\delta_{m}}^{n} - k_{m}k^{n}/k^{2})^{\ell}\!A_{n}(k)=0,            \eqno\ldots (4.28)
 $$
поэтому, для них имеем представление
$$
^{\ell}\!A_{m}(k)= k_{m}k^{n}\,^{\ell}\!A_{n}(k)/k^{2}.                  \eqno\ldots (4.29)
$$

Тогда, после введения обозначения
$$
\frac{ \theta(k^{0})k^{n}\,^{\ell}\!A_{n}(\pm k)}{\sqrt{2k^{0}}}= b^{\pm}(\vec{k}),       \eqno\ldots (4.30)
$$
где $k^{0}$  удовлетворяет (4.27),  (4.25)  переписывается в виде
$$
^{\ell}\!A_{m}(x)= \frac{{m_{2}}^{-2}}{(2\pi)^{3/2}}\int\!\frac{d\vec{k}}{\sqrt{2k^{0}}}\left( k_{m}b^{+}(\vec{k})e^{ikx}+k_{m}b^{-}(\vec{k})e^{-ikx}\right).          \eqno\ldots (4.31)
$$

Наконец, подстановка (4.20) и (4.31) в (2.2)--(2.4)  приводит к  следующим  трёхмерным импульсным представлениям ковариантных компонент соответствующих производных аффиноров  рассматриваемого  векторного поля.
\begin{multline*}
F_{mn}(x)=\frac{ i}{(2\pi)^{3/2}}\int\!\frac{d\vec{k}}{\sqrt{2k^{0}}}(( k_{m}a_{n}^{+}(\vec{k}) - k_{n}a_{m}^{+}(\vec{k}))e^{ikx}- ( k_{m}a_{n}^{-}(\vec{k}) -\\
 k_{n}a_{m}^{-}(\vec{k}))e^{-ikx}), \quad k^{0}= (\vec{k}^{2}+{m_{1}}^{2})^{1/2},    \qquad\ldots (4.32)
\end{multline*}
\begin{multline*}
  G_{mn}(x)=\frac{ i}{(2\pi)^{3/2}}\int\!\frac{d\vec{k}}{\sqrt{2k^{0}}}(( k_{m}a_{n}^{+}(\vec{k})+ k_{n}a_{m}^{+}(\vec{k}))e^{ikx}- ( k_{m}a_{n}^{-}(\vec{k}) + \\
   k_{n}a_{m}^{-}(\vec{k}))e^{-ikx})|_{k^{0}= (\vec{k}^{2}+{m_{1}}^{2})^{1/2}}+\\
  \frac{ i}{(2\pi)^{3/2}}\int\!\frac{d\vec{k}}{\sqrt{2k^{0}}}((\frac{2k_{m}k_{n}}{{m_{2}}^{2}}- \frac{ \textsl{g}_{mn}}{2})b^{+}(\vec{k})e^{ikx} - (\frac{2k_{m}k_{n}}{{m_{2}}^{2}}- \\
  \frac{ \textsl{g}_{mn}}{2})b^{-}(\vec{k})e^{-ikx})|_{k^{0}= (\vec{k}^{2}+{m_{2}}^{2})^{1/2}},       \qquad\ldots (4.33)
\end{multline*}
$$
H_{mn}(x)=\frac{ i}{(2\pi)^{3/2}}\int\!\frac{d\vec{k}}{\sqrt{2k^{0}}}\left(\frac{ \textsl{g}_{mn}}{2}b^{+}(\vec{k})e^{ikx}-
  \frac{ \textsl{g}_{mn}}{2}b^{-}(\vec{k})e^{-ikx})\right)|_{ k^{0}= (\vec{k}^{2}+{m_{2}}^{2})^{1/2}}. \eqno\ldots (4.34)
$$

В свою очередь, (3.14) и (3.15), с учётом (4.32)--(4.34), приводят к следующему выражению
 для 4-вектора  энергии--импульса  рассматриваемого векторного поля
$$
P'_{n}= \int \!T'_{0n}d\vec{x}= (1+a)\int\!d\vec{k}\,k_{n}\alpha_{\beta}^{+}(\vec{k})\alpha_{\beta}^{-}(\vec{k}) -
\frac{3a+b}{2{m_{2}}^{2}}\int \!d\vec{k}\,k'_{n}b^{+}(\vec{k})b^{-}(\vec{k}), \eqno\ldots (4.35)
$$
где
$$
\vec{a}\,^{\pm}(\vec{k})= \vec{e}_{\beta}\,\alpha_{\beta}^{\pm}(\vec{k}),         \eqno\ldots (4.36)
$$
$$
(\vec{e}_{\alpha}, \vec{e}_{\beta})= \delta_{\alpha\beta},   \quad \vec{e}_{3}=
\frac{\vec{k}}{|\vec{k}|}\,\frac{k^{0}}{m_{1}},            \eqno\ldots (4.37)
$$
$$
k_{n}= \{k_{0}, - \vec{k}\},              \eqno\ldots (4.38)
$$
$$
k'_{n}= \{k'_{0}, - \vec{k}\},              \eqno\ldots (4.39)
$$
и  $k_{0}$,  $k'_{0}$ определены, соответственно, уравнениями (4.22), (4.27).

Учитывая вещественность рассматриваемого поля, имеем
$$
(b^{\pm}(\vec{k}))^{\ast}= b^{\mp}(\vec{k}),            \eqno\ldots (4.40)
$$
то есть,   $b$-часть $P'_{0}$  является отрицательно определённой.

Однако, при этом важно отметить, что, из рассмотрения сокращения членов, содержащих $e^{\pm2ik_{0}x_{0}}$, $e^{\pm i(k_{0}-k'_{0})x_{0}}$   и  $e^{\pm i(k_{0}+k'_{0})x_{0}}$, находим, что сохраняющейся величиной является всё выражение $\int\!T'_{00}d\vec{x}$  в целом.

Следовательно, индефинитность   $b$-части $P'_{0}$, в рассматриваемой классической теории, не приводит мгновенно к соответствующему противоречию.

Действительно, в силу вышеуказанного замечания, мы вправе требовать положительную определённость только от всего выражения  $\int\!T'_{00}d\vec{x}$, но дефинитность полной энергии может быть обеспечена, если считать ${m_{2}}$  достаточно большим.

Наконец,  восстановление 4-вектора энергии-импульса $P_{n}$, соответствующее переходу к определяющему лагранжиану (2.1), приводит, в (4.35),  к  замене $(1+a)\rightarrow(\alpha+\beta)/2$, демонстрируя то, что, в положительно определённой части $P_{0}$, обсуждаемые $F$- и   $G$-поля вновь выступают совершенно равноправно и самостоятельно по отношению друг к другу.

Квантование, получение и свойства причинной функции,  а так же классическое взаимодействие со спинорным полем будут представлены в последующих двух главах.




\chapter{Квантование, пространство состояний и причинная функция}


\quad Продолжается построение теории действительного  векторного поля, начатое в главе I \cite{Al4}.

Обсуждено квантование, определено пространство состояний и найдена  причинная функция Грина  векторного поля в рамках обсуждаемой  теории.

Ключевые слова: квантование, пространство состояний, причинная функция.

\section*{\S \,1 Квантование}
\addcontentsline{toc}{section}{\S \,1 Квантование}

\qquad Условие совместности трансформационных свойств операторных полевых функций и, вытекающих из постулата квантования  \cite{BSh}, трансформационных свойств  амплитуд состояний, мы можем накладывать как на  $^{\tau}\!A_{m}(x)$, так и на  $^{\ell}\!A_{m}(x)$, в силу их ковариантного определения, с одной стороны, и полевой независимости, в силу ранее доказанной теоремы,  с другой.

  В результате, в случае бесконечно малых пространственно-временных трансляций, это условие приводит к уравнениям
$$
i\,\partial_{m}\,^{\tau}\!A_{n}(x)= [\,^{\tau}\!A_{n}(x), P_{m}],       \eqno\ldots (1.1)
$$
$$
i\,\partial_{m}\,^{\ell}\!A_{n}(x)= [\,^{\ell}\!A_{n}(x), P_{m}],       \eqno\ldots (1.2)
$$
которые, в свою очередь, приводят к следующим  перестановочным  соотношениям
$$
\frac{\alpha+\beta}{2}\,\left[\alpha_{\beta}^{+}(\vec{k}), \alpha_{\gamma}^{-}(\vec{q})\right]= - \delta_{\beta\gamma}\,\delta ( \vec{k}- \vec{q}\,),        \eqno\ldots (1.3)
$$
$$
\left[\alpha_{\beta}^{+}(\vec{k}), b^{-}(\vec{q})\right]=0,          \eqno\ldots (1.4)
$$
$$
\frac{3\beta+\gamma}{4{m_{2}}^{2}}\,\left[b^{-}(\vec{k}), b^{+}(\vec{q})\right]= - \delta ( \vec{k}- \vec{q}\,).  \eqno\ldots (1.5)    $$

Коммутатор  для   $b$-операторов, как видим, оказывается аналогичным коммутатору для “ временных фотонов”  электромагнитного поля.

В свою очередь, используя (1.3), получаем перестановочные соотношения для операторов $a_{m}^{\pm}(\vec{k})$
$$
\left[a_{m}^{-}(\vec{k}), a_{n}^{+}(\vec{q})\right]= \frac{2}{\alpha+\beta}\,\left(- \textsl{g}_{mn}+\frac{k_{m}k_{n}}{{m_{1}}^{2}}\right)\,\delta(\vec{k}-\vec{q}),       \eqno\ldots (1.6)
$$
которые, с точностью до мультипликативной постоянной, совпадают с соответствующими соотношениями для операторов массивного вещественного векторного поля  стандартной теории, с массой, определяемой посредством  (I.2.23).

Теперь можем найти коммутатор для операторов    в координатном представлении.

Для этого, следуя стандартной процедуре, разбиваем операторную функцию поля $A_{m}(x)$   на следующие положительно\,-  и отрицательно\,-частотные части
$$
A_{m}(x)= A_{m}^{+}(x)+ A_{m}^{-}(x),           \eqno\ldots (1.7)
$$
$$
A_{m}^{\pm}(x)=\;^{\tau}\!A_{m}^{\pm}(x)+ \;^{\ell}\!A_{m}^{\pm}(x),           \eqno\ldots (1.8)
$$
$$
^{\tau}\!A_{m}^{\pm}(x)= \frac{1}{(2\pi)^{3/2}}\int \!\frac{d\vec{k}}{\sqrt{2k^{0}}}\,e^{\pm ikx}\,a_{m}^{\pm}(\vec{k}),     \eqno\ldots (1.9)
$$
$$
^{\ell}\!A_{m}^{\pm}(x)= \frac{{m_{2}}^{-2}}{(2\pi)^{3/2}}\int \!\frac{d\vec{k}}{\sqrt{2k^{0}}}\,e^{\pm ikx}\,k'_{m}b^{\pm}(\vec{k}).     \eqno\ldots (1.10)
$$

Лоренц-инвариантность разложения (1.7) следует из соответствующей инвариантности представлений  (I.4.1), (I.4.20) и (I.4.25).

Тогда, учитывая перестановочность одночастотных операторов, имеем
$$
\left[ A_{m}(x), A_{n}(y) \right] = \left[ A_{m}^{+}(x), A_{n}^{-}(y) \right] +  \left[ A_{m}^{-}(x), A_{n}^{+}(y) \right],    \eqno\ldots (1.11)
$$
и, используя (1.6), получаем
$$
\left[\,^{\tau}\! A_{m}^{-}(x), \,^{\tau}\! A_{n}^{+}(y) \right]= \frac{2}{\alpha+\beta}\, (\textsl{g}_{mn}+\frac{\partial_{m}\partial_{n}}{{m_{1}}^{2}})\, iD^{-}(x-y; m_{1}),   \eqno\ldots (1.12)
$$
 где  $D^{-}(x; m)$ -- отрицательно-частотная часть функции Паули\,--\,Йордана.

Аналогично, используя (1.5), находим
$$
\left[\,^{\ell}\! A_{m}^{-}(x), \,^{\ell}\! A_{n}^{+}(y) \right]= - \frac{\partial_{m}\partial_{n}}{m^{2}}\, iD^{-}(x-y; m_{2}).   \eqno\ldots (1.14)
$$

В итоге, соответствующая подстановка (1.14) и (1.12) в (1.11) приводит к следующему выражению для коммутатора операторной функции $A_{m}(x)$  рассматриваемой теории
$$
\left[ A_{m}(x), A_{n}(y) \right] =\frac{2}{\alpha+\beta}\,\{(\textsl{g}_{mn}+\frac{\partial_{m}\partial_{n}}{{m_{1}}^{2}})\, iD (x-y; m_{1}) - \frac{\partial_{m}\partial_{n}}{m^{2}}\, iD (x-y; m_{2})\},   \eqno\ldots (1.15)
$$
где  $D(x;m)$ -- перестановочная функция Паули\,--\,Йордана.

В свою очередь, (1.15), с учётом явного вида  $D$-функции, представляется
\begin{multline*}
  \left[ A_{m}(x), A_{n}(y) \right] =\frac{2i}{\alpha+\beta}\,\{\textsl{g}_{mn}[\frac{1}{2\pi}\varepsilon(x^{0}-y^{0})\delta(\lambda) -\\ \frac{m_{1}}{4\pi\sqrt{\lambda}}\,\varepsilon(x^{0}-y^{0})\theta(\lambda)J_{1}(m_{1}\sqrt{\lambda})] -
  \partial_{m}\partial_{n}[\frac{1}{4\pi\sqrt{\lambda}}\,\varepsilon(x^{0}-y^{0})\theta(\lambda)\frac{1}{m_{1}}J_{1}(m_{1}\sqrt{\lambda})]+\\
  \frac{2(\alpha+\beta)}{3\beta+\gamma}\,\partial_{m}\partial_{n}[\frac{1}{4\pi\sqrt{\lambda}}\,\varepsilon(x^{0}-y^{0})
\theta(\lambda) \frac{1}{m_{2}}J_{1}(m_{2} \sqrt{\lambda})]\}.                \qquad\ldots (1.16)
\end{multline*}

(1.16) демонстрирует, что полученный коммутатор, во-первых, допускает предел при $m\rightarrow 0$, в отличие от соответствующего коммутатора стандартной теории \cite{BSh}, во-вторых, в этом пределе получаем, с точностью до мультипликативной постоянной, коммутационные соотношения для операторных функций электродинамического потенциала
$$
 \left[ A_{m}(x), A_{n}(y) \right]= \frac{2i}{\alpha+\beta}\,\textsl{g}_{mn}D_{0}(x-y),             \eqno\ldots (1.17)
$$
и, в третьих, так как в окрестности светового конуса
$$
D(x;m)= \frac{1}{2\pi}\,\varepsilon(x^{0})\delta(\lambda) - \frac{m^{2}}{8\pi}\,\varepsilon(x^{0})\theta(\lambda)+ 0(\lambda),      \eqno\ldots (1.18)
$$
то самая сильная особенность функции Паули\,--\,Йордана на световом конусе,  $\delta$-функция, исключена из-под знака производной второго порядка в обсуждаемом коммутаторе рассматриваемой теории.

Уместно так же отметить и важное достоинство коммутационных соотношений (1.15) заключающееся в следующем.

Как подчёркивается в \cite{BLOT}, к целому ряду удручающих следствий теоремы Хаага добавляется возможность более сингулярного характера перестановочных соотношений при каноническом лагранжевом подходе к квантовой теории поля.

В частности,  демонстрируется, что стандартные перестановочные соотношения уже для свободного векторного поля приводят к появлению, в канонических перестановочных соотношениях, усингуляренного члена пропорционального $\triangle\delta(x)$, что является основанием ожидать настолько сингулярного коммутатора взаимодействующих полей, что его необходимо сглаживать по всем четырём координатам и нельзя рассматривать, в фиксированный момент времени, как обобщённую функцию трех пространственных координат.

Проведя аналогичное рассмотрение на основе коммутатора (1.15), находим, что $\triangle\delta(x)$-члена не возникает.

А теперь, используя (1.15), можем получить коммутационные соотношения для основного объекта нашего внимания – операторных  полевых функций $G_{mn}(x)$  и $H_{mn}(x)$.

Для этого, дифференцируем (1.15)  по $x$  и $y$   и, используя    (I.2.2), (I.2.3) и (I.2.4), находим
\begin{multline*}
\left[ G_{kl}(x), G_{mn}(y)\right] = \frac{2}{\alpha+\beta}\,\{(\textsl{g}_{\ell n}\partial_{k} \partial_{m}+ \textsl{g}_{\ell m}\partial_{k} \partial_{n}+ \textsl{g}_{k n}\partial_{\ell} \partial_{m}+\textsl{g}_{km}\partial_{\ell} \partial_{n} -  \\
 \frac{4}{{m_{1}}^{2}}\,\partial_{k} \partial_{\ell} \partial_{m} \partial_{n})\,iD(x-y;m_{1})+ \frac{4}{{m_{1}}^{2}}\,\partial_{k} \partial_{\ell} \partial_{m} \partial_{n}\,iD(x-y;m_{2}) - \\
 (\textsl{g}_{kl}\partial_{m} \partial_{n}+\textsl{g}_{mn}\partial_{k} \partial_{\ell}+\frac{1}{4}\,\textsl{g}_{k\ell}\textsl{g}_{mn}\Box)[(1+
 \frac{1}{{m_{1}}^{2}}\,\Box )\,iD(x-y;m_{1})- \\
 \frac{1}{{m_{1}}^{2}}\,\Box \,iD(x-y; m_{2})]\},    \qquad\ldots (1.19)
\end{multline*}
\begin{multline*}
\left[ H_{kl}(x), H_{mn}(y)\right] = \frac{1}{2(\alpha+\beta)}\,\textsl{g}_{k\ell}\textsl{g}_{mn}\Box\{(1+\frac{1}{{m_{1}}^{2}}\Box)\,iD(x-y;m_{1})-   \\
\frac{1}{{m_{1}}^{2}}\Box\,iD(x-y;m_{2})\},     \qquad\ldots (1.20)
\end{multline*}
\begin{multline*}
 \left[ H_{kl}(x), G_{mn}(y)\right] = \frac{2}{\alpha+\beta}\,(\textsl{g}_{k\ell}\partial_{m}\partial_{n}+ \frac{1}{4}\textsl{g}_{k\ell}\textsl{g}_{mn}\Box) \{(1+\\
  \frac{1}{{m_{1}}^{2}}\Box)\,iD(x-y;m_{1})- \frac{ 1}{{m_{1}}^{2}}\Box\,iD(x-y;m_{2})\},          \qquad\ldots (1.21)
\end{multline*}
\begin{multline*}
 \left[ G_{kl}(x), F_{mn}(y)\right] = \frac{2}{\alpha+\beta}\,(\textsl{g}_{\ell n}\partial_{k}\partial_{m}- \textsl{g}_{\ell m}\partial_{k}\partial_{n} +\textsl{g}_{k n}\partial_{\ell}\partial_{m}-   \\
\textsl{g}_{km}\partial_{\ell}\partial_{n})\,iD(x-y;m_{1}), \qquad\ldots (1.22)
\end{multline*}
$$
\left[ H_{kl}(x), F_{mn}(y)\right] = 0.              \eqno\ldots (1.23)
$$

Соотношения (1.19)--(1.23) проявляют регулярные свойства, аналогичные свойствам коммутатора (1.15).

В частности, допускается  безмассовый  предел, в котором (1.19)--(1.22) представляются, соответственно,
\begin{multline*}
  \left[ G_{kl}(x), G_{mn}(y)\right] = \frac{2}{\alpha+\beta}\,\{(\textsl{g}_{\ell n}\partial_{k} \partial_{m}+ \textsl{g}_{\ell m}\partial_{k} \partial_{n}+ \textsl{g}_{k n}\partial_{\ell} \partial_{m}+\\
  \textsl{g}_{km}\partial_{\ell} \partial_{n})\,iD_{0}(x-y) - (\textsl{g}_{kl}\partial_{m} \partial_{n}+\textsl{g}_{m n}\partial_{k} \partial_{\ell})\,iD_{0}(x-y)\}, \qquad\ldots (1.24)
\end{multline*}
$$
\left[ H_{kl}(x), H_{mn}(y)\right] =0 ,        \eqno\ldots (1.25)
$$
$$
 \left[ H_{kl}(x), G_{mn}(y)\right] = \frac{2}{\alpha+\beta}\,\textsl{g}_{k\ell}\partial_{m}\partial_{n}\,iD_{0}(x-y),    \eqno\ldots (1.26)
$$
\begin{multline*}
 \left[ G_{kl}(x), F_{mn}(y)\right] = \frac{2}{\alpha+\beta}\,(\textsl{g}_{\ell n}\partial_{k} \partial_{m}- \textsl{g}_{\ell m}\partial_{k} \partial_{n}+ \textsl{g}_{k n}\partial_{\ell} \partial_{m}-  \\
  \textsl{g}_{km}\partial_{\ell} \partial_{n})\,iD_{0}(x-y).    \qquad\ldots (1.27)
\end{multline*}

Аналогично, используя (1.15), получаем следующие коммутационные соотношения для операторных полевых функций $F_{mn}(x)$
\begin{multline*}
 \left[ F_{kl}(x), F_{mn}(y)\right] = \frac{2}{\alpha+\beta}\,(\textsl{g}_{\ell n}\partial_{k} \partial_{m}- \textsl{g}_{\ell m}\partial_{k} \partial_{n}- \textsl{g}_{k n}\partial_{\ell} \partial_{m}+  \\
  \textsl{g}_{km}\partial_{\ell} \partial_{n})\,iD(x-y;m_{1}),    \qquad\ldots (1.28)
\end{multline*}
которые, как видим, с точностью до прежней мультипликативной постоянной, совпадают со стандартными.

То есть, учёт поля, определяемого симметрическими производными аффинорами, в рассматриваемой теории, не искажает стандартные коммутационные соотношения для операторных компонент стандартного антисимметрического производного аффинора.

В свою очередь, подобно тому, как стандартные лагранжианы вещественного векторного поля  являются частными случаями лагранжиана (I.2.1), определяемыми частными значениями коэффициентов $\alpha, \beta$ и $\gamma$, так и соответствующие стандартные коммутационные соотношения являются соответствующими частными случаями коммутационных соотношений (1.15) и (1.28).

При этом,  в каждом из подобных случаев, необходимо иметь в виду, что рассмотренная в разделе  (I.1)  инвариантность составляющих правой части (I.1.3) означает и аналогичную инвариантность   $F$-,  $G$-  и  $H$-составляющих определяющего лагранжиана (I.2.1), поэтому коэффициенты $\alpha, \beta$  и $\gamma$, в рассматриваемой теории, играют роль параметров инвариантного глобального “включения” или “выключения” \cite{BSh} полей, определяемых  $F$-,  $G$-  и  $H$-составляющими производного аффинора векторного поля.

В связи с этим, наиболее приемлемым, на наш взгляд, частным случаем является частный случай, определяемый выбором следующих ограничительных условий на рассматриваемые коэффициенты
$$
\alpha+\beta =2, \quad \alpha \neq0,  \quad \beta \neq 0,                           \eqno\ldots (1.29)
$$
упрощающий соответствующие мультипликативные постоянные как в коммутационных соотношениях (1.15)--(1.28), так и в векторе энергии-импульса $P_{n}$, и одновременно сохраняющий как наличие обсуждаемых инвариантных составляющих в определяющем лагранжиане (I.2.1), так и наличие равноправных вкладов  $F$-,  $G$-полей в положительно определённой части $P_{0}$.

С другой стороны, стандартный частный случай, определяемый соотношениями
$$
\alpha= 2,   \quad  \beta =0,              \eqno\ldots (1.30)
$$
так же упрощающий вышеназванные мультипликативные постоянные, означает не только инвариантное исключение реально существующего поля, определяемого  $G$-составляющей производного аффинора, но и соответствующее завышение энергетического вклада полей, определяемых $F$-составляющей данного аффинора, в выражении 4-вектора энергии-импульса $P$.

\section*{\S \,2  Вакуум, пространство состояний и причинная функция.}
\addcontentsline{toc}{section}{\S \,2  Вакуум, пространство состояний и причинная функция.}

\qquad И так, наша система состоит из невзаимодействующих квантованных полей, характеризуемых операторными функциями $\,^{\tau}\!A_{m}^{\pm}(x)$  и  $\,^{\ell}\!A_{m}^{\pm}(x)$.

Приступим к определению амплитуды вакуумного состояния рассматриваемой системы и последующему определению пространства состояний этой системы.

В силу релятивистской инвариантности представления (I.4.1) и динамической независимости продольной и поперечной  частей рассматриваемого поля, амплитуда вакуумного состояния $\Psi_{0}$   определяется операторами  $\,^{\tau}\!A_{m}^{\pm}(x)$  и  $\,^{\ell}\!A_{m}^{\pm}(x)$   независимо друг от друга.

В частности, если отдать определяющую роль операторам $^{\tau}\!A_{m}^{\pm}(x)$, то рамки стандартных требований приводят к стандартным тождествам
$$
^{\tau}\!A_{m}^{-}(x)\,\Psi_{0}=0,    \quad             \stackrel{\ast}{\Psi_{0}}\,^{\tau}\!A_{m}^{+}(x)=0,      \eqno\ldots (2.1)
$$
которые, после перехода к импульсному представлению,  приводят к следующим условиям на трёхмерном
гиперболоиде $k^{0}=({\vec{k}}^{2}+{m_{1}}^{2})^{1/2}$
$$
a_{m}^{-}(\vec{k}) \Psi_{0}=0,  \quad     \stackrel{\ast}{\Psi_{0}} a_{m}^{+}(\vec{k})=0,      \eqno\ldots (2.2)
$$
где $ a_{m}^{\pm}(\vec{k})$  определены согласно  (I.4.20) и (I.4.23).

При этом, амплитуда вакуумного состояния  подчиняется стандартному условию нормировки
$$
\Psi_{0}\stackrel{\ast}{\Psi_{0}}=1.         \eqno\ldots (2.3)
$$

Что же касается доопределения вакуума операторами  $\,^{\ell}\!A_{m}^{\pm}(x)$, то тут положение напоминает
 ситуацию с  “временными”  фотонами электромагнитного поля.

Однако, если в случае электромагнитного поля рамки релятивистской инвариантности определения вакуума предоставляют только одну  возможность
$$
a_{0}^{-}(\vec{k}) \Psi_{0}=0,  \quad     \stackrel{\ast}{\Psi_{0}} a_{0}^{+}(\vec{k})=0,      \eqno\ldots (2.4)
$$
которая, как общеизвестно, приводит к индефинитности норм соответствующих состояний, то в нашем случае, в силу вышеуказанной релятивистской инвариантности разложения (I.4.1) и динамической независимости его частей, представляются две возможности, которые приводят к существенно разным теориям в отношении математической структуры пространства состояний рассматриваемой системы.

Первое возможное доопределение вакуума \cite{HV} есть
$$
^{\ell}\!A_{m}^{+}(x)\,\Psi_{0}=0,    \quad             \stackrel{\ast}{\Psi_{0}}\,^{\ell}\!A_{m}^{-}(x)=0,      \eqno\ldots (2.5)
$$
которое, после соответствующего перехода к импульсному представлению, сводится к следующим
условиям на гиперболоиде $k^{0}=({\vec{k}}^{2}+{m_{2}}^{2})^{1/2}$
$$
b^{+}(\vec{k}) \Psi_{0}=0,  \quad     \stackrel{\ast}{\Psi_{0}}b^{-}(\vec{k})=0,      \eqno\ldots (2.6)
$$
где  операторы $ b^{\pm}(\vec{k})$   определены в соответствии с (I.4.31) и (I.4.40).

 $ a_{m}^{\pm}(\vec{k})$ и   $ b^{\pm}(\vec{k})$  представляем как операторы из минимального ядерного пространства регулярных состояний $\Omega$, формирующие соответствующие обобщённые состояния пространства $\Omega^{\ast}$.

В частности, в силу рассматриваемого определения вакуума, произвольное обобщённое  $n$-частичное  $b$-состояние c определёнными импульсами   $b$-частиц представляется  в виде  \cite{BLOT1},
$$
\Phi^{b}(p_{1},...,p_{n})= \frac{1}{\sqrt{n!}}\,\int \!\Phi(q_{1},...,q_{n})b^{-}({\vec{p}}_{1})...b^{-}({\vec{p}}_{n})
\,\frac{d{\vec{q}}_{1}...d{\vec{q}}_{n}}{\sqrt{q_{1}^{0}...q_{n}^{0}}}\,\Psi_{0}.          \eqno\ldots (2.7)
$$

Тогда, на соответствующих регулярных  $n$-частичных состояниях, формирующих произвольный регулярный
вектор состояния  $\Phi \in \Omega$,  соответственно сглаженные  $b$-операторы
$$
b^{\pm}(u) = \int\!b^{\pm}(\vec{p})u(\vec{p})\frac{d^{3}\vec{p}}{\sqrt{p^{0}}},   \quad u(p)\in S(R_{3}),    \eqno\ldots (2.8)
$$
имеют, в силу коммутационных соотношений (1.5), следующие представления
 $$
 (b^{-}(u)\Phi)_{n}(p_{1},...,p_{n})= \frac{1}{\sqrt{n}}\,\sum \limits_{k=1}^{n}u({\vec{p}}_{k})\Phi_{n-1}(p_{1},...,p_{k-1},p_{k+1}),...,p_{n}),     \eqno\ldots (2.9)
 $$
 $$
 (b^{+}(u)\Phi)_{n}(p_{1},...,p_{n})= \sqrt{n+1}\int \limits_{V^{+}_{m_{2}}}\!\!u(\vec{p})\,\Phi_{n+1}(p, p_{1},...,p_{n})\,\frac{d^{3}\vec{p}}{p^{0}},     \eqno\ldots (2.10)
 $$
которые говорят о том, что операторы $b^{-}(u)$  и  $b^{+}(u)$  являются, соответственно, операторами увеличения и уменьшения числа   $b$-частиц, при их действии на  $n$-частичные состояния рассматриваемой системы.

Следовательно, нормальное произведение  $b$-операторов имеет вид $b^{-}(\vec{p})\, b^{+}(\vec{p})$.

 При этом, как следует из  (2.9)--(2.10),  “ненормальное” произведение, в отличие от нормального, не определено ни для одного вектора из $\mathcal{H}$  \cite{BLOT1}.

 Но тогда является неопределённым и выражение  оператора энергии-импульса  $b$-частиц
$$
^{b}\!P_{n} = - \frac{3\beta+\gamma}{8{m_{2}}^{2}}\int\! d\vec{k}\,k_{n}(b^{-}(\vec{k})\,b^{+}(\vec{k})+b^{+}(\vec{k})\,b^{-}(\vec{k})). \eqno\ldots (2.11)
$$

Таким образом, не только условие исключения из теории псевдофизических величин типа нулевой энергии, нулевого импульса и так далее, но и математическая последовательность теории требуют переопределения динамических переменных, например, путём введения формы нормального произведения для этих переменных.

В частности, оператор (2.11) должен быть теперь переопределён либо непосредственно
 $$
 ^{b}\!P_{n} \rightarrow \,:^{b}\!P_{n} :\,\eqdef - \frac{3\beta+\gamma}{4{m_{2}}^{2}}\int\! d\vec{k}\,k_{n}b^{-}(\vec{k})\,b^{+}(\vec{k}), \eqno\ldots (2.12)
 $$
либо путём соответствующего переопределения исходных динамических величин, таких как лагранжиан, тензор энергии-импульса и так далее  \cite{BSh}, что автоматически приводит к выражению (2.12).

Аналогичное переопределение касается и динамических переменных, содержащих операторы   $a_{m}$-частиц.

Теперь приступим к анализу математической структуры пространства состояний в рамках  рассматриваемого  определения вакуума.

Отмечаем, прежде всего, что доопределение (2.5) приводит к положительной определённости квадрата нормы любого состояния нашей системы, то есть, теория формулируется, в этом случае, без обращения к индефинитной метрике.

Действительно, рассматривая, для простоты, амплитуду одночастичного   $b$-состояния с ненормированной функцией распределения по импульсу, $c(\vec{k})$,
$$
\Phi_{1} = \int\!c(\vec{k})b^{-}(\vec{k})\,d\vec{k}\,\Psi_{0},        \eqno\ldots (2.14)
$$
имеем, в силу (1.5), (2.3) и (I.4.40),
$$
\|\Phi_{1}\|^{2}= \left(\frac{3\beta+\gamma}{4m}\right)^{2}\int\! |c(\vec{k})|^{2}\,d\vec{k}.     \eqno\ldots (2.15)
$$

То есть, ядерное пространство регулярных одночастичных состояний $\Omega_{1}$  представляется как ядерное подпространство гильбертова пространства одночастичных состояний $\mathcal{H}_{1}$.

А это означает, что теория формулируется, в рассматриваемом случае, в оснащённом гильбертовом пространстве
 состояний $\Omega \subset \mathcal{H} \subset \Omega^{\ast}$.

Однако, как и следовало ожидать, при определении вакуума посредством (2.5), приходим к противоречию с уже слабой формой постулата спектральности, которая является достаточной для получения таких важных результатов релятивистской квантовой теории поля, как теорема  Холла\,--\,Уайтмана  и  TCP-теорема.

Действительно, например для обобщённого одночастичного  $b$-состояния,
$$
\Phi_{1}(\vec{q}\,) = b^{-}(\vec{q}\,)\Psi_{0},              \eqno\ldots (2.16)
$$
имеем
$$
:P_{n}:\Phi_{1}(\vec{q}\,)= - q_{n}\Phi_{1}(\vec{q}\,),          \eqno\ldots (2.17)
$$
где   $q^{0}= (\vec{q}\,^{2}+{m_{2}}^{2})^{1/2}$.

Однако, как будет показано ниже, в рамках стандартного взаимодействия рассматриваемого поля со спинорным полем,  $b$-частицы, являющиеся виновниками  данного противоречия, не являются наблюдаемыми, в связи с чем,  вышеуказанное нарушение спектральности оказывается лишь математическим и не приводит теорию, рассматриваемую даже на пространстве всех состояний, к предсказаниям, противоречащим экспериментальным данным.

В противном случае, подобно тому, как трудности с вероятностной интерпретацией состояний с отрицательным и нулевым квадратом нормы в теориях с индефинитной метрикой могут быть решены путём разложения пространства состояний на физическое и нефизическое подпространства, в рассматриваемой теории вышеуказанная трудность с постулатом спектральности может быть решена аналогичным образом.

 С другой стороны, подобно тому, как теории с индефинитной метрикой, направленные на решение “извечных” проблем с расходимостями, избегают и следствий теоремы Челлена\,--\,Лемана (согласно которой, в условиях релятивистской инвариантности, дефинитности метрики и спектральности, одетые пропагаторы содержат те же сингулярности, что и голые) путём отказа от дефинитности метрики на некотором  подпространстве пространства состояний системы, в случае данной теории, избегаем аналогичных следствий вышеуказанной теоремы, но за счёт нарушения условия спектральности на соответствующем подпространстве состояний.

И наконец, рассматриваемое нарушение условия спектральности позволяет избежать удручающих следствий теоремы Хаага и относительно обоснованно использовать представление взаимодействия.

В результате, в свете вышеприведённого ожидается, что нарушение спектральности в рассматриваемой теории играет скорее положительную роль, нежели отрицательную.

Уместно отметить так же, что вопрос о существовании нетривиального решения “аксиоматических” уравнений движения приводит к большому вниманию в отношении построения теорий, в которых выполняется лишь часть общих принципов релятивистской квантовой теории.

 В нашем случае, эта часть определяется самой теорией, в рамках рассматриваемого определения вакуума.

Вышерассмотренное нарушение спектральности выражается, в частности, в отрицательности средних
значений энергии $b$-состояний.

Действительно, в силу (2.17), для одночастичного обобщённого  $b$-состояния (1.16), имеем
 $$
 \langle :P_{n}: \rangle_{1}^{b} = -\, q_{n}.               \eqno\ldots (2.18)
 $$

В свою очередь, индефинитность средних значений энергии  $b$-состояний также не может привести к какому-либо противоречию с экспериментальными данными в силу вышеуказанной ненаблюдаемости $b$-частиц.

Наконец, рассмотрим роль операторов   $b^{\pm}(\vec{q}\,)$  в отношении увеличения и уменьшения энергии--импульса системы, находящейся в некотором состоянии содержащем   $b$-частицы, и, одновременно, установим непротиворечивость рассматриваемого определения вакуума и последующего определения формы нормального произведения полевых операторов, уравнениям (1.2), то есть, соответствующим условиям совместности трансформационных свойств операторных полевых функций и трансформационных свойств амплитуд состояний рассматриваемой системы.

После этого, с учётом вывода, сделанного ранее из уравнений (2.9) и (2.10), можем окончательно выяснить возможность интерпретации операторов   $b^{\pm}(\vec{k})$, как операторов рождения и уничтожения  $b$-частиц
 с 4-~импульсом  $k$  и массой ${m_{2}}$.

Согласно (I.4.31),  (1.2) приводит к уравнениям
$$
-\, k_{n}b^{+}(\vec{k}) = [\,b^{+}(\vec{k}), P_{n}],                   \eqno\ldots (2.19)
$$
$$
 k_{n}b^{-}(\vec{k}) = [\,b^{-}(\vec{k}), P_{n}],                   \eqno\ldots (2.20)
$$
где   $k^{2}= {m_{2}}^{2}$.

Тогда, для переопределённого оператора энергии--импульса, (2.20) приводит к соотношению
$$
:P_{n}:\,\Phi_{2}(\vec{k},\vec{q}\,) = -\,(q_{n}+k_{n})\,\Phi_{2}(\vec{k},\vec{q}\,),        \eqno\ldots (2.21)
$$
где $\Phi_{2}(\vec{k},\vec{q}\,)$  -- обобщённая амплитуда двухчастичного  $b$-состояния, то есть,
$$
\Phi_{2}(\vec{k},\vec{q}\,)= b^{-}(\vec{k})b^{-}(\vec{q}\,)\Psi_{0}.         \eqno\ldots (2.22)
$$

 Таким образом, оператор   $b^{-}(\vec{k}\,)$  действительно можно рассматривать как оператор рождения отрицательно-энергетической частицы массы $m_{2}$   и  4-импульса   $k$.

Аналогично, используя (2.19), для состояния
$$
\tilde{\Phi}_{2}(\vec{k}, \vec{q}\,) = b^{+}(\vec{k})b^{-}(\vec{q}\,)\Psi_{0},            \eqno\ldots (2.23)
$$
получаем уравнение
 $$
 :P_{n}:\,\tilde{\Phi}_{2}(\vec{k}, \vec{q}\,) = -\,(q_{n}- k_{n})\tilde{\Phi}_{2}(\vec{k}, \vec{q}\,),    \eqno\ldots (2.24)
 $$
которое  означает,  что  оператор  $b^{+}(\vec{k})$  можно  интерпретировать как оператор уничтожения соответствующей отрицательно-энергетической частицы.

Таким образом, рассматриваемое доопределение вакуума (2.5) допускает стандартную интерпретацию операторов рождения и уничтожения частиц, в согласии с энергетичностью этих частиц.

С другой стороны, соотношения (2.21) и (2.24) позволяют интерпретировать доопределение вакуума (2.5), как условие неположительности энергии  $b$-частиц, в то время как стандартное определение (2.1) означает неотрицательность энергии физических  $a_{m}$-частиц, то есть, имеем релятивистски-инвариантное неперекрывающееся разделение спектров энергий рассматриваемых частиц.

А это, в частности, приводит к тому, что в конкретных задачах рассеяния с конечным числом исходных частиц, вклад невзаимодействующих   $b$-частиц в общую энергию всегда чисто отрицателен, вносится конечным числом этих частиц и является постоянным.

Теперь приступим к получению пропагатора рассматриваемого векторного поля $A(x)$   в рамках определения вакуума (2.1), (2.3) и (2.5).

В отличие от  стандартной теории векторного поля, определяем причинную функцию исходя из описания причинной связи процессов рождения и уничтожения частиц в различных точках пространства--времени, подобно тому, как определена причинная функция Грина скалярного  поля \cite{BSh1}, то есть
 $$
 D^{c}_{mn}(x-y) = -\,i \stackrel{\ast }{\Psi}_{0}T(A_{m}(x) A_{n}(y))\Psi_{0}.            \eqno\ldots (2.25)
 $$

Хронологическое произведение операторов поля определено  соотношением
$$
T(A_{m}(x) A_{n}(y)) \eqdef \theta(x_{0}-y_{0})A_{m}(x) A_{n}(y) + \theta(y_{0}-x_{0})A_{n}(y) A_{m}(x),     \eqno\ldots (2.26)
$$
где   $\theta(x_{0})$ -- единичная ступень Хевисайда.

С другой стороны, рассматриваемое определение вакуума и коммутационные соотношения (1.12) и (1.14) приводят к уравнениям
$$
 \stackrel{\ast }{\Psi}_{0}A_{m}(x) A_{n}(y)\Psi_{0}= \frac{2}{\alpha+\beta}\,(\textsl{g}_{mn}+ \frac{\partial_{m} \partial_{n}}{{m_{1}}^{2}})\,iD^{-}(x-y;m_{1}) + \frac{\partial_{m} \partial_{n}}{m^{2}}\,iD^{-}(y-x;m_{2}).     \eqno\ldots (2.27)
$$

Тогда, (2.25), согласно (2.26) и (2.27), представляется в виде
\begin{multline*}
  D_{mn}^{c}(x) = \theta(x_{0})[\frac{2}{\alpha+\beta}\,(\textsl{g}_{mn}+ \frac{\partial_{m} \partial_{n}}{{m_{1}}^{2}})\,D^{-}(x;m_{1}) +\\   \frac{\partial_{m} \partial_{n}}{m^{2}}\,D^{-}(-x;m_{2})]+  \theta(-x_{0})[\frac{2}{\alpha+\beta}\,(\textsl{g}_{mn}+
   \frac{\partial_{m} \partial_{n}}{{m_{1}}^{2}})\,D^{-}(-x;m_{1})+ \\
   \frac{\partial_{m} \partial_{n}}{m^{2}}\,D^{-}(x;m_{2})].        \qquad\ldots (2.28)
\end{multline*}

В свою очередь, воспользовавшись известными свойствами частотных частей функции Паули\,--\,Йордана, находим
\begin{multline*}
  D^{c}_{mn}(x)= \frac{2}{\alpha+\beta}\,\textsl{g}_{mn}D^{c}(x;m_{1})+ \theta(x^{0})\frac{\partial_{m} \partial_{n}}{m^{2}}D^{-}(x;m_{1})-\\
\theta(-x^{0})\frac{\partial_{m} \partial_{n}}{m^{2}}D^{+}(x;m_{1}) -  \theta(x^{0})\frac{\partial_{m} \partial_{n}}{m^{2}}D^{+}(x;m_{2})+\\ \theta(-x^{0})\frac{\partial_{m} \partial_{n}}{m^{2}}D^{-}(x;m_{2}),        \qquad\ldots (2.29)
\end{multline*}
где $D^{c}(x)$  -- причинная функция Грина скалярного поля.

С другой стороны, так как
\begin{multline*}
   \theta(\pm x^{0})\,{\partial_{0}}^{2}D^{\pm}(x;m)= {\partial_{0}}^{2}( \theta(\pm x^{0})D^{\pm}(x;m))\mp\\
   \partial_{0}(\delta(x^{0})D^{\pm}(x;m)) \mp \delta(x^{0})\partial_{0}D^{\pm}(x;m),    \qquad\ldots (2.30)
\end{multline*}
то соотношение  (2.29) принимает вид
\begin{multline*}
  D^{c}_{mn}(x)= \frac{2}{\alpha+\beta}\, \textsl{g}_{mn}D^{c}(x;m_{1})+ \frac{1 }{m^{2}}\, \{\partial_{m}\partial_{n}D^{c}(x;m_{1})- \\
 \partial_{m}\partial_{n}[\theta(x^{0}) D^{+}(x;m_{2}) - \theta(-x^{0})D^{-}(x;m_{2})] -\\ \delta_{m0}\delta_{n0}\,[\partial_{0}(\delta(x^{0})D(x;m_{1})) +
 \delta(x^{0})\partial_{0}D(x;m_{1})]+\\ \delta_{m0} \delta_{n0}\,[\partial_{0}(\delta(x^{0})D(x;m_{2})) + \delta(x^{0})\partial_{0}D(x;m_{2})]\}.     \qquad\ldots (2.31)
\end{multline*}

(2.31), прежде всего,  демонстрирует, с учётом уравнений
$$
\delta(x^{0})D(x;m) = 0,                   \eqno\ldots (2.32)
$$
$$
\delta(x^{0}) \partial _{0}D(x;m) = \delta(x),                   \eqno\ldots (2.33)
$$
сокращение    $\delta_{m0} \delta_{n0}$-членов и приводит к причинной функции
\begin{multline*}
  D^{c}_{mn}(x)=\frac{ 2}{\alpha+\beta}\,\textsl{g}_{mn}D^{c}(x;m_{1})+\frac{1}{m^{2}}\,\{\partial_{m} \partial_{n}D^{c}(x;m_{1}) -  \\
  \partial_{m} \partial_{n}[\theta(x^{0})D^{+}(x;m_{2}) - \theta(-x^{0})D^{-}(x;m_{2})]\}.       \qquad\ldots (2.34)
\end{multline*}

В свою очередь, (2.34), с учётом соотношения
$$
\theta(x^{0})D^{+}(x;m) - \theta(-x^{0})D^{-}(x;m) = 2\,D^{s}(x;m) - D^{c}(x;m),      \eqno\ldots (2.35)
$$
где  $D^{s}(x;m)$  -- чётное решение соответствующего неоднородного уравнения Клейна -- Гордона \cite{BSh}, принимает окончательное выражение
\begin{multline*}
 D^{c}_{mn}(x)=\frac{ 2}{\alpha+\beta}\,\{(\textsl{g}_{mn}+\frac{ \partial_{m} \partial_{n}}{{m_{1}}^{2}})D^{c}(x;m_{1}) +  \\
  \frac{ \partial_{m} \partial_{n}}{{m_{1}}^{2}}D^{c}(x;m_{2}) - \frac{2\partial_{m} \partial_{n}}{{m_{1}}^{2}}D^{s}(x;m_{2})\}.   \qquad\ldots (2.36)
\end{multline*}

Из  (2.36), прежде всего, видим, что учёт поля определяемого симметрическими производными аффинорами мультипликативно изменяет поперечную часть пропагатора и аддитивно – продольную часть, одновременно определяя структуру последней.

С другой стороны, находим, что, в отличие от стандартной теории \cite{BSh}, определение причинной функции согласно (2.25), то есть, по аналогии с определением причинной функции скалярного поля,  {\em не приводит} в рассматриваемой теории к появлению нековариантного члена, пропорционального  $\delta_{m0} \delta_{n0} \delta(x)$.

Эта возможность установления общего “причинного” принципа определения причинных функций и скалярного, и векторного полей несомненно является важным положительным свойством теории.

Рассмотрим теперь влияние второго и третьего членов правой части  (2.36)  на поведение $ D^{c}_{mn}(x)$   в окрестности светового конуса.

Учитывая представления функций  $D^{c}(x;m)$  и  $D^{s}(x;m)$   в данной окрестности,
$$
D^{c}(x;m) \approx \frac{1}{4\pi}\,\delta (\lambda)+ \frac{1}{4\pi^{2}i\lambda}+\frac{im^{2}}{8\pi^{2}}\,\ln \,\frac{m|\lambda|^{1/2}}{2}- \frac{m^{2}}{6\pi}\,\theta(\lambda),         \eqno\ldots (2.37)
$$
$$
D^{s}(x;m) \approx \frac{1}{2\pi}\,\delta (\lambda)- \frac{m^{2}}{2}\,\theta(\lambda),         \eqno\ldots (2.37)
$$
видим, что вышеназванные члены не регуляризуют поперечную часть причинной функции (2.36).

Таким образом, в рамках рассматриваемого определения вакуума, данная теория, с точки зрения  регулярности голого пропагатора, не предпочтительнее стандартной теории.

Теперь рассмотрим второе возможное доопределение вакуума, то есть, вместо (2.5), считаем, что имеют силу стандартные соотношения
 $$
^{\ell}\!A_{m}^{-}(x)\,\Psi_{0}=0,    \quad             \stackrel{\ast}{\Psi_{0}}\,^{\ell}\!A_{m}^{+}(x)=0,      \eqno\ldots (2.39)
 $$
которые приводят к следующим условиям на гиперболоиде $k^{0}~=~({\vec{k}}^{2}+~{m_{2}}^{2} )^{1/2}$
 $$
b^{-}(\vec{k}) \Psi_{0}=0,  \quad     \stackrel{\ast}{\Psi_{0}}b^{+}(\vec{k})=0.      \eqno\ldots (2.40)
 $$

Совершенно очевидно, что возникающие при этом  проблемы и возможности их решения аналогичны соответствующим для   “временных фотонов”  электромагнитного поля.

Однако, важно отметить, что если в случае электромагнитного поля соответствующие проблемы возникают для компоненты 4-вектора, то, в данной теории, они касаются динамически независимого скалярного поля, что значительно упрощает преодоление возникающих трудностей и предоставляет большие возможности их устранения.

В итоге, ситуация с  $b$-квантами  лишь отдаленно напоминает ситуацию с   $a^{0}$- фотонами электромагнитного поля.

По-прежнему,  операторы  $a_{m}^{\pm}(\vec{k}\,)$   и  $b^{\pm}(\vec{k}\,)$   определяем как операторы  из $\Omega$
 в   $\Omega^{\ast}$ и, в  частности, произвольное  $n$-частичное   $b$-состояние есть теперь
 $$
 \Phi^{b}(p_{1},...,p_{n})= \frac{1}{\sqrt{n!}}\,\int \!\Phi(q_{1},...,q_{n})b^{+}({\vec{p}}_{1})...b^{+}({\vec{p}}_{n})
\,\frac{d{\vec{q}}_{1}...d{\vec{q}}_{n}}{\sqrt{q_{1}^{0}...q_{n}^{0}}}\,\Psi_{0}.          \eqno\ldots (2.41)
 $$

Тогда соответствующие представления   $b$-операторов имеют вид (2.9) и (2.10), с соответствующей заменой $b^{\mp}(u)\rightarrow b^{\pm}(u)$, из которых следует, что на этот раз оператор  $b^{+}(u)$   является оператором
 увеличения числа  $b$-частиц, а оператор  $b^{-}(u)$  -- оператором уменьшения числа этих частиц.

Поэтому, нормальное произведение  $b$-операторов имеет вид $b^{+}(\vec{p}\,)\,b^{-}(\vec{p}\,)$  и прежние причины приводят к необходимости переопределения соответствующих динамических переменных.

В частности, оператор (2.11) теперь должен быть переопределён
 $$
 ^{b}\!P_{n} \rightarrow \,:^{b}\!P_{n} :\,\eqdef - \frac{3\beta+\gamma}{4{m_{2}}^{2}}\int\! d\vec{k}\,k_{n}b^{+}(\vec{k})\,b^{-}(\vec{k}). \eqno\ldots (2.42)
 $$

 Как и в случае электромагнитного поля, доопределение вакуума согласно (2.40), в силу коммутационных соотношений (1.5), приводит к индефинитности квадрата нормы на пространстве  амплитуд  состояний рассматриваемой  системы.

Например, для одночастичного  $b$-состояния, соответствующего (2.14), имеем
 $$
 \|\Phi_{1}\|^{2}= -\,\left(\frac{3\beta+\gamma}{4m}\right)^{2}\int\! |c(\vec{k})|^{2}\,d\vec{k},     \eqno\ldots (2.43)
 $$
в результате чего, ядерное пространство регулярных одночастичных состояний,  $\Omega_{1}$, уже не является подпространством гильбертова пространства состояний.

Таким образом, в рамках доопределения вакуума согласно (2.40), рассматриваемая теория принадлежит к классу теорий с индефинитной метрикой \cite{N}, получивших особенно широкое распространение после  выхода работ \cite{K}, \cite{L}.

Следовательно, на этот раз мы так же автоматически избегаем следствий теоремы Челлена\,--\,Лемана, но уже за счёт индефинитности метрики, что, в свою очередь, также позволяет относительно обоснованно использовать представление взаимодействия, избегая следствий теоремы  Хаага.

Однако, как и все теории подобного рода, предлагаемая теория в этом случае сталкивается с такой проблемой, как проблема унитарности   $S$-матрицы.

Решение этой проблемы должно осуществляться для каждого конкретного  типа взаимодействия персонально.

 В главе III мы рассмотрим стандартное взаимодействие данного векторного поля со спинорным полем и покажем, что исследуемая модель приводит к свободе  $b$-поля при наличии данного взаимодействия, поэтому, вышеуказанная трудность в этом случае представляется относительно тривиально преодолимой тем  или иным способом \cite{N1}.

В свою очередь замечаем, что в согласии  со слабым условием спектральности имеем
$$
:P_{n}:\Phi_{1}(\vec{q}\,)=  q_{n}\Phi_{1}(\vec{q}\,),          \eqno\ldots (2.44)
$$
где   $q^{0}= (\vec{q}\,^{2}+{m_{2}}^{2})^{1/2}$.

Следовательно, для среднего значения энергии--импульса системы, находящейся в состоянии   $\Phi_{1}(\vec{q}\,)$, находим
$$
 \langle :P_{n}: \rangle_{1}^{b} = q_{n}.               \eqno\ldots (2.45)
 $$
то есть, средние значения энергий положительны.

В свою очередь, уравнения (2.19) и (2.20) приводят к стандартным соотношениям
 $$
:P_{n}:\,\Phi_{2}(\vec{k},\vec{q}\,) = (q_{n}+k_{n})\,\Phi_{2}(\vec{k},\vec{q}\,),        \eqno\ldots (2.46)
$$
 $$
 :P_{n}:\,\tilde{\Phi}_{2}(\vec{k}, \vec{q}\,) = (q_{n}- k_{n})\tilde{\Phi}_{2}(\vec{k}, \vec{q}\,),    \eqno\ldots (2.47)
 $$
где
 $$
\Phi_{2}(\vec{k},\vec{q}\,)= b^{+}(\vec{k})b^{+}(\vec{q}\,)\Psi_{0}, \quad  \tilde{\Phi}_{2}(\vec{k}, \vec{q}\,) = b^{-}(\vec{k})b^{+}(\vec{q}\,)\Psi_{0},         \eqno\ldots (2.48)
$$
то есть, получаем обычную интерпретацию операторов   $b^{\pm}(\vec{k}\,)$.

И так имеем, что, при отсутствии взаимодействия  $b$-частиц, в любой конкретной задаче рассеяния энергетический вклад этих частиц положителен, вносится их конечным числом и постоянен.

Следовательно, эксперимент безразличен к присутствию этих  частиц в теории рассматриваемого типа.

Однако при этом важно подчеркнуть, что для теории в целом присутствие этих частиц небезразлично не только ввиду их соответствующей положительной роли в коммутаторе поля, но и, как увидим ниже, не менее положительной их роли в пропагаторе рассматриваемой теории.

Как и в первом случае, определяем  $D^{c}_{mn}(x)$  посредством (2.25), то есть, в соответствии с описанием причинной связи процессов рождения и уничтожения частиц и, после проведения соответствующих преобразований, вместо (2.36) получаем
$$
 D^{c}_{mn}(x)=\frac{ 2}{\alpha+\beta}\,\{(\textsl{g}_{mn}+\frac{ \partial_{m} \partial_{n}}{{m_{1}}^{2}})D^{c}(x;m_{1}) -
  \frac{\partial_{m} \partial_{n}}{{m_{1}}^{2}}D^{c}(x;m_{2})\}.   \eqno\ldots (2.49)
$$

(2.49) демонстрирует, что, во-первых, учёт поля определяемого симметрическими производными аффинорами, так же как и в первом случае доопределения вакуума, мультипликативно изменяет поперечную часть пропагатора и аддитивно – продольную часть, одновременно определяя структуру последней.

Во-вторых, рассматриваемая теория естественным образом  приводит к причинной функции, необходимо использующейся в общей теории функций Грина взаимодействующих полей, когда продольным членом пренебрегать нельзя.

В-третьих, как и в первом случае доопределения вакуума находим, что, определение причинной функции согласно (2.25), то есть,   по  аналогии  с определением причинной  функции  скалярного поля, вновь не приводит, в рассматриваемой  теории, к появлению нековариантного члена, пропорционального   $\delta_{m0} \delta_{n0} \delta(x)$.

Таким образом, приходим к общему выводу: квантовая теория векторного поля {\em допускает} причинное определение причинной функции и при этом вне зависимости от рассмотренных способов определения  вакуума.

Теперь рассмотрим влияние продольного члена правой части (2.49) на поведение полученной причинной функции в окрестности светового конуса.

Учитывая представление (2.37) находим, что   $D^{c}(x;m_{2})$-член, обусловленный  наличием  в теории   $b$-поля, действует, в (2.49),  как  регуляризатор, исключая  из-под знака  производной  второго порядка  самые сильные
 особенности на световом конусе,  $\delta(\lambda)$ и   $\lambda^{-1}$.

Таким образом, рассматриваемая теория, с точки зрения  регулярности голого пропагатора,  в рамках данного  определения вакуума  предпочтительнее стандартной.

В импульсном представлении, (2.49) принимает вид
 $$
 D^{c}_{mn}(k)= \frac{2}{\alpha+\beta}\,\{(\textsl{g}_{mn}-\frac{ k_{m} k_{n}}{{m_{1}}^{2}})\,\frac{1}{{m_{1}}^{2}-k^{2}-i\varepsilon}\,+\frac{ k_{m} k_{n}}{{m_{1}}^{2}}\,\frac{1}{{m_{2}}^{2}-k^{2}-i\varepsilon}\}.      \eqno\ldots (2.50)
 $$

Структура (2.50) демонстрирует, в свою очередь, что полученный пропагатор автоматически (без обращения к условию сохранения тока) приводит к ренормируемой теории, например, стандартного взаимодействия векторного поля со спинорным полем и позволяет выполнить процедуру ренормализации последовательно \cite{BShir}, \cite{Fr},  без дополнительной перестройки пропагатора на промежуточном этапе теории.

Что же касается ренормализации стандартной теории вышеуказанного взаимодействия, то, как известно, она достигается только за счёт эффективного изменения спаривания, а, следовательно, за счёт изменения причинной функции, используя  условие сохранения тока, то есть, фактически на промежуточном этапе теории, осуществляется переход
 $$
(\textsl{g}_{mn}-\frac{ k_{m} k_{n}}{m^{2}})\,\frac{1}{k^{2}-m^{2}-i\varepsilon}\rightarrow \textsl{g}_{mn}\,\frac{1}{k^{2}-m^{2}-i\varepsilon}.
 $$

Тут следует напомнить об  известном успехе нестандартных теорий – калибровочных теорий со спонтанным
нарушением симметрии \cite{Bern}--\cite{LSch}, в рамках которых проблему ренормируемости можно считать решённой.

Однако, эти успехи не должны восприниматься как основание для отсутствия интереса к стандартной теории и её проблемам.

(2.50) можно переписать, выделив член, соответствующий калибровке Ландау,
$$
 D^{c}_{mn}(k)= \frac{2}{\alpha+\beta}\,\{(\textsl{g}_{mn}-\frac{ k_{m} k_{n}}{k^{2}})\,\frac{1}{{m_{1}}^{2}-k^{2}-i\varepsilon}\,+
 \frac{{m_{2}}^{2}}{{m_{1}}^{2}}\,\frac{ k_{m} k_{n}}{k^{2}}\,\frac{1}{{m_{2}}^{2}-k^{2}-i\varepsilon}\}.      \eqno\ldots (2.51)
 $$

Но, согласно (I.2.23) и (I.2.24), имеем следующее замечательное, для рассматриваемой теории, соотношение масс
 $$
  \frac{{m_{2}}^{2}}{{m_{1}}^{2}}= \frac{2(\alpha+\beta)}{3\beta+\gamma},     \eqno\ldots (2.52)
 $$
благодаря которому, (2.51) представляется в виде
\begin{multline*}
   D^{c}_{mn}(k)= \frac{2}{\alpha+\beta}\,\{(\textsl{g}_{mn}-\frac{ k_{m} k_{n}}{k^{2}})\,\frac{1}{{m_{1}}^{2}-k^{2}-i\varepsilon}\,+ \\
 \frac{2(\alpha+\beta)}{3\beta+\gamma}\,\frac{ k_{m} k_{n}}{k^{2}}\,\frac{1}{{m_{2}}^{2}-k^{2}-i\varepsilon}\}.   \qquad\ldots (2.53)
\end{multline*}

  (2.53) демонстрирует, в свою очередь, отсутствие, в пропагаторе рассматриваемой теории, члена $k_{m} k_{n}/m^{2}$, который обычно приводит к  “суровым”  ультрафиолетовым проблемам \cite{Bern}.

При этом, как видим, мы не прибегаем ни к использованию калибровочной инвариантности, ни к использованию условия сохранения тока  \cite{Lee}.

С другой стороны, вновь, в отличие от причинной функции стандартной теории и в отличие от пропагатора \cite{FM}, видим,  что (2.53) допускает предел при
$$
m^{2}\rightarrow 0,         \eqno\ldots (2.54)
$$
который, в силу соотношений (I.2.23) и (I.2.24), означает,  что
$$
{m_{1}}^{2}\rightarrow 0,   \quad   {m_{2}}^{2}\rightarrow 0,        \eqno\ldots (2.55)
$$
и, поэтому, интерпретируется как безмассовый предел пропагатора.

Следовательно, пропагатор данной теории не разделяет той осторожности перехода к безмассовой теории, которая сопровождает пропагатор Фельдмана\,--\,Метьюса \cite{FM}.

В рассматриваемом пределе (2.55), пропагатор (2.53) представляется
\begin{multline*}
   D^{c}_{mn}(k)= \frac{2}{\alpha+\beta}\,\{(\textsl{g}_{mn}-\frac{ k_{m} k_{n}}{k^{2}})\,\frac{1}{-k^{2}-i\varepsilon}\,+ \\
 \frac{2(\alpha+\beta)}{3\beta+\gamma}\,\frac{ k_{m} k_{n}}{k^{2}}\,\frac{1}{-k^{2}-i\varepsilon}\},   \qquad\ldots (2.56)
\end{multline*}
то есть, имеет вид фотонной функции Грина в общековариантной калибровке.

При этом, одновременно приходим к конкретному значению фиксирующего калибровку  коэффициента
$$
d_{\ell}= 2(\alpha+\beta)/(3\beta+\gamma).                      \eqno\ldots (2.57)
$$

Таким образом, рассматриваемая теория позволяет говорить о массивной электродинамике, в том же самом смысле, как и в \cite{FM}.

При этом представляется уместным провести сравнение пропагатора рассматриваемой теории, с эффективным пропагатором массивного фотона в теории, основанной на механизме  Хиггса.

В последней, замечательным является то  \cite{Bern}, что она достаточно  “помнит” своё электромагнитное происхождение, чтобы произвести калибровочный член  $k_{m} k_{n}/k^{2}$, который неплохо ведёт себя при высоких энергиях.

С другой стороны, фотон приобретает массу за счёт спонтанного нарушения калибровочной симметрии.

В итоге, в таких теориях, представляется  возможным избежать  как инфракрасной, так и ультрафиолетовой проблем.

В рассматриваемой теории, благодаря замечательному  для нашей теории  соотношению (2.52), приходим, без использования электромагнитных происхождений, к утрате члена $k_{m}k_{n}/m^{2}$   и к воспроизведению члена  $k_{m}k_{n}/k^{2}$, уходя тем самым и от соответствующих ультрафиолетовых проблем.

Возвращаясь к (2.50) видим, что допускается и второй предел при
$$
\beta \rightarrow 0,    \quad  \gamma \rightarrow 0,           \eqno\ldots (2.58)
$$
физически интерпретируемый как исключение материального полевого объекта, определяемого  симметрическими производными аффинорами рассматриваемого векторного поля.

При этом, в силу (I.2.24), имеем
 $$
 {m_{2}}^{2} \rightarrow \infty,             \eqno\ldots (2.59)
 $$
то есть, рассматриваемый предел  соответствует   $\xi$-предельному процессу Ли и Янга \cite{LY}--\cite{S}.

 В свою очередь, в пределе (2.58),  (2.50), при $\alpha = 2$, принимает стандартный вид
$$
 D^{c}_{mn}(k)= (\textsl{g}_{mn}-\frac{ k_{m} k_{n}}{m^{2}})\,\frac{1}{m^{2}-k^{2}-i\varepsilon},    \eqno\ldots (2.60)
$$
то есть, в рассматриваемом пределе получаем пропагатор векторного поля стандартной теории, приводящий к значительному ухудшению ренормируемости или к полной её утрате, к соответствующей ультрафиолетовой
проблеме и к отсутствию предела при   $m \rightarrow 0$.

Таким образом находим, что целый ряд трудностей стандартной теории имеет чисто субъективное происхождение, то есть, они порождены  либо неучётом, либо преждевременным исключением полей,  определяемых  симметрическими производными аффинорами векторного поля.

 Уместно также отметить, что само существование предела (2.58) остаётся, в общем случае, неясным, так как функции Грина являются неренормализуемыми в этом пределе \cite{Hsu}.

Поэтому мы склонны заключить, что, в общем случае,  нельзя исключать из теории векторного поля симметрические производные аффиноры также и с этой точки зрения.

 Подобно тому, как    $\xi$-теории сталкиваются с решением проблемы унитарности    $S$-матрицы при конечных ненулевых значениях параметра $\xi$ \cite{Hsu}, так и рассматриваемая теория сталкивается, в общем случае, с аналогичной проблемой при конечных ненулевых значениях параметров $\beta$  и  $\gamma$ однако, как было отмечено ранее, тривиально преодолимой в рамках стандартного взаимодействия со спинорным полем.

Что же касается других типов взаимодействий, приводящих к рассеянию  $b$-поля, то тут вопрос унитарности также может быть решен, например, использованием соответствующих методов безмассовой теории Янга\,--\,Миллса или квантовой электродинамики с нелинейными калибровочными условиями \cite{Hsu}.



\renewcommand{\refname}{ \sl Список литературы}

\chapter{\:Классическое взаимодействие, калибровочная инвариантность, определяющая система полевых уравнений и заключительные замечания к общей теории.}


\qquad Продолжается построение теории действительного векторного поля начатое в главах I и II \cite {Al1, Al2}.

Обсуждены классическое взаимодействие со спинорным полем, калибровочная инвариантность рассматриваемой теории и
определяющие полевые уравнения.

Представлены соотношения, определяющие как непосредственное, так и опосредованное силовые воздействия джейтонной составляющей электроджейтонного поля на электрически заряженные частицы.

Проведено сравнение данной теории с известными теориями векторного поля.

Ключевые слова: классическое взаимодействие, спинорное поле, калибровочная инвариантность.

\section*{\S \,1 Классическое взаимодействие и калибровочная инвариантность}
\addcontentsline{toc}{section}{\S \,1  Классическое взаимодействие и калибровочная инвариантность}

\qquad Рассмотрим классическую теорию взаимодействующих полей, представленных исследуемым векторным полем $A(x)$ и спинорным электрон-позитронным полем $\psi(x), \overline{\psi}(x)$, оставаясь в рамках модели, в которой осуществлен переход   (I.1.5).

Следуя \cite{BSh}, лагранжиан системы этих полей выберем в виде
$$
L(x)=L_{0A}(x) +L_{0\psi}(x)+L_{I}(x),                                \eqno\ldots (1.1)
$$
где
$$
L_{0\psi}(x)=\frac{i}{2}(\overline{\psi}(x)\hat{\gamma}\cdot\partial\psi(x)\\
-\partial\overline{\psi}(x)\cdot\hat{\gamma}\psi(x))-m_{\psi}\overline{\psi}(x)\psi(x),                    \eqno\ldots (1.2)
$$
$$
L_{I}(x)= e\overline{\psi}(x)\hat{\gamma}\psi(x)\cdot A(x)=-j(x)\cdot A(x),                        \eqno\ldots (1.3)
$$
однако, лагранжиан $L_{0A}(x)$ теперь определяем, в соответствии с (I.1.6), согласно (I.2.1),
$$
L_{0A}(x)= -\frac{\alpha}{2}\,F(x)^{2}-\frac{\beta}{2}\,G_{d}(x)^{2}-\frac{\gamma}{2}\,G_{\ast}(x)^{2}+\frac{m^{2}}{2}\,A(x)^{2}. \eqno\ldots (1.4)
$$

4-тензоры $F(x)$,  $G_{d}(x)$  и   $G_{\ast}(x)$, входящие в (1.4),  и тензор
$$
G(x)=G_{d}(x)+G_{\ast}(x)
$$
определены соотношениями
$$
F(x)= \widetilde{rot}\,A(x),                                         \eqno \ldots(1.5)
$$
$$
G_{d}(x)= \widetilde{def}_{d}\,A(x),                       \eqno\ldots (1.6)
$$
$$
G_{\ast}(x)= \widetilde{def}_{\ast}\,A(x),                   \eqno\ldots (1.7)
$$
$$
G(x)= \widetilde{def}\,A(x),                      \eqno\ldots (1.8)
$$
в которых  $ \widetilde{rot}$,  $\widetilde{def}_{d}$ ,  $ \widetilde{def}_{\ast}$  и  $\widetilde{def}$  -- тензорные дифференциальные операторы, имеющие вид
$$
 \widetilde{rot}= \partial\, \overset{a}{\otimes},                               \eqno\ldots (1.9)
$$
$$
\widetilde{def}_{d}= \partial  \overset{s}{\otimes}- \frac{\textsl{g}}{tr \textsl{g}}\,\partial\cdot  ,                 \eqno\ldots (1.10)
$$
$$
\widetilde{def}_{\ast}= \frac{\textsl{g}}{tr \textsl{g}}\,\partial\cdot  ,                            \eqno\ldots (1.11)
$$
$$
\widetilde{def}=  \partial  \overset{s}{\otimes}  .                               \eqno\ldots (1.12)
$$

$F(x)^{2}$,  $G_{d}(x)^{2}$ и  $G_{\ast}(x)^{2}$  в (1.4) представлены как скалярные квадраты соответствующих тензоров (как квадраты евклидовых норм этих тензоров) \cite{Gil}.

Прежние тензоры, представленные в (I.2.2)--(I.2.4), являющиеся удвоенными по отношению к данным, в дальнейшем будем обозначать как $F_{L}(x)$,  $G_{dL}(x)$,   $G_{\ast L}(x)$  и   $G_{L}(x)$ и называть {\em лоренцевыми\/} производными тензорами векторного поля $A(x)$.

Индексы {\em d}   и   $\ast$   в буквенных обозначениях девиатора  (1.6) и  дилататора  (1.7)  заимствованы у  \cite{Gil}.

Всюду в данном контексте тензор рассматривается как инвариантный, по классу преобразований координат {\em пассивного\/ } типа, объект, представляемый различными наборами компонент в различных системах отсчета \cite{Ko}.

Термин {\em векторное поле\/ } рассматривается всюду  в обсуждаемой теории  исключительно в математическом смысле (аспекте).

С самого начала следует подчеркнуть, что в {\em электродинамическом варианте  теории, \/ } то-есть, когда векторное поле представлено {\em электродинамическим \/ } 4-потенциалом $A(x)$, лагранжиан (1.1), включая его частный случай,  соответствующий выбору $\alpha= \beta=\gamma =1$, при котором, в силу инвариантного соотношения
$$
\partial A(x)^{2}= F(x)^{2}+  G_{d}(x)^{2} +  G_{\ast}(x)^{2},                 \eqno\ldots (1.13)
$$
лагранжиан $L_{0A}(x)$ принимает хорошо известный "диагональный"\  вид,
$$
L_{0A}(x)= -\frac{1}{2}\,\partial A(x)^{2},                                  \eqno \ldots(1.14)
$$
теперь рассматривается как лагранжиан системы взаимодействующих электрон--позитронного, электромагнитного и электроджейтонного полей, в котором 4-потенциал $A(x)$ является 4-потенциалом не только электромагнитного, но и электроджейтонного полей, а (1.3) выступает теперь как лагранжиан минимального взаимодействия рассматриваемого спинорного поля с электромагнитным и электроджейтонным полями.

Следуя традиционному изложению классической теории поля \cite{La}, обсуждаемая электродинамика
 ограничивается  рамками  электродинамики  вакуума и точечных электрических зарядов, а используемая система единиц очевидна из контекста и специально не оговаривается.

Как и в главах I и  II, действительные числовые коэффициенты $\alpha, \beta$ и $\gamma$, выступающие, в частности, в качестве "функций  глобального включения и выключения"   \cite{BSh} полевых переменных  определённых тензорами (1.5), (1.6) и (1.7), не ограничиваются, c самого начала, конкретными числовыми значениями с тем, чтобы проследить роль каждой из данных полевых переменных на отдельных этапах построения теории, в частности,  в соотношениях {\em явно \/ }не содержащих эти полевые переменные.

Уравнения Лагранжа\,--\,Эйлера, соответствующие варьированиям действия, определяемого лагранжианом (1.1), по
$A(x), \overline{\psi}(x)$ и $\psi(x)$, принимают вид
$$
\alpha\, \partial \!\cdot\! F(x)+ \beta\, \partial\!\cdot\! G_{d}(x)+ \gamma\, \partial\!\cdot\! G_{\ast}(x)+ m^{2}\,A(x)= j(x), \eqno\ldots (1.15)
$$
$$
i\hat{\gamma}\cdot\partial\psi(x)- m_{\psi}\psi(x)= -e\hat{\gamma}\psi(x)\!\cdot\! A(x),                \eqno\ldots (1.16)
$$
$$
i\partial\overline{\psi}(x)\cdot\hat{\gamma}+ m_{\psi}\overline{\psi}(x)= e\hat{\gamma}\overline{\psi}(x)\!\cdot \!A(x).  \eqno\ldots (1.17)
$$

Из (1.16) и (1.17) прежде всего следует, что плотность тока $ j(x) $  удовлетворяет уравнению непрерывности при любых $A(x)$, в том числе и при $A(x)$, являющихся решениями уравнения (1.15).

В электродинамическом случае, в отличие от соответствующего уравнения Максвелла, теперь имеем
$$
\alpha\, \partial\!\cdot\! F(x)+ \beta\, \partial \!\cdot\! G_{d}(x)+ \gamma\, \partial\!\cdot\! G_{\ast}(x)= j(x).    \eqno\ldots (1.18)
$$

Данное уравнение прежде всего явно демонстрирует, что ток $ j(x) $ создаёт, в общем случае, не только общеизвестное электромагнитное, но и электроджейтонное поля.

При этом уместно отметить, что, как известно, стандартное уравнение второй пары системы уравнений Максвелла,
$$
\partial \!\cdot\! F_{L}(x)= j(x),                          \eqno\ldots (1.19)
$$
приводит к трудностям при квантовании электромагнитного поля в произвольной калибровке в локальной теории
 \cite{G}.

 В частности, из уравнения (1.19) мгновенно следует, что для любой локальной наблюдаемой  $\varphi(x)$  имеет место соотношение
 $$
\left[Q, \varphi(x)\right]= \left[\int j^{0}(x)\,dx , \varphi(x)\right]= 0.                        \eqno\ldots (1.20)
$$

То есть, (1.19), при произвольных условиях квантования, не позволяет говорить о заряженных квантованных полях, удовлетворяющих соотношению
$$
[Q, \varphi(x)]= e\varphi(x).                                 \eqno\ldots (1.21)
$$

Уравнение (1.18), как видим, к подобному осложнению не приводит и, при этом, именно потому, что оно содержит {\em симметрические} производные аффиноры векторного потенциала.

Нетрудно убедиться в том, что рассматриваемый лагранжиан (1.1) и вытекающие из него уравнения движения (1.15)--(1.17), инвариантны, с точностью до 4-дивергенции, относительно "специализированных  калибровочных преобразований" \cite{BSh}
$$
A(x) \rightarrow A^{'}(x)= A(x)+ \partial f(x),                \eqno\ldots (1.22)
$$
$$
\psi(x)\rightarrow \psi(x)^{'}= e^{ief(x)}\psi(x),                       \eqno \ldots(1.23)
$$
$$
\overline{\psi}(x)\rightarrow \overline{\psi}(x)^{'}= e^{-ief(x)}\overline{\psi}(x),   \eqno\ldots (1.24)
$$

с $f(x)$, подчиняющейся уравнению
$$
(\Box - {m_{2}} ^{2})\, \partial f(x)= 0,    \eqno\ldots (1.25)
$$

где ${m_{2}} ^{2}$ определена согласно (I.2.24), то есть,
$$
{m_{2}} ^{2}= \frac{4m^{2}}{3\beta+\gamma}.                    \eqno\ldots (1.26)
$$

При этом,  $f(x)$ должна удовлетворять так же условию
$$
f(x) \neq - \partial_{x}D_{0} ^{c}(x-y)\cdot A(y),   \eqno\ldots (1.27)
$$
так как в противном случае ни лагранжиан, ни уравнения движения не инвариантны относительно рассматриваемых преобразований.

Следовательно, инвариантность при преобразованиях (1.22)--(1.24) не является непосредственным основанием для традиционного исключения продольной части векторного поля из полного лагранжиана и уравнений движения, если оставаться в классе функций  $f(x)$, подчиняющихся дополнительному условию (1.27).

С другой стороны находим, что при $m\neq0$ рассматриваемая теория обладает специализированной калибровочной инвариантностью  \cite{Hall} и, тем самым, представляется, в этом случае,  вариантом калибровочно-инвариантной формулировки теории нейтрального векторного поля с ненулевой массой покоя  \cite{BSh}, однако, без использования вспомогательных полей  \cite{St, Fu} и без специализированного выбора перестановочных соотношений.

При этом уместно отметить следующий нестандартный взгляд на калибровочное преобразование (1.22).

Как увидим ниже,  $^{l}A(x) $, при наличии рассматриваемого взаимодействия, остается свободным полем и подчиняется по-прежнему уравнению (I.4.9).

Тогда, переписав (1.22) в виде
$$
A(x)=\, ^{\tau}\!A(x)+\,  ^{l}\!A(x) \rightarrow A^{'}(x) = \, ^{\tau}\! A(x)+\, ^{l}\!A(x)+ \partial f(x)   \eqno\ldots (1.28)
$$
и учтя (1.25) находим, что калибровочное преобразование (1.22) имеет тривиальный математический смысл --- представляет собой добавление к общему решению его линейно зависимой части.

В свою очередь, выбор
$$
f(x) = - \partial_{x}D_{0}^{c}(x-y)\! \cdot\! A(y),   \eqno\ldots (1.29)
$$
обеспечивающий исключение  $\,^{l}A(x)$, приводит не только к неинвариантности лагранжиана и уравнений движения, но и к исключению из общего решения этих уравнений его линейно независимой части, что представляет собой потерю соответствующей математической общности теории.

 \section*{\S \,2 Локальные законы сохранения и калибровочная инвариантность}
 \addcontentsline{toc}{section}{\S \,2  Локальные законы сохранения и калибровочная инвариантность}

\qquad Ввиду принципиальной важности вопроса калибровочной инвариантности теории, предварительно определим процедуру получения локальных законов сохранения, сопровождающих калибровочную инвариантность, выделив  следующие три этапа этого процесса.

1. Следуя общему подходу к выводу локальных законов сохранения \cite{BSh}, рассматриваем вариацию действия, $\overline{\delta}S $, обусловленную изменением формы полевых функций и их первых производных при произвольном фиксированном объеме интегрирования, выделяя в этой вариации ее эйлерову часть, то есть,
 $$
\overline{\delta}S= \int_{V}\{ \delta L(x) /\delta u(x) \cdot \overline{\delta}u(x)+ \partial\!\cdot\!(\partial_{\partial u(x)}L(x)\cdot \overline{\delta}u(x))\}\,dx,     \eqno\ldots (2.1)
$$
где $\delta L(x)/\delta u(x)$ -- лагранжева производная от $L(x)$, имеющая, в рассматриваемом случае, вид
$$
\delta L(x) /\delta u(x)= \partial_{ u(x)}L(x) - \partial \!\cdot\!\partial_{\partial u(x)}L(x).     \eqno\ldots (2.2)
$$

(2.1) демонстрирует, что на экстремалях, то есть, на полевых функциях удовлетворяющих уравнению Лагранжа--Эйлера, $ \delta L(x) /\delta u(x)= \theta$, вариация (2.1) принимает "дивергентный"  вид
$$
\overline{\delta}S= \int_{V} \partial\!\cdot\!(\partial_{\partial u(x)}L(x)\cdot \overline{\delta}u(x))\ dx,    \eqno\ldots (2.3)
$$
"готовый" дать соответствующие законы сохранения.

2. Теперь, вместо традиционного "требуем"\, обращения $\overline{\delta}S$ в нуль, определим группу калибровочной инвариантности действия, на которой, при любой фиксированной области интегрирования,
$\overline{\delta}S=0$, что можно сделать, например, прямолинейным вычислением рассматриваемой вариации и последующим анализом полученного выражения.

3. В результате, на этой группе, на экстремалях, при любой фиксированной области интегрирования будем иметь тождественное равенство
$$
\int_{V} \partial\!\cdot\!(\partial_{\partial u(x)}L(x)\cdot \overline{\delta}u(x))\ dx= 0,   \eqno\ldots (2.4)
$$
которое, в силу произвольности области интегрирования, приводит, на этой группе калибровочных преобразований и на данных экстремалях, к локальному закону сохранения
$$
\partial\!\cdot\!(\partial_{\partial u(x)}L(x)\cdot \overline{\delta}u(x))\ = 0.      \eqno\ldots (2.5)
$$

Следуя данной последовательности действий, рассмотрим наиболее важный частный случай рассматриваемой теории, когда векторное поле представлено электродинамическим 4-потенциалом $A(x)$, а исходный лагранжиан определен выражением (1.1) при $m= 0$.

Для упрощения дальнейшего анализа данной полевой системы будем рассматривать частный случай, когда  $\gamma= \beta$, а тензор электроджейтонного поля представлен соотношением (1.8).

В этом случае, лагранжиан взаимодействующих электрон--позитронного, электромагнитного и электроджейтонного полей, через    очевидные обозначения, принимает вид
$$
L(x)= L_{0}(F(x))+  L_{0}(G(x))+  L_{0}(\psi(x),\overline{\psi}(x))+ L_{I}(A(x), \psi(x), \overline{\psi}(x)). \eqno\ldots (2.6)
$$

Определим вариацию соответствующего действия обусловленную изменением формы полевых функций и их первых производных, содержащихся в $L(x)$, при обобщенных инфинитезимальных калибровочных преобразованиях второго рода.

В этом случае, (2.6) приводит к соотношениям
$$
(\partial_{\partial\psi(x)}L(x))\overline{\delta}\psi(x)+ (\partial_{\partial\overline{\psi}(x)}L(x))\overline{\delta}\,\overline{\psi}(x)= j(x)f(x),                \eqno\ldots (2.7)
$$
$$
(\partial_{\partial A(x)}L(x))\cdot\overline{\delta}A(x)= - (\alpha F(x)+ \beta G(x))\cdot\partial f(x),    \eqno\ldots (2.8)
$$
в силу которых вариация (2.3) принимает вид
$$
\overline{\delta}S= \int_{V} \partial\!\cdot\!\{- (\alpha F(x)+ \beta G(x))\cdot \partial f(x)+j(x)f(x)  \}\, dx,    \eqno\ldots (2.9)
$$

Теперь, в соответствии с общей схемой, определим группу рассматриваемой калибровочной инвариантности рассматриваемого действия.

Для этого вычислим ту же вариацию, на той же произвольной но фиксированной области интегрирования, непосредственно по $L(x)$.

Так как сумма лагранжианов,  $L_{0}(F(x))+   L_{0}(\psi(x),\overline{\psi}(x))+ L_{I}(A(x), \psi(x), \overline{\psi}(x))$,
является инвариантной при обобщенных калибровочных преобразованиях второго рода, представляя собой лагранжиан традиционной классической электродинамики  \cite{BSh} с традиционной калибровочной группой $ \mathcal G$, то рассматриваемая вариация принимает вид
$$
\overline{\delta}S= \int_{V}\,\overline{\delta}L_{0}(G(x))\,dx= \int_{V}\,\{\beta G(x)\cdot\cdot\,\,\widetilde{\Box}f(x)\}\,dx, \eqno\ldots (2.10)
$$
где $\widetilde{\Box}$ -- тензорный дифференциальный оператор Даламбера, определенный соотношением
$$
\widetilde{\Box}= - \partial \otimes \partial.   \eqno\ldots (2.11)
$$

Из (2.10) следует, что при произвольной области интегрирования  вариация $\overline{\delta}S$ обращается в нуль при следующем условии на калибровочную функцию $f(x)$
$$
 \widetilde{\Box}f(x)= \widetilde{0}.   \eqno\ldots (2.12)
$$

Таким образом, рассматриваемое действие калибровочно инвариантно на группе обобщенных градиентных преобразований второго рода    $\mathcal G_{c}: \{f(x)= C_{\mu}x^{\mu}+ C\}$, в которой подгруппа $R_{c} : \{f(x)= C_{\mu}x^{\mu}\}$, представляет собой известную группу $R$-калибровочных преобразований полевых функций и связанную с ней  $R$-инвариантность  \cite{Br}.

При этом следует отметить, что данная группа калибровочной инвариантности рассматриваемого действия соответствует группе калибровочной инвариантности наблюдаемых полевых переменных, а таковыми теперь являются полевые переменные, определяемые не только тензором электромагнитного поля, но и тензором электроджейтонного поля.

Следуя далее вышеуказанной общей схеме находим, что на группе $\mathcal G_{c}$ и на экстремалях выполняется локальный закон сохранения (2.5), принимающий, в данном случае, вид
$$
 \partial\!\cdot\!\{- (\alpha F(x)+ \beta G(x))\cdot \partial f(x)+j(x)f(x)  \}= 0,  \quad f(x)\in \mathcal G_{c}.       \eqno\ldots (2.13)
$$

В свою очередь, (2.13) можно переписать
$$
\{-\alpha\partial\!\cdot\! F(x)-\beta\partial\!\cdot\! G(x)+ j(x)\}\cdot\partial f(x)+ \beta G(x)\cdot\cdot\,\, \widetilde{\Box}f(x)+ \partial\!\cdot\! j(x) f(x)= 0.         \eqno\ldots (2.14)
$$

Так как в уравнении (2.14) калибровочная функция $f(x)$ принадлежит  $\mathcal G_{c}$, то в этом уравнении, в силу (2.12), выполняется соотношение
$$
\beta\, G(x)\cdot\cdot\,\, \widetilde{\Box}f(x)= 0.       \eqno\ldots (2.15)
$$

Таким образом, калибровочная инвариантность, которой обладает рассматриваемое действие, определяемое лагранжианом (2.6), не требует обращения $\beta$ в нуль и, тем самым, не требует исключения тензора электроджейтонного поля из лагранжиана $L(x)$.

Последующий учет  (2.15)  представляет закон сохранения (2.14) в виде
$$
\{-\alpha\, \partial\!\cdot\! F(x)-\beta\, \partial\!\cdot\! G(x)+ j(x)\}\!\cdot\!\partial f(x)+ \partial \!\cdot\! j(x)f(x)= 0.     \eqno\ldots (2.16)
$$

В силу того, что (2.16) выполняется на экстремалях, выражение в фигурных скобках левой части этого уравнения равно нулевому 4-вектору, в результате чего данное уравнение принимает вид соотношения
$$
 \partial \!\cdot\! j(x)f(x)= 0,    \quad f(x)\in \mathcal G_{c},       \eqno\ldots (2.17)
$$
которое, в силу соответствующей произвольности калибровочной функции $f(x)$, при любых фиксированных значениях полевых переменных входящих в (2.17), приводит к дифференциальному уравнению закона сохранения электрического заряда, $\partial \!\cdot\! j(x)= 0$.

В итоге, вышеуказанная калибровочная инвариантность второго рода, которой обладают и лагранжиан, и рассматриваемое  действие рассматриваемой полевой системы, также приводит к уравнению непрерывности для плотности электрического тока.

В заключение данного раздела вновь следует подчеркнуть, что частный случай электродинамического лагранжиана (2.6),  соответствующий выбору коэффициентов $\alpha=\beta=1$ и  последующему использованию инвариантного соотношения
$$
\partial A(x)^{2}= F(x)^{2}+ G(x)^{2},        \eqno\ldots (2.18)
$$
представлен хорошо известным выражением \cite{Schw}
$$
L(x)= - \frac{1}{2}\, \partial A(x)^{2}+ L_{0\psi}(x)+ L_{I}(x),      \eqno\ldots (2.19)
$$
однако которое, в силу (2.18), теперь выступает в качестве лагранжиевой плотности для взаимодействия поля Дирака с электромагнитным и электроджейтонным полями, а 4-потенциал $A(x)$ входящий в него теперь является электромагнитоджейтонным потенциалом.

В связи с этим, (2.19) предпочтительнее представлять в виде
\begin{multline*}
L(x)= -\frac{1}{2}\, F(x)^{2}-\frac{1}{2}\, G(x)^{2}+ \frac{i}{2}\, (\overline{\psi}(x)\hat{\gamma}\cdot\partial\psi(x)  - \partial\overline{\psi}(x)\cdot\hat{\gamma}\psi(x)) - m_{\psi}\overline{\psi}(x)\psi(x)\\ +  e\overline{\psi}(x)\hat{\gamma}\psi(x)\cdot A(x),       \quad\ldots (2.20)
\end{multline*}
явно отражая факт равноправного существования электромагнитного и электроджейтонного полей.

\section*{\S \,3 Полевые уравнения}
 \addcontentsline{toc}{section}{\S \,3  Полевые уравнения}

\qquad Возвращаясь к исходной системе полевых уравнений (1.15)--(1.17), проведем, как и в случае свободного поля, лоренц-инвариантное разложение $A(x)$ согласно (I.4.1) и найдем уравнения движения для составляющих $\,^{\tau}A(x)$ и $\,^{\ell}A(x)$ в рамках рассматриваемого взаимодействия.

Для этого, используя (1.6)--(1.8), представим (1.15) в виде
$$
\alpha\, \partial\!\cdot\! F(x)+ \beta\, \partial\!\cdot\! G(x)+ (\gamma - \beta)\, \partial\!\cdot\! G_{\ast}(x)+ m^{2}A(x)= j(x).       \eqno\ldots (3.1)
$$

(3.1), в свою очередь, после использования соотношений
$$
\partial\!\cdot\! F(x)= - \frac{1}{2}\,\,  \Box  \, ^{\tau}\!A(x),    \eqno\ldots (3.2)
$$
$$
\partial \!\cdot\! G(x)=  - \frac{1}{2}\,\,  \Box \,  ^{\tau}\! A(x) - \Box \,^{\ell}\! A(x),     \eqno\ldots (3.3)
$$
$$
\partial\!\cdot\! G_{\ast}(x)= \frac{1}{4}\,\,  \partial \,\partial \!\cdot\! A(x),         \eqno\ldots (3.4)
$$
приводит к уравнению
$$
- \frac{\alpha+\beta}{2}\,\, \Box \,^{\tau}\!A(x) - \beta \,\,\Box \,^{\ell}\!A(x)+\frac{\gamma-\beta}{4}\partial \,\partial \cdot\! A(x)+ m^{2}\,^{\tau}\!A(x)+ m^{2}\,^{\ell}\!A(x) = \jmath (x),                  \eqno\ldots (3.5)
$$
которое, после учета равенства $\partial \, \partial\!\cdot\! A(x)= -\, \Box\,^{\ell}\!A(x)$, принимает вид
$$
- \frac{\alpha+\beta}{2}\,\,\{(\Box - {m_{1}}^{2})\,^{\tau}\!A(x)+ \frac{{m_{1}}^{2}}{{m_{2}}^{2}}\,(\Box - {m_{2}}^{2})\,^{\ell}\!A(x)\}= \jmath (x),                      \eqno\ldots (3.6)
$$
где $ {m_{1}}^{2} $ и    $ {m_{2}}^{2} $ по-прежнему определены согласно (I.2.23) и    (I.2.24), то-есть,
$$
{ m_{1}}^{2}= \frac{2m^{2}}{\alpha+ \beta},             \eqno\ldots (3.7)
$$
а ${m_{2}}^{2}$ представлена соотношением (1.26).

В свою очередь, внутреннее умножение тензорного оператора Даламбера (2.11) на обе части уравнения (3.6) с последующим учетом равенств
$$
\widetilde{\Box}\cdot\, ^{\tau}\!A(x)= \theta = \widetilde{\Box}\cdot \jmath(x),             \eqno\ldots (3.8)
$$
приводит, при $ 3\beta + \gamma \neq 0 $, к уравнению
$$
(\Box -  {m_{2}}^{2})\,\widetilde{\Box}\cdot\, ^{\ell}\!A(x)= \theta.                        \eqno\ldots (3.9)
$$

С другой стороны, в силу соотношений
$$
 ^{\ell}\!A(x)= - D_{0}^{c}(x- y)\,\widetilde{\Box}\cdot\, ^{\ell}\!A(y),          \eqno\ldots (3.10)
$$
$$
\Box \,^{\ell}\!A(x)= - \Box_{x}D_{0}^{c}(x- y)\,\widetilde{\Box}\cdot\, ^{\ell}\!A(y)= - D_{0}^{c}(x- y)\,\Box_{y}\widetilde{\Box}\cdot\, ^{\ell}\!A(y),                            \eqno\ldots (3.11)
$$
имеем
$$
(\Box - m_{2}^{2})\,^{\ell}\!A(x)= - D_{0}^{c}(x- y)\,\{(\Box -{ m_{2}}^{2})\,\widetilde{\Box}\cdot\, ^{\ell}\!A(y)\}.        \eqno\ldots (3.12)
$$

Последующее использование (3.9) в (3.12) приводит к искомому уравнению для $ ^{\ell}A(x)$,
$$
(\Box - {m_{2}}^{2})\,^{\ell}\!A(x)= \theta.                            \eqno\ldots (3.13)
$$

В свою очередь, (3.6) после учета  (3.13) приводит к уравнению для поперечной части рассматриваемого векторного поля, $^{\tau}A(x)$ , которое, при прежнем  условии (II.1.29), имеет вид
$$
(\Box - {m_{1}}^{2})\,^{\tau}\!A(x)= - \jmath(x).                      \eqno\ldots (3.14)
$$

(3.13) демонстрирует, что 4-продольная составляющая векторного поля $A(x)$, приводящая, в частности, к индефинитности энергии, продолжает описываться уравнением движения свободного поля и при наличии рассматриваемого взаимодействия.

При этом легко видеть, что $^{\ell}A(x)$ не принимает участия в данном взаимодействии и согласно уравнений движения для спинорного поля.

Действительно, после подстановки (I.4.1) в (1.3) лагранжиан взаимодействия представляется в виде
$$
L_{I}(x)= - \jmath(x)\cdot\, ^{\tau}\!A(x) - \jmath(x)\cdot\, ^{\ell}\!A(x),                   \eqno\ldots (3.15)
$$
а второе скалярное произведение правой части (3.15), в силу уравнения непрерывности для плотности тока  $\jmath(x)$, сводится к несущественной, для уравнений движения, 4-дивергенции, то есть,
$$
 \jmath(x)\!\cdot\!\, ^{\ell}\!A(x)= \partial\!\cdot\!\{\jmath(x)D_{0}^{c}(x- y)\,\partial \!\cdot\! A(y)\}.        \eqno\ldots (3.16)
$$

Следовательно, уравнения движения (1.16) и (1.17) представляются, соответственно,
$$
  i\hat{\gamma}\cdot\partial\psi(x)- m_{\psi}\psi(x)= -e\hat{\gamma}\psi(x)\cdot \,^{\tau}\! A(x),                  \eqno\ldots (3.17)
$$
$$
 i\partial\overline{\psi}(x)\cdot\hat{\gamma}+ m_{\psi}\overline{\psi}(x)= e\hat{\gamma}\overline{\psi}(x)\cdot\,^{\tau}\! A(x). \quad \eqno\ldots (3.18)
$$

(3.17) и (3.18) совместно с (3.13) демонстрируют, что 4-продольная составляющая рассматриваемого векторного поля, $ ^{\ell}\!A(x)$, является, в рассматриваемом случае взаимодействия, "полностью"  \,свободным полем.

Поэтому, динамические переменные рассматриваемой полевой системы определяемые этой составляющей являются сохраняющимися, при данном взаимодействии, величинами.

В частности, часть общей энергии, обусловленная полем $ ^{\ell}\!A(x)$, является не меняющейся в этом случае величиной, то есть, представляет собой аддитивную константу не влияющую на результат физического процесса.

В итоге, нет необходимости ни в наложении соответствующих дополнительных условий на $ ^{\ell}\!A(x)$, ни в переопределении энергии векторного поля.

Более того, с учетом замечания в конце главы I, не является очевидной и сама последовательность такого переопределения энергии в общем случае.

Наконец, следуя \cite{Feld} можно показать, что все $S$-матричные элементы, включающие $b$-кванты как внешние частицы, исчезают тождественно в силу сохранения тока, без каких-либо дополнительных условий на пространство амплитуд состояний рассматриваемой системы.

С другой стороны, так же в силу сохранения тока, члены пропагатора содержащие $ k_{m}k_{n},$ не дают вклада в элементы матрицы рассеяния, однако, этот факт нельзя считать основанием для предварительного исключения этих членов из пропагатора, напротив, следует рассматривать данный факт как основание для сохранения этих членов вплоть до конечного этапа теории.

Далее вновь возвращаемся к наиболее важному частному случаю теории, когда векторное поле представлено электродинамическим потенциалом $A(x)$, а сама обсуждаемая теория выступает как теория электромагнитного и электроджейтонного полей.

Этот случай отличается прежде всего тем, что в классическом варианте теории на первое место выступают уже сами производные тензоры (производные аффиноры) 4-потенциала электродинамики, в связи с чем, далее уделяется главное внимание полевым уравнениям для этих полевых переменных.

В $\xi$-формализме электродинамики 4-скаляр $\partial \! \cdot \! A(x)$  рассматривают как вспомогательную величину определяющую фиктивный ток \cite{BLOT}.

В рассматриваемой теории эта величина содержится в тензоре $G_{\ast}(x)$, вследствие чего фиктивные токи будут пропорциональными 4-дивергенции $ \partial\!\cdot\! G_{\ast}(x) $.

Уравнение (1.18), в рамках данного формализма, принимает вид
$$
\alpha\, \partial\!\cdot\! F(x)+ \beta\, \partial\!\cdot\! G_{d}(x)= j(x)+ j_{f}(x) .    \eqno\ldots (3.19)
$$
где фиктивный ток $ j_{f}(x)$ определен соотношением
$$
 j_{f}(x)= (\beta - \gamma )\,\partial\!\cdot \!G_{\ast}(x).                     \eqno\ldots (3.20)
$$

В свою очередь, уравнение (3.19), после использования соотношения
$$
\partial\!\cdot\! G(x)= \partial\!\cdot\! F(x)+ 4\, \partial\!\cdot\! G_{\ast}(x),             \eqno\ldots (3.21)
$$
можно переписать в виде
$$
(\alpha + \beta)\, \partial\!\cdot\! F(x)= j(x)+ j_{fF}(x),                        \eqno\ldots (3.22)
$$
в котором фиктивный ток $ j_{fF}(x)$ представлен
$$
 j_{fF}(x)= - (3\beta + \gamma)\,\partial\!\cdot\! G_{\ast}(x).               \eqno\ldots (3.23)
$$

Уравнение (3.22), при прежнем условии (II.1.29), приводит к модифицированному уравнению второй пары системы уравнений Максвелла в $\xi -$ формализме электродинамики \cite{BLOT},
$$
\partial\!\cdot\! F_{L}(x)=  j(x)+ j_{fF}(x).              \eqno (3.24)
$$

Фиктивный ток (3.23), после учета соотношения (3.4), принимает вид
$$
 j_{fF}(x)= - \frac{3\beta+ \gamma}{4}\partial\,\partial \!\cdot\! A(x),        \eqno\ldots (3.25)
$$
демонстрирующий, что параметр определяющий калибровку, в данном случае, представлен выражением
$$
\xi_{F}= \frac{4}{3\beta+\gamma}.                \eqno\ldots (3.26)
$$

В свою очередь, (3.24), при стандартном дополнительном условии классической электродинамики, условии Лоренца, приводит к общеизвестному немодифицированному уравнению второй пары системы уравнений для электромагнитного поля (1.19).

Таким образом, уравнение (1.19), как и его $\xi$-модификация (3.24), выступают в качестве следствий уравнения (1.18) рассматриваемой теории.

Однако, с таким же успехом, можно использовать соотношение (3.21) и для получения другого следствия уравнения (1.18).

Действительно, (3.21) позволяет представить уравнение (3.19) так же и в виде
$$
(\alpha + \beta)\, \partial\!\cdot\! G(x)= j(x)+ j_{fG}(x),         \eqno\ldots (3.27)
$$
где фиктивный ток $ j_{fG}(x)$ теперь определен соотношением
$$
 j_{fG}(x)=  (4\alpha + \beta - \gamma)\,\partial\!\cdot\! G_{\ast}(x).          \eqno\ldots (3.28)
$$

В свою очередь, (3.27), при том же условии (II.1.29), приводит к модифицированному уравнению второй пары системы уравнений для электроджейтонного поля в рассматриваемом $\xi$-формализме электродинамики,
$$
 \partial\!\cdot\! G_{L}(x)= j(x)+ j_{fG}(x),         \eqno\ldots (3.29)
$$
где фиктивный ток $ j_{fG}(x)$ теперь представлен
$$
 j_{fG}(x)= - \frac{3\beta+\gamma - 8}{4}\partial\,\partial \!\cdot\! A(x),        \eqno\ldots (3.30)
$$
а параметр определяющий калибровку теперь имеет вид
$$
\xi_{G}= \frac{4}{3\beta+\gamma - 8}\, .                \eqno\ldots (3.31)
$$

(3.29), при прежнем традиционном дополнительном условии классической электродинамики, приводит к немодифицированному уравнению второй пары системы уравнений для электроджейтонноо поля,
$$
\partial\!\cdot \!G_{L}(x)= j(x).              \eqno\ldots (3.32)
$$

Таким образом, в квантовом варианте теории имеем, что "в физическом гильбертовом пространстве квантовой электродинамики выполняется"\,\cite{BLOT}  {\em не  только\/} уравнение Максвелла
$$
\partial \!\cdot\! \mathfrak{F} _{L}(x)=   \mathfrak{J} (x),         \eqno\ldots (3.33)
$$
но и уравнение
$$
\partial\! \cdot\! \mathfrak{G}_{L}(x)=   \mathfrak{J} (x),         \eqno\ldots (3.34)
$$
где $ \mathfrak{F} _{L}(x) $, $\mathfrak{G}_{L}(x)$ и $\mathfrak{J} (x) $ -- соответствующие операторные обобщенные функции.

При этом, в этом пространстве состояний по-прежнему имеем соотношение $\partial \!\cdot \!\mathfrak{F} _{L}(x) = - \Box\, \mathfrak{A}(x) =  \partial \!\cdot \!\mathfrak{G}_{L}(x) $, означающее,что (3.33) и (3.34) эквивалентны друг другу как уравнения определяющие свойства оператора   $\mathfrak{A}(x)$, но не эквивалентны как уравнения на операторы $\mathfrak{F}_{L}(x)$ и $\mathfrak{G}_{L}(x)$ -- каждое из этих уравнений представляет свои соотношения соответствующих полевых переменных.

Классические уравнения (1.19) и (3.32), как уравнения на 4-потенциал $A(x)$, представляют собой, в рамках все того же дополнительного условия классической электродинамики, одно и то же уравнение, являющееся стандартным неоднородным уравнением Даламбера
$$
\Box \, A(x)= - j(x).                               \eqno\ldots (3.35)
$$

Но это означает лишь только то, что решения этого уравнения, $ A(x)$, являются функциями, подстановка которых в (1.5)--(1.8) позволяет определить тензоры напряженностей как электромагнитного, так и электроджейтонного полей.

При этом вновь следует подчеркнуть, что уравнения (1.19) и (3.32), в классическом электродинамическом случае, когда на первое место выступают $ \mathit F$-  и $ \mathit G$-  поля, не являются эквивалентными друг другу -- каждое из этих уравнений определяет свои свойства своих классических полевых переменных.

Наглядно демонстрируют данное различие данных уравнений их представления в трехмерном евклидовом пространстве $R^{3}$.

Действительно, если (1.19) в этом случае имеет хорошо известный вид
$$
\partial ^{0}\vec{E}_{F_{L}}(x) + \vec{\triangledown}\cdot \mathcal{F}_{L}(x)= - \vec{j}(x),
$$
$$
 div\, \vec{E}_{F_{L}}(x)= j^{0}(x),                                   \eqno\ldots (3.36)
$$
$$
 \vec{E}_{F_{L}}(x)=  \vec{E}_{A}(x) + \vec{E}_{\varphi}(x),
$$
то (3.32) представлено в  $R^{3}$ следующей системой уравнений
$$
\partial^{0}\vec{E}_{G_{L}}(x) + \vec{\triangledown}\cdot \mathcal{G}_{L}(x)= - \vec{j}(x),
$$
$$
 div\, \vec{E}_{G_{L}}(x)= -  j^{0}(x) +\partial^{0}G_{L}^{00}(x),                                   \eqno\ldots (3.37)
$$
$$
\vec{E}_{G_{L}}(x)=  \vec{E}_{A}(x) - \vec{E}_{\varphi}(x).
$$

$ \mathcal{F}_{L}(x) $ и  $ \mathcal{G}_{L}(x)$, в (3.36) и (3.37), -- лоренцевы 3-тензоры магнитного и джейтонного полей, определенные соотношениями
$$
\mathcal{F}_{L}(x) = 2\, \tilde{rot}\, \vec{A}(x),                                      \eqno\ldots (3.38)
$$
$$
 \mathcal{G}_{L}(x) =  2\, \tilde{def}\, \vec{A}(x),                                      \eqno\ldots (3.39)
$$
в которых $\tilde{rot}\,$  и   $\tilde{def}\,$ -- 3-тензорные дифференциальные операторы, представленные равенствами \cite{Pob}
$$
\tilde{rot} =  \vec{\triangledown}\, \overset{a}{\otimes},                                              \eqno\ldots (3.40)
$$
$$
 \tilde{def} = \vec{\triangledown} \overset{s}{\otimes}.                                              \eqno\ldots (3.41)
$$

Система уравнений (3.37), прежде всего, демонстрирует соответствующие {\em математические } отличительные свойства компонент тензора электроджейтонного поля, $ \mathit G_{L}(x) $, вследствие чего эта система не является гипотетической уже в {\em математическом }  отношении.

Дальнейшие комментарии к данным системам уравнений идентичны комментариям к системам уравнений (I.2.15), (I.2.16) и тут не приводятся.

 \section*{\S \,4 Уравнения классической электродинамики при наличии гравитационного поля}
  \addcontentsline{toc}{section}{\S \,4  Уравнения классической электродинамики при наличии гравитационного поля}

\qquad И так, в псевдоевклидовом пространстве событий уравнения вторых пар электродинамических полевых уравнений имеют вид
$$
\partial\!\cdot\! F_{L}(x)= j(x),
$$
$$
\partial\!\cdot \!G_{L}(x)= j(x).
$$

Осуществляя традиционную замену обычных производных, присутствующих в этих уравнениях, на абсолютные производные в пространстве аффинной связности $L_{n}$  \cite{Rash}, получаем
$$
\bigtriangledown \cdot \tilde{F}_{L}(x) = J(x),                              \eqno\ldots (4.1)
$$
$$
\bigtriangledown \cdot \tilde{G}_{L}(x) = J(x),                              \eqno\ldots (4.2)
$$
где
$$
\tilde{F}_{L}(x) = 2\, Rot\, A(x),                        \eqno\ldots (4.3)
$$
$$
\tilde{G}_{L}(x) = 2\, Def\, A(x),                        \eqno\ldots (4.4)
$$
в которых, $ Rot $ и $ Def $ -- тензорные дифференциальные операторы пространства  $L_{n}$, представленные соотношениями
$$
Rot = \bigtriangledown \, \overset{a}{\otimes} ,                           \eqno\ldots (4.5)
$$
$$
Def = \bigtriangledown \overset{s}{\otimes} .                              \eqno\ldots (4.6)
$$

Плотность 4-тока $  \mathit J(x) $, в (4.1) и (4.2), определена соответствующим обобщением 4-вектора плотности тока в псевдоевклидовом пространстве \cite{La}.

Далее, временно перейдем от тензорного языка к традиционному языку тензорного исчисления \cite{Gil}.

Так как в $L_{n}$ ковариантные компоненты абсолютного градиента векторного поля $ \mathit A (x)$ представлены соотношениями \cite{Rash}
$$
\bigtriangledown_{m}A_{n}(x) = \partial_{m}A_{n}(x) - \mathit \Gamma^{k}_{\,mn}(x)A_{k}(x),        \eqno\ldots (4.7)
$$
то соответствующие компоненты тензоров  $ \tilde{ \mathit F}(x) $  и    $  \tilde{ \mathit G}(x) $, в пространстве аффинной связности    $ \mathit L_{n}$, принимают вид
$$
\tilde{ \mathit F}_{mn}(x) = \mathit F_{mn}(x) - \mathit \Gamma^{k}_{\,[mn]}(x)A_{k}(x),        \eqno\ldots (4.8)
$$
$$
\tilde{ \mathit G}_{mn}(x) = \mathit G_{mn}(x) - \mathit \Gamma^{k}_{\,(mn)}(x)A_{k}(x).        \eqno\ldots (4.9)
$$

Последующее использование соотношения
$$
\mathit \Gamma^{k}_{\,mn}(x) = \frac{1}{2}\,\mathit C^{k}_{\,mn}(x) +\tilde{ \mathit \Gamma}^{k}_{\,mn}(x),   \eqno\ldots (4.10)
$$
в котором $\mathit C^{k}_{\,mn}(x) =\mathit \Gamma^{k}_{\,[mn]}(x) $ -- компоненты тензора картанова кручения пространства аффинной связности ($ \mathit C $ -- кручения ), а  $\tilde{ \mathit \Gamma}^{k}_{\,mn}(x) =\mathit \Gamma^{k}_{\,(mn)}(x) $, позволяет переписать (4.8) и (4.9) в виде
$$
\tilde{ \mathit F}_{mn}(x) = \mathit F_{mn}(x) - \frac{1}{2}\,\mathit C^{k}_{\,mn}(x)A_{k}(x),        \eqno\ldots (4.11)
$$
$$
\tilde{ \mathit G}_{mn}(x) = \mathit G_{mn}(x) - \tilde{ \mathit \Gamma}^{k}_{\,mn}(x)A_{k}(x).        \eqno\ldots (4.12)
$$

(4.11) и (4.12) демонстрируют, что тензор $ \tilde{ \mathit F}(x) $ аддитивно "реагирует"\,  на  {\em кручение } пространства  $ \mathit L_{n}$, а тензор  $  \tilde{ \mathit G}(x) $ -- на {\em симметрическую } связность данного пространства.

Таким образом, имеем {\em принципиальное }  различие электромагнитного и электроджейтонного полей и в данном отношении.

В свою очередь, используя аппарат абсолютного дифференцирования в пространстве аффинной связности, легко находим соотношения для контравариантных компонент абсолютной дивергенции тензора  $ \tilde{ \mathit F}(x) $  в данном пространстве
$$
\bigtriangledown_{m}\tilde{ \mathit F}^{mn}(x) = \partial_{m}\tilde{ \mathit F}^{mn}(x) + \mathit \Gamma^{k}_{\,km}(x)\tilde{ \mathit F}^{mn}(x) +  \mathit \Gamma^{n}_{\,kl}(x)\tilde{ \mathit F}^{kl}(x).          \eqno\ldots (4.13)
$$

Последующее использование (4.10) представляет (4.13) в виде
\begin{multline*}
\bigtriangledown_{m}\tilde{ \mathit F}^{mn}(x) = \partial_{m}\tilde{ \mathit F}^{mn}(x) +\frac{1}{2}\, \mathit C^{k}_{\,kl}(x)\tilde{ \mathit F}^{ln}(x) +  \tilde{ \mathit \Gamma}^{k}_{\,kl}(x)\tilde{ \mathit F}^{ln}(x) \\+ \frac{1}{2}\, \mathit C^{n}_{\,kl}(x)\tilde{ \mathit F}^{kl}(x) + \tilde{ \mathit \Gamma}^{n}_{\,kl}(x)\tilde{ \mathit F}^{kl}(x).       \quad\ldots (4.14)
\end{multline*}

В свою очередь, определив $ \textsl{g}_{mn}(x)$ как величину, удовлетворяющую соотношению
$$
\tilde{ \mathit \Gamma}^{k}_{\,mn}(x)= \frac{1}{2}\,\textsl{g}^{kl}(x)(\partial_{n}g_{lm}(x)+ \partial_{m}\textsl{g}_{ln}(x) - \partial_{l}\textsl{g}_{mn}(x) ),               \eqno\ldots (4.15)
$$
приходим к известному представлению для ковариантных компонент следа симметрической части связности пространства  $ \mathit L_{n}$,
$$
\tilde{ \mathit \Gamma}^{k}_{\,kl}(x) = \frac{1}{\sqrt{-\textsl{g}(x)}}\,\partial_{l}\,\sqrt{-\textsl{g}(x)},                 \eqno\ldots (4.16)
$$
которое позволяет представить сумму первого и третьего слагаемых правой части (4.14) в известной форме, то есть,
$$
\partial_{m}\tilde{ \mathit F}^{mn}(x) + \tilde{ \mathit \Gamma}^{k}_{\,kl}(x)\tilde{ \mathit F}^{ln}(x)=  \frac{1}{\sqrt{-\textsl{g}(x)}}\,\partial_{m}(\sqrt{-\textsl{g}(x)}\, \tilde{ \mathit F}^{mn}(x)).       \eqno\ldots (4.17)
$$

Последующее использование (4.17) в (4.14), с учетом обращения в нуль последнего слагаемого правой части (4.14) и с последующим возвратом к лоренцевым полевым переменным, приводит к следующей ковариантной форме уравнения (4.1)
$$
\frac{1}{\sqrt{-\textsl{g}(x)}}\,\partial_{m}(\sqrt{-\textsl{g}(x)}\, \tilde{ \mathit F}_{L}^{mn}(x)) + \frac{1}{2}\, (\mathit C^{k}_{\,kl}(x)\tilde{ \mathit F}_{L}^{ln}(x) + \mathit C^{n}_{\,kl}(x)\tilde{ \mathit F}_{L}^{kl}(x) ) =  J^{n}(x).               \eqno\ldots (4.18)
$$

Уравнение (4.18) является модификацией традиционного обобщения уравнения второй пары системы уравнений Максвелла в присутствии гравитационного поля \cite{La}.

Определяя возможную физическую роль второго члена левой части уравнения (4.18), обозначаемого ниже как $ \mathit K_{F}^{n}(x)$, рассмотрим два варианта представления этого уравнения.

После введения величины ${m_{F}}^{2}$, определяемой соотношением
$$
K_{F}(x) ={ m_{F}}^{2}A(x),                        \eqno\ldots (4.19)
$$
имеющим решение, при $ A(x)^{2} \neq 0$,
$$
 {m_{F}}^{2}= \frac{K_{F}(x)\cdot A(x)}{A(x)^{2}},                      \eqno\ldots (4.20)
$$
инвариантная форма уравнения (4.18) принимает вид неоднородного уравнения Прока,
$$
\frac{1}{\sqrt{-\textsl{g}(x)}}\,\partial \!\cdot\! (\sqrt{-\textsl{g}(x)}\, \tilde{ \mathit F}_{L}(x)) + {m_{F}}^{2}A(x) = J(x).               \eqno\ldots (4.21)
$$

Таким образом, в результате взаимодействия электромагнитного поля с кручением пространства $ \mathit L_{n}$, электромагнитное поле приобретает "массу"\,, определяемую соотношением (4.20).

Инвариантная форма уравнения (4.18) может быть представлена так же в виде
$$
\frac{1}{\sqrt{-\textsl{g}(x)}}\,\partial \!\cdot\! (\sqrt{-\textsl{g}(x)}\, \tilde{ \mathit F}_{L}(x))=  J(x) + J_{F}(x),       \eqno\ldots (4.22)
$$
где $J_{F}(x)= -K_{F}(x) $  -- картанов  4-ток  смещения,  обусловленный наличием картанова кручения пространства  $ \mathit L_{n}$, в котором рассматривается электромагнитное поле.

Наличие материальных тел, в общем случае, не только искривляет пространство событий, но и закручивает его, в связи с этим, уравнение (4.18) не представляется гипотетическим, вне зависимости от малости соответствующих эффектов обусловленных кручением.

Следуя \cite{La}, уместно представить и соответствующее обобщение уравнения движения электрически заряженной частицы  в гравитационном и электромагнитном полях на случай  пространства $ \mathit L_{n}$,
$$
m_{0}c\, \left(\frac{du^{n}(x)}{ds} + \tilde{ \mathit \Gamma}^{n}_{\,kl}(x)u^{k}(x)u^{l}(x)\right) = \frac{e}{c}\,F_{L\, k}^{n}(x) u^{k}(x) -  \frac{e}{c}\,C^{k,nl}(x)A_{k}(x)u_{l}(x).                \eqno\ldots (4.23)
$$

Таким образом, кручение пространства $ \mathit L_{n}$ приводит не только к дополнительному полевому току смещения в уравнении вторых пар системы уравнений Максвелла, но и к дополнительной силе, действующей на электрически заряженную частицу, представленной вторым слагаемым правой части (4.23).

Переходя  к уравнению (4.2), вместо (4.13)  теперь имеем
$$
\bigtriangledown_{m}\tilde{ \mathit G}^{mn}(x) = \partial_{m}\tilde{ \mathit G}^{mn}(x) + \mathit \Gamma^{k}_{\,km}(x)\tilde{ \mathit G}^{mn}(x) +  \mathit \Gamma^{n}_{\,kl}(x)\tilde{ \mathit G}^{kl}(x).          \eqno\ldots (4.24)
$$

Последующее использование (4.10) представляет (4.24) в виде
\begin{multline*}
\bigtriangledown_{m}\tilde{ \mathit G}^{mn}(x) = \partial_{m}\tilde{ \mathit G}^{mn}(x) +\frac{1}{2}\, \mathit C^{k}_{\,kl}(x)\tilde{ \mathit G}^{ln}(x) +  \tilde{ \mathit \Gamma}^{k}_{\,kl}(x)\tilde{ \mathit G}^{ln}(x) + \\ \frac{1}{2}\, \mathit C^{n}_{\,kl}(x)\tilde{ \mathit G}^{kl}(x) + \tilde{ \mathit \Gamma}^{n}_{\,kl}(x)\tilde{ \mathit G}^{kl}(x).       \quad\ldots (4.25)
\end{multline*}

В свою очередь, сумма первого и третьего членов правой части (4.25), после использования (4.16), представляется
$$
\partial_{m}\tilde{ \mathit G}^{mn}(x) + \tilde{ \mathit \Gamma}^{k}_{\,kl}(x)\tilde{ \mathit G}^{ln}(x)=  \frac{1}{\sqrt{-\textsl{g}(x)}}\,\partial_{m}(\sqrt{-\textsl{g}(x)}\, \tilde{ \mathit G}^{mn}(x)).       \eqno\ldots (4.26)
$$

Последующее использование (4.26) в (4.25), с учетом обращения в $\theta$  четвертого слагаемого правой части (4.25) и с последующим возвратом к лоренцевым полевым переменным, приводит к следующей ковариантной форме уравнения (4.2)
$$
\frac{1}{\sqrt{-\textsl{g}(x)}}\,\partial_{m}(\sqrt{-\textsl{g}(x)}\, \tilde{ \mathit G}_{L}^{mn}(x)) + \frac{1}{2}\, \mathit C^{k}_{\,kl}(x)\tilde{ \mathit G}_{L}^{ln}(x) + \tilde{ \mathit \Gamma}^{n}_{\,kl}(x)\tilde{ \mathit G}_{L}^{kl}(x)  =  J^{n}(x).               \eqno\ldots (4.27)
$$

Как и ранее, обозначив бездивергентную составляющую левой части уравнения (4.27) как $ K_{G}^{n}(x)$, рассмотрим два варианта представления этого уравнения.

После введения величины ${m_{G}}^{2}$, определяемой соотношением
$$
K_{G}(x) = {m_{G}}^{2}A(x),                        \eqno\ldots (4.28)
$$
имеющим решение, при $ A(x)^{2} \neq 0$,
$$
 {m_{G}}^{2}= \frac{K_{G}(x)\cdot A(x)}{A(x)^{2}},                      \eqno\ldots (4.29)
$$
инвариантная форма уравнения (4.27)  принимает вид
$$
\frac{1}{\sqrt{-\textsl{g}(x)}}\,\partial\! \cdot\! (\sqrt{-\textsl{g}(x)}\, \tilde{ \mathit G}_{L}(x)) + {m_{G}}^{2}A(x) = J(x).               \eqno\ldots (4.30)
$$

Таким образом, в результате взаимодействия электроджейтонного поля с кручением и с симметрической составляющей связности пространства $ \mathit L_{n}$, электроджейтонное поле приобретает  "массу"\,, определяемую соотношением (4.29).

Инвариантная форма уравнения (4.27) может быть представлена так же и  в виде
$$
\frac{1}{\sqrt{-\textsl{g}(x)}}\,\partial \!\cdot\! (\sqrt{-\textsl{g}(x)}\, \tilde{ \mathit G}_{L}(x))=  J(x) + J_{G}(x),       \eqno\ldots (4.31)
$$
где $J_{G}(x)= -K_{G}(x) $  -- 4-ток смещения, обусловленный наличием картанова кручения и симметрической части связности пространства $ \mathit L_{n}$, в котором рассматривается электроджейтонное поле.

Сравнивая ${m_{F}}^{2}$ и ${m_{G}}^{2}$, при $\mathit C^{k}_{\,kl}(x) = 0 $, находим, что первая "масса" \,обусловлена взаимодействием электромагнитного поля {\em  только с кручением} пространства $ \mathit L_{n}$, а вторая "масса"\, определяется взаимодействием электроджейтонного поля {\em только с симметрической }частью связности этого пространства.

Таким образом, электромагнитное и электроджейтонное поля, в пространстве $ \mathit L_{n}$, обладают разными, по своей природе, "массами"\,, что вновь демонстрирует очередное  принципиальное различие этих полей.

Теперь рассмотрим традиционное псевдориманово пространство, отождествляемое с пространством событий стандартной общей теорией относительности, интерпретируя все последующие уравнения как уравнения в традиционном гравитационном поле.

То есть, теперь стартуем с риманова пространства $V_{n}$  \cite{Rash}, в котором метрический тензор взаимно  однозначно определяет связность этого пространства, $\mathit \Gamma^{k}_{\,mn}(x)$, представленную соответствующими символами Кристоффеля.

В таком пространстве событий, вместо (4.11), (4.12) теперь имеем соотношения
$$
\tilde{ \mathit F}_{mn}(x) = \mathit F_{mn}(x) ,        \eqno\ldots (4.32)
$$
$$
\tilde{ \mathit G}_{mn}(x) = \mathit G_{mn}(x) -  \mathit \Gamma ^{k}_{\,mn}(x)A_{k}(x),        \eqno\ldots (4.33)
$$
которые демонстрируют, что в то время как напряженности электромагнитного поля  аддитивно "не реагируют"\, на присутствие гравитационного поля, напряженности  электроджейтонного поля приобретают, в этом случае, аддитивные "добавки"\,, представленные вторым членом правой части (4.33), изменяющие как связь $G(x)$ c потенциалом $A(x)$, так и вид уравнений первой пары системы уравнений  электроджейтонного поля.

(4.18) и (4.27), в рассматриваемом частном случае пространства событий, принимают, соответственно, вид
$$
\frac{1}{\sqrt{-\textsl{g}(x)}}\,\partial_{m}(\sqrt{-\textsl{g}(x)}\,  \mathit F_{L}^{mn}(x)) =  J^{n}(x),               \eqno\ldots (4.34)
$$
$$
\frac{1}{\sqrt{-\textsl{g}(x)}}\,\partial_{m}(\sqrt{-\textsl{g}(x)}\, \tilde{ \mathit G}_{L}^{mn}(x))  + \mathit \Gamma^{n}_{\,kl}(x)\tilde{ \mathit G}_{L}^{kl}(x)  =  J^{n}(x).               \eqno\ldots (4.35)
$$

(4.34) представляет собой хорошо известное уравнение электродинамики, обобщающее уравнение второй пары системы уравнений Максвелла на случай наличия гравитационного поля \cite{La}.

Уравнение (4.35), в отличие от (4.34), по-прежнему допускает два варианта своего представления, определяемых уравнениями (4.30) и (4.31), при соответствующих переопределениях величин $ {m_{G}}^{2} $ и   $ J_{G}(x)$.

Таким образом, в присутствии традиционного гравитационного поля, электроджейтонное поле, в отличие от электромагнитного, становится "массивным" в рассматриваемом отношении.

 \section*{\S \,5 Заключительные замечания к общей теории и сравнение с известными теориями векторного поля}
 \addcontentsline{toc}{section}{\S \,5  Заключительные замечания к общей теории и сравнение с известными теориями векторного поля}

\qquad В отношении исходной посылки, рассматриваемая теория отдаленно напоминает нелинейную электродинамику Борна\,--\,Инфельда \cite{BI}.

Однако, изначально, преследуя цель объединения электромагнетизма и гравитации и следуя предположению А. Эйнштейна, в этой теории симметрическая часть тензора объединённого поля отождествляется с метрическим полем.

Тем самым, вопрос о существовании неметрического поля, определяемого симметрическими производными аффинорами векторного потенциала и наиболее естественным образом замыкающим математическую структуру электродинамики, не возникает с самого начала.

С точки зрения уравнений движения, предлагаемая теория напоминает многочисленные в прошлом  попытки построения теорий, исходящих из лагранжианов, содержащих вторые производные от полевых переменных \cite{PU}--\cite{Gr}.

Такие теории автоматически лишены основных сингулярностей стандартных теорий, однако приводят к непреодоленным пока трудностям согласования требований ковариантности, положительной определенности энергии и причинного поведения векторов состояния.

Предлагаемая теория  не содержит в лагранжиане производных от полевых функций выше первых, однако, приводит, подобно тому как это происходит в механике сплошных сред \cite{Nov}, к уравнениям движения четвертого порядка.

Именно это обстоятельство позволяет последовательно получить теорию ренормируемого типа, с одной стороны, и лишить стандартную теорию целого рада проблем, с другой стороны, однако избежав при этом основных трудностей теорий с высшими производными.

При этом, принципиальным является и то, что ренормируемость рассматриваемой теории достигается не за счет исключения, тем или иным способом, зависящей от калибровки части пропагатора, а за счет функции Грина поля $\,^{l}\!A(x)$.

 Следует при этом отметить, что подобный механизм достижения ренормируемости не является новым \cite{L},\cite{LY} однако, до сих пор он рассматривается как вспомогательный {\em математический \,} прием, заключающийся в {\em формальном\,} введении в лагранжиан калибровочно  неинвариантного члена $\xi\,\partial \!\!\cdot\!\! A(x)^{2}$ и последующего перехода к пределу $\xi \rightarrow 0 $ на соответствующих этапах построения теории.

 Однако, эта процедура не является удовлетворительной даже в математическом отношении, так как не ясно, существует или нет сам предел $\xi \rightarrow 0 $, ибо функции Грина неренормируемы в этом пределе.

 Именно это обстоятельство и явилось основанием для рассмотрения в той же мере формальной $\xi $-процедуры с ненулевым значением параметра $\xi$ \cite{Hsu}.

 Предлагаемая теория, как видим, строится исходя из принципиально иных позиций -- учета, предполагаемого к  {\em реальному \,} существованию, самостоятельного физического поля, определяемого симметрическими производными аффинорами 4-потенциала электродинамики, а ренормируемость, как и все прочие свойства, являются сопровождающим, а не определяющим теорию фактом.

 При этом уместно отметить соответствующие успехи спонтанно нарушенных калибровочных теорий, однако и в них ренормируемость является определяющей.

 По внешнему виду пропагатора, предлагаемая теория, в рамках второго доопределения вакуума, напоминает массивную электродинамику Фельдмана\,--\,Метьюса \cite{Feld}, однако не разделяет той осторожности перехода к безмассовой теории, которая сопровождает пропагатор последней.

 И так, исключение математической асимметрии, обсуждаемой во введении главы  I  \cite{Al4}, приводит, уже в рамках монографических методов построения теории, в целом, к существенному улучшению теории векторного поля, с одной стороны, и проливает свет на природу целого ряда трудностей, сопровождающих стандартную теорию, с другой стороны.

 Однако, важно еще раз подчеркнуть, что {\em  радикальной\,} отличительной особенностью (определяющей идеей) настоящей теории является постановка вопроса о реальном существовании и о последующем равноправном учете материального полевого объекта, определяемого (ассоциированного с) симметрическим производным аффинором классического электродинамического 4-потенциала.

 Последнее, в свою очередь, представляет собой главное {\em идеологическое\,} отличие рассматриваемой теории от всех теорий, предложенных ранее  \cite{BG}--\cite{Nak}.

При этом, является уместным  привести так же и краткие {\em предварительные} замечания о физическом содержании поля, определяемого симметрическими производными аффинорами 4-векторного потенциала классической электродинамики.

В электродинамическом случае, (3.14) и (3.13) принимают, соответственно, вид
$$
\Box \,\,^{\tau}\!A(x)= - \jmath(x),                      \eqno\ldots (5.1)
$$
$$
\Box \,\,^{\ell}\!A(x)= \theta.                            \eqno\ldots (5.2)
$$

В свою очередь, (5.1) и (5.2), с учетом (I.4.1.), приводят к стандартному неоднородному уравнению Даламбера (3.35).

Последующее действие на обе части уравнения (3.35) дифференциальных операторов (1.9) и (1.12) приводит к следующей определяющей системе релятивистски-инвариантных и {\em линейно независимых}, по отношению к друг другу, неоднородных волновых уравнений
$$
\square\, F(x) = - \widetilde{rot}\,j(x),                 \eqno\ldots (5.3)
$$
$$
\square\, G(x) = - \widetilde{def}\,j(x),                 \eqno\ldots (5.4)
$$
из которых, в частности, следуют соответствующие неоднородные волновые уравнения для магнитной и джейтонной составляющих электромагнитного и электроджейтонного полей,
$$
\square\,\mathcal{F}(x) = - \tilde{rot}\,\vec{j}(x),             \eqno\ldots (5.5)
$$
$$
\square\,\mathcal{G}(x) = - \tilde{def}\,\vec{j}(x),             \eqno\ldots (5.6)
$$
где тензорные дифференциальные операторы $\tilde{rot}$ и $\tilde{def}$ определены соотношениями (3.40)
и (3.41), соответственно.

Уравнения (5.3) и (5.4) имеют, прежде всего, фундаментальное {\em физиечское\,} значение, демонстрируя те локально-независимые физические  процессы,  в источнике излучения, которые могут генерировать, соответственно, локально независимые электромагнитное и электроджейтонное поля.

В частности, так как в рамках классического описания источника электродинамического поля \cite{La}, векторное поле плотности тока представляет собой векторное поле, имеющее кинематический смысл,  то уравнение (5.4), в соответствии с  \cite{Rash},  демонстрирует, что классическое электроджейтонное поле, определяемое симметрическими производными аффинорами классического векторного потенциала, генерируется, в частности, {\em деформационными\,} токами (процессами) в $\delta$-окрестности рассматриваемой  точки источника поля.

Соответственно, уравнение (5.3) демонстрирует известную генерацию электромагнитного поля, обусловленную {\em вихревыми\,} токами (процессами) в $\delta$-окрестности данной  точки источника поля.

Таким образом, электроджейтонное поле создается реальными и локально-независимыми физическими процессами в источнике поля и поэтому имеет не только чисто математическую, но и физическую основу для своего реального и самостоятельного существования.

Физическая реальность электроджейтонного поля определяется так же и как непосредственным, так и опосредованным силовым воздействием составляющих этого поля на электрически заряженные частицы \cite{Al2}, \cite{Al3}, \cite{Al6}, относительно подробно рассмотренным в  части II.

 Как и в \cite{Al6}, тут лишь  кратко (декларативно)  коснемся  вопросов математического описания данных силовых воздействий в
 рамках обсуждаемой модели классической электродинамики.

 По-прежнему, следуя  \cite{La} ограничиваемся  рамками электродинамики вакуума и точечных электрических зарядов.

 Джейтонная сила, определяющая непосредственное силовое воздействие классического джейтонного поля на движущуюся точечную электрически заряженную частицу, представлена соотношением
 $$
 \vec{F}_{\mathcal{G}}(x) = e\,  \mathcal{G}(x)\!\cdot\! \vec{v}(x),          \eqno\ldots (5.7)
 $$
 где $\mathcal{G}(x)$ -- 3\,-тензор джейтонного поля,
 $$
  \mathcal{G}(x) = \tilde{def}\vec{A}(x),                \eqno\ldots (5.8)
 $$
 в котором $\tilde{def}$ -- тензорный дифференциальный оператор, определенный равенством (3.41).

 В соответствии с взаимно однозначным разложением данного тензора, $ \mathcal{G}(x)$, на линейно независимые части,   $\mathcal{G}_{d}(x)$    и $\mathcal{G}_{\ast}(x)$, представляющие собой {\em девиатационное \,} джейтонное поле и {\em дилатационное\,} джейтонное поле, соответственно, имеем следующие выражения для джейтонных сил, определяющих непосредственные силовые воздействия каждого из этих полей на движущиеся электрически заряженные частицы, \cite{Al6},
 $$
 \vec{F}_{\mathcal{G}_{d}}(x) = e\,  \mathcal{G}_{d}(x)\!\cdot\! \vec{v}(x),          \eqno\ldots (5.9)
 $$
$$
 \vec{F}_{\mathcal{G}_{\ast}}(x) = e\,  \mathcal{G}_{\ast}(x)\!\cdot\! \vec{v}(x),          \eqno\ldots (5.10)
 $$
 где
 $$
  \mathcal{G}_{d}(x) = \tilde{def}_{d}\,\vec{A}(x),                \eqno\ldots (5.11)
 $$
 $$
  \mathcal{G}_{\ast}(x) = \tilde{def}_{\ast}\,\vec{A}(x),                \eqno\ldots (5.12)
 $$
 $$
 \tilde{def}_{d} = \vec{\bigtriangledown} \overset{s}{\otimes} - \frac{g}{tr g}\,\vec{\bigtriangledown}\,\cdot ,     \eqno\ldots (5.13)
 $$
$$
 \tilde{def}_{\ast} = \frac{g}{tr g}\,\vec{\bigtriangledown}\,\cdot ,   \qquad  g = \vec{e}_{\alpha}\otimes\vec{e}\,^{\alpha}.                     \eqno\ldots (5.14)
 $$

 В полевых системах, созданных стационарными токами, $\mathcal{G}_{\ast}(x) =\tilde{0} $, поэтому силовое воздействие джейтонного поля определяется, в таких случаях, только {\em девиатационной \,} джейтонной силой (5.9), то есть, силовым воздействием {\em девиатационного \,} джейтонного поля, представленного тензором $\mathcal{G}_{d}(x)$.

 Магнитное поле теперь определено 3-тензором
 $$
 \mathcal{F}(x) = \tilde{rot}\,\vec{A}(x),            \eqno\ldots (5.15)
 $$
 где $ \tilde{rot}$ -- тензорный дифференциальный оператор, определенный равенством (3.40), а непосредственное силовое воздействие, так определенного магнитного поля, представлено соотношением стандартного вида
 $$
 \vec{F}_{ \mathcal{F}}(x) = e\, \mathcal{F}(x)\!\cdot\! \vec{v}(x).         \eqno\ldots (5.16)
 $$

 В силу очевидного соотношения между 3-тензором $\mathcal{F}(x)$ и стандартным лоренцевым  3-тензором магнитного поля,  $\mathcal{F}_{L}(x)$, сила (5.16) составляет {\em половину \,} от традиционной силы Хевисайда\,--\,Лоренца,
 $$
 \vec{F}_{L}(x) = e\, \mathcal{F}_{L}(x)\!\cdot\! \vec{v}(x).         \eqno\ldots (5.17)
 $$

Линейная независимость составляющих джейтонного поля, $\mathcal{G}_{d}(x)$   и   $\mathcal{G}_{\ast}(x) $, позволяет получить, как следствие (5.6), систему двух линейно независимых, как по отношению к друг другу, так и к (5.5), неоднородных волновых уравнений
$$
\square\, \mathcal{G}_{d}(x) = -  \tilde{def}_{d}\,\,\vec{j}(x),               \eqno\ldots (5.18)
$$
 $$
\square\, \mathcal{G}_{\ast}(x) = -  \tilde{def}_{\ast}\,\,\vec{j}(x),               \eqno\ldots (5.19)
$$
демонстрирующих, в частности, что в качестве источников волнового девиатационного джейтонного поля и волнового дилатационного джейтонного поля могут выступать структуры плотности тока, определяемые правыми частями этих уравнений.

Дифференциальные законы {\em электроджейтонной \,} индукции, определяющие  {\em опосредованное \,} силовое воздействие классических джейтонных полей на {\em системы \,} электрически заряженных частиц, представлены соотношениями \cite{Al6}
$$
\partial^{\,0}\, \mathcal{G}_{d}(x) = -  \tilde{def}_{d}\,\,\vec{E}_{A}(x),               \eqno\ldots (5.20)
$$
 $$
\partial^{\,0}\, \mathcal{G}_{\ast}(x) = -  \tilde{def}_{\ast}\,\,\vec{E}_{A}(x),               \eqno\ldots (5.21)
$$
в то время как соответствующий дифференциальный закон электромагнитной индукции Фарадея\,--\,Максвелла, представляющий  опосредованное силовое воздействие классического магнитного поля на системы электрически заряженных частиц, имеет вид
$$
\partial^{\,0}\, \mathcal{F}(x) = -  \tilde{rot}\,\,\vec{E}_{A}(x).               \eqno\ldots (5.22)
$$

Напряженность электрического поля, $\vec{E}_{A}(x)$, содержащаяся в правых частях уравнений (5.20)--(5.22), определена известным соотношением \cite{La}
$$
\vec{E}_{A}(x) = - \partial^{\,0}\vec{A}(x),                         \eqno\ldots (5.23)
$$
представляющим собой напряженность электрического поля, обусловленного изменяющимся во времени векторным потенциалом $\vec{A}(x)$.

Уравнения (5.20) и (5.21) являются линейно независимыми как по отношению к друг другу, так и по отношению к уравнению (5.22) и выполняются, как и уравнение (5.22), тождественно лишь в силу определений полевых переменных входящих в них, то есть, эти уравнения имеют более универсальный характер по отношению к полевым уравнениям, определяемым посредством соответствующих вариационных процедур.

С другой стороны, действие оператора $\partial^{\,0}$ на обе части уравнений (5.5), (5.18) и (5.19) и последующее использование (5.20)--(5.22) приводит к следующей системе неоднородных волновых уравнений
$$
\square\,\tilde{rot}\,\,\vec{E}_{A}(x) =  \tilde{rot}\,\partial^{\,0}\,\vec{j}(x),             \eqno\ldots (5.24)
$$
$$
\square\,\tilde{def}_{d}\,\,\vec{E}_{A}(x)  =  \tilde{def}_{d}\,\,\partial^{\,0}\,\vec{j}(x),               \eqno\ldots (5.25)
$$
 $$
\square\,\tilde{def}_{\ast}\,\,\vec{E}_{A}(x)  =  \tilde{def}_{\ast}\,\,\partial^{\,0}\,\vec{j}(x),               \eqno\ldots (5.26)
$$
демонстрирующих, в частности, что {\em структуры \,} электрических полей, определяемые правыми частями законов электродинамической индукции (5.20)--(5.22), обладают волновым характером распространения в пространстве, при этом, их источником могут выступать соответствующие структуры плотности тока, генерирующие $\mathcal{F}(x)$- ,
   $ \mathcal{G}_{d}(x)$-   и    $\mathcal{G}_{\ast}(x)$- поля.

Таким образом, данные {\em структуры \,} плотности тока могут одновременно генерировать как волновые магнитное и джейтонное поля, так и {\em структуры \,} электрических полей, порождаемых последними в соответствии с законами электродинамической индукции (5.20)--(5.22).

Подобные процессы наблюдаются и практически используются, в частности, при возбуждении соответствующих типов волн в соответствующих волноводах, демонстрируя физическую значимость вышеуказанных уравнений.

 В связи с этим находим, что рассматриваемые тензорные уравнения, в отличие от векторных соотношений, описывают поведение геометрических образов тензоров, представленных соответствующими геометрическими структурами соответствующих векторных линий (соответствующими конгруэнциями  векторных линий  соответствующих векторных полей), в результате чего тензорный язык  \cite{Gil} на трехмерных  и двухмерных многообразиях становится наполненным конкретным как геометрическим, так и физическим содержанием.

 И так, обсуждаемая теория, в основе которой заложен равноправный и явный учет как электромагнитного, так и электроджейтонного полей, приобретает, по сравнению с традиционной классической электродинамикой, более симметричный и законченный вид как в математическом, так и в физическом отношениях.

(5.3) и (5.4) вне источника поля принимают вид волновых уравнений Даламбера
$$
\square\, F(x) = \widetilde{0},                 \eqno\ldots (5.27)
$$
$$
\square\, G(x) = \widetilde{0}.                 \eqno\ldots (5.28)
$$

(5.28) демонстрирует, что электроджейтонное поле обладает волновым характером распространения в 3-пространстве, представляя собой, в этом случае, {\em  электроджейтонную\,} волну \cite{Al5}, которая как и электромагнитные волны, распространяется в открытом пространстве со скоростью света и является носителем как энергии и импульса, так и вышеуказанной информации об источнике излучения, а так же носителем соответствующих силовых воздействий, отмеченных выше.

С другой стороны, уравнения (5.27) и (5.28), так же как и (5.3) и (5.4),  представляют собой два инвариантных линейно независимых уравнения, описывающих волновое распространение, соответственно, самостоятельных и локально линейно независимых электромагнитного и электроджейтонного полей, тем самым, эти уравнения, как и уравнения (5.20)--(5.22), снимают своеобразное вырождение вкладов полей, определяемых симметрическим производным аффинором $ \mathit G(x) $, и полей, определяемых антисимметрическим производным аффинором  $ \mathit F(x) $, которое наблюдается в полевых уравнениях (I.2.7), (I.2.8) и, соответственно, в (1.15), (1.18), при соответствующем выборе значений коэффициентов входящих в них, и которое происходит в результате использования стандартных процедур варьирования и соответствующих дополнительных условий.

В заключение данной части, автор выражает признательность Боголюбову Н.Н. (ст.), Ширкову Д.В. и Блохинцеву Д.И. за обсуждение определяющей идеи рассматриваемой теории, а так же благодарит дирекцию Лаборатории Теоретической Физики Объединенного Института Ядерных Исследований (г. Дубна), за предоставленную, в прошлом, возможность выполнения, в данной Лаборатории, курсового (1968 г.), а затем и дипломного (1970 г.) проектов, в которых была представлена вышеуказанная определяющая идея о {\em  физическом\,} существовании классического поля, определяемого симметрическим производным аффинором классического  электродинамического 4-потенциала,  а  так же продемонстрирован отличный от нуля вклад данного поля  в тензор энергии--импульса произвольной классической электродинамической полевой системы и, тем самым, определена материальность этого поля.

Особая признательность  выражается Нестерину В. А. за постоянное внимание и поддержку в ходе последующих
исследований автора в рассматриваемой области электродинамики.

\renewcommand{\bibname}{\em Литература к части I}

\part* {Часть II}
\addcontentsline{toc}{part}{Часть II}

\chapter{ Явление электроджейтонной индукции}

\qquad Продолжается построение теории действительного векторного поля, представленной в \cite {Al1}--\cite {Al0}.

Обсуждается, в рамках этой теории, явление электроджейтонной индукции на основе дифференциальных форм законов, описывающих данное явление.

Введено понятие фазового портрета тензора второго ранга на двух и трехмерных  многообразиях (над двух и трехмерным векторным пространством), позволяющее с  {\em единой} геометрической точки зрения  продемонстрировать принципиальное различие тензоров второго ранга, обладающих разным типом симметрии.

Определены топологические структуры электрических полей  определяемых вышеуказанными законами электроджейтонной индукции  и найдена специфика силового воздействия этих полей на системы электрически заряженных частиц.

Представлена система дифференциальных уравнений, устанавливающих дифференциальные связи джейтонного и сопровождающего его электрического полей в электроджейтонном поле.

Исходя из этих уравнений, развивается представление об электро\-джейтонных волнах (как о поперечных, так и продольных) и в частности рассматривается вопрос о продольном электроджейтонном излучении электрически заряженных частиц.

Продемонстрированы наличие и фундаментальная роль электро\-джейтонного поля в фундаментальных классических полевых системах.

Ключевые слова: электроджейтонная индукция, электроджейтонная волна,  электроджейтонное излучение.

\section*{\S \,1 Дифференциальные законы электроджейтонной индукции и их интерпретация}
\addcontentsline{toc}{section}{\S \,1  Дифференциальные законы электроджейтонной индукции и их интерпретация}

\qquad Теория векторного поля с симметрическими аффинорами, \cite {Al1}--\cite {Al0}, базируясь на математическом равноправии и самостоятельности производных тензоров векторного поля  $ \mathit A(x) $, $\mathit F(x) = \partial \! \overset{a}{\otimes}\!\mathit A(x)$  и $\mathit G(x) =\partial  \overset{s}{\otimes}\mathit A(x)$, рассматривает поле, определяемое симметрическим производным 4-тензором 4-потенциала классической электродинамики, как самостоятельное и равноправное, по отношению к электромагнитному полю, физическое  поле, названное  электроджейтонным полем \cite {Al2}.

Для доказательства утверждения о физической реальности и самостоятельности электроджейтонного поля является достаточным продемонстрировать физическую реальность и самостоятельность {\em джейтонной \,} составляющей электроджейтонного поля, определяемой 3-тензором
$$
\mathcal{G}(x) = \vec{\bigtriangledown} \overset{s}{\otimes}\vec{A}(x).                    \eqno\ldots (1.1)
$$

В целях сопоставления свойств джейтонной составляющей электроджейтонного поля и свойств магнитной составляющей электромагнитного поля, будем рассматривать обе эти составляющие одновременно, представляя последнюю, в данном контексте, 3-тензором
$$
\mathcal{F}(x) = \vec{\bigtriangledown} \overset{a}{\otimes}\vec{A}(x).                    \eqno\ldots (1.2)
$$

В (1.1) и (1.2) и далее традиционным символом $\vec{\bigtriangledown}$ обозначен классический дифференциальный оператор Гамильтона, а символы $a$ и $s$ означают, соответственно, операции альтернирования и симметризации соответствующих внешних произведений.

Для дальнейшего является удобным использовать тензорные дифференциальные операторы \cite{Pob}
$$
\tilde{rot} =  \vec{\triangledown}\, \overset{a}{\otimes},                                              \eqno\ldots (1.3)
$$
$$
 \tilde{def} = \vec{\triangledown}\, \overset{s}{\otimes},                                               \eqno\ldots (1.4)
$$
с помощью которых соотношения (1.1) и (1.2) можно переписать в более компактном виде
$$
  \mathcal{G}(x) = \tilde{def}\vec{A}(x),                \eqno\ldots (1.5)
 $$
 $$
 \mathcal{F}(x) = \tilde{rot}\,\vec{A}(x).            \eqno\ldots (1.6)
 $$

 Тензор джейтонного поля, $ \mathcal{G}(x)$, в соответствии с его разложением на девиатор и дилататор, имеет две локально линейно независимые, по отношению к друг другу и к тензору $ \mathcal{F}(x)$, составляющие,  (I.5.11)--(I.5.14),
 $$
  \mathcal{G}_{d}(x) = \tilde{def}_{d}\,\vec{A}(x),                \eqno\ldots (1.7)
 $$
 $$
  \mathcal{G}_{\ast}(x) = \tilde{def}_{\ast}\,\vec{A}(x),                \eqno\ldots (1.8)
 $$
 $$
 \tilde{def}_{d} = \vec{\bigtriangledown} \overset{s}{\otimes} - \frac{g}{tr g}\,\vec{\bigtriangledown}\,\cdot ,     \eqno\ldots (1.9)
 $$
$$
 \tilde{def}_{\ast} = \frac{g}{tr g}\,\vec{\bigtriangledown}\,\cdot ,   \qquad  g = \vec{e}_{\alpha}\otimes\vec{e}\,^{\alpha}, \qquad tr g = g\cdot\cdot g.                     \eqno\ldots(1.10)
 $$

 Поля, представленные этими составляющими джейтонного поля, будем называть {\em девиатационным \,} джейтонным полем и {\em дилатационным \,} джейтонным полем, соответственно.

 Всюду в данном контексте тензор рассматривается как инвариантный, по классу преобразований координат пассивного типа, объект, представляемый различными, в общем случае, наборами компонент в разных системах отсчета \cite{Ko}.

 Следуя  \cite{La}, тут и далее рассматривается \,"электродинамика вакуума и точечных электрических зарядов"\,, а используемая система единиц очевидна из контекста.

 С тем, чтобы минимизировать использование "жирных"\,  или "полужирных"\, символов, 3-векторы евклидова пространства обозначаются "обычными"\,  символами содержащими соответствующие стрелки.

 В силу вышеуказанной линейной независимости $\mathcal{F}(x)$- , $  \mathcal{G}_{d}(x)$- и $ \mathcal{G}_{\ast}(x)$-полей, совокупность магнитного, девиатационного джейтонного и дилатационного джейтонного полей представляет собой систему {\em самостоятельных \,} полевых объектов, обладающих, как увидим ниже, {\em принципиально  \,} разными как математическими, так и физическими свойствами.

 Отвечая на вопрос, что порождает каждое из этих полей изменяясь во времени в некоторой фиксированной точке $\mathit M$ евклидова пространства, подействуем на обе части каждого из уравнений (1.6), (1.7) и (1.8) дифференциальным оператором $\mathit \partial ^{0}$, в результате чего приходим к следующей системе трех локально {\em линейно независимых\,}, по отношению к друг другу, уравнений, не содержащих явно векторный потенциал  $\vec{ \mathit A}(x)$,
$$
\partial  ^{0}\, \mathcal{F}(x) = -  \tilde{rot}\,\vec{E}_{A}(x),               \eqno\ldots (1.11)
$$
$$
\partial  ^{0}\, \mathcal{G}_{d}(x) = -  \tilde{def}_{d}\,\vec{E}_{A}(x),               \eqno\ldots (1.12)
$$
$$
\partial ^{0}\, \mathcal{G}_{\ast}(x) = -  \tilde{def}_{\ast}\,\vec{E}_{A}(x).              \eqno\ldots (1.13)
$$

Напряженность электрического поля,  $ \vec{\mathit E}_{A}(x) $, содержащаяся в правых частях уравнений (1.11)--(1.13), определена известным соотношением  \cite{La}
$$
\vec{E}_{A}(x)  \eqdef  - \partial ^{0}\vec{A}(x) ,           \eqno\ldots (1.14)
$$
представляющим собой напряженность электрического поля, обусловленного изменяющимся во времени векторным потенциалом $\vec{\mathit A}(x)$.

Первое из этих уравнений, как видим, есть общеизвестный дифференциальный  закон электромагнитной индукции Фарадея--Максвелла, представленный в тензорном виде (в пространстве $ V\!\otimes\!V$), тогда  два другие, естественно, рассматривать как соответствующие дифференциальные законы  {\em электроджейтонной \,} индукции.

Интерпретация соотношений (1.12) и (1.13) очевидна и идентична традиционной интерпретации уравнения (1.11), то есть,  если (1.11) говорит о том, что магнитное поле, представленное в (1.11) тензором $ \mathcal{F}(x)$, в процессе своего изменения во времени в некоторой фиксированной точке $\mathit M $, "порождает"\,, в окрестности этой точки, электрическое поле $ \vec{E}_{A}(x)$, имеющее структуру, определяемую тензором $  \tilde{rot}\,\,\vec{E}_{A}(x)$, то уравнения (1.12) и (1.13) говорят о том, что джейтонные поля, представленные в этих уравнениях тензорами $\mathcal{G}_{d}(x) $ и $ \mathcal{G}_{\ast}(x)$, в процессе своего изменения во времени в некоторой фиксированной точке  $\mathit M $, "порождают"\,,  в окрестности этой точки, электрические поля  $ \vec{E}_{A}(x)$, имеющие структуры, определяемые тензорами $  \tilde{def}_{d}\,\,\vec{E}_{A}(x)$ и $  \tilde{def}_{\ast}\,\,\vec{E}_{A}(x)$, соответственно.

И так, система уравнений (1.11)--(1.13) демонстрирует, что наряду с известным явлением электромагнитной индукции, представленным дифференциальным уравнением (1.11), существует и явление, описывающееся линейно независимыми, по отношению к (1.11), дифференциальными уравнениями (1.12) и (1.13), названное явлением электроджейтонной индукции
\footnote{Закон электроджейтонной индукции, представленный в виде $ \partial^{\,0}\, \mathcal{G}(x) = -\tilde{def}\,\,\vec{E}_{A}(x) $   и записанный в форме
 $ \partial^{\,0}\, \mathcal{G}(x) = -\tilde{def}\,\,\vec{E}_{\mathcal{F}L}(x) - \tilde{def}\,\,\vec{E}_{\varphi}(x) $, где $ \vec{E}_{\mathcal{F}L}(x) = \vec{E}_{A}(x) + \vec{E}_{\varphi}(x) $ -- традиционная лоренцева напряженность электрического поля, впервые представлен и предварительно проанализирован, в традиционном ковариантном виде, в работе  \cite{Al4}, когда джейтонное поле, определяемое тензором $ \mathcal{G}_{L}(x)$, называлось ещё "джи-полем".}.

 С другой стороны,  легко видеть, что уравнения (1.12) и (1.13), так же как и уравнение (1.11), выполняются тождественно в рамках определений величин входящих в них, вследствие чего эти уравнения носят более универсальный характер, по отношению к уравнениям второй пары системы уравнений Максвелла \cite{La}, связанных с тем или иным видом лагранжевых процедур.

 В свою очередь, воспользовавшись известным выражением для псевдовектора $ \vec{\mathcal{H}}(x) $,
 сопровождающего тензор $ \mathcal{F}(x) $ в трехмерном пространстве $ E\,^{3} $,
 $$
  \vec{\mathcal{H}}(x) = - \frac{1}{2}\,\, \varepsilon \cdot\cdot\, \mathcal{F}(x),                       \eqno\ldots (1.15)
 $$
 и последующим учетом соотношения
 $$
 \varepsilon\cdot\cdot\, \tilde{rot}\,\,\vec{E}_{A}(x) = - rot\,\vec{E}_{A}(x),                 \eqno\ldots (1.16)
 $$
 легко получить, как непосредственное следствие уравнения (1.11), и традиционную  дифференциальную форму закона электромагнитной индукции  \cite{La}.

 Псевдовектор $ \vec{\mathcal{H}}(x) $,  определенный посредством (1.15), связан с традиционной  "лоренцевой"\, напряженностью магнитного поля, обозначаемой в данном контексте как  $ \vec{\mathcal{H}}_{L}(x) $, очевидным
  соотношением  $ \vec{\mathcal{H}}(x)=~\displaystyle\frac{1}{2}\, \vec{\mathcal{H}}_{L}(x)$.

  Аналогичным соотношением связан тензор $ \mathcal{F}(x) $  с традиционным лоренцевым тензором лоренцева магнитного поля, обозначаемым, в дальнейшем, как  $ \mathcal{F}_{L}(x) $.

  Суммируя уравнения (1.11)--(1.13), приходим к~"объединенному"\, закону электродинамической индукции
  $$
  \partial\,^{0}\partial A(x) = - \,\partial E_{A}(x),                           \eqno\ldots (1.17)
  $$
  в котором, символами $ \partial A(x) $ и $ \partial E_{A}(x) $ обозначены 3-градиенты векторных
   полей $ \vec{A}(x) $ и $ \vec{E}_{A}(x) $, соответственно, то есть,
   $$
  \partial A(x)  \eqdef  \vec{\bigtriangledown}\,\otimes\vec{A}(x) ,                   \eqno\ldots (1.18)
   $$
  $$
  \partial E_{A}(x)  \eqdef  \vec{\bigtriangledown}\,\otimes\vec{E}_{A}(x) .                   \eqno\ldots (1.19)
   $$

   (1.17) так же представляет собой тождество по  $ \vec{A}(x) $, демонстрируя, тем самым, свою универсальность в вышеуказанном отношении, и обладает соответствующей физической интерпретацией, идентичной вышеуказанным
   интерпретациям соотношений (1.11)--(1.13).

   Стартуя с (1.17) легко придти обратно к системе уравнений (1.11)--(1.13), демонстрируя объединяющий характер объединенного закона электродинамической индукции.

Действительно, используя инвариантные соотношения
$$
\partial A(x) = \mathcal{F}(x)+ \mathcal{G}_{d}(x)+  \mathcal{G}_{\ast}(x),               \eqno\ldots (1.20)
$$
$$
 \partial E_{A}(x) = \tilde{rot}\,\,\vec{E}_{A}(x)+ \tilde{def}_{d}\,\,\vec{E}_{A}(x)+ \tilde{def}_{\ast}\,\,\vec{E}_{A}(x), \eqno\ldots (1.21)
$$
уравнение (1.17) можно переписать в развернутом виде, содержащим
явно $ \mathcal{F}(x)  $- , $ \mathcal{G}_{d}(x) $-  и $\mathcal{G}_{\ast}(x) $-поля,
\begin{multline*}
 \partial\,^{0} \mathcal{F}(x)+\partial\,^{0} \mathcal{G}_{d}(x)+ \partial\,^{0} \mathcal{G}_{\ast}(x) =\\
  - \tilde{rot}\,\,\vec{E}_{A}(x)- \tilde{def}_{d}\,\,\vec{E}_{A}(x)-
                                                                  \tilde{def}_{\ast}\,\,\vec{E}_{A}(x).        \qquad\ldots (1.22)
\end{multline*}

Соотношения (1.20) и (1.21) соответствуют известному представлению линейного пространства тензоров ранга (0,2) над действительным векторным пространством $V$, $T(V)=V\otimes V$, в виде прямой суммы~\cite{LS} попарно ортогональных линейных подпространств $ T_{a}(V) = V \overset{a}{\otimes} V $, $T_{d}(V) = V \overset{d}{\otimes} V$ и $T_{\ast}(V)=V \overset{\ast}{\otimes} V  $.

В качестве элементов пространства $V$, в данном случае, выступают векторы, представленные соответствующими инвариантными геометрическими объектами, при этом, само пространство $V$ рассматривается евклидовым.

Ортогональность элементов вышеуказанных подпространств определена по отношению к скалярному произведению тензоров ранга (0,2) \cite{Gil},
$$
(A(x), B(x))\, \eqdef A(x)\,\cdot\cdot\,B(x)^{T}\,,                   \eqno\ldots (1.23)
$$
индуцирующему евклидову норму этих элементов,
$$
\parallel A(x)\parallel \;\; \eqdef +\,\sqrt{ A(x)\,\cdot\cdot\,A(x)^{T}}\,.      \eqno\ldots (1.24)
$$

При этом, любой ортонормированный базис векторного пространства $V$, $ e_{V}= \{\,\vec{e}_{\alpha}\, \}$\,,
сопровождается ортонормированными, по отношению к (1.23) и (1.24), базисами пространств $T(V)$, $ T_{a}(V)$, $T_{d}(V)$ и $T_{\ast}(V)$\,, представленных, соответственно, соотношениями
$$
 e = \{\,\vec{e}_{\alpha}\,\otimes\,\vec{e}_{\beta} \}\,,                 \eqno\ldots (1.25)
$$
$$
 e_{a} = \{\,\vec{e}_{\alpha}\,\overset{a}{\otimes}\,\vec{e}_{\beta} \}\,,                 \eqno\ldots (1.26)
$$
$$
 e_{d} = \{\,\vec{e}_{\alpha}\,\overset{s} {\otimes}\,\vec{e}_{\beta}- \frac{g_{\alpha\beta}}{tr g}\,g \}\,,  \eqno\ldots (1.27)
$$
$$
 e_{\ast} = \{ \frac{g_{\alpha\beta}}{tr g}\,g  \},   \quad    g_{\alpha\beta}=(\vec{e}_{\alpha},\vec{e}_{\beta})\,.              \eqno\ldots (1.28)
$$

В рамках определений (1.23) и (1.24), соотношение (1.20) приводит к следующему представлению квадрата нормы производного тензора $\partial A(x)$
$$
\parallel \partial A(x)\parallel ^{2}\, =\,\, \parallel F(x)\parallel ^{2} + \parallel G_{d}(x)\parallel ^{2} + \parallel G_{\ast}(x)\parallel ^{2}.
$$

Возвращаясь к (1.22), умножим скалярно обе части этого уравнения на произвольный тензор $A_{a}(x)$ из пространства $T_{a}(V)$, в результате чего приходим к соотношению
$$
(\partial\, ^{0}\,\mathcal{F}(x) , A_{a}(x) ) =( -  \tilde{rot}\,\,\vec{E}_{A}(x) , A_{a}(x) )\,,           \eqno\ldots (1.29)
$$
которое, в силу произвольности тензора $A_{a}(x)$, означает выполнение равенства (1.11).

Совершенно аналогично, проведя скалярные умножения обеих частей уравнения (1.22) последовательно на произвольные тензоры  $A_{d}(x)$ и  $A_{\ast}(x)$ из пространств $T_{d}(V)$ и $T_{\ast}(V)$, приходим, соответственно, к равенствам $$
(\partial^{\,0}\, \mathcal{G}_{d}(x) , A_{d}(x)) = ( -  \tilde{def}_{d}\,\,\vec{E}_{A}(x) ,  A_{d}(x))\,,    \eqno\ldots (1.30)
$$
$$
(\partial^{\,0}\, \mathcal{G}_{\ast}(x) , A_{\ast}(x)) = (-  \tilde{def}_{\ast}\,\,\vec{E}_{A}(x) , A_{\ast}(x) ) ,                       \eqno\ldots (1.31)
$$
которые, в силу произвольности тензоров  $A_{d}(x)$ и  $A_{\ast}(x)$, означают выполнение соотношений (1.12) и (1.13).

Можно придти к этим же результатам и более прямолинейным способом (не прибегая к использованию вспомогательных тензоров  $A_{a}(x)$,  $A_{d}(x)$ и  $A_{\ast}(x)$).

В этом случае, переписав (1.22) в виде
\begin{multline*}
  \partial\,^{0} \mathcal{F}(x)+ \tilde{rot}\,\,\vec{E}_{A}(x) + \partial\,^{0} \mathcal{G}_{d}(x)+\tilde{def}_{d}\,\,\vec{E}_{A}(x)+ \\
  \partial\,^{0} \mathcal{G}_{\ast}(x)+\tilde{def}_{\ast}\,\,\vec{E}_{A}(x) = \tilde{0},     \qquad\ldots (1.32)
\end{multline*}
получаем
\begin{multline*}
 \parallel \partial\,^{0} \mathcal{F}(x)+ \tilde{rot}\,\,\vec{E}_{A}(x) + \partial\,^{0} \mathcal{G}_{d}(x)+\tilde{def}_{d}\,\,\vec{E}_{A}(x)+  \\
  \partial\,^{0} \mathcal{G}_{\ast}(x)+\tilde{def}_{\ast}\,\,\vec{E}_{A}(x) \parallel ^{2} = 0.                        \qquad\ldots (1.33)
\end{multline*}

В силу ортогональности соответствующих пар слагаемых левой части (1.32), уравнение (1.33) принимает вид соотношения
\begin{multline*}
\parallel \partial\,^{0} \mathcal{F}(x)+ \tilde{rot}\,\,\vec{E}_{A}(x)\parallel^{2} + \parallel \partial\,^{0} \mathcal{G}_{d}(x)+\tilde{def}_{d}\,\,\vec{E}_{A}(x)\parallel^{2}\\+ \parallel \partial\,^{0} \mathcal{G}_{\ast}(x)+\tilde{def}_{\ast}\,\,\vec{E}_{A}(x) \parallel ^{2} = 0,     \qquad\ldots(1.34)
\end{multline*}
которое вновь приводит к системе (1.11)--(1.13).

К этому же результату можно так же легко придти, если представить каждое слагаемое левой части (1.32) в виде соответствующих линейных комбинаций элементов базисов (1.26)--(1.28) и затем воспользоваться линейной независимостью этих базисов по отношению к друг другу.

\section*{\S \,2 Геометрическое  представление произвольного тензора ранга (0,2) в действительном векторном пространстве $V$}
\addcontentsline{toc}{section}{\S \,2 Геометрическое  представление произвольного тензора ранга (0,2) в действительном векторном пространстве $V$}

\qquad Возвращаясь к интерпретации уравнений (1.11)--(1.13), найдем конкретные виды структур
 электрических полей, определяемых правыми частями этих уравнений.

В целях общности полученных результатов рассмотрим общий случай геометрических объектов, представленных произвольным производным тензором  (1.18), в котором действительное дифференцируемое векторное поле  $ \vec{ \mathit A}(x)$ может иметь любое происхождение.

Данный тензор, являясь градиентом данного векторного поля, определяет, в каждый фиксированный момент времени, структуру неоднородной части векторного поля $ \vec{ \mathit A}(x)$ в  $\delta$-окрестности точки дифференциирования, в линейном приближении.

Действительно, согласно формулы Тейлора с дополнительным членом в форме Пеано, в каждый фиксированный момент времени $t$, в  $\delta$-окрестности некоторой точки  $ \mathit M $ евклидова пространства, данное векторное поле представляется соотношением
$$
\vec{ \mathit A}(N) = \vec{ \mathit A}_{0}(N) +  \partial A(M)^{T}\!\cdot\! \vec{\rho}\,(N) + \vec{0} (\rho), \quad  N \in \delta_{\mathit M},               \eqno\ldots (2.1)
$$
где $ \vec{ \mathit A}_{0}(N)\stackrel{\delta_{\mathit M}}{\equiv}\vec{ \mathit A}(M) $,  $ \vec{\rho}(N) =\vec{r}(N) - \vec{r}(M)$, а индекс $\mathit T$ означает операцию транспонирования соответствующего тензора.

Ради краткости записей  время $ t$  в аргументах соответствующих функций и выражений часто будем опускать без специальных оговорок.

Обратив главное внимание теперь на то, что в (2.1) точка $ \mathit N $ -- любая, принадлежащая  $\delta_{\mathit M}$, находим, что {\em  тензор} $ \partial A(M)$, заданный в точке $ \mathit M$, {\em порождает}, в рассматриваемой окрестности точки $ \mathit M $, {\em векторное поле }
$$
\vec{ \mathit A}_{\mathit \partial A}(N) =   \partial A(M)^{T}\!\cdot\! \vec{\rho}\,(N) , \quad  N \in \delta_{\mathit M},               \eqno\ldots (2.2)
$$
являющееся линейной частью неоднородной составляющей поля $ \vec{ \mathit A}(x)$ в данной  $\delta_{M}$, в данный момент времени $t$ .

В связи с этим, (2.2) теперь рассматривается не как традиционное отображение $V\rightarrow V$  \cite{Rash},  а как отображение  $V\otimes V\,\rightarrow V$, то есть, если в качестве операнда в (2.2) традиционно выступает $\vec{\rho}(N)$, то теперь операндом в данном соотношении рассматривается тензор  $ \partial A(\mathit M)$, в результате чего, этому тензору, как объекту из $V\otimes V$, ставится в соответствие, посредством (2.2), геометрический объект в $V$, представленный векторным полем $ \vec{ \mathit A}_{\partial A}(N)$.

Геометрически, векторное поле, определяемое тензором $ \partial A(\mathit M)$ посредством (2.2), в целом можно представить, при соответствующих условиях на это поле, в виде конгруэнции векторных линий этого поля, наглядно демонстрирующей его структуру в рассматриваемой области.

Эту конгруэнцию предлагается рассматривать в качестве геометрического объекта, представляющего
 производный тензор  $ \partial A(\mathit M)$ в векторном пространстве $V$.

 Данное геометрическое представление непосредственно отвечает вышеуказанному прямому
 назначению тензора  $ \partial A(\mathit M)$ как градиента векторного поля  $ \vec{\mathit A}(x)$.

Ограничение малости  $\delta_{\mathit M}$  при этом можно снять, то есть, можно изображать этот геометрический объект в некоторой не бесконечно малой окрестности $\triangle_{\mathit M}$, но при этом иметь в виду, что при $\triangle_{\mathit M}\rightarrow\delta_{\mathit M}$ рассматриваемый образ тензора  $ \partial A(\mathit M)$ представляет собой конгруэнцию векторных линий соответствующей части векторного поля  $ \vec{\mathit A}(x)$ в бесконечно малой окрестности  $\delta_{\mathit M}$.

С другой стороны, при таком геометрическом представлении тензора  $ \partial A(\mathit M)$,  {\em компоненты } этого тензора, в некотором базисе  $e_{\mathit V}\otimes e_{\mathit V} $  пространства $V\otimes V$, приобретают следующее геометрическое содержание.

Так как, данные компоненты представлены частными производными, $\partial_{\alpha}A^{\beta}(\mathit M)$, являющимися, по их определению, производными по направлениям соответствующих векторов  базиса $e_{\mathit V}$, то они, определяя структуру неоднородной части векторного поля $ \vec{\mathit A}(x)$ в $\delta$-окрестности точки дифференцирования именно по этим направлениям, представляют, тем самым, структуру рассматриваемой {\em конгруэнции } по направлениям векторов данного базиса $e_{\mathit V}$.

В связи с этим, различие компонент в разных базисах означает наличие {\em анизотропии } в конгруэнции векторных линий данного геометрического образа тензора $ \partial A(\mathit M)$.

И так, в рамках определения тензора как геометрического объекта \cite{Ko}, имеем, что тензор   $ \partial A(\mathit M)$ -- это объект пространства  $V\otimes V$, представленный в $V$ рассматриваемой конгруэнцией векторных линий векторного поля определяемого этим тензором, компоненты которого в каждом
базисе $ \{\,\vec{e}_{\alpha}\,\otimes\,\vec{e}_{\beta} \}$ пространства  $V\otimes V$ характеризуют структуру данной конгруэнции (данного геометрического объекта) вдоль векторов базиса $ \{\vec{e}_{\alpha}\}$ пространства $V$.

Закон преобразования компонент данного тензора является непосредственным следствием инвариантности этого тензора как рассматриваемого геометрического объекта.

Наконец, важно подчеркнуть, что данный способ геометрического представления рассматриваемых тензоров второго ранга применим к тензорам любой симметрии, что позволяет {\em с единой } геометрической точки зрения оценить степень различия таких тензоров.

Это обстоятельство имеет принципиальное значение при решении вопроса о математическом различии магнитного и джейтонного полей, отличающихся, прежде всего, типом симметрии тензоров представляющих
 эти поля.

 В рамках терминологии качественной теории дифференциальных уравнений \cite{Nem} рассматриваемую конгруэнцию можно называть {\em фазовым портретом  тензора } $ \partial A(\mathit M)$, в силу того, что данная конгруэнция, как и конгруэнция векторных линий любого векторного поля, определяется решениями автономной нормальной системы дифференциальных уравнений (динамической системы)
 $$
 \frac{d\, \vec{r}}{d \,\tau} = \vec{A}\,_{\mathit \partial A}(\vec{r}\,),       \eqno\ldots (2.3)
 $$
 которая, в силу (2.2), принимает вид линейной однородной автономной нормальной системы дифференциальных уравнений (однородной линейной динамической системы)
$$
 \frac{d\, \vec{r}}{d \,\tau} = \partial A(\mathit M)^{T}\!\cdot  \vec{r}.       \eqno\ldots (2.4)
 $$

 Таким образом, наиболее естественным геометрическим объектом, представляющим тензор второго ранга $ \partial A(\mathit M)$, является фазовый портрет этого тензора, определяемый динамической системой (2.4).

 В свою очередь, воспользовавшись соотношением (1.20), формулу Тейлора (2.1) можно переписать в виде
\begin{multline*}
 \vec{ \mathit A}(N) = \vec{ \mathit A}_{0}(N) +  \mathcal{F}(\mathit M)^{T}\!\cdot\! \vec{\rho}\,(N) +\mathcal{G}_{d}(\mathit M) \!\cdot\! \vec{\rho}\,(N) +  \\
  \mathcal{G}_{\ast}(\mathit M)\!\cdot\! \vec{\rho}\,(N) +  \vec{0}\, (\rho), \quad  N \in \delta_{\mathit M}.               \qquad\ldots (2.5)
\end{multline*}

 Представление (2.5) прежде всего демонстрирует, что тензоры $ \mathcal{F}(\mathit M)$, $\mathcal{G}_{d}(\mathit M)$ и $\mathcal{G}_{\ast}(\mathit M)$  {\em равноправно }   участвуют в формировании линейной части структуры векторного поля $\vec{\mathit A}(\mathit N)$ в $\delta_{\mathit M}$.

 С другой стороны, в полной аналогии с теоремой Коши -- Гельмгольца о разложении векторного поля скоростей точек бесконечно малой области сплошной среды, соотношение (2.5), в его линейной части, можно рассматривать, в электродинамическом случае, как теорему о линейном разложении векторного потенциала электродинамики и говорить, что электродинамический потенциал $\vec{\mathit A}(\mathit N)$ в  любой точке $N\in \delta_{\mathit M}$, в каждый фиксированный момент времени $t$, представлен, в линейном приближении, суперпозицией векторного потенциала однородного поля $\vec{ \mathit A}_{0}(\mathit N) \stackrel{\delta_{\mathit M} }{\equiv}\vec{A}(\mathit M)$ и векторных потенциалов $\vec{A}_{\mathcal{F}}(\mathit N)$, $\vec{A}_{\mathcal{G}_{d}}(\mathit N)$ и $\vec{A}_{\mathcal{G}_{\ast}}(\mathit N)$, обусловленных наличием, в точке $\mathit M$, магнитного, девиатационного джейтонного и дилатационного джейтонного полей, определяемых, соответственно. тензорами $\mathcal{F}(\mathit M)$, $\mathcal{G}_{d}(\mathit M)$ и $\mathcal{G}_{\ast}(\mathit M)$.

 Уместно представить тут и теорему Коши -- Гельмгольца  \cite{Sed} в форме, соответствующей (2.5), демонстрируя возникающую теперь аналогию в поведении базовых тензорных полевых переменных классической электродинамики и классической механики сплошных сред,
 \begin{multline*}
   \vec{ v}(\mathit N) \doteq \vec{ v}_{0}(\mathit N) + \omega (\mathit M)^{T} \!\cdot\!\vec{\rho}(\mathit N) +e_{d}(\mathit M) \!\cdot\! \vec{\rho}(\mathit N) + \\
  e_{\ast}(\mathit M)\!\cdot\! \vec{\rho}(\mathit N),     \quad  N \in \delta_{\mathit M}.                              \qquad\ldots (2.6)
 \end{multline*}

 В (2.6) тензоры $\omega (\mathit M)$, $e_{d}(\mathit M)$ и $e_{\ast}(\mathit M)$ определены, соответственно, соотношениями, идентичными (1.6), (1.7) и (1.8), а $\delta_{\mathit M}$ представляет собой бесконечно малую {\em сопровождающую область } (или {\em индивидуальный объем }) сплошной среды, то есть область состоящую из одних и тех же частиц данной среды \cite{Sed}.

 Данная теорема, представленная соотношением (2.6), утверждает  \cite{Rash}, что скорость $ \vec{ v}(\mathit N) $ любой точки $N$ бесконечно малой области $\delta_{\mathit M}$ сплошной среды складывается, в линейном приближении, из скорости $ \vec{ v}_{0}(\mathit N)$ соответствующего поступательного движения этой области, скорости  $ \vec{ v}_{\omega}(\mathit N)$ вращательного движения  данной области как абсолютно твердого тела, скорости  $ \vec{ v}_{e_{d}}(\mathit N)$ чистой деформации этой области без изменения её объема и скорости $ \vec{ v}_{e_{\ast}}(\mathit N)$ чистой деформации данной области без изменения её формы.

 С другой стороны, динамические системы (2.8), (2.11) и (2.12), представленные ниже, позволяют рассматривать полевые электродинамические переменные  $\mathcal{F}(\mathit M)$,  $\mathcal{G}_{d}(\mathit M)$  и
  $\mathcal{G}_{\ast}(\mathit M)$  как тензоры, определяющие соответствующие движения {\em  сопровождающей области}  (или {\em сопровождающего "объема"\,}, или {\em индивидуального "объема"\,})  \cite{Nem} в   {\em фазовом } пространстве, придав вышеуказанной теореме о разложении векторного потенциала электродинамики {\em тесное  }
   соответствие с теоремой Коши -- Гельмгольца.

  В свою очередь (2.5) демонстрирует, что наличие тензора $\mathcal{F}(\mathit M)$,  в данной точке $\mathit M$ рассматриваемого векторного поля $\vec{A}(x)$, говорит о том, что в  $\delta_{\mathit M}$ существует неоднородное векторное поле  $\vec{A}_{\mathcal{F}}(\mathit N)$, определяемое соотношением
  $$
  \vec{A}_{\mathcal{F}}(\mathit N)  \eqdef  \mathcal{F}(\mathit M)^{T}\!\cdot\!\vec{\rho}(N),  \quad  N \in \delta_{\mathit M},               \eqno\ldots (2.7)
  $$
  конгруэнция векторных линий которого определяется фазовым портретом тензора $\mathcal{F}(\mathit M)$, то есть фазовым портретом однородной линейной динамической системы
 $$
 \frac{d\, \vec{r}}{d \,\tau} = \mathcal{F}(\mathit M)^{T}\!\cdot\! \vec{r}.       \eqno\ldots (2.8)
 $$

 Совершенно аналогично, из (2.5) находим так же, что наличие тензоров  $\mathcal{G}_{d}(\mathit M)$  и
  $\mathcal{G}_{\ast}(\mathit M)$,  в данной точке $\mathit M$ рассматриваемого векторного поля $\vec{A}(x)$, говорит о том, что в  $\delta_{\mathit M}$ существуют неоднородные векторные поля
   $\vec{A}_{\mathcal{G}_{d}}(\mathit N)$ и  $\vec{A}_{\mathcal{G}_{\ast}}(\mathit N)$, определяемые, соответственно,  соотношениями
   $$
  \vec{A}_{\mathcal{G}_{d}}(\mathit N)  \eqdef  \mathcal{G}_{d}(\mathit M) \!\cdot\!\vec{\rho}\,(N),  \quad  N \in \delta_{\mathit M},               \eqno\ldots (2.9)
  $$
 $$
  \vec{A}_{\mathcal{G}_{\ast}}(\mathit N)  \eqdef  \mathcal{G}_{\ast}(\mathit M) \!\cdot\!\vec{\rho}\,(N),  \quad  N \in \delta_{\mathit M},               \eqno\ldots (2.10)
  $$
  конгруэнции векторных линий которых определяются фазовыми портретами тензоров
   $\mathcal{G}_{d}(\mathit M)$  и   $\mathcal{G}_{\ast}(\mathit M)$, то есть, фазовыми портретами динамических
    систем
 $$
 \frac{d\, \vec{r}}{d \,\tau} = \mathcal{G}_{d}(\mathit M)\!\cdot\! \vec{r},       \eqno\ldots (2.11)
 $$
$$
 \frac{d\, \vec{r}}{d \,\tau} = \mathcal{G}_{\ast}(\mathit M)\!\cdot\! \vec{r}.       \eqno\ldots (2.12)
 $$

 Определим свойства каждого из векторных полей (2.7), (2.9) и (2.10) на предмет
 их соленоидальности и потенциальности.

Использование соотношения
$$
\vec{\bigtriangledown}\cdot(T\cdot\vec{r}\,) = T\cdot\cdot\,g = tr\,T,             \eqno\ldots (2.13)
$$
где $T$ -- постоянный тензор второго ранга, мгновенно приводит к равенствам
$$
div\,\vec{A}_{\mathcal{F}}(\vec{r}\,) = \vec{\bigtriangledown}\cdot(\mathcal{F}(\mathit M)^{T}\cdot\vec{r}\,) = tr\,\mathcal{F}(\mathit M)\equiv 0,                 \eqno\ldots (2.14)
$$
$$
div\,\vec{A}_{\mathcal{G}_{d }}(\vec{r}\,) = \vec{\bigtriangledown}\cdot(\mathcal{G}_{d}(\mathit M)\cdot\vec{r}\,) = tr\,\mathcal{G}_{d}(\mathit M) \equiv 0,                 \eqno\ldots (2.15)
$$
$$
div\,\vec{A}_{\mathcal{G}_{\ast }}(\vec{r}\,) = \vec{\bigtriangledown}\cdot(\mathcal{G}_{\ast}(\mathit M)\cdot\vec{r}\,) = tr\,\mathcal{G}_{\ast}(\mathit M) = tr\,\partial A(\mathit M) = div\,\vec{A}(\mathit M),                 \eqno\ldots (2.16)
$$
в любой точке $\mathit N\in \triangle_{\mathit M} $.

В свою очередь, воспользовавшись соотношением
$$
\vec{\bigtriangledown}\times(T\cdot\vec{r}\,) =  - \varepsilon\cdot\cdot\,T\,^{T},             \eqno\ldots (2.17)
$$
где $T$ -- так же постоянный тензор второго ранга, получаем
$$
rot\,\vec{A}_{\mathcal{F}}(\vec{r}\,) = \vec{\bigtriangledown}\times(\mathcal{F}(\mathit M)^{T}\cdot\vec{r}\,) =  - \varepsilon\cdot\cdot\,\mathcal{F}(\mathit M),             \eqno\ldots (2.18)
$$
из которого, с учетом (1.15), находим
$$
rot\,\vec{A}_{\mathcal{F}}(\vec{r}\,) = rot\,\vec{A}(\vec{r}\,).             \eqno\ldots (2.19)
$$

Аналогичное использование (2.17) для  $\vec{A}_{\mathcal{G}_{d}}(\mathit N)$ и  $\vec{A}_{\mathcal{G}_{\ast}}(\mathit N)$ дает
$$
rot\,\vec{A}_{\mathcal{G}_{d}}(\vec{r}\,) = \vec{\bigtriangledown}\times(\mathcal{G}_{d}(\mathit M)\cdot\vec{r}\,) =  - \varepsilon\cdot\cdot\,\mathcal{G}_{d}(\mathit M) \equiv \vec{0},             \eqno\ldots (2.20)
$$
$$
rot\,\vec{A}_{\mathcal{G}_{\ast}}(\vec{r}\,) = \vec{\bigtriangledown}\times(\mathcal{G}_{\ast}(\mathit M)\cdot\vec{r}\,) =  - \varepsilon\cdot\cdot\,\mathcal{G}_{\ast}(\mathit M) \equiv \vec{0}.             \eqno\ldots (2.21)
$$

Соотношения (2.14)--(2.16) и (2.19)--(2.21) демонстрируют, что фазовые портреты тензоров  $\mathcal{F}(\mathit M)$, $\mathcal{G}_{d}(\mathit M)$  и   $\mathcal{G}_{\ast}(\mathit M)$ представлены конгруэнциями векторных линий соленоидального (вихревого), лапласова (одновременно и соленоидального и потенциального) и потенциального (коллинеарного эйлеровому полю  \cite{Arn}) полей, соответственно.

С другой стороны, возвращаясь к (2.5) находим, что любое дифференцируемое векторное поле, в его неоднородной линейной части, представлено, в общем случае, суперпозицией соленоидального, лапласова и коллинеарного эйлеровому полей.

Потенциальные функции (или скалярные потенциалы) \cite{Fich} векторных полей  $\vec{A}_{\mathcal{G}_{d}}(\vec{r})$ и  $\vec{A}_{\mathcal{G}_{\ast}}(\vec{r})$ определены, соответственно,
$$
\Phi_{\mathcal{G}_{d}}(\vec{r}\,) = \frac{1}{2}\,\vec{r}\cdot\mathcal{G}_{d}(\mathit M)\cdot\vec{r} = \frac{1}{2}\,\mathcal{G}_{d}(\mathit M)\cdot\cdot \,\vec{r}\otimes\vec{r},        \eqno\ldots (2.22)
$$
$$
\Phi_{\mathcal{G}_{\ast}}(\vec{r}\,) = \frac{1}{2}\,\vec{r}\cdot\mathcal{G}_{\ast}(\mathit M)\cdot\vec{r} = \frac{1}{2}\,\mathcal{G}_{\ast}(\mathit M)\cdot\cdot \,\vec{r}\otimes\vec{r}.        \eqno\ldots (2.23)
$$

\section*{\S \,3 Фазовые портреты тензоров ранга (0,2) обладающих разным типом симметрии}
\addcontentsline{toc}{section}{\S \,3 Фазовые портреты тензоров ранга (0,2) обладающих разным типом симметрии}

\qquad Теперь определим конкретные виды конгруэнций векторных линий каждого из векторных полей,  $\vec{A}_{\mathcal{F}}(\vec{r})$,   $\vec{A}_{\mathcal{G}_{d}}(\vec{r})$ и  $\vec{A}_{\mathcal{G}_{\ast}}(\vec{r})$, то есть, конкретные виды фазовых портретов тензоров  $\mathcal{F}(\mathit M)$, $\mathcal{G}_{d}(\mathit M)$  и   $\mathcal{G}_{\ast}(\mathit M)$, представленных фазовыми портретами динамических систем (2.8), (2.11) и (2.12), соответственно.

При этом, тут и далее будем ограничиваться, без специальных оговорок, случаями динамических систем на плоскости \cite{Arr}, рассмотрение которых является достаточным для получения основных результатов в области физических приложений обсуждаемой теории.

Прежде всего следует отметить, что, в силу того, что тензоры  $\mathcal{F}(\mathit M)$, $\mathcal{G}_{d}(\mathit M)$  и   $\mathcal{G}_{\ast}(\mathit M)$ несингулярны, то есть, динамические системы (2.8), (2.11) и (2.12) являются простыми динамическими системами \cite{Arr}, точка $\mathit M$ является единственной изолированной особой точкой для каждой из этих систем. Это является очевидным так же и непосредственно из соотношений (2.7), (2.9) и (2.10).

Более того, простые динамические системы  (2.8) и (2.12), представленные в ковариантном виде, например, изначально являются каноническими \cite{Arr}, а динамическая система (2.11) становится таковой простым переходом к собственному базису тензора $\mathcal{G}_{d}(\mathit M)$.

Классификация возможных топологически неэквивалентных типов фазовых портретов простых линейных динамических систем на плоскости традиционно осуществляется, исторически следуя классификации А.~Пуанкаре, по характеру
 спектра, определяющего эту систему  \cite{Nem, Arr, Mat}.

 В   данном  контексте, классификация линейных динамических систем проводится в зависимости от типа симметрии классов тензоров второго ранга, в соответствии с поставленной физической задачей  и, в частности, в соответствии с задачей определения геометрических образов тензоров, входящих в систему (1.11)--(1.13).

 Наиболее ярко демонстрируется определяющее влияние свойств симметрии тензоров  $\mathcal{F}(\mathit M)$, $\mathcal{G}_{d}(\mathit M)$  и   $\mathcal{G}_{\ast}(\mathit M)$  на структуру соответствующих конгруэнций соответствующих векторных линий, в процессе получения уравнений векторных линий этих
  конгруэнций исходя из  {\em симметричных} форм нормальных динамических систем (2.8), (2.11) и (2.12).

  В связи с этим, рассмотрим ниже именно эти формы данных нормальных систем с последующим определением уравнений векторных линий, определяющих вид фазовых портретов рассматриваемых динамических систем.

  В прямоугольной декартовой системе координат симметричная форма динамической системы (2.8),
  $$
  \frac{d\,x}{A_{\mathcal{F}}^{x}(\mathit N)} =  \frac{d\,y}{A_{\mathcal{F}}^{y}(\mathit N)},         \eqno\ldots (3.1)
  $$
  после использования соотношений, следующих из (2.7),
  $$
 A_{\mathcal{F}}^{x}(\mathit N) = \mathcal{F}^{yx}(\mathit M)\,y,               \eqno\ldots (3.2)
  $$
  $$
 A_{\mathcal{F}}^{y}(\mathit N) = \mathcal{F}^{xy}(\mathit M)\,x,               \eqno\ldots (3.3)
  $$
  принимает вид
  $$
  \frac{d\,x}{\mathcal{F}^{yx}(\mathit M)\,y} =  \frac{d\,y}{\mathcal{F}^{xy}(\mathit M)\,x}.         \eqno\ldots (3.4)
  $$

  Последующее использование в (3.4) антисимметричности тензора $\mathcal{F}(\mathit M)$ приводит к дифференциальному уравнению
  $$
  \frac{d\,x}{y} = -  \frac{d\,y}{x},         \eqno\ldots (3.5)
  $$
  означающему, что фазовый портрет тензора  $\mathcal{F}(\mathit M)$, в $E^{\,2}$, представлен
   конгруэнцией векторных линий векторного поля $\vec{A}_{\mathcal{F}}(\mathit N)$, имеющих
   вид концентрических
  окружностей с общим центром в точке $\mathit M$.

  Таким образом, фазовый портрет антисимметричного тензора  $\mathcal{F}(\mathit M)$
   представлен {\em классическим центром }[13, 15--17], имеющим вид изображенный на рис. 3.1.

 \setcounter{section}{3}
	\setcounter{figure}{0}

\begin{figure}[!h]
\begin{center}
\vspace{-6pt}
\includegraphics[width=48mm]{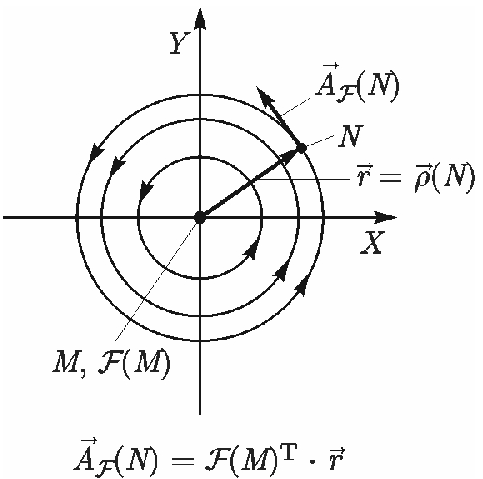}
\vspace{1pt}
\caption{Фазовый портрет антисимметричного тензора ${\mathcal F}(M)$  в евклидовом пространстве $E^{2}.$}
\end{center}
\end{figure}

Направление векторных линий определяется стандартными методами качественной
теории дифференциальных уравнений (см., например, \cite{Mat}) и тут не обсуждается.

Так как, в случае евклидова пространства $E\,^{3}$, антисимметричный тензор
 второго ранга, $\mathcal{F}(\mathit M)$, связан с сопровождающим его
  псевдовектором  $\vec{\mathcal{H}}(\mathit M)$ легко проверяемым соотношением
$$
\mathcal{F}(\mathit M)\cdot\vec{a} = \vec{a}\times \vec{\mathcal{H}}(\mathit M),               \eqno\ldots (3.6)
$$
где $\vec{a}$ -- произвольный вектор, то (2.7) можно переписать, в этом случае, так же в виде соотношения
$$
  \vec{A}_{\mathcal{F}}(\mathit N)  =\vec{\mathcal{H}}(\mathit M)\times \vec{r},    \quad  N \in \delta_{\mathit M},               \eqno\ldots (3.7)
$$
которое, при заданном $\vec{\mathcal{H}}(\mathit M)$, так же позволяет определить
направление векторных линий фазового портрета данного тензора  $\mathcal{F}(\mathit M)$.

В свою очередь, (3.7) позволяет переписать уравнение (2.8) в форме
$$
 \frac{d\, \vec{r}}{d \,\tau} = \vec{\mathcal{H}}(\mathit M)\times \vec{r},       \eqno\ldots (3.8)
 $$
 соответствующей формуле Эйлера, определяющей закон распределения мгновенных
 скоростей точек   абсолютно твердого тела, обусловленных вращательным движением этого тела.

 В результате, если рассматривать решения динамической системы (2.8), как уравнения движения
 индивидуальных точек в фазовом пространстве по их фазовым кривым, то (3.8) будет означать, что любая сопровождающая область (или индивидуальный "объем"\,) \cite{Nem, Sed} в фазовом   пространстве движется (вращается) как абсолютно   твердое тело, при этом, $\vec{\mathcal{H}}(\mathit M)$ будет выступать   в качестве  угловой скорости вращательного движения этой   области, а $ \vec{A}_{\mathcal{F}}(\vec{r}\,)$ -- в качестве соответствующей мгновенной линейной  скорости точки данной области.

Таким образом  можно "вдохнуть\," \, кинематический смысл в любое непрерывно
 дифференцируемое векторное поле $ \vec{A}_{\mathcal{F}}(\vec{r}\,)$, не обязательно являющееся
 векторным полем скоростей точек некоторой сплошной среды.

 При таком подходе, псевдовектор   $\vec{\mathcal{H}}(\mathit M)$, сопровождающий
 тензор $\mathcal{F}(\mathit M)$ в   $E\,^{3}$, выступает в качестве соответствующей характеристики
 вышеуказанного вращательного движения рассматриваемой сопровождающей области
  фазового пространства, тем самым, одновременно устанавливается и иерархия отношений данных математических объектов.

Симметричная форма динамической системы (2.11), в декартовой системе
 координат, имеет вид, аналогичный (3.1),
  $$
  \frac{d\,x}{A_{\mathcal{G}_{d}}^{x}(\mathit N)} =  \frac{d\,y}{A_{\mathcal{G}_{d}}^{y}(\mathit N)}.         \eqno\ldots (3.9)
  $$

  Выбрав декартову систему координат, орты которой представлены собственным базисом тензора $\mathcal{G}_{d}(\mathit M)$, из (2.9) теперь имеем
  $$
 A_{\mathcal{G}_{d}}^{x}(\mathit N) = \mathcal{G}_{d}^{xx}(\mathit M)\,x,               \eqno\ldots (3.10)
  $$
  $$
 A_{\mathcal{G}_{d}}^{y}(\mathit N) = \mathcal{G}_{d}^{yy}(\mathit M)\,y.               \eqno\ldots (3.11)
  $$

В результате, (3.9), в такой системе координат, принимает вид соотношения
$$
  \frac{d\,x}{\mathcal{G}_{d}^{xx}(\mathit M)\,x} =  \frac{d\,y}{\mathcal{G}_{d}^{yy}(\mathit M)\,y},         \eqno\ldots (3.12)
$$
которое, после учета равенства $ tr\,\mathcal{G}_{d}(\mathit M) = 0 $, приводит к
 дифференциальному уравнению
 $$
  \frac{d\,x}{x} =  - \frac{d\,y}{y},         \eqno\ldots (3.13)
 $$
означающему, что фазовый портрет тензора   $\mathcal{G}_{d}(\mathit M)$  представлен,
 в   $E\,^{2}$, конгруэнцией векторных линий векторного поля $\vec{A}_{G_{d}}(N)$, каждая из которых
 является равносторонней гиперболой или её вырождением.

Таким образом, фазовый портрет симметричного бесследового тензора  $\mathcal{G}_{d}(\mathit M)$
 представлен {\em классическим седлом} (см., например, \cite{Arr, Ob}\,), имеющим вид,
  изображенный на рис. 3.2 {\em a})\,.

 \begin{figure}[!h]
\begin{center}
\vspace{-6pt}
\includegraphics[width=128mm]{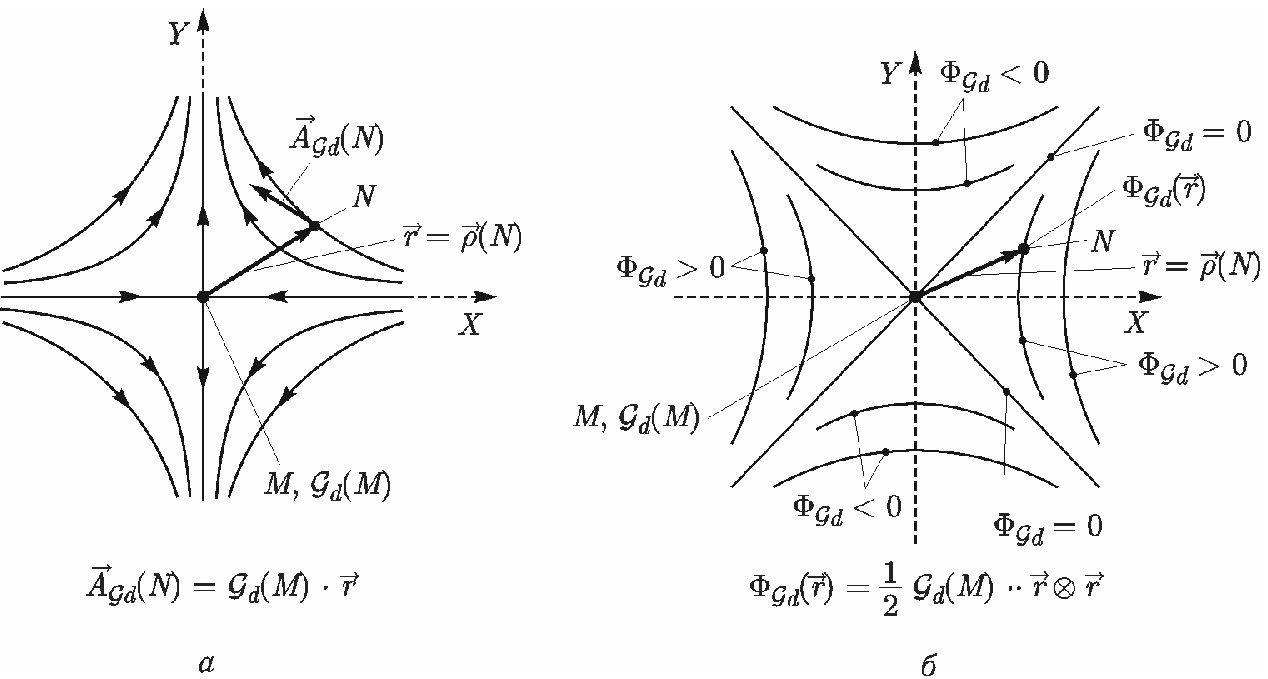}
\vspace{1pt}
\caption{Фазовый портрет симметричного тензора ${\mathcal G}_{d}(M)$ и сопровождающий его континуум "поверхностей"\, Коши (эквипотенциальных "поверхностей"\,) в евклидовом пространстве  $E^{2}.$}
\end{center}
\end{figure}

Сопровождающим геометрическим объектом  симметричного тензора  $\mathcal{G}_{d}(\mathit M)$
теперь является континуум поверхностей Коши (или эквипотенциальных поверхностей),
 потенциал каждой из которых определен соотношением (2.22), а сами эквипотенциальные
  поверхности представлены, в $E\,^{2}$, конгруэнцией эквипотенциальных линий, каждая
   из которых так же является равносторонней гиперболой или её вырождением (рис. 3.2~{\em б})\,).

Действительно, уравнение, определяющее некоторую эквипотенциальную поверхность потенциала (2.22),
$$
\Phi_{\mathcal{G}_{d}}(\vec{r}\,) =  \frac{1}{2}\,\mathcal{G}_{d}(\mathit M)\cdot\cdot \,\vec{r}\otimes\vec{r} = C,        \eqno\ldots (3.14)
$$
в вышеуказанной системе координат, принимает следующий вид
$$
\mathcal{G}_{d}^{xx}(\mathit M)\,x^{2} + \mathcal{G}_{d}^{yy}(\mathit M)\,y^{2} = 2C,           \eqno\ldots (3.15)
$$
который, после учета равенства  $ tr\,\mathcal{G}_{d}(\mathit M) = 0 $, приводит к уравнению
$$
x\,^{2} - y\,^{2} = \frac{2C}{\mathcal{G}_{d}^{xx}(\mathit M)},       \eqno\ldots (3.16)
$$
являющемуся уравнением равносторонней гиперболы, асимптоты которой представлены
 биссектрисами прямых углов, образованных сепаратрисами рассматриваемого седла.

Симметричная форма динамической системы (2.12), в декартовой системе координат,
 представлена соотношением
$$
  \frac{d\,x}{A_{\mathcal{G}_{\ast}}^{x}(\mathit N)} =  \frac{d\,y}{A_{\mathcal{G}_{\ast}}^{y}(\mathit N)}.    \eqno\ldots (3.17)
$$

В силу определений (1.8) и (1.10), тензор  $\mathcal{G}_{\ast}(\mathit M)$  имеет вид
$$
\mathcal{G}_{\ast}(\mathit M) = \frac{div\,\vec{A}(M)}{tr g}\,g,              \eqno\ldots  (3.18)
$$
означающий, что
$$
\mathcal{G}_{\ast}^{xy}(\mathit M) = 0 = \mathcal{G}_{\ast}^{yx}(\mathit M),      \eqno\ldots  (3.19)
$$
вследствие чего (2.10) приводит к равенствам
$$
 A_{\mathcal{G}_{\ast}}^{x}(\mathit N) = \mathcal{G}_{\ast}^{xx}(\mathit M)\,x,               \eqno\ldots (3.20)
$$
$$
 A_{\mathcal{G}_{\ast}}^{y}(\mathit N) = \mathcal{G}_{\ast}^{yy}(\mathit M)\,y.               \eqno\ldots (3.21)
$$
подстановка которых в (3.17) приводит данную симметричную форму к виду
$$
  \frac{d\,x}{\mathcal{G}_{\ast}^{xx}(\mathit M)\,x} =  \frac{d\,y}{\mathcal{G}_{\ast}^{yy}(\mathit M)\,y}.     \eqno\ldots (3.22)
$$

В свою очередь, (3.22), после учета соотношения
 $$
\mathcal{G}_{\ast}^{xx}(\mathit M) =  \mathcal{G}_{\ast}^{yy}(\mathit M) = \frac{div\,\vec{A}(M)}{tr g},      \eqno\ldots  (3.23)
$$
приводит к дифференциальному уравнению
$$
  \frac{d\,x}{x} =  \frac{d\,y}{y},         \eqno\ldots (3.24)
$$
которое означает, что фазовый портрет симметричного тензора  $\mathcal{G}_{\ast}(\mathit M)$  представлен, в  $E\,^{2}$, {\em дикритическим } ({\em звездным}) {\em  узлом} (см., например, \cite{Arr}).

Пример такого фазового портрета, представленного неустойчивым дикритическим узлом, дан на рис. 3.3 {\em а})\,.
\\

\begin{figure}[!h]
\begin{center}
\vspace{-6pt}
\includegraphics[width=91mm]{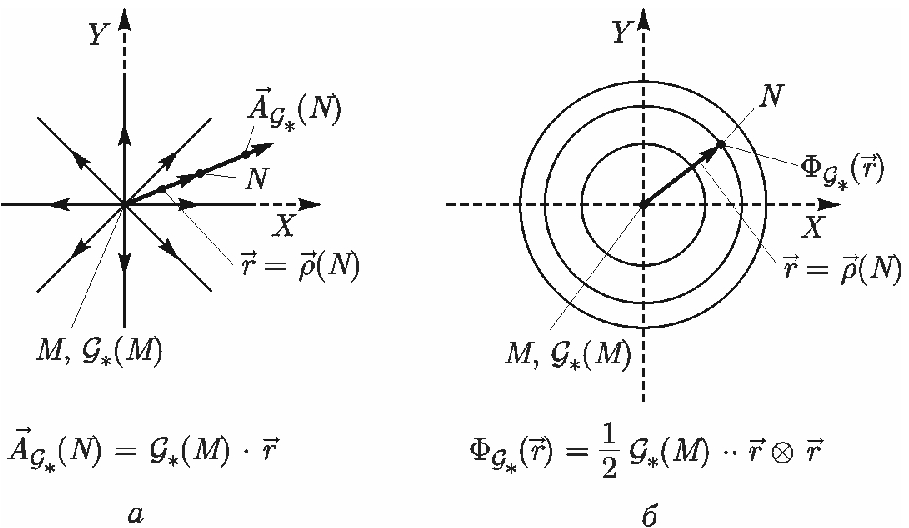}
\vspace{1pt}
\caption{Фазовый портрет симметричного тензора ${\mathcal G}_{*}(M)$ при $div\; {\vec A}(M) > 0$ и сопровождающий его континуум "поверхностей"\, Коши (или эквипотенциальных "поверхностей"\,) в евклидовом пространстве  $E^{2}.$}
\end{center}
\end{figure}

Сопровождающим геометрическим объектом симметричного тензора $\mathcal{G}_{\ast}(\mathit M)$  является континуум поверхностей Коши (или эквипотенциальных поверхностей), потенциал каждой из которых определен соотношением (2.23), а сами эти поверхности представлены, в $E\,^{2}$, конгруэнцией эквипотенциальных линий в виде континуума концентрических окружностей с общим центром в точке $\mathit M$  (рис. 3.3\,{\em б})).

Действительно, уравнение, определяющее некоторую эквипотенциальную поверхность потенциала (2.23),
$$
\Phi_{\mathcal{G}_{\ast}}(\vec{r}\,) = \frac{1}{2}\,\mathcal{G}_{\ast}(\mathit M)\cdot\cdot \,\vec{r}\otimes\vec{r} = C,          \eqno\ldots (3.25)
$$
в рассматриваемой системе координат, принимает вид, аналогичный (3.15), то есть,
$$
\mathcal{G}_{\ast}^{xx}(\mathit M)\,x\,^{2} + \mathcal{G}_{\ast}^{yy}(\mathit M)\,y\,^{2} = 2C.           \eqno\ldots (3.26)
$$

Последующий учет соотношения (3.23) сводит (3.26) к уравнению
$$
x\,^{2} + y\,^{2} = \frac{2\mathit C\,tr g}{div\,\vec{A}(\mathit M)},        \eqno\ldots (3.27)
$$
являющимся уравнением окружности с центром в точке $\mathit M$.

К аналогичному результату можно так же легко придти и без обращения к некоторой системе координат.

Действительно, согласно (3.18), имеем соотношения
$$
\mathcal{G}_{\ast}(\mathit M)\cdot\cdot \,\vec{r}\otimes\vec{r} = \frac{div\,\vec{A}(M)}{tr g}g\,\cdot\cdot\,\vec{r}\otimes\vec{r} = \frac{div\,\vec{A}(M)}{tr g}\,r\,^{2},           \eqno\ldots (3.28)
$$
использование которых в (3.25) мгновенно приводит к уравнению
$$
r\,^{2} =  \frac{2\mathit C\,tr g}{div\,\vec{A}(\mathit M)},        \eqno\ldots (3.29)
$$
соответствующему (3.27).

Данный результат заранее очевиден, так как структура векторных линий дикритического узла говорит о том, что эквипотенциальные линии (по-прежнему рассматриваем векторное поле в евклидовом пространстве $E\,^{2}$) представлены концентрическими окружностями с общим центром в точке $\mathit M$.

В свою очередь, (2.23), после учета (3.28), принимает вид
$$
\Phi_{\mathcal{G}_{\ast}}(\vec{r}\,) =\frac{div\,\vec{A}(M)}{2\,tr g}\,r\,^{2},           \eqno\ldots (3.30)
$$
из которого следует, что знак потенциала $\Phi_{\mathcal{G}_{\ast}}(\vec{r}\,)$ определяется знаком $div\,\vec{A}(M)$,
в связи с чем, константа $C$ в (3.25) положительна при $div\,\vec{A}(M)\,>\,0$ и отрицательна при $div\,\vec{A}(M)\,<\,0$, что приводит к согласованности правой и левой частей уравнения(3.29).

Как в случае тензора  $\mathcal{G}_{d}(\mathit M)$, так и в случае тензора  $\mathcal{G}_{\ast}(\mathit M)$,
 конгруэнции их эквипотенциальных линий (или эквипотенциальных поверхностей,
  в случае  $E\,^{3}$) можно называть {\em потенциальными} портретами данных тензоров.

Фазовые портреты тензоров  $\mathcal{F}(\mathit M)$ и  $\mathcal{G}_{\ast}(\mathit M)$ можно так же легко получить и инвариантной процедурой (без обращения к соответствующей системе координат и к соответствующим симметричным формам нормальных динамических систем), то есть,  как и в основной части настоящего контекста, отдать предпочтение использованию тензорного языка \cite{Gil}.

Действительно, рассмотрим некоторое произвольное решение $\vec{r}\,(\tau)$ динамической
 системы (2.8),  определяющее соответствующую векторную линию $\mathfrak{L}_{\mathcal{F}}$ векторного поля (2.7).

 Тогда, для любой точки этой линии, для любого значения параметра $\tau\,\in\,I$, будем иметь "удовлетворение"\, уравнения (2.8) на этом решении, то есть,
 $$
 \frac{d\, \vec{r}\,(\tau)}{d \,\tau} = \mathcal{F}(\mathit M)^{T}\!\cdot\, \vec{r}\,(\tau),   \quad \tau\,\in\,I.     \eqno\ldots (3.31)
 $$

 Умножая (3.31) скалярно на данное решение  $\vec{r}\,(\tau)$, получаем
 $$
\vec{r}\,(\tau)\cdot \frac{d\, \vec{r}\,(\tau)}{d \,\tau} = \vec{r}\,(\tau)\cdot\mathcal{F}(\mathit M)^{T}\,\cdot \vec{r}\,(\tau),   \quad \tau\,\in\,I.     \eqno\ldots (3.32)
 $$

 Но, в силу антисимметричности тензора  $\mathcal{F}(\mathit M)$, правая часть (3.32) тождественно  обращается в нуль, вследствие чего (3.32) принимает вид
  $$
\vec{r}\,(\tau)\cdot \frac{d\, \vec{r}\,(\tau)}{d \,\tau} = 0,   \quad \tau\,\in\,I,     \eqno\ldots (3.33)
 $$
 означающий, что рассматриваемая векторная линия  $\mathfrak{L}_{\mathcal{F}}$ является окружностью с центром в точке $\mathit M$.

 Таким образом, и таким способом явно демонстрируется, что {\em антисимметрия} тензора второго ранга $\mathcal{F}(\mathit M)$ однозначно определяет вид фазового портрета этого тензора.

 Вид фазового портрета тензора $\mathcal{G}_{\ast}(\mathit M)$ так же просто определяется непосредственно из уравнения (2.12).

 Действительно, как и выше, рассмотрим некоторое произвольное решение $\vec{r}\,(\tau)$
 динамической системы (2.12), определяющее соответствующую векторную
  линию $\mathfrak{L}_{\mathcal{G}_{\ast}}$ векторного поля (2.10).

Тогда, для любой точки этой линии, для любого значения $\tau\in I$, будем иметь, согласно (2.12),
  $$
 \frac{d\, \vec{r}\,(\tau)}{d \,\tau} = \mathcal{G}_{\ast}(\mathit M)\cdot \vec{r}\,(\tau),   \quad \tau\,\in\,I.     \eqno\ldots (3.34)
 $$

 Дальнейшее использование "специфического"\, вида тензора $\mathcal{G}_{\ast}(\mathit M)$, определяемого соотношением (3.18), позволяет представить (3.34) в виде
  $$
 \frac{d\, \vec{r}\,(\tau)}{d \,\tau} =  \frac{div\,\vec{A}(\mathit M)}{tr g}\, \vec{r}\,(\tau),   \quad \tau\,\in\,I,     \eqno\ldots (3.35)
 $$
 означающим "звездность"\, фазового портрета тензора $\mathcal{G}_{\ast}(\mathit M)$.

Одновременно, (3.35) демонстрирует, что устойчивость или неустойчивость данного звездного узла  \cite{Arr} определяется знаком $div\,\vec{A}(\mathit M)$.

С другой стороны, (2.10) в итоге принимает вид
$$
\vec{A}_{\mathcal{G}_{\ast}}(\vec{r}) =  \frac{div\,\vec{A}(\mathit M)}{tr g}\vec{r},     \eqno\ldots (3.36)
$$
демонстрирующий, что векторное поле, порождаемое тензором  $\mathcal{G}_{\ast}(\mathit M)$, является коллинеарным
эйлеровому векторному полю, специфическая отличительная черта которого, по отношению к кулоновскому векторному полю точечного источника, состоит в том, что с удалением от его "источника"\, абсолютные значения этого поля возрастают, а  "густота"\,  векторных линий данного векторного поля при этом уменьшается.

Сопоставление полученных фазовых портретов тензоров  $\mathcal{F}(\mathit M)$, $\mathcal{G}_{d}(\mathit M)$  и   $\mathcal{G}_{\ast}(\mathit M)$, представленных на рис. 3.1--3.3, наглядно демонстрирует как {\em кардинальное } различие, так и локальную линейную независимость этих тензоров по отношению к друг другу,  поэтому, нельзя рассматривать поле, представленное одним из этих тензоров, в качестве составляющей поля, представленного любым другим из них.

Данное обстоятельство является особенно важным в электродинамическом случае, когда тензорами $\mathcal{F}(\mathit M)$, $\mathcal{G}_{d}(\mathit M)$  и   $\mathcal{G}_{\ast}(\mathit M)$ представлены, соответственно, магнитное, девиатационное джейтонное и дилатационное джейтонное поля.

С другой стороны, в завершение анализа соотношения (2.5), констатируем, что, в силу взаимно однозначного соответствия между рассматриваемыми тензорами и их фазовыми портретами, наличие того или иного тензора в некоторой точке $\mathit M$ евклидова пространства означает наличие, в $\delta$-окрестности этой точки, векторного поля, представленного фазовым портретом данного тензора, и обратно, наличие в $\delta_{\mathit M}$ векторного поля, обладающего структурой фазового портрета одного из рассматриваемых тензоров, означает наличие в точке $\mathit M$ этого тензора.

\section*{\S \,4 Конгруэнтные разложения векторных полей в евклидовом пространстве $E\,^{2}$}
\addcontentsline{toc}{section}{\S \,4 Конгруэнтные разложения векторных полей в евклидовом пространстве $E\,^{2}$}

\qquad Вышеуказанное соответствие позволяет {\em  аналитическому} разложению (1.20) поставить, во взаимно однозначное соответствие, {\em  геометрическое} разложение произвольной конгруэнции векторных линий на локально линейно независимые составляющие конгруэнции векторных линий векторных полей, определяемых тензорами правой части (1.20).

Подобные геометрические разложения векторных полей, в которых элементами разложения являются кнгруэнции векторных линий соответствующих полей. в дальнейшем будем называть  {\em конгруэнтными} разложениями этих ролей.

В итоге, в соответствии с (2.5), конгруэнтное разложение произвольного  дифференцируемого
плоского векторного поля $\vec{A}(\mathit N), \mathit N\in \delta_{\mathit M}$, содержащего, в общем случае, и однородную составляющую данного поля, $\vec{A_{0}}(\mathit N)$, качественно имеет вид, представленный на рис. 4.1.
\\

\setcounter{section}{4}
\setcounter{figure}{0}

\begin{figure}[!h]
\begin{center}
\vspace{-6pt}
\includegraphics[width=135mm]{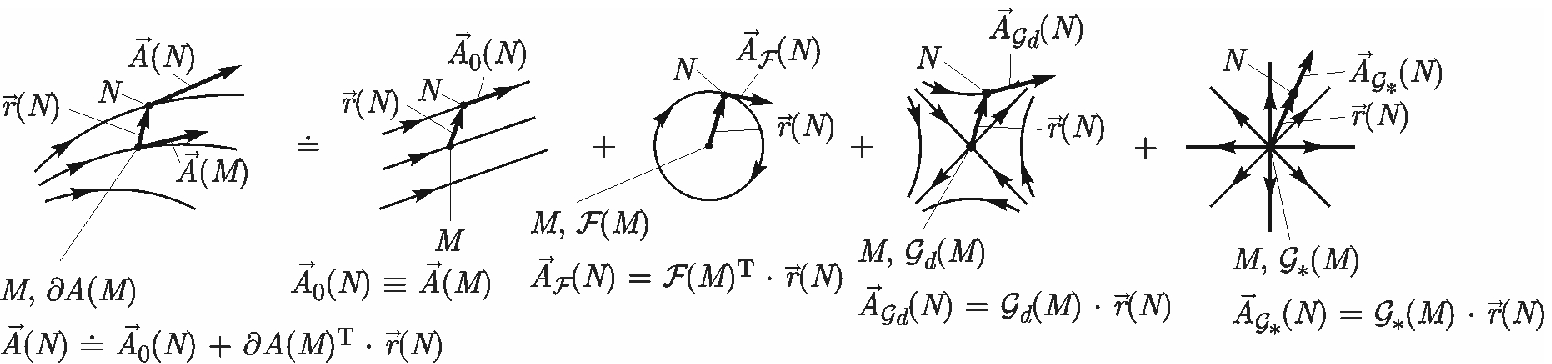}
\vspace{1pt}
\caption{Качественная картина конгруэнтного разложения векторного поля ${\vec A}(N)$, $N \in \delta_{M}$, на составляющие базисные конгруэнции в евклидовом пространстве $E^{2}.$}
\end{center}
\end{figure}
Рис. 4.1 демонстрирует, что, в $\delta_{\mathit M}$, конгруэнция векторных линий любого дифференцируемого векторного поля  $\vec{A}(\mathit N)$ представлена, в линейном приближении, в общем случае, суперпозицией четырех базовых конгруэнций, представленных центром, классическим седлом, дикритическим узлом и конгруэнцией векторных линий однородного векторного поля, определяемого значением данного векторного поля в точке $\mathit M$.

То есть, совокупность данных четырех топологических объектов представляет собой топологический базис в классе объектов, представленных конгруэнциями векторных линий дифференцируемых  векторных полей  в  $\delta$-окрестности рассматриваемой точки $\mathit M$ евклидова пространства $E\,^{2}$.

При этом имеем, что в качестве "основных"\, объектов рассматриваемого топологического базиса выступают фазовые портреты тензоров $\mathcal{F}(\mathit M)$, $\mathcal{G}_{d}(\mathit M)$  и   $\mathcal{G}_{\ast}(\mathit M)$, что вновь демонстрирует   {\em  фундаментальную} роль  {\em  всех} этих тензоров в рассматриваемых математических процессах.

Конкретные примеры качественных конгруэнтных разложений конкретных видов векторных полей представлены на рис.~4.2--4.5.

Всюду на этих рисунках рассматриваются конгруэнтные разложения в евклидовом пространстве $E\,^{2}$ (или в соответствующих плоскостях евклидова пространства $E\,^{3}$) в пределах $\delta$-окрестности
 заданной точки  $\mathit M$ этого пространства.

 В общем случае изменяющегося во времени векторного поля,  $\vec{A}(\vec{r},t)$, рассматриваемое разложение этого поля соответствует некоторому фиксированному моменту времени $t$, то есть, время  $t$ выступает, в этом случае, как параметр, фиксирующий вид конгруэнции векторных линий этого поля  в данный момент времени.
\\
\\
\begin{figure}[!h]
\begin{center}
\vspace{-6pt}
\includegraphics[width=125mm]{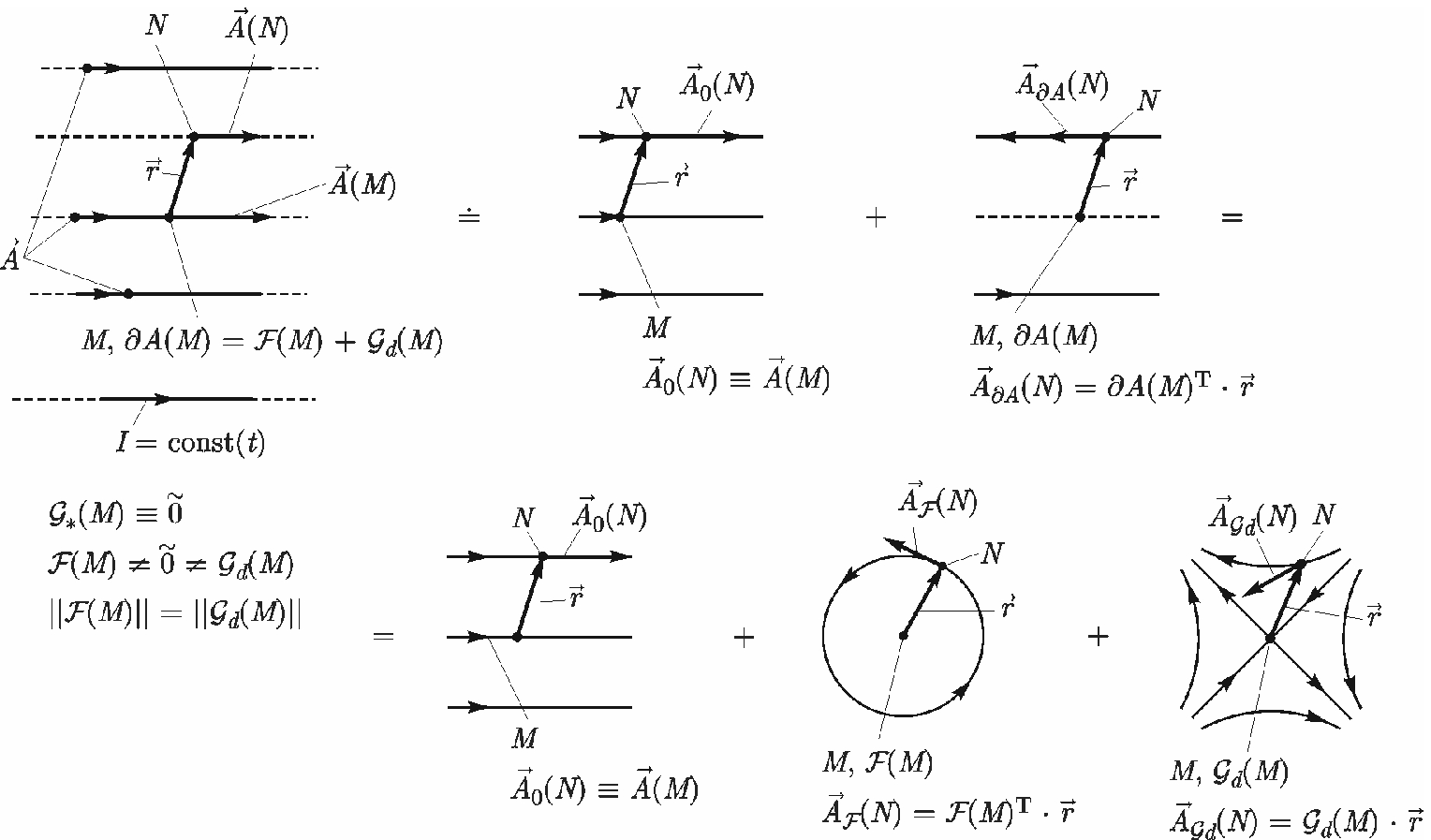}
\vspace{1pt}
\caption{Конгруэнтное разложение векторного поля векторного потенциала прямолинейного бесконечно длинного неподвижного проводника с постоянным током.}
\end{center}
\end{figure}

\begin{figure}[!h]
\begin{center}
\vspace{-6pt}
\includegraphics[width=135mm]{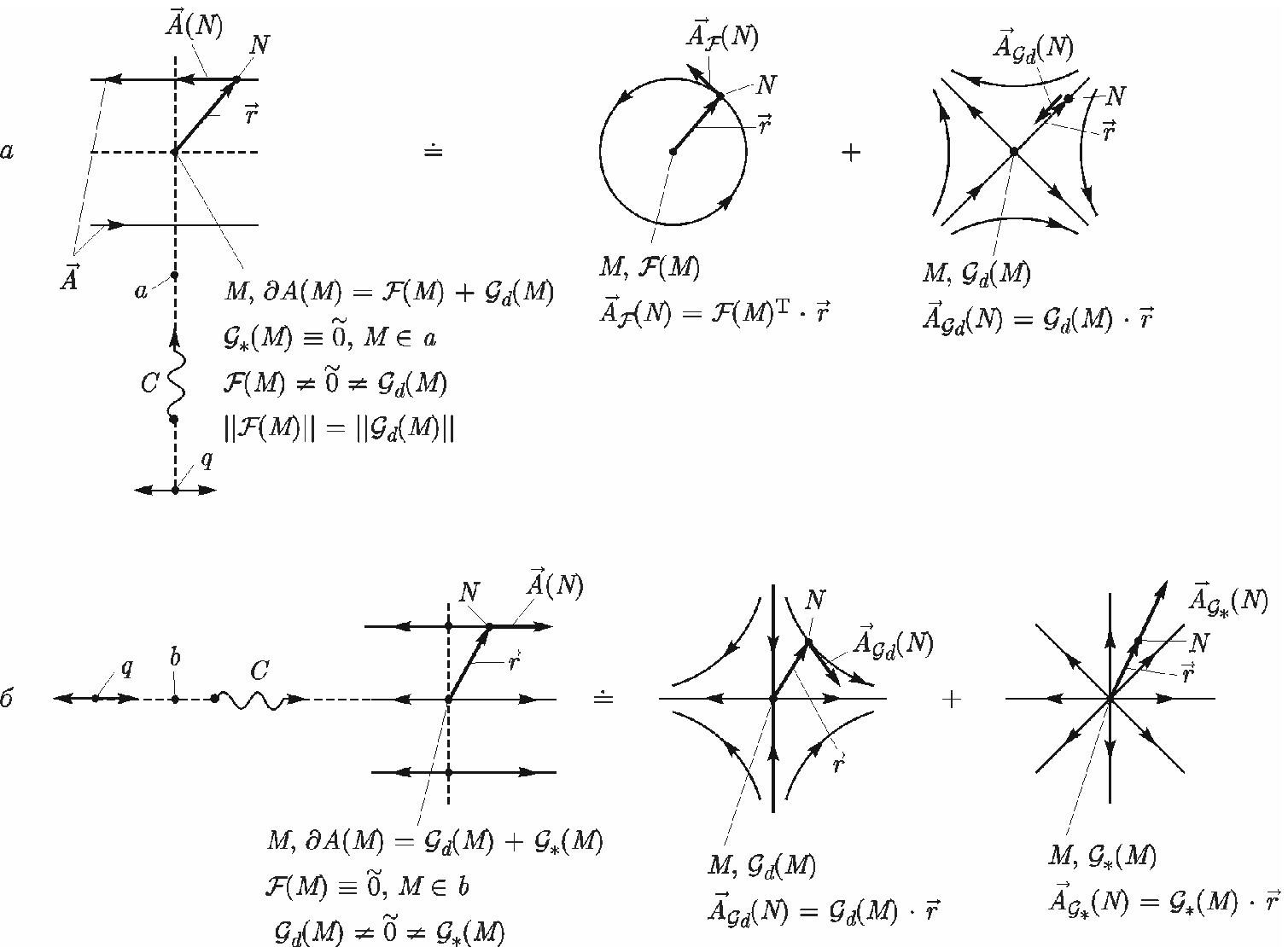}
\vspace{1pt}
\caption{Конгруэнтные разложения векторного поля векторного потенциала Лиенара\,--\,Вихерта~[5] в меридиональной плоскости линейно вибрирующей точечной электрически заряженной частицы (поперечное и продольное направления волновой зоны). Штрихованными линиями обозначены особые линии соответствующих непростых динамических систем~[15]. Случай, когда точка $M$ находится вне особой линии, сводится лишь к дополнительному учёту соответствующего однородного поля.}
\end{center}
\end{figure}

\begin{figure}[!h]
\begin{center}
\vspace{-6pt}
\includegraphics[width=135mm]{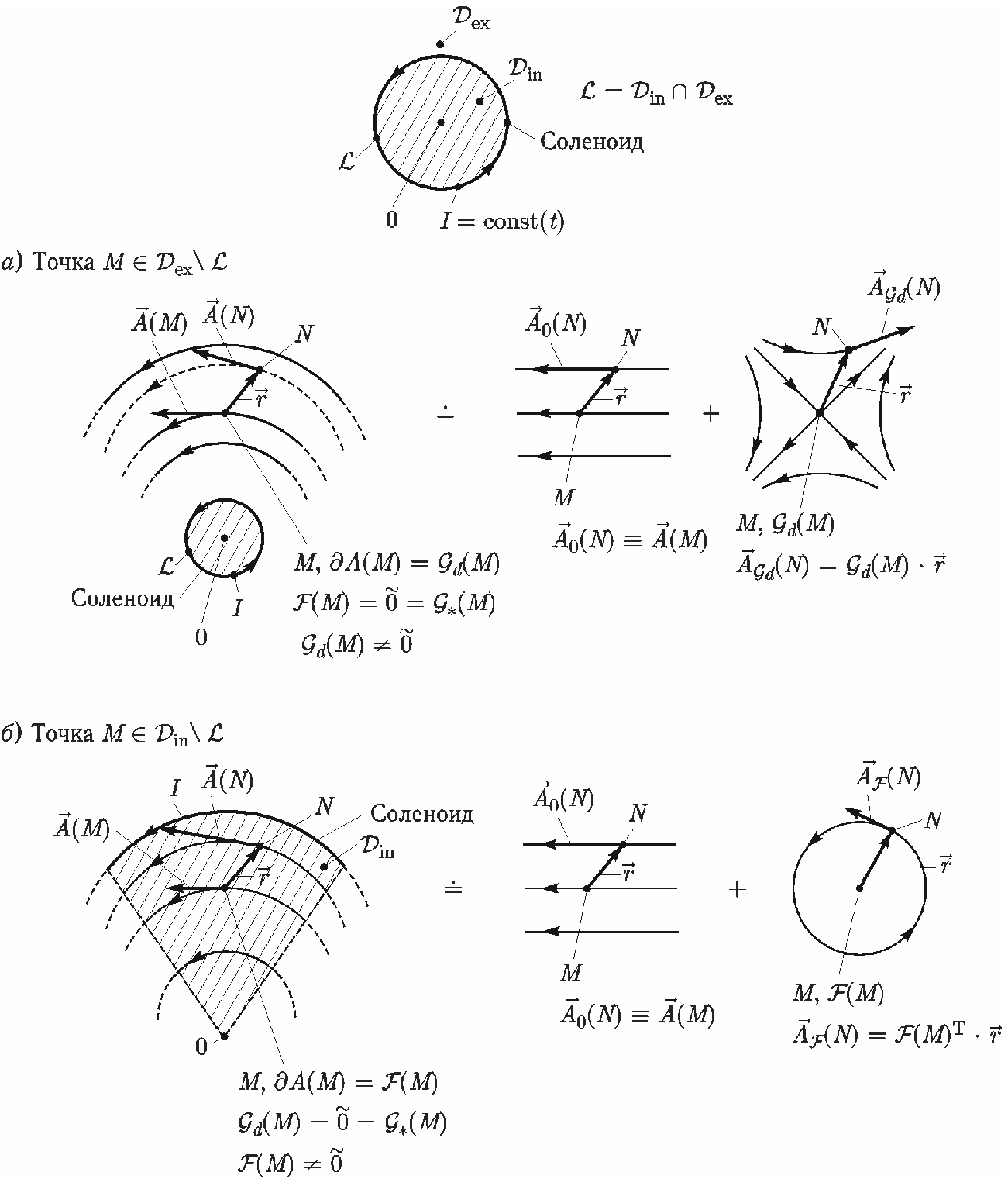}
\vspace{1pt}
\caption{Конгруэнтные разложения векторного поля векторного потенциала неподвижного бесконечно длинного цилиндрического соленоида с постоянным током. Индексы, отличающие векторные потенциалы вне и внутри соленоида, в целях упрощения записей, опущены. Рассматриваемая плоскость нормальна к оси соленоида.}
\end{center}
\end{figure}

\begin{figure}[!h]
\begin{center}
\vspace{-6pt}
\includegraphics[width=135mm]{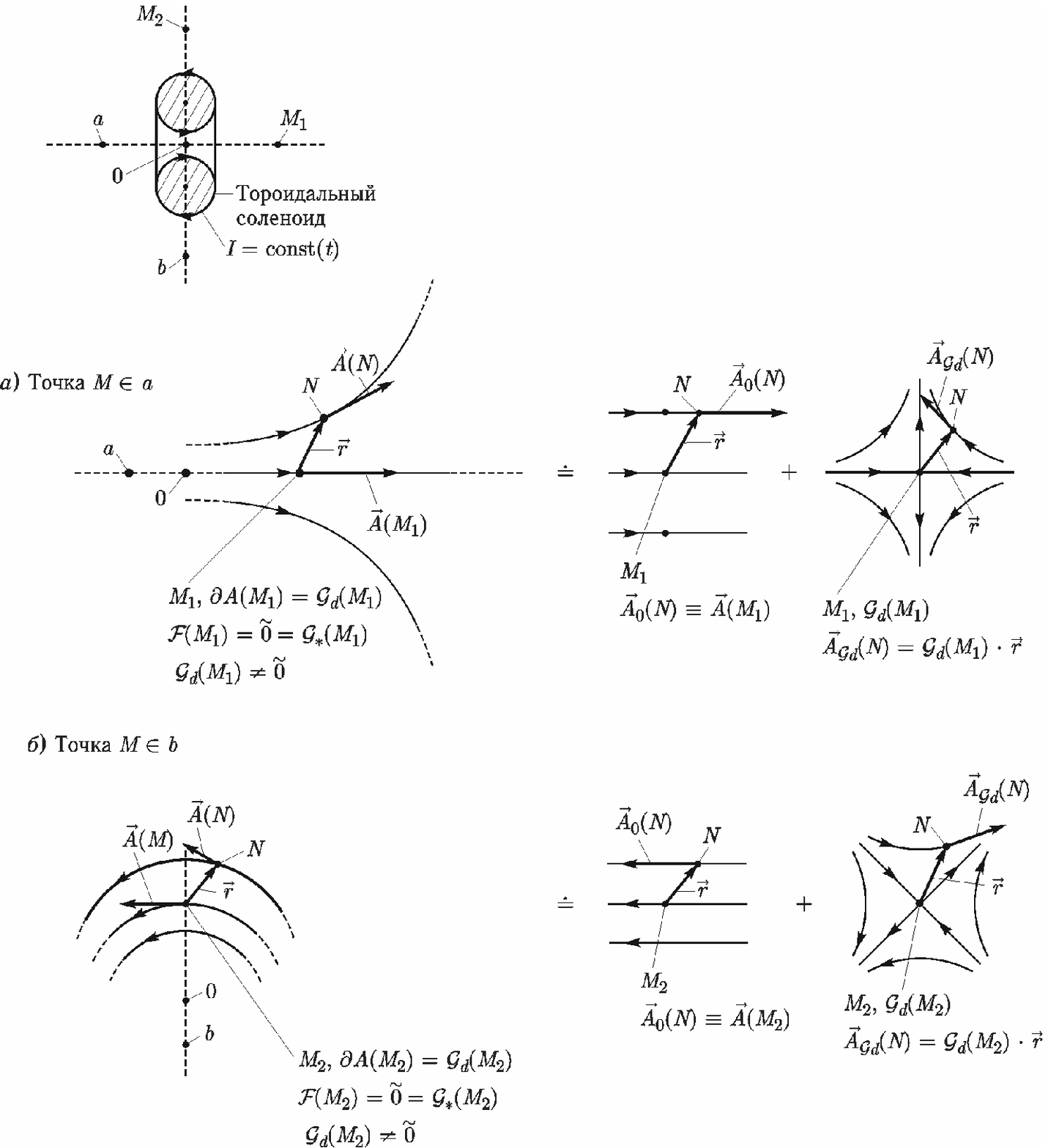}
\vspace{1pt}
\caption{Конгруэнтные разложения векторного поля векторного потенциала внешней области меридиональной плоскости неподвижного тороидального соленоида с постоянным током.}
\end{center}
\end{figure}
Рис. 4.2 и рис. 4.3, прежде всего, демонстрируют, что рассматриваемые векторные поля,
представленные непростыми (сингулярными) динамическими системами  \cite{Arr},
раскладываются на сумму векторных полей, описывающихся  {\em простыми} динамическими системами.

В этом отношении, рис. 4.4 и рис. 4.5, в свою очередь, демонстрируют, что рассматриваемые
 поля векторного потенциала цилиндрического и тороидального соленоидов "изначально"\,
 представлены несингулярными динамическими системами.

С другой стороны, рис.4.2 наглядно демонстрирует, что рассматриваемый прямолинейный
проводник с постоянным током создает, в окружающем его пространстве, не только магнитное
поле, представленное тензором  $\mathcal{F}(\mathit M)$, но и девиатационное джейтонное поле,
определяемое тензором  $\mathcal{G}_{d}(\mathit M)$, каждое из которых геометрически  представлено
 локально линейно независимой конгруэнцией соответствующих векторных линий
 соответствующих векторных полей.

Аналогичные  {\em  физические} заключения относятся и к {\em фундаментальным}
 классическим полевым системам представленным на рис. 4.3--4.5.

\clearpage

\section*{\S \,5 Аналитический анализ полевой системы, созданной стационарным электрическим током}
\addcontentsline{toc}{section}{\S \,5 Аналитический анализ полевой системы, созданной стационарным электрическим током}

\quad \, Проведем и общий  {\em аналитический}   анализ произвольной полевой системы, созданной стационарным током.

 В качестве такой полевой системы, рассмотрим класс векторных потенциалов, являющихся
  решениями уравнения Пуассона,
 $$
 \bigtriangleup\, \vec{A}\,(\vec{r}\,) = -  \vec{j}\,(\vec{r}\,),             \eqno\ldots (5.1)
 $$
 и удовлетворяющих традиционному условию, $div\,\vec{A}\,(\vec{r}) \equiv 0$, приводящему, в силу (3.18), к тождественному равенству
 $$
 \mathcal{G}_{\ast}(\vec{r}\,) \equiv \tilde{0}.                            \eqno\ldots (5.2)
 $$

 В этом случае, (5.1), после использования соотношения
 $$
 \bigtriangleup\, \vec{A}\,(\vec{r}\,) = \vec{\bigtriangledown}\cdot\partial A(\vec{r}),             \eqno\ldots (5.3)
 $$
 и последующего учета (1.20) и (5.2), приводит к уравнению
$$
  \vec{\bigtriangledown}\cdot \mathcal{F}(\vec{r}\,) + \vec{\bigtriangledown}\cdot \mathcal{G}_{d}(\vec{r}\,)  = -  \vec{j}\,(\vec{r}\,),             \eqno\ldots (5.4)
 $$
  {\em  аналитически}  демонстрирующему, что рассматриваемые стационарные электрические
   токи, в общем случае,    порождают, в окружающем их 3-~пространстве,  не только
    стационарное магнитное поле,   представленное в этом уравнении тензором $ \mathcal{F}(\vec{r})$, но
    и стационарное девиатационное джейтонное поле,
   определяемое тензором $\mathcal{G}_{d}(\vec{r})$.

В силу того, что в обсуждаемом классе потенциалов при одновременном присутствии
в рассматриваемой точке 3-пространства магнитного и девиатационного джейтонного полей
выполняется соотношение
 $$
  \vec{\bigtriangledown}\cdot \mathcal{F}(\vec{r}\,)\equiv \vec{\bigtriangledown}\cdot \mathcal{G}_{d}(\vec{r}\,),      \eqno\ldots (5.5)
 $$
 уравнение (5.4), в свою очередь, приводит к {\em двум} уравнениям, математически
 эквивалентным (5.1)  как уравнениям на векторный потен-циал~$\vec{A}\,(\vec{r})$,
 $$
  \vec{\bigtriangledown}\cdot \mathcal{F}_{\mathit L}(\vec{r}\,)   = -  \vec{j}\,(\vec{r}\,),             \eqno\ldots (5.6)
 $$
 $$
  \vec{\bigtriangledown}\cdot \mathcal{G}_{\mathit L}(\vec{r}\,)  = -  \vec{j}\,(\vec{r}\,).             \eqno\ldots (5.7)
 $$

 Первое из этих уравнений, после использования соотношения, связывающего дивергенцию антисимметричного 3-тензора второго ранга c  сопровождающим его псевдовектором,
 $$
  \vec{\bigtriangledown}\cdot \mathcal{F}_{\mathit L}(\vec{r}\,)   = -  rot\,\vec{\mathcal{H}_{\mathit L}}(\vec{r}\,),             \eqno\ldots (5.8)
 $$
 приводит к дифференциальному уравнению
 $$
  rot\,\vec{\mathcal{H}}_{\mathit L}(\vec{r}\,) = \vec{j}\,(\vec{r}\,),         \eqno\ldots (5.9)
 $$
 представляющему собой общеизвестную дифференциальную форму закона (теоремы) Ампера
  о циркуляции псевдовектора $\vec{\mathcal{H}_{\mathit L}}(\vec{r}\,)$, сопровождающего лоренцев
   3-тензор магнитного поля, $ \mathcal{F}_{\mathit L}(\vec{r}\,)$.

 В итоге, уравнение (5.4) выступает в качестве соответствующего {\em физического} обобщения закона Ампера, представленного соотношением (5.9).

 В свою очередь, последовательное действие, на обе части уравнения (5.1), операторов (1.3), (1.9) и (1.10)
 приводит, в общем случае, заранее не предопределяющим выполнение равенства (5.2),
 к системе линейно независимых, по отношению к друг другу, уравнений {\em для каждого}
  из рассматриваемых тензорных полей,
  $$
\bigtriangleup\,\mathcal{F}(\vec{r}\,)   = - \, \tilde{rot}\, \vec{j}\,(\vec{r}\,),             \eqno\ldots (5.10)
 $$
 $$
\bigtriangleup\,\mathcal{G}_{d}(\vec{r}\,)   = - \, \tilde{def}_{d}\, \,\vec{j}\,(\vec{r}\,),             \eqno\ldots (5.11)
 $$
  $$
\bigtriangleup\,\mathcal{G}_{\ast}(\vec{r}\,)   = - \, \tilde{def}_{\ast}\, \,\vec{j}\,(\vec{r}\,).             \eqno\ldots (5.12)
 $$

 Уравнение (5.10) представляет собой дифференциальную форму известного закона
  Био \!--\! Савара \!--\! Лапласа в пространстве $V\otimes V$.

Уместно подчеркнуть, что уравнения системы (5.10)--(5.12), как и все другие подобные соотношения, определены (имеют смысл) как в  $V_{3}\otimes V_{3}$, так и в $V_{2}\otimes V_{2}$.

Соответствующая традиционная дифференциальная форма закона Био \!--\! Савара \!--\! Лапласа
 следует из уравнения (5.10), после двойного скалярного умножения обеих частей этого
  уравнения на псевдотензор третьего ранга Леви-Чивиты, $\varepsilon$, и последующего использования соотношения (1.15) и тождества $\varepsilon \cdot\cdot \,\tilde{rot}\,\vec{j}\,(\vec{r}\,) \equiv rot \,\vec{j}\,(\vec{r}\,)$, идентичного (1.16).

  Таким образом, в качестве следствия (5.10), имеем
 $$
\triangle \vec{\mathcal{H}}_{\mathit L} (\vec{r}\,)   = - rot\, \vec{j}(\vec{r}\,).             \eqno\ldots (5.13)
 $$

 При этом очевидно так же, что (5.13) является, одновременно, и мгновенным следствием уравнения (5.1), после действия на обе части этого уравнения дифференциального оператора $rot$ и  последующего использования определения лоренцевой напряженности магнитного поля, $ \vec{\mathcal{H}}_{\mathit L} (\vec{r}\,)  \eqdef rot\,\vec{A}(\vec{r}) $.

 Обратно, от (5.13) можно легко перейти к уравнению (5.10), если скалярно умножить обе части уравнения (5.13) на псевдотензор $\varepsilon$ и затем воспользоваться соотношениями
 $$
 \varepsilon \cdot \vec{\mathcal{H}} (\vec{r}\,) = \mathcal{F}(\vec{r}\,),         \eqno\ldots (5.14)
 $$
 $$
 \varepsilon \cdot rot\,\vec{j}(\vec{r}\,) = 2\,\tilde{rot}\,\vec{j}(\vec{r}\,).             \eqno\ldots (5.15)
 $$

 Решения уравнений (5.10)--(5.12), соответствующие полям созданными {\em  системой} \cite{La, Jack}, определяются соотношениями
 $$
\mathcal{F}(\vec{r}\,) = \frac{1}{4\pi}\int_{\mathit V}\frac{\tilde{rot}\,\vec{j}(\vec{r}\,^{'})}{|\, \vec{r} -
\vec{r}\,^{'}|}\,d\,V\,^{'},                                 \eqno\ldots (5.16)
$$
$$
\mathcal{G}_{d}(\vec{r}) =  \frac{1}{4\pi}\int_{\mathit V}\frac{\tilde{def}_{d}\,\,\vec{j}(\vec{r}\,^{'})}{|\, \vec{r} - \vec{r}\,^{'}|}\,d\,V\,^{'},                          \eqno\ldots (5.17)
$$
$$
\mathcal{G}_{\ast}(\vec{r}) =  \frac{1}{4\pi}\int_{\mathit V}\frac{\tilde{def}_{\ast}\,\,\vec{j}(\vec{r}\,^{'})}{|\, \vec{r} - \vec{r}\,^{'}|}\,d\,V\,^{'}.                           \eqno\ldots (5.18)
$$

Дилататор плотности тока в подинтегральной функции правой части (5.18), согласно (1.10), имеет вид
$$
\tilde{def}_{\ast}\,\,\vec{j}(\vec{r}\,) = \frac{div\,\vec{j}(\vec{r}\,)}{tr g}\,g,      \eqno\ldots (5.19)
$$
который, в силу уравнения непрерывности стационарных токов, приводит к равенству
$$
\tilde{def}_{\ast}\,\,\vec{j}(\vec{r}\,) \equiv \tilde{0},                  \eqno\ldots (5.20)
$$
возвращающему нас, в соответствии с (5.18), к (5.2).

С другой стороны, так как тензоры    $\tilde{rot}\,\vec{j}(\vec{r}\,)$ и $\tilde{def}_{d}\,\,\vec{j}(\vec{r}\,)$, в соответствии с их геометрическими образами (в соответствии с их фазовыми портретами), определяют соответствующие конгруэнции
векторных линий плотности тока $\vec{j}(\vec{r}\,)$, представленные, соответственно, центром и классическим седлом, то соотношения (5.16) и (5.17) демонстрируют, одновременно, что магнитное и девиатационное джейтонное поля могут генерироваться, соответственно, и такими структурами плотности тока.

Используя соотношение
$$
\vec{\bigtriangledown}\otimes(\varphi \,(\vec{r}\,)\,\,\vec{a}\,(\vec{r}\,)) =\vec{\bigtriangledown}\varphi \,(\vec{r}\,)\otimes \vec{a}\,(\vec{r}\,)+\varphi (\vec{r}\,)\,\vec{\bigtriangledown}\otimes\vec{a}\,(\vec{r}\,),     \eqno\ldots (5.21)
$$
в котором $\vec{a}\,(\vec{r}\,) $ и $\varphi\, (\vec{r}\,)$ --~произвольные дифференцируемые векторное и скалярное поля, представим подинтегральные функции интегралов правых частей (5.16) и (5.17), соответственно, в видах
$$
\frac{\tilde{rot}\,\vec{j}(\vec{r}\,^{'})}{|\, \vec{r} - \vec{r}\,^{'}|}= - \frac{\vec{r} - \vec{r}\,^{'}}{|\, \vec{r} - \vec{r}\,^{'}|\,^{3}}\stackrel{ a}{\otimes}\vec{j}(\vec{r}\,^{'}) + \tilde{rot}_{\vec{r}\,^{'}}\frac{\vec{j}(\vec{r}\,^{'})}{|\, \vec{r} - \vec{r}\,^{'}|},                         \eqno\ldots (5.22)
$$
$$
\frac{\tilde{def}_{d}\,\,\vec{j}(\vec{r}\,^{'})}{|\, \vec{r} - \vec{r}\,^{'}|} = - \frac{\vec{r} - \vec{r}\,^{'}}{|\, \vec{r} - \vec{r}\,^{'}|\,^{3}}\stackrel{ d}{\otimes}\vec{j}(\vec{r}\,^{'}) + \tilde{def}_{d \,\,\vec{r}\,^{'}}\frac{\vec{j}(\vec{r}\,^{'})}{|\, \vec{r} - \vec{r}\,^{'}|}.                         \eqno\ldots (5.23)
$$

(5.22) и (5.23), в рамках соответствующих условий на границах токовой системы,
приводят (5.16) и (5.17), соответственно, к виду
$$
\mathcal{F}(\vec{r}\,) = - \frac{1}{4\pi}\int_{\mathit V}\frac{(\vec{r} - \vec{r}\,^{'})\stackrel{ a}{\otimes}\vec{j}(\vec{r}\,^{'})}{|\, \vec{r} - \vec{r}\,^{'}|\,^{3}}\,d\,V\,^{'},                           \eqno\ldots (5.24)
$$
$$
\mathcal{G}_{d}(\vec{r}) =  - \frac{1}{4\pi}\int_{\mathit V} \frac{(\vec{r} - \vec{r}\,^{'})\stackrel{ d}{\otimes}\vec{j}(\vec{r}\,^{'})}{|\, \vec{r} - \vec{r}\,^{'}|\,^{3}}\,d\,V\,^{'}.                           \eqno\ldots (5.25)
$$

(5.24) представляет собой интегральную форму закона Био\! --\! Савара~\!--\! Лапласа в
 соответствующем тензорном пространстве.

 В частном случае тензорного пространства над 3-мерным евклидовым
  векторным пространством  $\mathit V_{3}$, двойное внутреннее произведение тензора $\varepsilon$ и
  соответствующих тензоров уравнения (5.24), с последующим использованием определения (1.15)
   и учетом соотношения
$$
\varepsilon\,\cdot\cdot\,(\vec{a}\stackrel{ a}{\otimes}\vec{b}) = - \vec{a}\times\vec{b},          \eqno\ldots (5.26)
$$
приводит, в качестве следствия (5.24), к стандартной интегральной форме данного
закона в векторном пространстве $V_{3}$,
$$
\vec{\mathcal{H}}_{\mathit L} (\vec{r}\,) = \frac{1}{4\pi}\int_{\mathit V}\frac{\vec{j}(\vec{r}\,^{'})\times (\vec{r} - \vec{r}\,^{'})}{|\, \vec{r} - \vec{r}\,^{'}|\,^{3}}\,d\,V\,^{'}.      \eqno\ldots (5.27)
$$

Обратный переход то (5.27) к (5.24) осуществляется так же просто, как и переход от (5.13) к (5.10).

Подобные переходы для законов (5.11) и (5.25) не имеют места в силу того, что векторный инвариант любого симметричного тензора второго ранга равен нуль-вектору векторного пространства $V_{3}$, над которым определен этот тензор.

Поэтому, эти законы остаются представленными, в данном контексте, в виде соотношений (5.11) и (5.25).

Вернемся вновь к анализу примеров конгруэнтных разложений классических векторных полей, представленных на рис. 4.3 и 4.4.

Рис. 4.3\,{\em a})\,, на примере излучения линейно вибрирующей точечной электрически заряженной частицы, геометрически демонстрирует, что и традиционное поле электродинамической плоской волны \cite{La} в действительности представлено, помимо соответствующего электрического поля, не только магнитным полем, определенным тензором $\mathcal{F}(x)$, но и девиатационным джейтонным полем, представленным тензором $\mathcal{G}_{d}(x)$, то есть, рассматриваемая  {\em фундаментальная} волна является не электромагнитной, а {\em  электромагнитоджейтонной} волной.

Рис. 4.3\,{\em б})\,, в свою очередь, геометрически демонстрирует наличие, на продольной оси линейно вибрирующей точечной электрически заряженной частицы, волновых джейтонных полей, представленных тензорами $\mathcal{G}_{d}(x)$ и $\mathcal{G}_{\ast}(x)$, при тождественно отсутствии, на этой оси, магнитного поля.

Конгруэнтные разложения, представленные на рис. 4.4\, {\em a})\, и 4.4\, {\em б})\,, геометрически демонстрируют известный факт наличия классического магнитного поля {\em  внутри}  рассматриваемого  соленоида, при полном отсутствии, в этой области, джейтонных полей, и факт наличия девиатационного джейтонного поля {\em  вне} данного соленоида, при полном отсутствии, в этой области, магнитного и дилатационного джейтонного полей.

В связи с этим, {\em  классический} векторный потенциал рассматриваемого соленоида, во внешней области этого соленоида, теперь перестает быть "нулевым"\, потенциалом в вышеуказанном {\em  классическом} отношении.

\section*{\S \,6 Аналитический анализ полевой системы, созданной бесконечно длинным цилиндрическим соленоидом с постоянным электрическим током}
\addcontentsline{toc}{section}{\S \,6 Аналитический анализ полевой системы, созданной бесконечно длинным цилиндрическим соленоидом с постоянным электрическим током}

\qquad Проведем более подробный {\em  анлитический} анализ данной классической полевой системы,
 представленной на рис. 6.1.

 \setcounter{section}{6}
\setcounter{figure}{0}

\begin{figure}[!h]
\begin{center}
\vspace{-6pt}
\includegraphics[width=135mm]{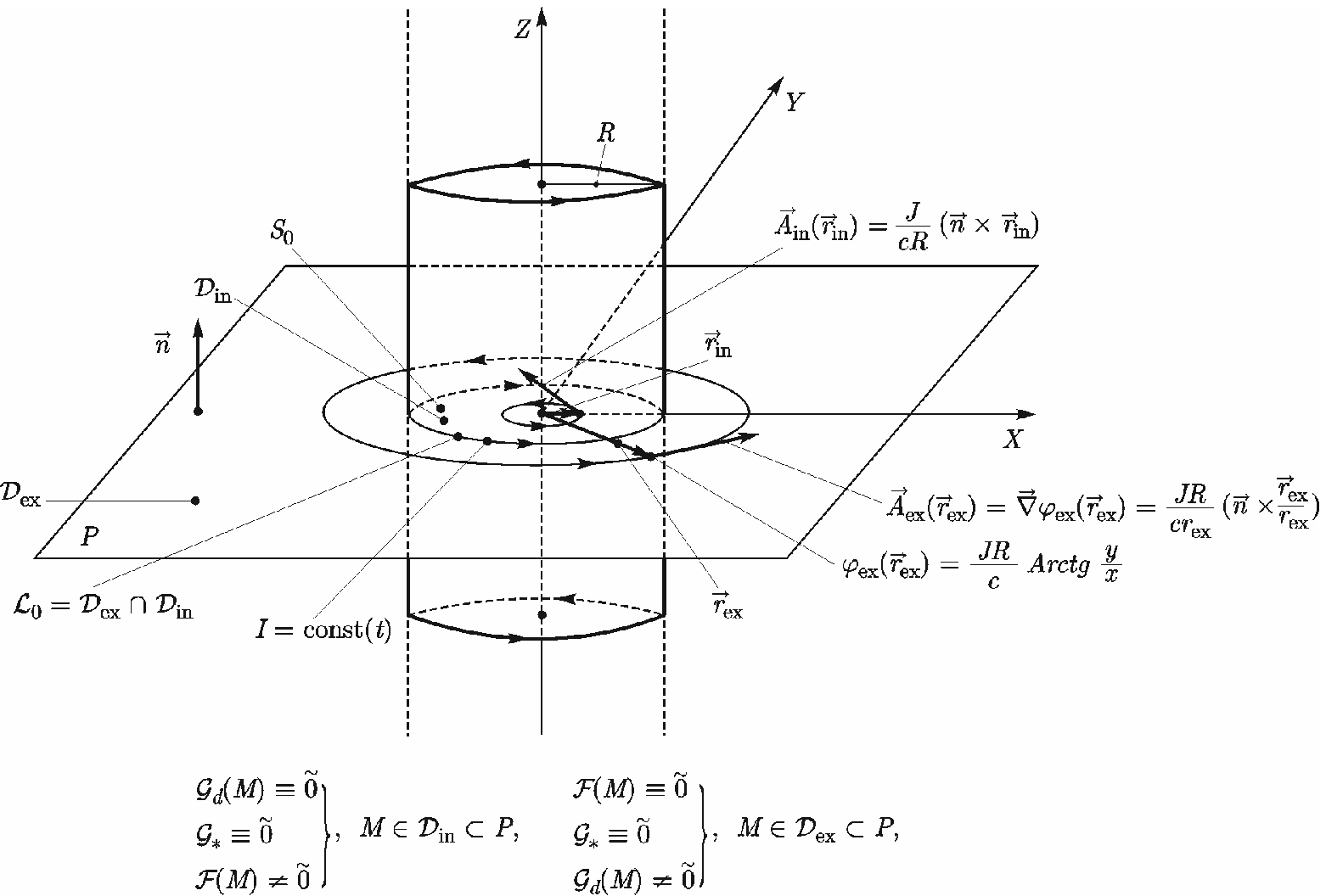}
\vspace{1pt}
\caption{Полевая система бесконечно длинного цилиндрического соленоида с постоянным током. Рассматриваемая  плоскость нормальна к продольной оси соленоида.  $S_{0}$~--- кусочно гладкая поверхность, натянутая на ${\mathcal L}_{0}$, $J = 2\pi Rn_{1} I$, $n_{1} = \frac{N}{l}.$}
\end{center}
\end{figure}

Тут и далее, в данном разделе, используется гауссова система единиц.

Внутри данного соленоида векторный потенциал определен соотношением \cite{Ag}
$$
\vec{A}_{in}(\vec{r}\,) = \frac{J}{c\,R}\,\,(\vec{n}\times \vec{r}\,),                    \eqno\ldots (6.1)
$$
которое, согласно (1.6), приводит к следующему выражению для тензора магнитного
 поля, $ \mathcal{F}_{in}(\vec{r}\,)$,
$$
\mathcal{F}_{in}(\vec{r}\,) = \frac{J}{c\,R}\,\,\tilde{rot}\,\,(\vec{n}\times \vec{r}\,).                     \eqno\ldots (6.2)
$$

С другой стороны, имеем легко проверяемое соотношение
$$
\vec{\bigtriangledown}\otimes (\vec{n}\times \vec{r}\,) = \vec{n}\,\cdot\,\varepsilon ,              \eqno\ldots (6.3)
$$
демонстрирующее, что  тензор $\vec{\bigtriangledown}\otimes (\vec{n}\times \vec{r}\,)$
антисимметричен, вследствие чего
$$
\tilde{rot}\,\,(\vec{n}\times \vec{r}\,) = \vec{\bigtriangledown}\otimes (\vec{n}\times \vec{r}\,),      \eqno\ldots (6.4)
$$
в итоге, (6.2)  принимает вид
$$
\mathcal{F}_{in}(\vec{r}\,) = \frac{J}{c\,R}\,\,\vec{n}\,\cdot\,\varepsilon.                     \eqno\ldots (6.5)
$$

В свою очередь, (6.5), совместно с (1.15), приводит к выражению для псевдовектора $ \vec{\mathcal{H}}_{in}(\vec{r}\,)$, сопровождающего тензор   $ \mathcal{F}_{in}(\vec{r}\,)$,
 $$
\vec{\mathcal{H}}_{in}(\vec{r}\,) = \frac{J}{c\,R}\,\,\vec{n}.                     \eqno\ldots (6.6)
$$

Соотношения (6.5) и (6.6) демонстрируют факт наличия магнитного поля внутри рассматриваемого соленоида и его однородность.

Используя (6.5), легко получить выражение для евклидовой нормы тензора  $ \mathcal{F}_{in}(\vec{r}\,)$
$$
|| \,\mathcal{F}_{in}(\vec{r}\,)\, ||= \frac{J\,\sqrt{2}}{c\,R}\,.                     \eqno\ldots (6.7)
$$

С другой стороны, согласно (6.6) имеем
$$
\mathcal{H}_{in}(\vec{r}\,) = \frac{J}{c\,R}\,.                     \eqno\ldots (6.8)
$$

(6.7) и (6.8), в свою очередь, позволяют выразить токовый параметр $\mathit J $ через полевые переменные,
$$
J = \frac{1}{\sqrt{2}}\,c\,R\, || \,\mathcal{F}_{in}(\vec{r}\,)\, || = c\,R\,\,\mathcal{H}_{in}(\vec{r}\,)\,,      \eqno\ldots (6.9)
$$
и представить выражение для векторного потенциала (6.1) в стандартном виде \cite{Ryd}, "изгнав"\,, из выражения (6.1), самого  {\em  создателя} рассматриваемых классических полей,
$$
 \vec{A}_{in}(\vec{r}\,) =\mathcal{H}_{in}(\vec{r}\,)\,(\vec{n}\times \vec{r}\,) = \frac{\mathcal{H}_{L\,in}}{2}\,(\vec{n}\times \vec{r}\,),     \eqno\ldots (6.10)
$$
$$
A_{in}(\vec{r}\,) =\frac{\mathcal{H}_{L\,in}\,r}{2},                   \eqno\ldots (6.11)
$$
$$
 \vec{A}_{in}(\vec{r}\,) = \frac{|| \,\mathcal{F}_{in}\,||}{\sqrt{2}}\,(\vec{n}\times \vec{r}\,).     \eqno\ldots (6.12)
$$

Тензор {\em   дилатационного} джейтонного поля, определенный соотношением (1.8), для потенциала (6.1) принимает вид
$$
\mathcal{G}_{\ast \,in}(\vec{r}\,) = \frac{J}{c\,R}\,\,\tilde{def}_{\ast}\,\,(\vec{n}\times \vec{r}\,).                     \eqno\ldots (6.13)
$$

Но, в силу (1.10), имеем тождественное равенство
$$
\tilde{def}_{\ast}\,\,(\vec{n}\times \vec{r}\,) = \frac{g}{tr g}\vec{\bigtriangledown}\cdot (\vec{n}\times \vec{r}\,) \equiv \tilde{0}, \eqno\ldots (6.14)
$$
использование которого в  (6.13) приводит к соотношению
$$
\mathcal{G}_{\ast \,in}(\vec{r}\,) \equiv \tilde{0}.                     \eqno\ldots (6.15)
$$

Данный результат заранее ожидаем и соответствует (5.2).

Тензор  {\em  девиатационного} джейтонного поля, определенный соотношением (1.7), для потенциала (6.1), с учетом (6.15),  принимает вид
$$
\mathcal{G}_{d \,in}(\vec{r}\,) =\mathcal{G}_{in}(\vec{r}\,) =\frac{J}{c\,R}\,\,\tilde{def}\,\,(\vec{n}\times \vec{r}\,).                     \eqno\ldots (6.16)
$$

Но, в силу антисимметричности тензора $\vec{\bigtriangledown}\otimes (\vec{n}\times \vec{r}\,)$, имеем
$$
\tilde{def}\,\,(\vec{n}\times \vec{r}\,) \equiv \tilde{0},           \eqno\ldots (6.17)
$$
вследствие чего из (6.16) следует,что и
$$
\mathcal{G}_{d \,in}(\vec{r}\,) \equiv \tilde{0}.             \eqno\ldots (6.18)
$$

Полученные соотношения (6.15) и (6.18) демонстрируют факт отсутствия как дилатационного, так и девиатационного джейтонных полей  {\em  внутри} рассматриваемого соленоида, то есть, поле, связанное с 3-градиентами векторного потенциала, представлено, в данной области, исключительно магнитным полем.

Во внешней области рассматриваемого соленоида, векторный потенциал представлен соотношением \cite{Ag}
$$
\vec{A}_{ex}(\vec{r}\,) = \frac{J R}{c}\,\,\left (\vec{n}\times \frac{\vec{r}\,}{r\,^{2}}\right ),                    \eqno\ldots (6.19)
$$
из которого, прежде всего следует, что, при неизменной силе тока соленоида, $ \vec{A}_{ex}(\vec{r}\,)$, в отличие от $\vec{A}_{in}(\vec{r}\,)$, с ростом радиуса соленоида -- {\em  возрастает }.

С другой стороны, векторное поле  $ \vec{A}_{ex}(\vec{r}\,)$, представленное соотношением (6.19), определено всюду кроме оси соленоида, то есть, это векторное поле занимает двусвязную область представляя собой очередной пример потенциального векторного поля, обладающего многозначным потенциалом.

Аналогичное поведение демонстрирует, например, векторное поле, представленное напряженностью магнитного поля бесконечного прямолинейного проводника с постоянным током \cite{Bor}.

Тензор магнитного поля, $ \mathcal{F}(\vec{r}\,)$, определенный соотношением (1.6), для векторного потенциала (6.19) теперь имеет вид
$$
\mathcal{F}_{ex}(\vec{r}\,) = \frac{J R}{c}\,\,\tilde{rot}\,\,\left (\vec{n}\times \frac{\vec{r}\,}{r\,^{2}}\right ).            \eqno\ldots (6.20)
$$

Однако на этот раз, используя соотношение
$$
\vec{\bigtriangledown}\otimes \left (\,\vec{n}\times \frac{\vec{r}\,}{r\,^{2}}\,\right ) = n^{i}\varepsilon_{ik\alpha}\left (\,\frac{g^{\beta k}}{r\,^{2}} - \frac{2\,x^{\beta}x^{k}}{r\,^{4}}\,\right)\,\vec{e}\,^{\alpha}\otimes \vec{e}_{\beta},               \eqno\ldots (6.21)
$$
находим, что тензор $\vec{\bigtriangledown}\,\otimes\,\left (\vec{n}\times \frac{\displaystyle\vec{r}\,}{\displaystyle r\,^{2}}\,\right )  $  {\em  симметричен}, вследствие чего
$$
\tilde{rot}\,\,\left (\vec{n}\times \frac{\vec{r}\,}{r\,^{2}}\right ) \equiv \tilde{0}.            \eqno\ldots (6.22)
$$

Последующее использование (6.22) в (6.20) означает, что в области определения векторного поля (6.19) выполняется тождественное равенство
$$
\mathcal{F}_{ex}(\vec{r}\,) \equiv \tilde{0},            \eqno\ldots (6.23)
$$
демонстрирующее факт отсутствия магнитного поля вне рассматриваемого соленоида, с одной стороны, и выступающее в качестве необходимого условия потенциальности (или необходимого условия интегрируемости  \cite{Fich}) векторного поля (6.19), с другой стороны.

Тензор дилатационного джейтонного поля, $\mathcal{G}_{\ast}(\vec{r}\,)$, определенный соотношением (1.8), для векторного потенциала (6.19) принимает вид
$$
\mathcal{G}_{\ast \,ex}(\vec{r}\,) = \frac{J R}{c}\,\,\tilde{def}_{\ast}\,\,\left (\vec{n}\times \frac{\vec{r}\,}{r\,^{2}}\right ).     \eqno\ldots (6.24)
$$

Как и в (6.14), использование определения (1.10) приводит к соотношению
$$
\tilde{def}_{\ast}\,\,\left (\vec{n}\times \frac{\vec{r}\,}{r\,^{2}}\right ) = \frac{g}{tr g}\vec{\bigtriangledown} \cdot \left (\vec{n}\times \frac{\vec{r}}{r\,^{2}}\,\right ) \equiv \tilde{0},                               \eqno\ldots (6.25)
$$
учет которого в (6.24) означает, что и
$$
\mathcal{G}_{\ast \,ex}(\vec{r}\,) \equiv \tilde{0}.     \eqno\ldots (6.26)
$$

Таким образом, в соответствии с (5.2), полученные соотношения, (6.18) и (6.26), демонстрируют отсутствие {\em  дилатационного} джейтонного поля как внутри, так и вне рассматриваемого соленоида.

Тензор {\em  девиатационного} джейтонного поля, $\mathcal{G}_{d}(\vec{r}\,)$, определенный соотношением (1.7), для векторного потенциала (6.19), с учетом (6.26), имеет вид
$$
\mathcal{G}_{d \,ex}(\vec{r}\,) =\mathcal{G}_{ex}(\vec{r}\,) =\frac{J R}{c}\,\,\tilde{def}\,\,\left (\vec{n}\times \frac{\vec{r}}{r\,^{2}}\,\right ).                     \eqno\ldots (6.27)
$$

Но, в силу вышеуказанной симметрии тензора  $\vec{\bigtriangledown}\,\otimes\,\left (\vec{n}\times \frac{\displaystyle\vec{r}\,}{\displaystyle r\,^{2}}\,\right )$,
$$
\tilde{def}\,\,\left (\vec{n}\times \frac{\vec{r}}{r\,^{2}}\,\right ) = \vec{\bigtriangledown}\,\otimes\,\left (\vec{n}\times \frac{\displaystyle\vec{r}\,}{\displaystyle r\,^{2}}\,\right ),                \eqno\ldots (6.28)
$$
вследствие чего (6.27) представляется соотношением
$$
\mathcal{G}_{d \,ex}(\vec{r}\,) =\frac{J R}{c}\,\,\vec{\bigtriangledown}\otimes\,\,\left (\vec{n}\times \frac{\vec{r}}{r\,^{2}}\,\right ),                     \eqno\ldots (6.29)
$$
демонстрирующим, в частности, факт наличия {\em  девиатационного} джейтонного поля в рассматриваемой области.

С другой стороны, (6.29) демонстрирует так же, что с увеличении радиуса рассматриваемого соленоида, при постоянной силе тока в этом соленоиде, девиатационное джейтонное поле возрастает, то есть, для получения более ярко выраженных эффектов, обусловленных данным полем, нужно использовать соленоиды возможно большего диаметра.

Таким образом, во {\em  внешней} области данного соленоида, поле, связанное с градиентами векторного потенциала, представлено исключительно {\em  девиатационным} джейтонным полем, определяемым соотношением (6.29).

В свою очередь, в силу того, что векторный потенциал (6.1) является носителем исключительно магнитного поля, естественно называть этот потенциал векторным потенциалом {\em  магнитного} поля, а в силу того, что векторный потенциал (6.19) выступает в качестве носителя исключительно девиатационного джейтонного поля, столь же естественно представлять данный  потенциал векторным потенциалом  {\em  девиатационного джейтонного} поля.

При этом, рассматриваемый "магнитный"\, векторный потенциал является  {\em  соленоидальным} векторным полем, а "джейтонный"\, векторный потенциал представлен  {\em потенциальным} векторным полем.

Аналогичное относится и к полю векторного потенциала тороидального соленоида с постоянным электрическим током (полевая система, представленная на рис. 4.5).

Таким образом имеем, что целые  {\em  области} 3-пространства (внешние области бесконечно длинного  цилиндрического соленоида, тороидального соленоида, с постоянными электрическими токами, и соответствующие области других, подобных им,  источников полевых систем) "заняты"\, исключчительно {\em девиатационным} джейтонным полем, что является демонстрацией автономного, по отношению к магнитному полю,  существования последнего, с последующими  {\em  автономными} физическими эффектами, обусловленными этим полем.

В тех случаях, когда векторный потенциал является носителем одновременно и магнитного, и джейтонного полей (полевая система, представленная на рис. 5, например), так же естественно представлять его  {\em  магнтоджейтонным} векторным потенциалом.

Таким образом констатируется, что векторный потенциал классической электродинамики далеко не всегда является векторным потенциалом магнитного или электромагнитного, в общем случае, поля.

Возвращаясь к рассматриваемому соленоиду, находим, что использование (6.29), с учетом (6.21), приводит к следующему выражению для евклидовой нормы тензора  $\mathcal{G}_{d\,ex}(\vec{r}\,)$
$$
|| \,\mathcal{G}_{d\,ex}(\vec{r}\,)\, ||= \frac{J R\,\sqrt{2}}{c\,r\,^{2}}\,,                     \eqno\ldots (6.30)
$$
откуда, в частности, имеем
$$
|| \,\mathcal{G}_{d\,ex}(R\,)\, ||= \frac{J\,\sqrt{2}}{c\,R}\,.                     \eqno\ldots (6.31)
$$

Последующее сопоставление (6.31) и (6.7) приводит к равенству норм
$$
|| \,\mathcal{G}_{d\,ex}(R\,)\, ||= || \,\mathcal{F}_{in}(\vec{r}\,)\, ||,                     \eqno\ldots (6.32)
$$
которое означает, что норма тензора $\mathcal{G}_{d\,ex}(\vec{r}\,)$ непрерывно "переходит"\, в норму тензора
 $\mathcal{F}_{in}(\vec{r}\,)$, а это, в свою очередь, означает, что плотность энергии магнитного и джейтонного полей рассматриваемого соленоида распределена в 3-пространстве непрерывно.

 (6.31) позволяет, в свою очередь, выразить токовый параметр $ J $ через норму тензора $\mathcal{G}_{d\,ex}(\vec{r}\,)$ на $ \mathfrak{L}_{0}$ (см. рис. 6.1)
$$
J = \frac{1}{\sqrt{2}}\,c\, \mathit R\, || \,\mathcal{G}_{d\,ex}( \mathit R\,)\,||,      \eqno\ldots (6.33)
$$
и переписать соотношение (6.19) в виде
$$
\vec{A}_{ex}(\vec{r}\,) = \frac{ || \,\mathcal{G}_{d\,ex}( \mathit R\,)\,||\,\mathit R\,^{2}}{r\,\sqrt{2}}\,\,\left (\vec{n}\times \frac{\vec{r}\,}{r}\right ).                    \eqno\ldots (6.34)
$$

С другой стороны, использование (6.9) позволяет представить $ \vec{A}_{ex}(\vec{r}\,)$ в традиционном
 \cite{Ryd}, {\em  математически} эквивалентном виде
 $$
\vec{A}_{ex}(\vec{r}\,) = \mathcal{H}_{in}R\,^{2}\,\,\left (\vec{n}\times \frac{\vec{r}\,}{r\,^{2}}\right ) = \frac{ \mathcal{H}_{in\,L}R\,^{2}}{2\,r}\,\,\left (\vec{n}\times \frac{\vec{r}\,}{r}\right ),                    \eqno\ldots (6.35)
$$
$$
A_{ex}(r) = \frac{ \mathcal{H}_{in}R\,^{2}}{r} =  \frac{ \mathcal{H}_{in\,L}R\,^{2}}{2\,r}.       \eqno\ldots (6.36)
$$

В настоящем контексте предпочтение отдается представлениям (6.1) и (6.19), подчеркивающим, что  {\em  создателем} векторных полей $ \vec{A}_{in}(\vec{r}\,)$ и $ \vec{A}_{ex}(\vec{r}\,)$ является, в данном случае, электрический ток рассматриваемого соленоида, а все остальные формы представления этих полей являются эквивалентными исходным лишь в {\em  математическом}  отношении.

Наконец, можно выразить токовый параметр $J$ и через поток $  \Phi _{S_{0}}$ псевдовектора $ \vec{\mathcal{H}}_{in}$
через кусочно-гладкую поверхность $ S_{0}$, имеющую своей границей контур $ \mathcal{L}_{0}$  \cite{Fich}, получив другие, часто использующиеся представления рассматриваемых векторных потенциалов.

Действительно, из соотношений
$$
\Phi _{S_{0}} = \int \!\! \! \int_{ S_{0}}\vec{\mathcal{H}_{in}}\cdot \vec{n}\,\,d S = \frac{\pi J R}{c},      \eqno\ldots (6.37)
$$
мгновенно получаем искомое выражение
$$
J = \frac{c\,\Phi _{S_{0}}}{\pi R},                  \eqno\ldots (6.38)
$$
использование которого в (6.1) и (6.19) приводит к соотношениям
$$
\vec{A}_{in}(\vec{r}\,) = \frac{\Phi _{S_{0}}}{\pi R\,^{2}}\,\,(\vec{n}\times \vec{r}\,) =  \frac{\Phi _{S_{0}L}}{2\,\pi R\,^{2}}\,\,(\vec{n}\times \vec{r}\,),                    \eqno\ldots (6.39)
$$
$$
\vec{A}_{ex}(\vec{r}\,) = \frac{\Phi _{S_{0}}}{\pi r}\,\,\left (\vec{n}\times \frac{\vec{r}\,}{r}\right ) =  \frac{\Phi _{S_{0}L}}{2\,\pi r}\,\,\left (\vec{n}\times \frac{\vec{r}\,}{r}\right ),                    \eqno\ldots (6.40)
$$
в которых $ \Phi _{S_{0}L}$ -- поток через поверхность  $ S_{0}$ лоренцевой напряженности магнитного поля, $ \vec{\mathcal{H}}_{ \mathit L}(\vec{r}\,)$.

С другой стороны, хорошо известный интеграл Гаусса, $ \textsl{g }$   \cite{Fich}, представленный в виде криволинейного интеграла второго типа, в инвариантной векторной форме имеет вид
$$
\textsl{g}\,  \eqdef  \int_{\mathcal{L}} \vec{A}_{\, \textsl{g }}\!(\vec{r}\,) \cdot d \vec{r},        \eqno\ldots (6.41)
$$
где плоское потенциальное векторное поле $ \vec{\mathit A}_{\textsl{g}}(\vec{r}\,)$ (векторное поле Гаусса) и его потенциальная функция $ \varphi_{\textsl{g}}(\vec{r}\,) $ определены соотношениями
$$
 \vec{A}_{\, \textsl{g }}\!(\vec{r}\,) \eqdef \vec{\triangledown}\, \varphi_{\textsl{g}}(\vec{r}\,),              \eqno\ldots (6.42)
$$
$$
  \varphi_{\textsl{g}}(\vec{r}\,)  \eqdef  \arctan \frac{y}{x}.                        \eqno\ldots (6.43)
$$

Подстановка (6.43) в (6.42) позволяет представить векторное поле Гаусса в виде
$$
 \vec{A}_{\, \textsl{g }}\!(\vec{r}\,) = \frac{1}{r\,^{2}}\,(\vec{n}\times \vec{r}\,).          \eqno\ldots (6.44)
$$

В свою очередь, (6.44) позволяет переписать (6.19) в виде
$$
\vec{A}_{ex}(\vec{r}\,) = \frac{J R}{c}\, \vec{A}_{\, \textsl{g }}\!(\vec{r}\,),                    \eqno\ldots (6.45)
$$
демонстрирующим, что векторный потенциал бесконечно длинного цилиндрического соленоида с постоянным током, в области $ \mathcal{D}_{ex}$, представляет собой, по существу, векторное поле Гаусса, свойства которого хорошо представлены в математической литературе \cite{Fich}.

В частности имеем, что циркуляции векторного поля $ \vec{A}_{ex}(\vec{r}\,)$ по всевозможным простым кусочно-гладким контурам, охватывающих рассматриваемый соленоид и принадлежащих $ \mathcal{D}_{ex}\subset P $ (рис. 6.1), отличны от нуля и равны между собой.

Это "весьма замечательное"\,  \cite{Fich} свойство рассматриваемого векторного поля, наиболее прямолинейно демонстрирует, например, соотношение, непосредственно вытекающее из векторной формы теоремы Стокса, примененной к кусочно-гладкой поверхности $ S$, ограниченной двумя кусочно-гладкими контурами $ \mathcal{L}_{1}$ и $ \mathcal{L}_{2}$, представленными ниже на рис.~6.2~{\em a})\,,
$$
\oint_{\mathcal{L}_{1}^{+}}\vec{A}(\vec{r}\,)\cdot d \vec{r} = - \int\!\!\!\int_{S}\left(\varepsilon\cdot\cdot\mathcal{F}(\vec{r}\,)\right)\cdot \vec{n}(\vec{r}\,)\,dS + \oint_{\mathcal{L}_{2}^{-}}\vec{A}(\vec{r}\,)\cdot d \vec{r}.             \eqno\ldots (6.46)
$$

\begin{figure}[!h]
\begin{center}
\vspace{-6pt}
\includegraphics[width=135mm]{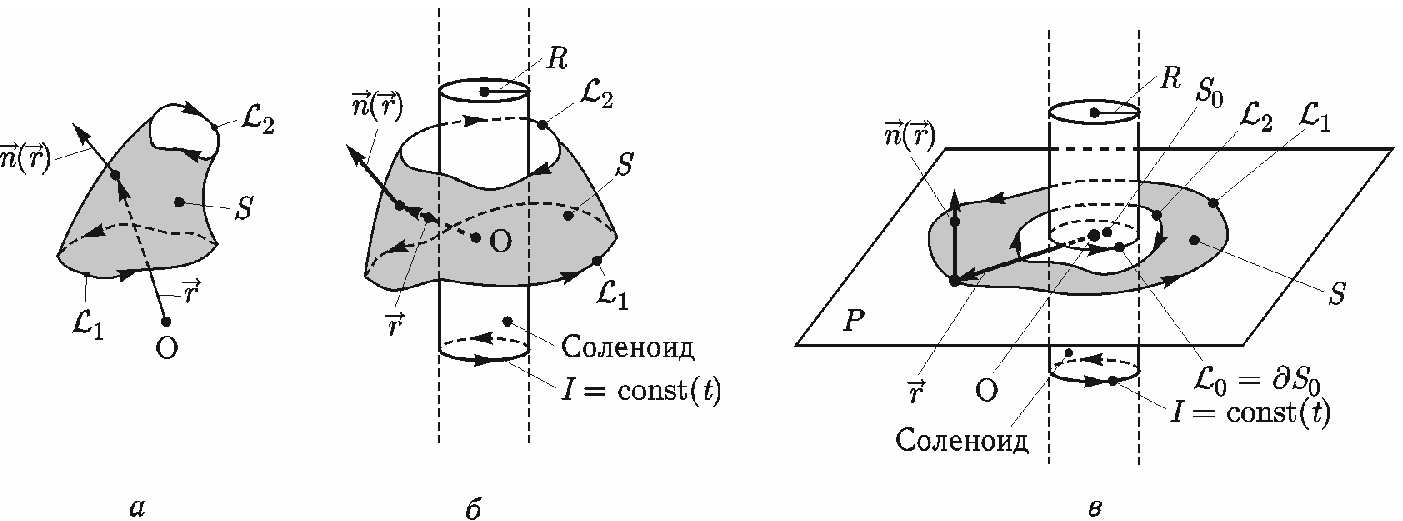}
\vspace{1pt}
\caption{К использованию теоремы Стокса в случае кусочно-гладкой поверхности, для которой два кусочно-гладких контура являются границей этой поверхности.}
\end{center}
\end{figure}

(6.46) можно рассматривать и как общее соотношение, определяющее связь циркуляций векторного поля $ \vec{A}(\vec{r}\,)$ по контурам $ \mathcal{L}_{1}$ и $ \mathcal{L}_{2}$, находящихся, совместно с $ S$, в области определения векторного поля  $ \vec{A}(\vec{r}\,)$, удовлетворяющего соответствующим условиям применимости теоремы Стокса.

Использование (6.46) для контуров $ \mathcal{L}_{1}$ и $ \mathcal{L}_{2}$, охватывающих рассматриваемый соленоид, рис. 6.2 {\em б})\,, приводит, в силу (6.23), к равенству циркуляций векторного поля $ \vec{A}_{ex}(\vec{r}\,)$ по любым таким контурам,
$$
\oint_{\mathcal{L}_{1}^{+}}\vec{A}_{ex}(\vec{r}\,)\cdot d \vec{r} =  \oint_{\mathcal{L}_{2}^{-}}\vec{A}_{ex}(\vec{r}\,)\cdot d \vec{r}.             \eqno\ldots (6.47)
$$

Общее значение всех таких интегралов, то есть, циклическая постоянная $ \sigma$, принимает вид
$$
\sigma = \frac{2\pi JR}{c},                        \eqno\ldots (6.48)
$$
демонстрирующий, в частности, что, при заданной силе тока в соленоиде, увеличение радиуса соленоида приводит к увеличению всех рассматриваемых циркуляций.

С другой стороны, (6.1) и (6.19) демонстрируют выполнение равенства $ \vec{A}_{ex}(\vec{r}\,) =  \vec{A}_{in}(\vec{r}\,)$, при $ \vec{r}=\vec{R}$, в результате чего находим, что векторный потенциал рассматриваемого соленоида представлен  непрерывным (но не непрерывно дифференцируемым) векторным полем.

Указанная непрерывность, совместно с (6.47), приводит к соотношениям
$$
\oint_{\mathcal{L}}\vec{A}_{ex}(\vec{r}\,)\cdot d \vec{r} \stackrel{\mathcal{L}}{\equiv}\oint_{\mathcal{L}_{0}}\vec{A}_{ex}(\vec{r}\,)\cdot d \vec{r}=  \oint_{\mathcal{L}_{0}}\vec{A}_{in}(\vec{r}\,)\cdot d \vec{r},             \eqno\ldots (6.49)
$$
где $ \mathcal{L}$~--~любой кусочно-гладкий контур однократно охватывающий рассматриваемый соленоид.

С другой стороны, использование теоремы Стокса позволяет представить циркуляцию правой части (6.49) в виде
$$
 \oint_{\mathcal{L}_{0}}\vec{A}_{in}(\vec{r}\,)\cdot d \vec{r}=  \int \!\! \! \int_{ S_{0}}\vec{\mathcal{H}}_{in\,L}(\vec{r})\cdot \vec{n}\,\,d S = \Phi_{S_{0}L}.             \eqno\ldots (6.50)
$$

В итоге, вышеуказанное замечательное свойство векторного потенциала $ \vec{A}_{ex}(\vec{r}\,) $, совместно с рассматриваемой непрерывностью, приводят к следующему {\em  нелокальному} соотношению
$$
\oint_{\mathcal{L}}\vec{A}_{ex}(\vec{r}\,)\cdot d \vec{r} \stackrel{\mathcal{L}}{\equiv} \Phi_{S_{0}L}.                              \eqno\ldots (6.51)
$$

(6.50) и (6.51), в частности демонстрируют, что циклическая постоянная $ \sigma$ представлена потоком псевдовекторного поля $ \vec{\mathcal{H}}_{in\,L}(\vec{r})$ через поверхность  натянутую на $ \mathcal{L}_{0}$, то есть, $ \sigma =  \Phi_{S_{0}L}$.

В случае нестационарного векторного поля, $ \vec{A}(\vec{r},t)$, уравнение (6.51) имеет аналогичный вид, при этом, как было отмечено ранее, время $t$  рассматривается как параметр, фиксирующий структуру данного векторного поля в заданный момент времени \cite{Bor}.

В таком случае, взяв частную производную по времени от соответствующего аналога (6.51), при фиксированном $ \mathcal{L}$, приходим к соответствующему закону М. Фарадея для  недеформируемых неподвижных контуров $ \mathcal{L}$
$$
\oint_{\mathcal{L}}\vec{E}_{A\,ex}(\vec{r},t)\cdot d \vec{r}\, \stackrel{\mathcal{L}}{\equiv} - \partial ^{0} \Phi_{S_{0}L}(t),                             \eqno\ldots (6.52)
$$
однако, в отличие от стандартных интерпретаций, соотношение (6.52) теперь рассматривается лишь только как {\em  нелокальное} соотношение, возникающее как непосредственное  следствие (6.49), между классическим  лоренцевым магнитным потоком соленоида и циркуляцией электрического поля
$ \vec{E}_{A\,ex}(\vec{r},t) \!=\! - \partial ^{0}\vec{A}_{ex}(\vec{r},t)$ по контуру однократно охватывающему  соленоид.

И так, выше рассмотренные примеры полевых систем (рис. \!4.2--4.5) демонстрируют, что джейтонное поле равноправно и равновелико, по отношению к магнитному полю, присутствует в важнейших классических полевых системах, созданных фундаментальными классическими источниками.

\section*{\S \,7 Топологические структуры электрических полей, определяемых дифференциальными законами электромагнитной и электроджейтонной индукции}
\addcontentsline{toc}{section}{\S \,7 Топологические структуры электрических полей, определяемых дифференциальными законами электромагнитной и электроджейтонной индукции}

\qquad Возвращаясь к вопросу интерпретации правых частей уравнений (1.11)--(1.13), воспользуемся основными результатами выше рассмотренной общей схемы анализа топологической структуры произвольного векторного поля для соответствующего анализа  топологической структуры  электрических полей, определяемых правыми частями данных уравнений.

Аналогом (2.5) теперь является соотношение
\begin{multline*}
   \vec{ \mathit E}_{\mathit A}(\mathit N) = \vec{ \mathit E}_{A0}(\mathit N) + \tilde{rot}\,\vec{ \mathit E}_{\mathit A}(\mathit M)^{T} \cdot \vec{\rho}\,(\mathit N) +\tilde{def}_{d}\, \vec{ \mathit E}_{A}(\mathit M)\cdot \vec{\rho}\,(\mathit N) + \\
   \tilde{def}_{\ast}\,\vec{ \mathit E}_{\mathit A}(\mathit M) \cdot \vec{\rho}\,(\mathit N) +  \theta \, (\rho), \quad \mathit N \in \delta_{\mathit M},           \qquad \ldots (7.1)
\end{multline*}
из которого следует, что наличие тензора $ \tilde{rot}\,\vec{ \mathit E}_{\mathit A}(\mathit M)$ в данной точке $ \mathit M$ говорит о том, что в $ \delta _{ \mathit M}$ существует неоднородное электрическое поле $ \vec{\mathit E}_{\mathit A\mathcal{F}}( \mathit N)$, определяемое соотношением
$$
 \vec{\mathit E}_{\mathit A\mathcal{F}}( \mathit N) =   \tilde{rot}\,\vec{ \mathit E}_{\mathit A}(\mathit M)^{T} \cdot \vec{\rho}\,(\mathit N), \quad \mathit N \in \delta_{\mathit M},         \eqno\ldots (7.2)
$$
конгруэнция векторных линий которого определяется фазовым портретом тензора $  \tilde{rot}\,\vec{ \mathit E}_{\mathit A}(\mathit M)^{T}$, то есть, фазовым портретом однородной линейной динамической системы
$$
  \frac{d\,\vec{r}}{d\,\tau}= \tilde{rot}\,\vec{ \mathit E}_{\mathit A}(\mathit M)^{T} \cdot \vec{r}.        \eqno\ldots (7.3)
$$

Аналогично, (7.1) демонстрирует так же, что наличие тензоров
 $ \tilde{def}_{d}\, \vec{ \mathit E}_{A}(\mathit M)$  и  $ \tilde{def}_{\ast}\,\vec{ \mathit E}_{\mathit A}(\mathit M) $ в данной точке $ \mathit M$ рассматриваемого электрического поля говорит о том, что  в $  \delta_{\mathit M}$ существуют и неоднородные электрические поля $ \vec{\mathit E}_{\mathit A\mathcal{G}_{d}}( \mathit N)$  и $ \vec{\mathit E}_{\mathit A\mathcal{G}_{\ast}}( \mathit N)$, определяемые соотношениями
$$
  \vec{\mathit E}_{\mathit A\mathcal{G}_{d}}( \mathit N) =  \tilde{def}_{d}\, \vec{ \mathit E}_{A}(\mathit M)\cdot \vec{\rho}\,(\mathit N),  \quad \mathit N \in \delta_{\mathit M},           \eqno\ldots (7.4)
$$
$$
  \vec{\mathit E}_{\mathit A\mathcal{G}_{\ast}}( \mathit N) =  \tilde{def}_{\ast}\, \vec{ \mathit E}_{A}(\mathit M)\cdot \vec{\rho}\,(\mathit N),  \quad \mathit N \in \delta_{\mathit M},           \eqno\ldots (7.5)
$$
 конгруэнции векторных линий которых определяются, соответственно,  фазовыми портретами тензоров $ \tilde{def}_{d}\, \vec{ \mathit E}_{A}(\mathit M)$ и $ \tilde{def}_{\ast}\, \vec{ \mathit E}_{A}(\mathit M)$, то есть, фазовыми портретами однородных линейных динамических систем
 $$
  \frac{d\,\vec{r}}{d\,\tau}= \tilde{def}_{d}\,\vec{ \mathit E}_{\mathit A}(\mathit M) \cdot \vec{r},        \eqno\ldots (7.6)
 $$
 $$
 \frac{d\,\vec{r}}{d\,\tau}= \tilde{def}_{\ast}\,\vec{ \mathit E}_{\mathit A}(\mathit M) \cdot \vec{r}.        \eqno\ldots (7.7)
 $$

 Теперь, в соответствии с результатами выше рассмотренного общего случая произвольного дифференцируемого действительного векторного поля $ \vec{A}(x)$, находим, что фазовые портреты производных
 тензоров $  \tilde{rot}\,\vec{ \mathit E}_{\mathit A}(x)$, $ \tilde{def}_{d}\, \vec{ \mathit E}_{A}(x)$  и
  $ \tilde{def}_{\ast}\,\vec{ \mathit E}_{\mathit A}(x) $, в каждый фиксированный момент времени $ t$, представлены, соответственно, конгруэнциями векторных линий электрических полей (7.2), (7.4) и (7.5), то есть, классическим центром, седлом и дикритическим узлом (по-прежнему ограничиваемся случаями динамических систем на плоскости), представленными  на рис. 7.1 {\em a})\,--\,{\em в})\,.

 \setcounter{section}{7}
\setcounter{figure}{0}

\begin{figure}[!h]
\begin{center}
\vspace{-6pt}
\includegraphics[width=118mm]{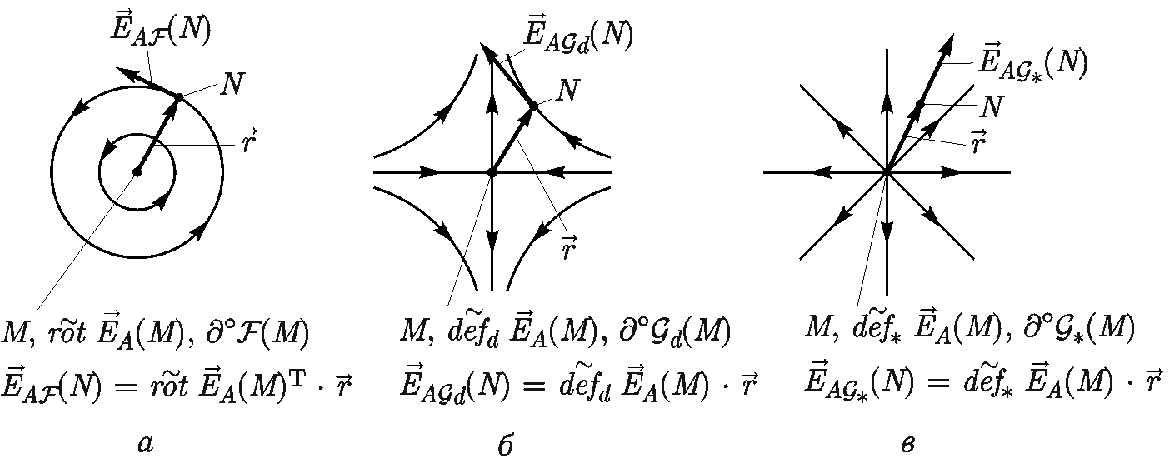}
\vspace{1pt}
\caption{Фазовые портреты тензоров $\tilde{rot}\,{\vec E}_{A}(M)$, $\tilde{def}_{d}\,{\vec E}_{A}(M)$ и $\tilde{def}_{*}\,{\vec E}_{A}(M)$ в пространстве $E^{2}.$}
\end{center}
\end{figure}

  Таким образом, наличие в некоторой точке $ \mathit M$ рассматриваемых тензоров, $  \tilde{rot}\,\vec{ \mathit E}_{\mathit A}(\mathit M)$, $ \tilde{def}_{d}\, \vec{ \mathit E}_{A}(\mathit M)$  и
  $ \tilde{def}_{\ast}\,\vec{ \mathit E}_{\mathit A}(\mathit M) $, означает наличие, в $ \delta$-~окрестности этой точки, электрических полей, конгруэнции векторных линий которых представлены вышеуказанными фазовыми портретами данных тензоров.

  При этом, так же в соответствии с выше указанной общей схемой, имеем, в качестве аналогов уравнений (2.14)--(2.16) и (2.19)--(2.21), следующую систему соотношений
$$
    div\,\vec{E}_{A\mathcal{F}}(\vec{r}\,) =  \vec{\bigtriangledown}\cdot\,\left(\tilde{rot}\,\vec{E}_{A}(\mathit M)\,^{T}\cdot\,\vec{r}\, \right) = tr\,\tilde{rot}\,\vec{E}_{A}(\mathit M) \equiv 0,                                  \eqno\ldots (7.8)
$$
$$
    div\,\vec{E}_{A\mathcal{G}_{d}}(\vec{r}\,) = \vec{\bigtriangledown}\cdot\,\left(\tilde{def}_{\ast}\,\vec{E}_{A}(\mathit M)\,\cdot\,\vec{r}\, \right) = tr\,\tilde{def}_{\ast}\,\vec{E}_{A}(\mathit M) \equiv 0,                                   \eqno\ldots (7.9)
$$
\begin{multline*}
 div\,\vec{E}_{A\mathcal{G}_{\ast}}(\vec{r}\,) = \vec{\bigtriangledown}\cdot\,\left(\tilde{def}_{d}\,\vec{E}_{A}(\mathit M)\,\cdot\,\vec{r}\, \right) = tr\,\tilde{def}_{d}\,\vec{E}_{A}(\mathit M) =  \\
  div\,\vec{E}_{A}(\mathit M),        \qquad\ldots (7.10)
\end{multline*}
\begin{multline*}
 rot\,\vec{E}_{A\mathcal{F}}(\vec{r}\,) =  \vec{\bigtriangledown}\times\,\left(\tilde{rot}\,\vec{E}_{A}(\mathit M)\,^{T}\cdot\,\vec{r}\, \right) =  - \varepsilon\cdot\cdot\,\tilde{rot}\,\vec{E}_{A}(\mathit M) =  \\
  rot\,\vec{E}_{A}(\mathit M),            \qquad\ldots (7.11)
\end{multline*}
$$
    rot\,\vec{E}_{A\mathcal{G}_{d}}(\vec{r}\,) = \vec{\bigtriangledown}\times\,\left(\tilde{def}_{d}\,\vec{E}_{A}(\mathit M)\,\cdot\,\vec{r}\, \right) = - \varepsilon\cdot\cdot\,\tilde{def}_{d}\,\vec{E}_{A}(\mathit M) \equiv \vec{0} ,                                   \eqno\ldots (7.12)
$$
$$
    rot\,\vec{E}_{A\mathcal{G}_{\ast}}(\vec{r}\,) = \vec{\bigtriangledown}\times\,\left(\tilde{def}_{\ast}\,\vec{E}_{A}(\mathit M)\,\cdot\,\vec{r}\, \right) = - \varepsilon\cdot\cdot\,\tilde{def}_{\ast}\,\vec{E}_{A}(\mathit M) \equiv \vec{0},                                   \eqno\ldots (7.13)
$$
которая говорит о том, что фазовые портреты тензоров  $  \tilde{rot}\,\vec{ \mathit E}_{\mathit A}(\mathit M)$, $ \tilde{def}_{d}\, \vec{ \mathit E}_{A}(\mathit M)$  и
  $ \tilde{def}_{\ast}\,\vec{ \mathit E}_{\mathit A}(\mathit M) $  представлены, как и фазовые портреты тензоров $\mathcal{F}(\mathit M)$, $\mathcal{G}_{d}(\mathit M)$  и   $\mathcal{G}_{\ast}(\mathit M)$, конгруэнциями  соответствующих векторных (силовых) линий соленоидального (вихревого), лапласова и эйлерова электрических полей, соответственно.

Данное обстоятельство обеспечивает математическую согласованность правой и левой частей каждого из уравнений системы (1.11)--(1.13) в рассматриваемом геометрическом отношении.

Соответствующие {\em  сопровождающие} геометрические объекты рассматриваемых тензоров и их интерпретация соответствуют  аналогичным объектам рассмотренного выше общего случая произвольного  векторного поля и тут не приводятся.

Ввиду {\em  деформирующего} силового воздействия электрических полей, представленных на рис. 7.1 {\em б})\, и {\em в})\,, на протяженные  {\em  системы} электрически заряженных частиц, будем называть эти электрические поля, в дальнейшем, девиатационным деформирующим электрическим полем и дилатационным деформирующим электрическим полем, соответственно.

В итоге имеем, что каждое из  {\em  тензорных} электрических полей, представленных на рис. 7.1, оказывает, на протяженные системы электрически заряженных частиц,  {\em  своё} специфическое силовое  воздействие, а именно, вращательное (без изменения формы и объема системы), деформирующее, сопровождающееся изменением формы без изменения объема, и деформирующее воздействие, которое сопровождается изменением объема без изменения формы электрической системы, то есть, данные тензорные поля вызывают в рассматриваемых системах типичные процессы рассматриваемые в  классической механике сплошных сред.

Таким образом,  в качестве  {\em  физических} воздействий тензорных полей,  $  \tilde{rot}\,\vec{ \mathit E}_{\mathit A}(\mathit M)$, $ \tilde{def}_{d}\, \vec{ \mathit E}_{A}(\mathit M)$  и   $ \tilde{def}_{\ast}\,\vec{ \mathit E}_{\mathit A}(\mathit M) $, выступают силовые воздействия их электрических полей, представленных на рис. 7.1 {\em a})\,--\,{\em в})\,.

Рассматриваемые силовые воздействия данных полевых объектов на протяженные  системы электрически заряженных частиц будем называть {\em тензорными} воздействиями.

И так имеем, что электрические поля, порождаемые переменными во времени магнитным и джейтонными полями,  обладают принципиально разными   тензорными силовыми воздействиями на системы электрически заряженных частиц.

В этом отношении, каждый из рассматриваемых законов имеет  самостоятельное {\em  физическое} значение, что позволяет перевести  уравнения,  представляющие эти законы, из класса математических соотношений в класс фундаментальных {\em  физических} уравнений.

 С другой стороны имеем, что, в соответствии с явлением электроджейтонной индукции представленным уравнениями (1.12) и (1.13), джейтонная составляющая электроджейтонного поля (джейтонное поле) оказывает, посредством порождаемого ей (им) деформирующих электрических полей, специфическое {\em  тензорное} силовое воздействие на системы электрически заряженных частиц.

В свою очередь, наличие самостоятельных  физических воздействий джейтонного поля  позволяет перевести уже само джейтонное  поле, а с ним и электроджейтонное поле,  из класса линейно независимых {\em математических} объектов в класс фундаментальных {\em  физических} полевых систем.

Теперь можем завершить {\em  физическую} интерпретацию уравнений (1.11)--(1.13), констатируя, что если уравнение (1.11), представляющее известный закон электромагнитной индукции, говорит о том, что переменное во времени магнитное поле, представленное в (1.11) тензором $ \mathcal{ F}(x)$, порождает, в $ \delta$-окрестности точки $ \mathit M$, электрическое поле $ \vec{\mathit E}_{\mathit A \mathcal{ F}}(x)$, имеющее конгруэнцию силовых линий представленных на рис. 7.1{\em a} )\, и вызывающее в протяженных электрических системах {\em  вихревые} процессы (движения), то законы электроджейтонной индукции, представленные уравнениями (1.12) и (1.13), говорят о том, что переменные во времени джейтонные поля, представленные в (1.12) и (1.13) тензорами $ \mathcal{G}_{d}(x)$ и $  \mathcal{G}_{\ast}(x)$, порождают, в $ \delta$-окрестности точки $ \mathit M$, электрические поля
 $ \vec{\mathit E}_{\mathit A \mathcal{ G}_{d}}(x)$  и  $ \vec{\mathit E}_{\mathit A \mathcal{ G}_{\ast}}(x)$, обладающие принципиально иными видами конгруэнций силовых линий представленных на рис. 7.1 {\em б})\, и {\em в})\, и вызывающие,  в протяженных электрических системах,  {\em деформационные} процессы (движения), представленные выше.

 С другой стороны, согласно (7.1), как математического аналога (2.5), имеем конгруэнтное разложение плоского электрического поля $ \vec{\mathit E}_{\mathit A}(x)$,   топологически идентичное представленному на рис. 4.1, то есть,  конгруэнция силовых линий {\em  произвольного} плоского электрического поля $ \vec{\mathit E}_{\mathit A}(x)$ всегда может быть представлена, в общем случае,  суперпозицией четырех базовых конгруэнций, являющихся классическим центром, седлом,  дикритическим узлом и конгруэнцией силовых линий соответствующего однородного электрического поля.

 С учетом этого, тогда имеем, что система (1.11)--(1.13) является  {\em  полной} системой в том отношении, что она представляет (описывает) "рождение"\,, соответствующими переменными во времени полями,  {\em  всех} возможных базовых конгруэнций неоднородной составляющей произвольного электрического поля $ \vec{\mathit E}_{\mathit A}(x)$.

 В этом отношении, описание явления электродинамической индукции только теперь можно считать математически завершенным.

 Рассматриваемую систему (1.11)--(1.13) можно переписать, используя определения (1.6)--(1.8), в виде соотношений, содержащих векторный потенциал $ \vec{\mathit A}(x)$, традиционно "изгоняемый"\, из подобных уравнений  \cite{La},
 $$
\partial^{\,0} \tilde{rot}\,\vec{A}(x) = -  \tilde{rot}\,\,\vec{E}_{\mathit A}(x),               \eqno\ldots (7.14)
$$
$$
\partial^{\,0} \tilde{def}_{d}\,\vec{A}(x) = -  \tilde{def}_{d}\,\,\vec{E}_{\mathit A}(x),               \eqno\ldots (7.15)
$$
$$
\partial^{\,0} \tilde{def}_{\ast}\,\vec{A}(x) = -  \tilde{def}_{\ast}\,\,\vec{E}_{\mathit A}(x),              \eqno\ldots (7.16)
$$
которые непосредственно демонстрируют, с учетом (1.14), тождественный характер каждого из этих уравнений, как уравнений на векторный потенциал  $ \vec{\mathit A}(x)$, что было отмечено в самом начале настоящего  контекста.

С другой стороны, эти уравнения демонстрируют, что структура каждого из рассматриваемых электрических  полей
 повторяет топологию геометрического образа соответствующего тензора,
 $ \mathcal{ F}(x)$, $ \mathcal{G}_{d}(x)$ или $  \mathcal{G}_{\ast}(x)$,  вследствие чего, дифференциальные законы электродинамической индукции, заданные в аналитическом виде уравнениями (1.11)--(1.13), могут быть  качественно  представлены в виде  геометрических соотношений, представленных ниже на рис.\, 7.2.

\begin{figure}[!h]
\begin{center}
\vspace{-6pt}
\includegraphics[width=90mm]{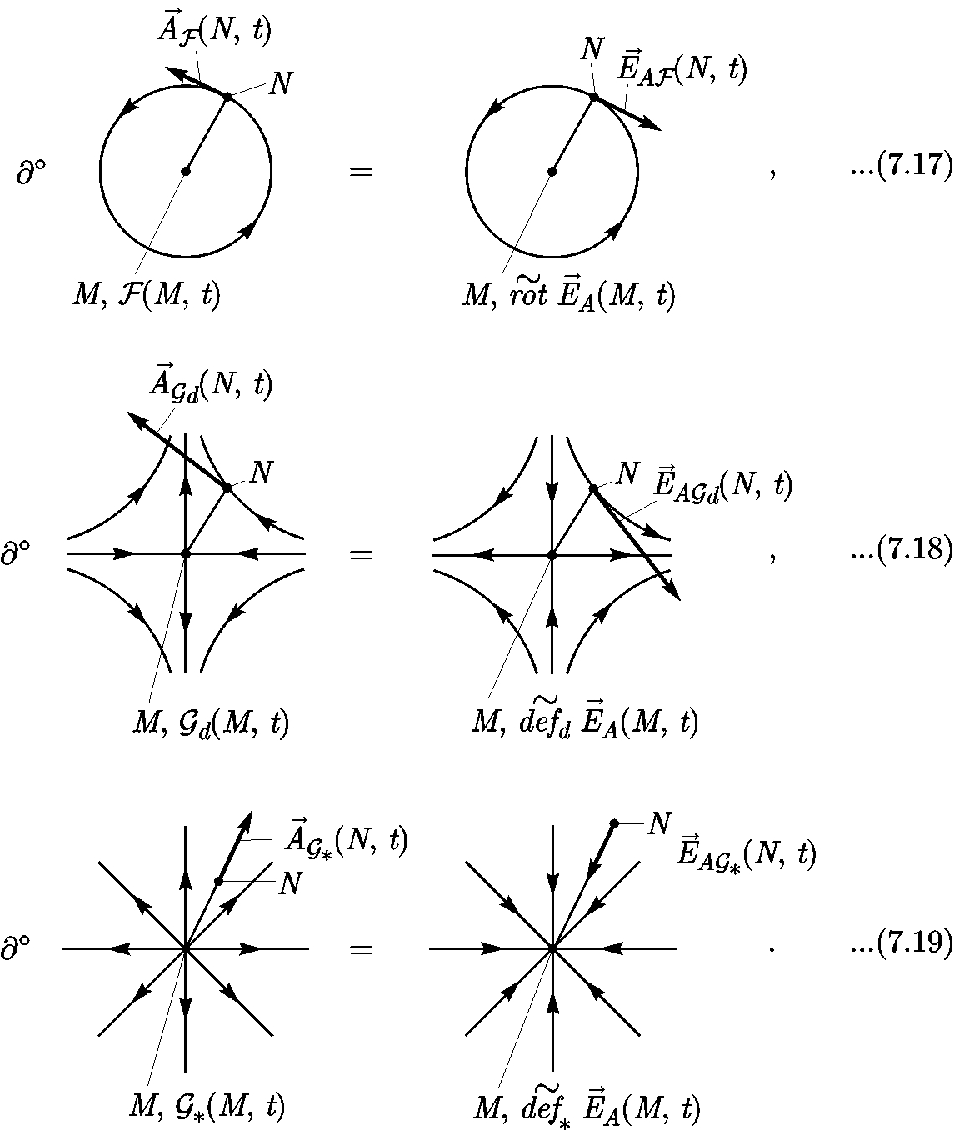}
\vspace{1pt}
\caption{Качественное графическое представление аналитических  соотношений (1.11)~\!--~\!(1.13) для случая возрастающих, во времени, векторных потенциалов.}
\end{center}
\end{figure}

Соотношение (7.18), как и его аналитический аналог (1.12), демонстрируют, с другой стороны, что электрическое поле квадрупольного вида может иметь так же и полевое происхождение, а именно, может порождаться изменяющимся во времени девиатационным джейтонным полем.

Наблюдаемость электрических полей, порождаемых изменяющимися во времени джейтонными полями, представленными тензорами   $ \mathcal{G}_{d}(x)$ и $  \mathcal{G}_{\ast}(x)$, означает, что законы электроджейтонной индукции, представленные уравнениями (1.12) и (1.13), относятся к рангу {\em  физических} законов, а сами джейтонные поля относятся, как уже было подчеркнуто выше,  к классу {\em  физических} полей.

  Далее рассмотрим  конкретные примеры полевых систем, в которых "действуют"\, обсуждаемые явления  электродинамической  индукции.

Как и прежде, время $ t$ рассматривается как параметр, определяющий структуру (состояние) нестационарного векторного поля в данный момент времени.

Ради краткости записей, этот параметр,  в аргументах соответствующих функций и выражений, по-прежнему часто будем опускать без специальных оговорок.

\section*{\S \,8 Топологические структуры конгруэнтных составляющих электрического поля плоской волны Ландау}
\addcontentsline{toc}{section}{\S \,8 Топологические структуры конгруэнтных составляющих электрического поля плоской волны Ландау}

\qquad Рассмотрим электродинамическую полевую систему, созданную линейно вибрирующим (совершающим финитное одномерное гармоническое колебательное движение) точечным электрическим зарядом, векторный и скалярный потенциалы которой представлены хорошо известными потенциалами Лиенара\,--\,Вихерта \cite{La, Jack}, представляющими собой точные решения соответствующего неоднородного уравнения Даламбера для 4-потенциала $ \mathit A(x)$.

Ниже изложенное выступает, одновременно, и как анализ полевой системы, созданной ускоренно движущимся точечным электрическим зарядом.

Ограничимся рассмотрением полей волновой зоны излучения данного электрического вибратора в его экваториальной плоскости, соответствующей максимуму диаграммы направленности электродинамического излучения данного вибратора, что автоматически (без использования соответствующих дополнительных условий на электродинамический потенциал $ \mathit A(x)$) обеспечивает выполнение как соотношения $ \vec{\mathit E}(x) = \vec{\mathit E}_{ \mathit A}(x)$, характерного для плоской электродинамической волны Ландау  \cite{La}, так и равенство $  \mathcal{G}_{\ast}(x) = \tilde{0}$ в данной области волновой зоны.

Относительно подробный геометрический анализ данной фундаментальной полевой системы представлен на рис. 8.1.

\setcounter{section}{8}
\setcounter{figure}{0}

\begin{figure}[!h]
\begin{center}
\vspace{-6pt}
\includegraphics[width=90mm]{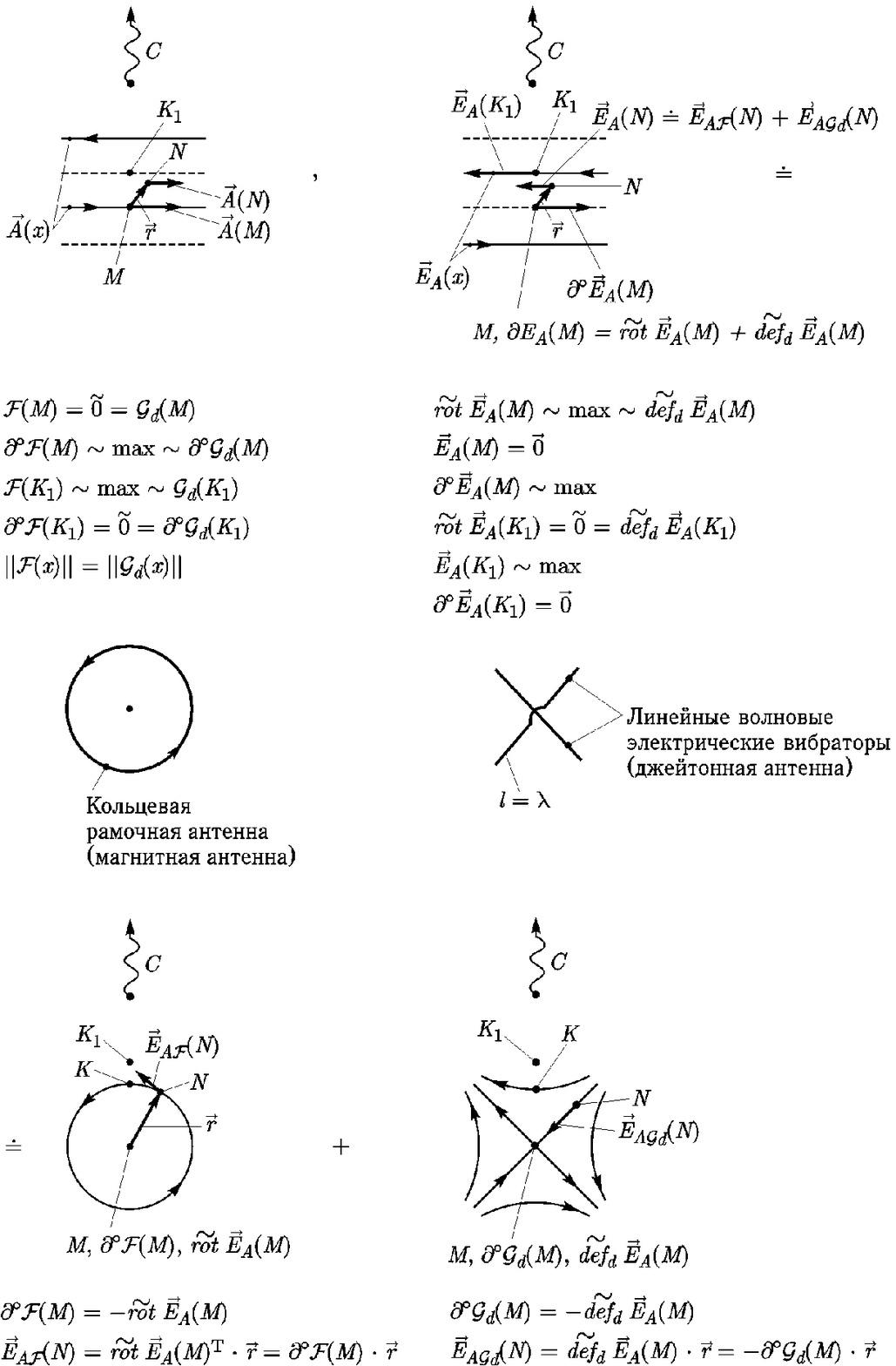}
\vspace{1pt}
\caption{Конгруэнтное разложение электрического поля плоской волны Ландау созданной линейно вибрирующей точечной электрически заряженной частицей, дифференциальные законы электромагнитной и электроджейтонной индукции, действующие в этой волне, и их графическая (геометрическая) интерпретация. Случай когда точка $M$ находится вне особой линии электрического поля ${\vec E}_{A}(M)$  сводится к дополнительному учету соответствующего однородного электрического поля, что содержательной стороны рассматриваемых процессов в рассматриваемой волне не затрагивает. Анализ произвольного направления излучения идентичен представленному на данном рисунке с элементарной заменой ${\vec A}(x) \to \vec {A}^{\tau}(x)$ и $\vec{ E}_{A}(x) \to \vec {E}_{A}^{\tau}(x).$}
\end{center}
\end{figure}

Рис. 8.1, совместно с рис. 4.3 {\em a})\,, демонстрируют, что, во-первых, рассматриваемая плоская волна представлена, помимо электрического поля с напряженностью  $  \vec{\mathit E}_{ \mathit A}(x)$, не только магнитным полем, определенным тензором $ \mathcal{ F}(x)$, но и девиатационным джейтонным полем, представленным тензором $ \mathcal{G}_{d}(x)$,
то есть, данная фундаментальная волна является, в действительности, электромагнитоджейтонной волной, что уже было констатировано в комментариях к рис. 4.3 {\em a})\,.

Во-вторых, магнитное и  девиатационное  джейтонное поля данной волны, в соответствии с законами электромагнитной и электроджейтонной индукции, "порождают"\,, в $ \delta _{\mathit M}$, электрические поля, конгруэнции силовых линий которых, в результате их сложения, составляют, в линейном приближении, конгруэнцию исходного электрического поля, с напряженностью  $  \vec{\mathit E}_{ \mathit A}(x)$, в этой окрестности $ \delta _{\mathit M}$.

Таким образом, и магнитное и девиатационное джейтонное  поля {\em  совместно}  участвуют в "создании"\, {\em  наблюдаемого} электрического поля данной волны.

При этом важно еще раз подчеркнуть, что магнитное поле   ассоциируется  теперь с тензором $ \mathcal{ F}(x)$ и сопровождающим его псевдовектором $ \vec{\mathcal{H}}(x)$, в отличие от традиционной классической электродинамики, где магнитное поле связывается с соответствующими {\em  лоренцевыми} величинами, $ \mathcal{ F}_{\mathit L}(x)$   и $ \vec{\mathcal{H}}_{\mathit L}(x)$.

Как следствие такого определения и равенства норм, $ ||\,\tilde{rot}\,\vec{ \mathit E}_{ \mathit A}(\mathit M)\,|| = ||\,\tilde{def}_{d}\,\vec{ \mathit E}_{ \mathit A}(\mathit M)\,||$, выполняющегося в данном случае,  в любой точке $ \mathit K$, принадлежащей $ \delta_{\mathit M}$ и находящейся на прямой $a$ (рис. 8.1), имеем
$$
\vec{E}_{\mathit A \mathcal{ F}}( \mathit K) = \frac{1}{2}\,\vec{E}_{\mathit A}( \mathit K) =
 \vec{E}_{\mathit A \mathcal{ G}_{d}}( \mathit K).                              \eqno\ldots (8.1)
$$

То есть, в таких точках, принадлежащих экваториальной плоскости максимума диаграммы направленности рассматриваемого излучения,   {\em  половина} напряженности электрического поля данной волны "создается"\, переменным во времени магнитным полем, представленным тензором $ \mathcal{ F}(x)$, а вторая  {\em  половина} --
переменным во времени девиатационным джейтонным полем, определенным тензором $ \mathcal{G}_{d}(x)$, тем самым, демонстрируется равноправная  роль магнитного и девиатационного джейтонного полей в процессах, происходящих в {\em  фундаментальной}  полевой системе, представленной рассматриваемой плоской волной Ландау.

Традиционная теория, определяющая магнитное поле посредством лоренцева тензора $ \mathcal{ F}_{\mathit L}(x)$, рассматривает электрическое поле в точке $ \mathit K$ как поле рожденное исключительно  {\em  магнитным} полем, в результате чего  {\em  кратно} завышается роль магнитного поля, при полном игнорировании девиатационного джейтонного поля и связанного с ним явления электроджейтонной индукции.

Волновой характер распространения полей, составляющих рассматриваемую полевую систему, является непосредственным следствием волнового характера распространения векторного потенциала   $\vec{\mathit A}(x)$, выступающего в качестве "носителя"\, данного классического полевого комплекса.

Другие,  не менее фундаментальные полевые системы, созданные, например, элементарным электрическим излучателем или полуволновым линейным электрическим вибратором, принципиально идентичны, в рассматриваемой области волновой зоны, выше рассмотренной полевой системе линейно вибрирующего точечного электрического заряда, поэтому, все результаты выше рассмотренного анализа автоматически распространяются и на эти полевые системы этих фундаментальных источников электродинамического излучения.

\section*{\S \,9 Топологические структуры конгруэнтных составляющих электрического поля прямолинейного бесконечно длинного проводника с переменным электрическим током квазистационарной частоты}
\addcontentsline{toc}{section}{\S \,9 Топологические структуры конгруэнтных составляющих электрического поля прямолинейного бесконечно длинного проводника с переменным электрическим током квазистационарной частоты}

\qquad В силу квазистационарности данной полевой системы, топологии конгруэнций силовых линий электрического поля, определяемого напряженностью $ \vec{ \mathit E}_{ \mathit A}(x)$,  повторяют соответствующие топологии конгруэнций векторных линий векторного потенциала $ \vec{ \mathit A}(x)$,  представленные на рис.~4.2, в связи с чем, конгруэнтное разложение данного электрического поля данной полевой системы имеет топологически идентичный вид, представленный на рис. 9.1.
\\
\\
\setcounter{section}{9}
\setcounter{figure}{0}

\begin{figure}[!h]
\begin{center}
\vspace{-6pt}
\includegraphics[width=120mm]{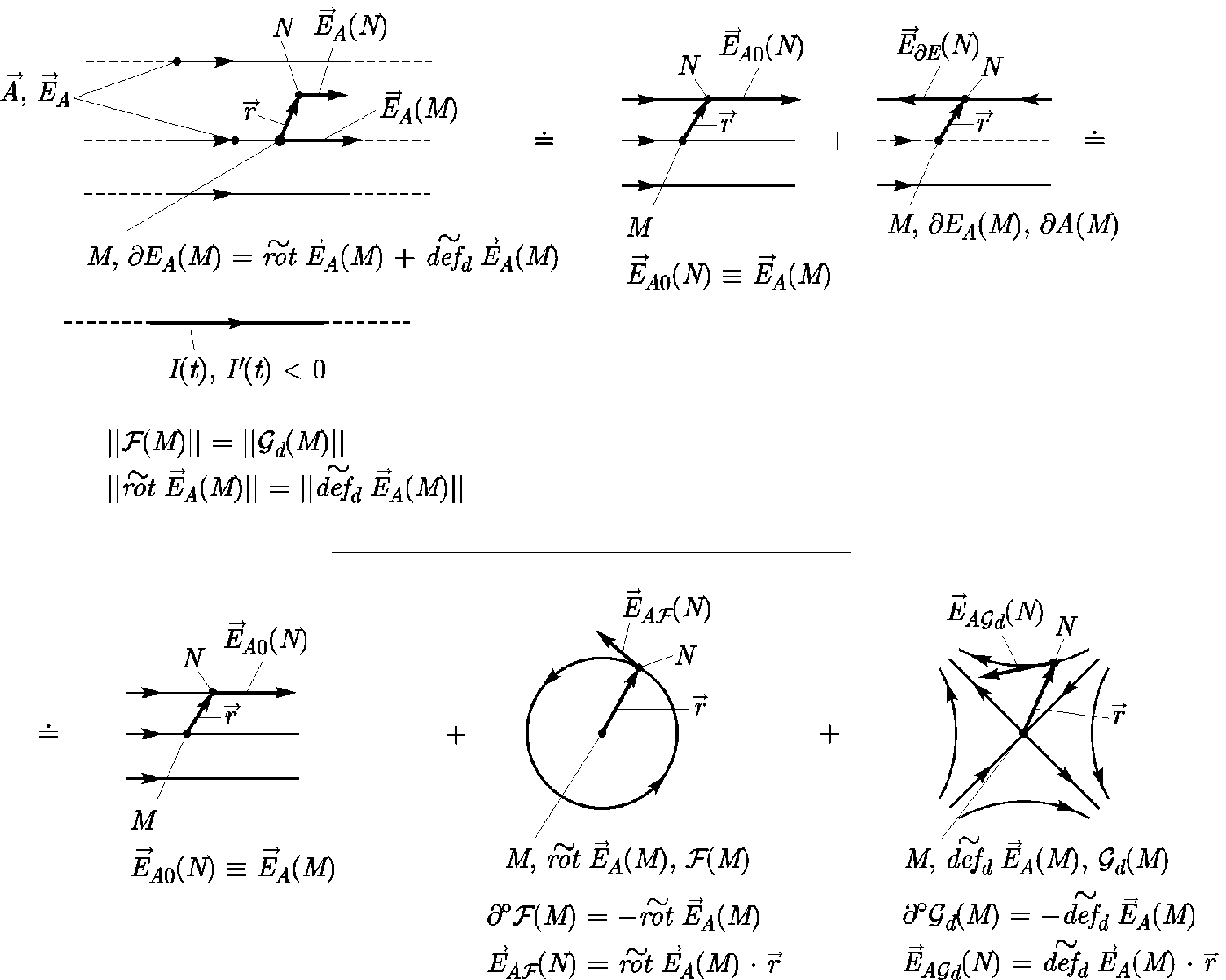}
\vspace{1pt}
\caption{Конгруэнтное разложение квазистационарного электрического поля ${\vec E}_{A}(x)$ прямолинейного бесконечно длинного проводника с переменным током квазистационарной частоты, дифференциальные законы электромагнитной и электроджейтонной индукции и их геометрическая интерпретация.}
\end{center}
\end{figure}

Рис. 9.1, совместно с рис. 4.2, демонстрируют, что и полевая система, порождаемая рассматриваемым переменным током, представлена, в действительности, электромагнитоджейтонным полем, в котором, так же как и в полевой системе рис.~8.1, действуют законы электромагнитной и электроджейтонной индукции, представленные уравнениями (1.11) и (1.12), в соответствии с которыми переменные во времени магнитное и девиатационное джейтонное поля данной системы порождают, в $ \delta_{\mathit M}$, соответствующие электрические поля, которые, совместно с однородной частью данного поля, $\vec{\mathit E}_{\mathit A\,0}(\mathit N)$, составляют, в линейном приближении, {\em  наблюдаемое} электрическое поле $ \vec{\mathit E}_{\mathit A}(\mathit N)$ в рассматриваемой $ \delta$-окрестности точки  $ \mathit M$.

При этом, девиатационное джейтонное поле ведет себя, в этом процессе, так же равноправно, по отношению к соответствующему магнитному полю, как и в полевой системе рис. 8.1.

Более того, в силу топологической неэквивалентности  \cite{Arr} порождаемых ими электрических полей, роль девиатационного джейтонного поля в этом процессе не может быть сведена к роли магнитного поля и наоборот, что является очередным фактом, демонстрирующим самостоятельность этих полей по отношению к друг другу.

\section*{\S \,10 Топологические структуры конгруэнтных составляющих электрического поля бесконечно длинного цилиндрического соленоида с переменным электрическим током квазистационарной частоты}
\addcontentsline{toc}{section}{\S \,10 Топологические структуры конгруэнтных составляющих электрического поля бесконечно длинного цилиндрического соленоида с переменным электрическим током квазистационарной частоты}

\qquad  Вновь, в силу квазистационарности рассматриваемой классической полевой системы, конгруэнтное разложение векторного поля $ \vec{ \mathit E}_{ \mathit A}(x)$, в некоторой $ \delta_{\mathit M}$, повторяет топологию конгруэнтного разложения векторного потенциала $ \vec{ \mathit A}(x)$, в той же $ \delta_{\mathit M}$, продемонстрированного на рис. 4.4.

Данная полевая система, представленная ниже на рис. 10.1, интересна прежде всего тем, что она, являясь примером автономного существования магнитного и девиатационного джейтонного полей, одновременно  представляет собой пример существования пространственно автономных, по отношению к друг другу, явлений электромагнитной и электроджейтонной индукции, а так же пример пространственно автономного непосредственного силового воздействия данных полей на движущиеся, в этих полях, электрически заряженные частицы и, в частности, на проводники с электрическим током.

\setcounter{section}{10}
\setcounter{figure}{0}

\begin{figure}[!h]
\begin{center}
\vspace{-6pt}
\includegraphics[width=100mm]{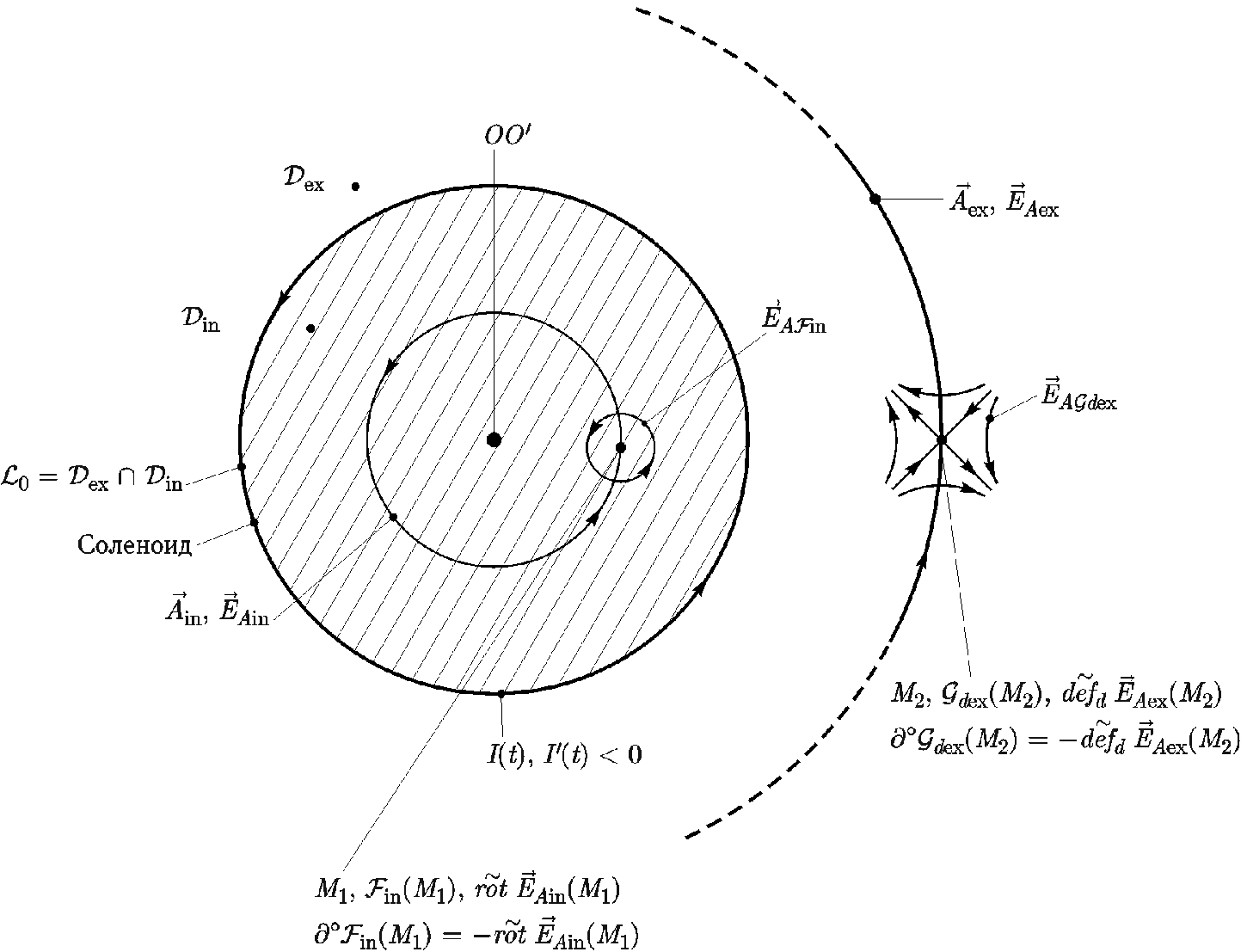}
\vspace{1pt}
\caption{Дифференциальные законы электродинамической индукции внутри и вне рассматриваемого соленоида и их геометрическая интерпретация. В любой точке $M_{1} \in {\mathcal D}_{in}\backslash {\mathcal L}_{0}$ действует только закон электромагнитной индукции, а в любой точке  $M_{2} \in {\mathcal D}_{ex}\backslash {\mathcal L}_{0}$~--- только закон электроджейтонной индукции. Рассматриваемая плоскость нормальна к продольной оси $OO^{\prime}$ данного соленоида.}
\end{center}
\end{figure}

 И так,  рис. 10.1  демонстрирует, что {\em  внутри} данного соленоида, ввиду отсутствия в этой области джейтонных полей, "действует"\, только общеизвестный закон электромагнитной индукции, представленный, как и ранее, тензорным дифференциальным уравнением~(1.11),
 $$
\partial^{\,0}\, \mathcal{F}_{in}(\mathit M) = -  \tilde{rot}\,\,\vec{E}_{\mathit A\,in}(\mathit M), \quad  \mathit M\in \mathcal{D}_{in}\backslash \mathcal{L}_{0},              \eqno\ldots (10.1)
$$
вследствие чего в $ \forall\,\,\delta_{ \mathit M}\subset \mathcal{D} _{in}$ имеем, помимо соответствующего однородного электрического поля, только классический центр векторного поля $ \vec{E}_{\mathit A \mathcal{F}in}(\mathit N)$, "порожденный"\,, согласно (10.1), изменяющимся во времени магнитным полем, представленным тензором $ \mathcal{F}_{in}(\mathit M)$.

Таким образом,  {\em  внутри} рассматриваемого соленоида электрическое поле, "рожденное"\, изменяющимся во времени магнитным полем, совместно с соответствующим однородным электрическим полем, образуют (формируют)  {\em  наблюдаемую} конгруэнцию силовых линий электрического поля $ \vec{E}_{\mathit A in}(\mathit N)$, в $ \forall\,\,\delta_{ \mathit M}\subset \mathcal{D}_{in}$, что является одним из фактов, относящих магнитное поле к классу физических полей.

Вне данного соленоида, ввиду отсутствия  в этой области  магнитного и дилатационного джейтонного полей, "действует"\, только закон электроджейтонной индукции, представленный дифференциальным уравнением~(1.12),
$$
\partial^{\,0}\, \mathcal{G}_{d\,ex}(\mathit M) = - \, \tilde{def}_{d}\,\,\vec{E}_{A\,ex}(\mathit M), \quad  \mathit M\in \mathcal{D}_{ex}\backslash \mathcal{L}_{0},                 \eqno\ldots (10.2)
$$
вследствие чего в $ \forall\,\,\delta_{ \mathit M}\subset\mathcal{D}_{ex}$ имеем, помимо соответствующего однородного электрического поля, только классическое седло векторного поля $ \vec{E}_{\mathit A \mathcal{G}_{d}ex}(\mathit N)$,
"порожденное"\,, согласно (10.2), изменяющимся во времени девиатационным джейтонным полем, представленным тензором $ \mathcal{G}_{d\,ex}(\mathit M)$.

То есть,  {\em  вне} данного соленоида, электрическое поле "рожденное"\,  изменяющимся во времени девиатационным джейтонным полем, совместно с соответствующим однородным электрическим полем, образуют (формируют)  {\em  наблюдаемую} конгруэнцию силовых линий  электрического  поля $ \vec{E}_{\mathit A\, ex}(\mathit N)$ в  $ \forall\,\,\delta_{ \mathit M}\subset \mathcal{D}_{ex}$, что, как и в выше рассмотренном случае  магнитного поля, так же следует рассматривать одним из фактов, относящих данное джейтонное поле к классу {\em  физических} полей.

\section*{\S \,11 Топологические структуры конгруэнтных составляющих электрического поля тороидального соленоида с переменным электрическим током квазистационарной частоты}
\addcontentsline{toc}{section}{\S \,11 Топологические структуры конгруэнтных составляющих электрического поля тороидального соленоида с переменным электрическим током квазистационарной частоты}

\qquad По-прежнему,    в силу квазистационарности данной  полевой системы, конгруэнтное разложение векторного поля $ \vec{ \mathit E}_{ \mathit A}(x)$, в некоторой $ \delta_{\mathit M}$, повторяет топологию конгруэнтного разложения векторного потенциала $ \vec{ \mathit A}(x)$, в той же $ \delta_{\mathit M}$, продемонстрированного на рис. 4.5.

Данная классическая полевая система, представленная ниже на рис. 11.1, по отношению к выше рассмотренной полевой системе цилиндрического соленоида, специфична еще и тем, что источник этой полевой системы имеет конечные размеры, в связи с чем, {\em  практическую} пространственную автономность магнитного и девиатационного джейтонного полей в этом случае осуществить проще.

Как и в полевой системе  цилиндрического соленоида, в данной полевой системе имеем, что вне тора электрическое поле "рожденное"\, изменяющимся во времени девиатационным джейтонным полем, совместно с соответствующим однородным электрическим полем, образуют (формируют) {\em  наблюдаемую} конгруэнцию векторных линий векторного поля $ \vec{E}_{\mathit A\, ex}(\mathit N)$ в  $ \forall\,\,\delta_{ \mathit M}\subset \mathcal{D}_{ex}$, что так же относит девиатационное джейтонное поле тороида к классу физических полей.

При этом уместно так же отметить, что фазовые портреты тензоров $ \mathcal{G}_{d}(\mathit M)$ и
$ \tilde{def}_{d}\,\vec{E}_{\mathit A}( \mathit M )$, в случае полевых систем рассмотренных на рис.~4.3~{\em б})\,, 4.5 {\em a})\, и 11.1, представлены  {\em  трехмерными} стандартными (классическими) седлами с усами
  $ \mathit R^{m_{+}} = \mathit R\,^{2}$  и $ \mathit R^{m_{-}} = \mathit R\,^{1}$, внеусовые  фазовые кривые которых являются семействами {\em  неравнобочных}  гипербол с показателем  $ k = - 1/2$  \cite{Arn}, в отличие от полевых систем рассмотренных на рис. 4.2, 4.3 {\em а})\,, 4.4 {\em a})\,, 4.5 {\em б})\,, и 9.1, для которых фазовые портреты соответствующих тензоров представлены {\em  двухмерными} стандартными седлами, обладающими усами  $ \mathit R^{m_{+}} = \mathit R\,^{1}$  и $ \mathit R^{m_{-}} = \mathit R\,^{1}$, а внеусовые фазовые кривые которых являются "настоящими"\,~гиперболами~с~показателем~$k = - 1$.
 \\
 \\
\setcounter{section}{11}
\setcounter{figure}{0}
\begin{figure}[!h]
\begin{center}
\vspace{-6pt}
\includegraphics[width=135mm]{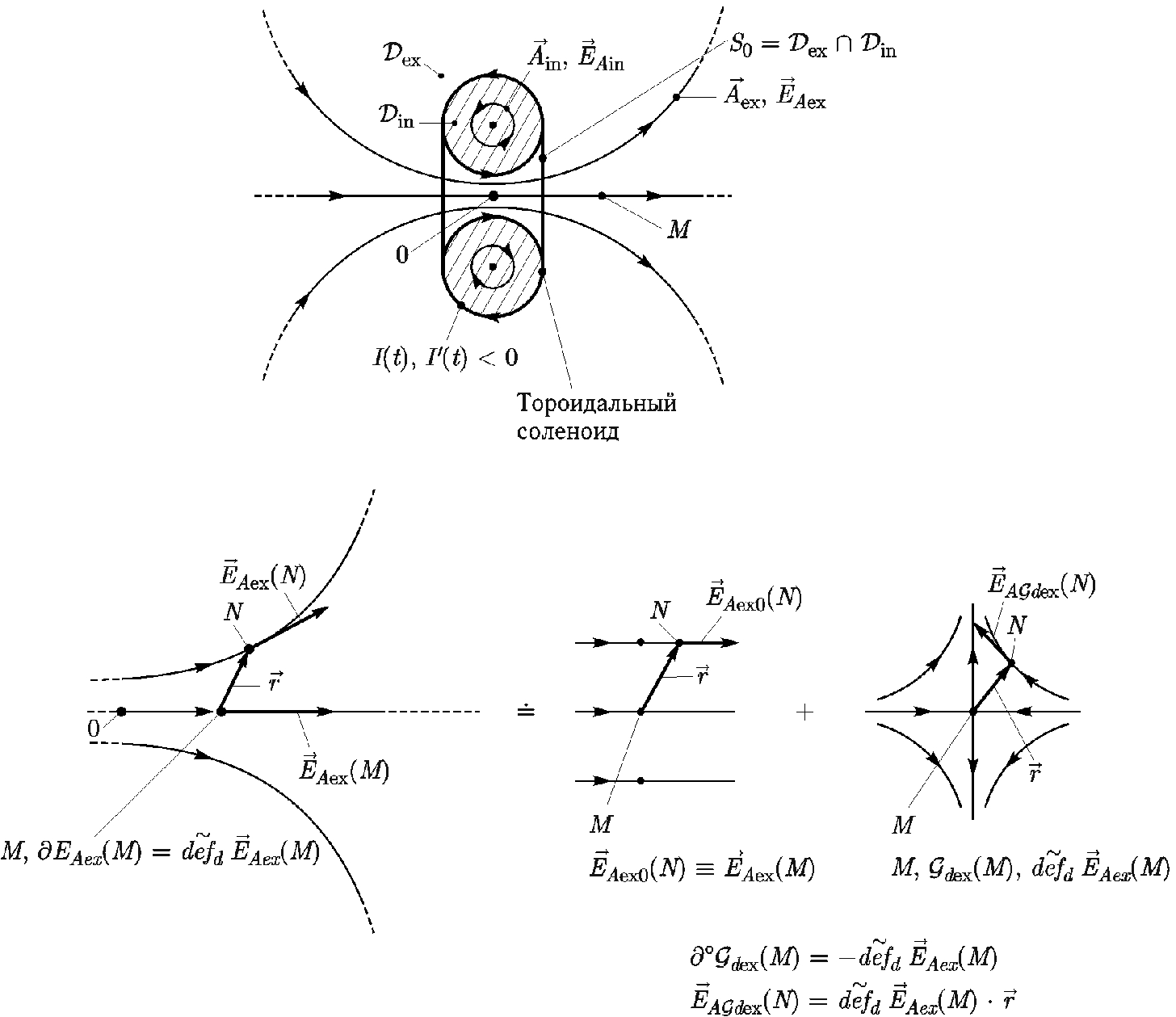}
\vspace{1pt}
\caption{Конгруэнтное разложение квазистационарного электрического поля ${\vec E}_{A}(x)$ в меридиональной плоскости неподвижного тороидального соленоида с переменным током квазистационарной частоты, дифференциальный закон электроджейтонной индукции и его геометрическая интерпретация.}
\end{center}
\end{figure}

  \section*{\S \,12 Топологические структуры электрических полей волноводов и законы электромагнитной и электроджейтонной индукции в соответствующих полевых системах}
\addcontentsline{toc}{section}{\S \,12 Топологические структуры электрических полей волноводов и законы электромагнитной и электроджейтонной индукции в соответствующих полевых системах}

\qquad В качестве очередных полевых систем, демонстрирующих действие законов электродинамической индукции (1.11)--(1.13), выступают полевые системы соответствующих волноводов.

  В рамках концепции Л. Бриллюэна эти полевые системы  можно рассматривать и  как результат суперпозиции выше рассмотренных плоских волн Ландау, то есть, плоских  электромагнитоджейтонных волн.

В этом случае, результат данной суперпозиции будет так же электромагнитоджейтонной волной, что и демонстрируют структуры электрических полей различных типов волн в рассматриваемых направляющих системах.

Таким образом имеем, что и поля подобных хорошо известных электрических систем представлены, в действительности, электромагнитоджейтонными полями, а соответствующие волны этих систем являются, в общем случае, электромагнитоджейтонными волнами.

Конкретные примеры автономного действия законов электродинамической индукции, (1.11)--(1.13), в полевых системах волноводов представлены на рис. 12.1--12.3.
\\
\setcounter{section}{12}
\setcounter{figure}{0}

\begin{figure}[!h]
\begin{center}
\vspace{-6pt}
\includegraphics[width=125mm]{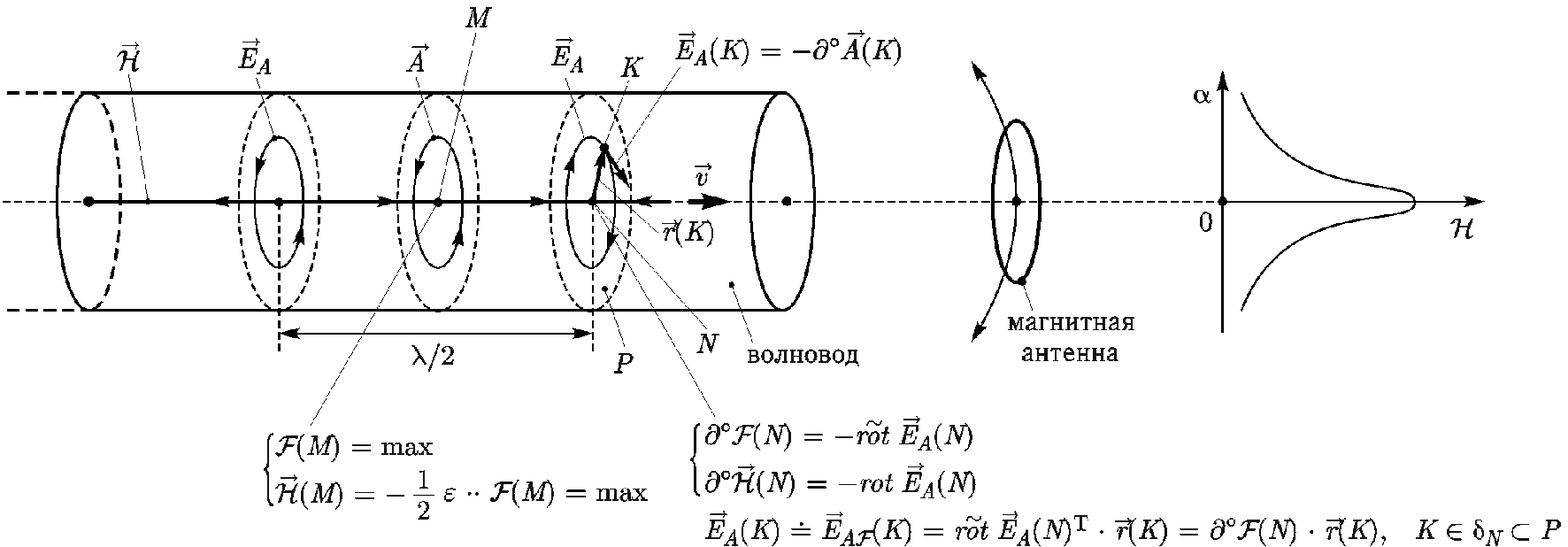}
\vspace{1pt}
\caption{$H_{01}$-волна волновода круглого поперечного сечения: полевая система на продольной оси и в ее $\delta$-окрестности. Конгруэнция векторных линий псевдовекторного поля $\vec{\mathcal {H}}(x)$ вне продольной оси, в целях упрощения рисунка, не указана.}
\end{center}
\end{figure}

Наличие центра электрического поля $ \vec{\mathit E}_{\mathit A}(\mathit K), \mathit K\in \delta_{\mathit N}\subset \mathit P $, в представленной на рис. 12.1 $ \mathit H_{01}$-волне, говорит о наличии в точке $  \mathit N$  производного тензора
 $ \tilde{rot}\,\vec{\mathit E}_{\mathit A}(\mathit N)$, а наличие последнего означает, в силу закона электромагнитной индукции  представленного дифференциальным уравнением (1.11),
 $$
 \partial\, ^{0}\,\mathcal{F}( \mathit N ) = - \tilde{rot}\,\vec{\mathit E}_{\mathit A}(\mathit N),           \eqno\ldots (12.1)
 $$
 наличие, на продольной оси рассматриваемого волновода,  переменного во времени магнитного поля, представленного
  тензором   $ \mathcal{F}( \mathit N )$,  создавшего, в $ \delta_{\mathit N}$, данный центр
  электрического поля $  \vec{\mathit E}_{\mathit A\mathcal{F}}(\mathit K)$.

  Данное магнитное поле волновым образом распространяется вдоль данной продольной оси представляя собой продольную ($\vec{\mathcal{H}}(x)\,coll\,\vec{v}\,$) "чисто"\, {\em  магнитную} волну.

  Наличие седла электрического поля $ \vec{\mathit E}_{\mathit A}(\mathit K), \mathit K\in \delta_{\mathit N}\subset \mathit P $, в представленной на рис. 12.2 $ \mathit H_{11}$-волне, говорит о наличии в точке $  \mathit N$  производного тензора
 $ \tilde{def}_{d}\,\vec{\mathit E}_{\mathit A}(\mathit N)$, а наличие последнего означает, в силу закона электоджейтонной индукции, представленного дифференциальным уравнением (1.12),
 $$
 \partial\, ^{0}\,\mathcal{G}_{d}( \mathit N ) = - \tilde{def}_{d}\,\vec{\mathit E}_{\mathit A}(\mathit N),           \eqno\ldots (12.2)
 $$
 наличие на продольной оси рассматриваемого волновода, с рассматриваемым типом волны, переменного во времени плоского девиатационного джейтонного поля, представленного тензором $ \mathcal{G}_{d}(\mathit N)$,   создавшего, в $ \delta_{\mathit N}$, данное седло электрического поля $  \vec{\mathit E}_{\mathit A\mathcal{G}_{d}}(\mathit K)$.
 \\
  \begin{figure}[!h]
\begin{center}
\vspace{-6pt}
\includegraphics[width=125mm]{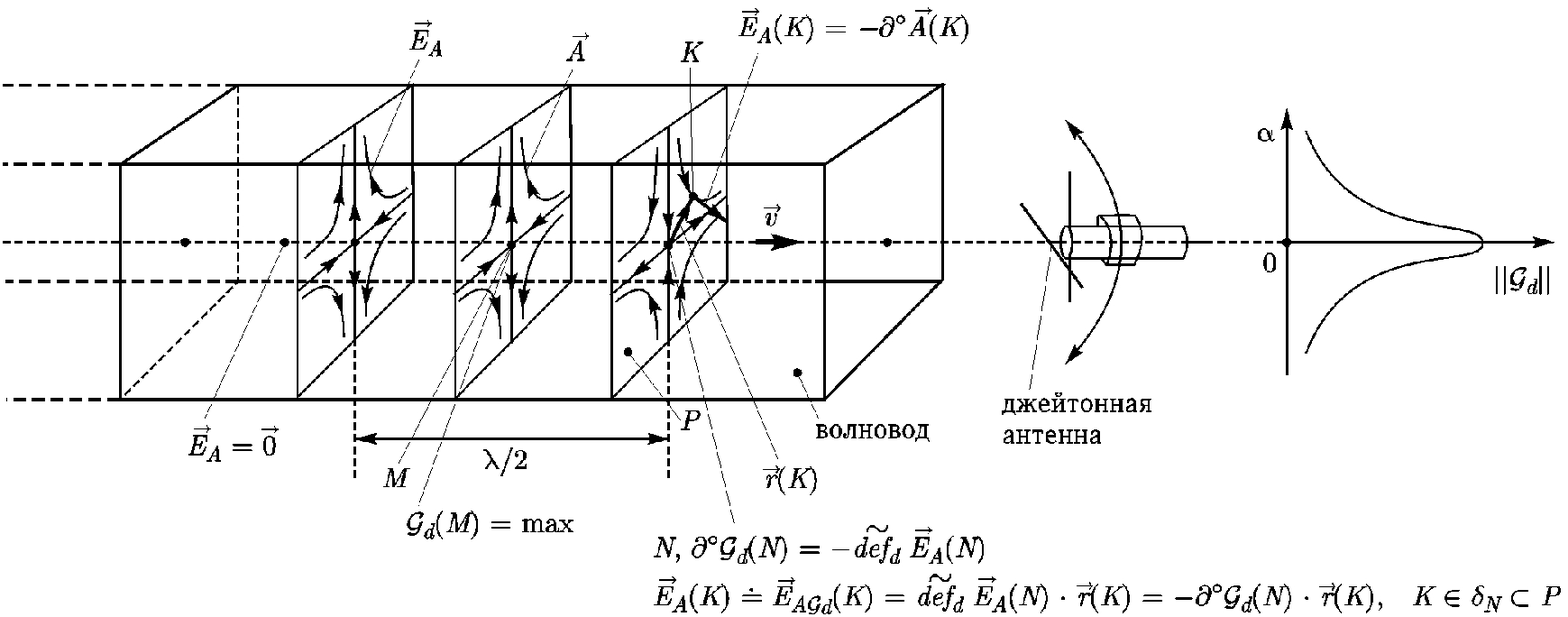}
\vspace{1pt}
\caption{$H_{11}$-волна волновода прямоугольного поперечного сечения: полевая система на продольной оси и в ее $\delta$-окрестности. Конгруэнция векторных линий псевдовекторного поля $\vec{\mathcal {H}}(x)$ и конгруэнция векторных линий векторных полей $\vec{ A}(x)$ и $\vec{ E}_{A}(x)$ вне рассматриваемых поперечных плоскостей на данном рисунке не показаны.}
\end{center}
\end{figure}

  Данное джейтонное поле волновым образом распространяется вдоль данной продольной оси представляя собой "чисто"\, {\em  джейтонную} волну.

В полной аналогии с двумя предыдущими полевыми системами, в случае $ \mathit E_{01}$-волны, представленной на рис.12.3,  наличие плоского дикритического узла электрического поля $ \vec{\mathit E}_{\mathit A}(\mathit K), \mathit K\in \delta_{\mathit N}\subset \mathit P$  говорит о наличии в точке $  \mathit N$ производного тензора $ \tilde{def}_{\ast}\,\vec{\mathit E}_{\mathit A}(\mathit N)$, а наличие последнего означает, в силу закона электоджейтонной индукции, представленного дифференциальным уравнением (1.13),
$$
 \partial\, ^{0}\,\mathcal{G}_{\ast}( \mathit N ) = - \tilde{def}_{\ast}\,\vec{\mathit E}_{\mathit A}(\mathit N),       \eqno\ldots (12.3)
 $$
 наличие на продольной оси рассматриваемого волновода, с рассматриваемым типом волны, переменного во времени плоского дилатационного джейтонного поля, представленного тензором $ \mathcal{G}_{\ast}(\mathit N)$,   создавшего, в $ \delta_{\mathit N}$, данный дикритический узел электрического поля $  \vec{\mathit E}_{\mathit A\mathcal{G}_{\ast}}(\mathit K)$.

 \begin{figure}[!h]
\begin{center}
\vspace{-6pt}
\includegraphics[width=125mm]{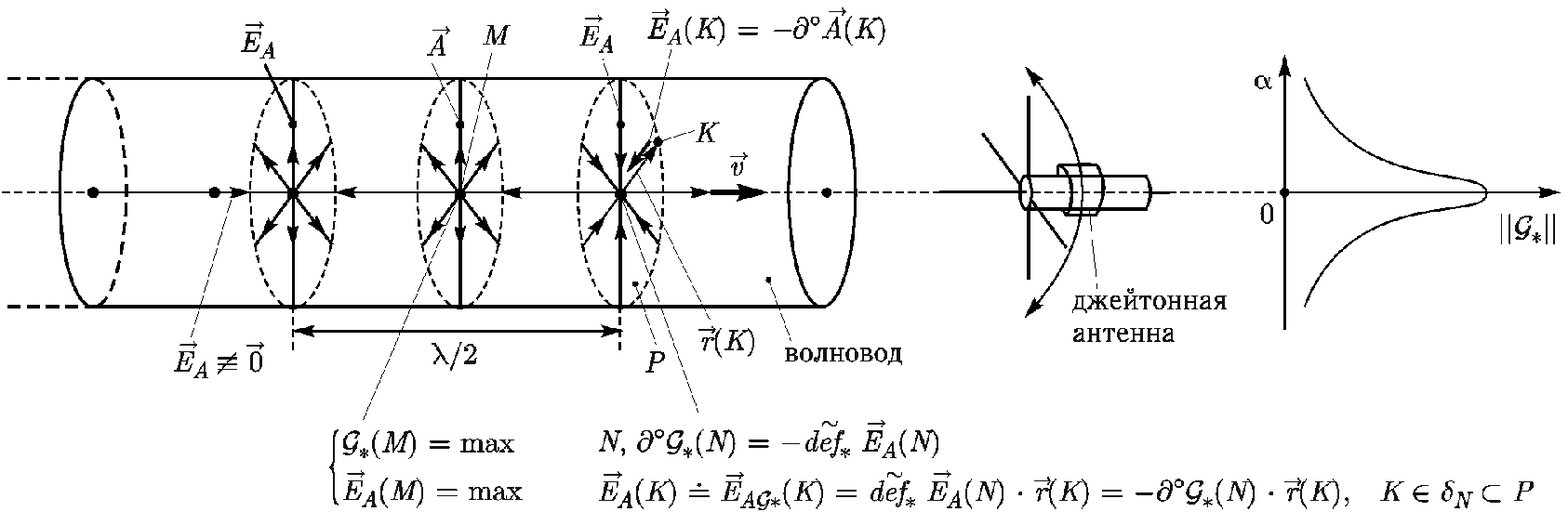}
\vspace{1pt}
\caption{$E_{01}$-волна волновода круглого поперечного сечения: полевая система на продольной оси и в ее  $\delta$-окрестности. Конгруэнция векторных линий псевдовекторного поля $\vec{\mathcal {H}}(x)$ и трехмерных сёдел векторных полей ${\vec A}(x)$ и ${\vec E}_{A}(x)$ на данном рисунке не указаны.}
\end{center}
\end{figure}

 Данное джейтонное поле волновым образом распространяется вдоль рассматриваемой оси данного волновода, представляя, совместно с $ \vec{\mathit E}_{\mathit A}(x)$,
 {\em  продольную}  ($\vec{\mathit{E}_{\mathit A}}(x)\,coll\,\vec{v}\,$) электроджейтонную волну, представленную полевым комплексом \{$\vec{\mathit E}_{\mathit A}(x), \mathcal{G}_{\ast}(x)$\}.

При этом нужно иметь в виду, что в данном случае имеем  трехмерные {\em седла} векторных линий соответствующих векторных полей, поперечные части  (поперечные усы) которых представлены, в рассматриваемых $ \delta$-окрестностях, дикритическими узлами.

Так же уместно отметить, что классический полевой комплекс \{$\vec{\mathit A}(x),  \vec{\mathit E}_{\mathit A}(x)$\}, исключающий из рассмотрения магнитное и джейтонные поля, нельзя рассматривать как волну, так как $\vec{\mathit A}(x)$ и $ \vec{\mathit E}_{\mathit A}(x)$, в обсуждаемых волновых процессах, несинфазны, поэтому соотношение (1.14) с самого начала рассматривается, и в данном контексте, традиционно  \cite{La}, то есть, как {\em определение} напряженности электрического поля, обусловленного векторным потенциалом $\vec{\mathit A}(x)$.

\section*{\S \,13 Основные динамические характеристики классической полевой системы, представленной плоской волной Ландау}
\addcontentsline{toc}{section}{\S \,13 Основные динамические характеристики классической полевой системы, представленной плоской волной Ландау}

\qquad Рассмотрим основные динамические характеристики, плотность энергии и плотность потока энергии, а так же соответствующий дифференциальный закон сохранения для фундаментальной классической полевой системы, представленной плоской электродинамической волной Ландау (рис. 8.1).

Следуя хорошо известной процедуре  \cite{La}, выражения для этих динамических величин, через решения уравнений движения данной системы, легко получить непосредственно из самих уравнений движения этой  системы.

Рассматриваемая полевая  система  состоит из электрического поля, представленного напряженностью
 $\vec{\mathit E}_{\mathit A}(x)$, магнитного поля, определяемого тензором $\mathcal{F}(x)$, и девиатационного джейтонного поля, представленного тензором $\mathcal{G}_{d}(x)$, а в качестве соответствующих уравнений движения, связывающих эти величины, выступают, в частности, рассматриваемые законы электродинамической индукции (1.11)--(1.12) -- уравнения из "первых пар"\,  \cite{La}  электродинамических уравнений.

 Необходимое для нашей цели уравнение движения из "вторых пар"\, электродинамических уравнений, связывающее эти же полевые переменные, является непосредственным следствием стандартного уравнения Даламбера для векторного потенциала и имеет, в данном случае, вид
$$
  \vec{\bigtriangledown}\cdot\mathcal{F}(x) +  \vec{\bigtriangledown}\cdot\mathcal{G}_{d}(x) = - \partial ^{0}\vec{\mathit E}_{\mathit A}(x).                 \eqno\ldots (13.1)
$$

Это уравнение так же говорит о том, что переменное во времени электрическое поле данной полевой системы "порождает"\, не только магнитное поле, представленное в этом уравнении тензором $\mathcal{F}(x)$, но и девиатационное джейтонное поле, представленное тензором $\mathcal{G}_{d}(x)$, что соответствует тому, что в этой волне присутствуют оба эти поля.

Так как в рассматриваемой полевой системе выполняется соотношение идентичное (5.5), то из данного уравнения следуют {\em два} уравнения, математически эквивалентные друг другу как уравнения на векторный потенциал $ \vec{\mathit A}(x)$, то есть,
 $$
 \vec{\bigtriangledown}\cdot \mathcal{F}_{\mathit L}(x) =  - \partial ^{0}\vec{\mathit E}_{\mathit A}(x).             \eqno\ldots (13.2)
 $$
$$
 \vec{\bigtriangledown}\cdot \mathcal{G}_{d \mathit L}(x) =  - \partial ^{0}\vec{\mathit E}_{\mathit A}(x).          \eqno\ldots (13.3)
$$
первое из которых, после использования соотношения (5.8),  приводит к дифференциальному уравнению для псевдовектора $ \vec{\mathcal{H}}_{\mathit L}(x)$, сопровождающего тензор $ \mathcal{F}_{ \mathit L}(x)$,
$$
rot\,\vec{\mathcal{H}}_{\mathit L}(x) = \partial ^{0}\vec{\mathit E}_{\mathit A}(x),          \eqno\ldots (13.4)
$$
представляющему собой известную дифференциальную  форму закона Ампера\,--\,Максвелла о
циркуляции псевдовектора  $ \vec{\mathcal{H}}_{\mathit L}(x)$ для рассматриваемого случая плоской волны.

При этом, как и в магнитоджейтонностатическом случае (уравнение (5.9)), находим, что из уравнения (13.4) нельзя делать традиционного вывода о том, что рассматриваемое электрическое поле рассматриваемой плоской волны "порождает"\, (сопровождает) только магнитное поле.

Аналогичное замечание относится и к интерпретации уравнения (13.3).

Таким образом, (13.4) говорит лишь о том, что ток смещения рассматриваемой плоской волны, порождает, помимо девиатационного джейтонного поля, представленного тензором $ \mathcal{G}_{d}(x)$,  поле псевдовектора $ \vec{\mathcal{H}}_{\mathit L}(x)$, для которого выполняется данная дифференциальная форма закона Ампера\,--\,Максвелла.

В связи с этим, уравнение (13.1) выступает в качестве обобщения закона Ампера\,--\,Максвелла (13.4), принимающего во внимание факт существования как магнитного, так и девиатационного джейтонного полей.

Таким образом, система уравнений, описывающая процессы происходящие в рассматриваемой плоской волне, представлена уравнениями (1.11), (1.12) и (13.1).

Продолжая вышеуказанную процедуру, умножаем обе части уравнения (13.1) скалярно на
 $ \vec{\mathit E}_{\mathit A}(x)$, а обе части уравнений (1.11) и (1.12) на $ \mathcal{F}(x)$ и $ \mathcal{G}_{d}(x)$, соответственно, получаем следующую систему трех скалярных уравнений
 $$
 \frac{1}{2}\,\partial^{0}\,\vec{E}_{ \mathit A}(x)\,^{2} = - \left(\vec{\bigtriangledown}\cdot  \mathcal{F}(x), \vec{\mathit E}_{\mathit A}(x) \right) - \left(\vec{\bigtriangledown}\cdot  \mathcal{G}_{d}(x), \vec{\mathit E}_{\mathit A}(x) \right),    \eqno\ldots (13.5)
 $$
 $$
 \frac{1}{2}\,\partial^{0}\,\mathcal{F}(x)\,^{2} = - \left(\tilde{rot}\, \vec{\mathit E}_{\mathit A}(x),  \mathcal{F}(x)\right),   \eqno\ldots (13.6)
 $$
  $$
 \frac{1}{2}\,\partial^{0}\,\mathcal{G}_{d}(x)\,^{2} = - \left(\tilde{def}_{d}\, \vec{\mathit E}_{\mathit A}(x),  \mathcal{G}_{d}(x)\right),   \eqno\ldots (13.7)
 $$
 в которой, скалярные произведения правых частей уравнений (13.6) и (13.7) и соответствующие скалярные квадраты левых частей этих уравнений определены посредством (1.23).

 Сложение уравнений (13.5)--(13.7),  с последующим учетом соотношений
 $$
\left(\vec{\bigtriangledown} \cdot  \mathcal{F}(x), \vec{\mathit E}_{\mathit A}(x) \right) = \vec{\bigtriangledown}\cdot
\left(\mathcal{F}(x) \cdot \vec{\mathit E}_{\mathit A}(x) \right) - \left(\mathcal{F}(x), \vec{\bigtriangledown}\otimes \,\vec{\mathit E}_{\mathit A}(x) \right),      \eqno\ldots (13.8)
$$
$$
\left(\vec{\bigtriangledown} \cdot  \mathcal{G}_{d}(x), \vec{\mathit E}_{\mathit A}(x) \right) = \vec{\bigtriangledown}\cdot
\left(\mathcal{G}_{d}(x) \cdot \vec{\mathit E}_{\mathit A}(x) \right) - \left(\mathcal{G}_{d}(x), \vec{\bigtriangledown}\otimes \,\vec{\mathit E}_{\mathit A}(x) \right),                                                 \eqno\ldots (13.9)
$$
где

$$
\left(\mathcal{F}(x), \vec{\bigtriangledown}\otimes \,\vec{\mathit E}_{\mathit A}(x) \right) \equiv \left(\mathcal{F}(x), \tilde{rot}\,\vec{\mathit E}_{\mathit A}(x) \right),                                    \eqno\ldots (13.10)
$$

$$
\left(\mathcal{G}_{d}(x), \vec{\bigtriangledown}\otimes \,\vec{\mathit E}_{\mathit A}(x) \right) \equiv \left(\mathcal{G}_{d}(x), \tilde{def}_{d}\,\vec{\mathit E}_{\mathit A}(x) \right),                                    \eqno\ldots (13.11)
$$
приводит к следующему дифференциальному закону сохранения энергии для рассматриваемой полевой системы
\begin{multline*}
  \frac{1}{2}\,\partial^{0}\left(\vec{E}_{ \mathit A}(x)\,^{2} + \mathcal{F}(x)\,^{2}+ \mathcal{G}_{d}(x)\,^{2} \right) = \\
  -  \vec{\bigtriangledown}\cdot \left(\mathcal{F}(x) \cdot \vec{\mathit E}_{\mathit A}(x) \right) -  \vec{\bigtriangledown}\cdot
\left(\mathcal{G}_{d}(x) \cdot \vec{\mathit E}_{\mathit A}(x) \right).          \qquad\ldots(13.12)
\end{multline*}

Уравнение (13.12) демонстрирует, что плотность энергии и плотность потока энергии в данной полевой системе представлены, соответственно,
$$
W(x) = \frac{\vec{E}_{ \mathit A}(x)\,^{2} + \mathcal{F}(x)\,^{2}+ \mathcal{G}_{d}(x)\,^{2}}{2},      \eqno\ldots (13.13)
$$
$$
\vec{S}(x) = \mathcal{F}(x) \cdot \vec{\mathit E}_{\mathit A}(x) +
 \mathcal{G}_{d}(x) \cdot \vec{\mathit E}_{\mathit A}(x).                                     \eqno\ldots (13.14)
$$

В свою очередь, уравнения (13.13)  и (13.14) демонстрируют, что девиатационное джейтонное поле, по отношению к магнитному полю, представленному тензором $ \mathcal{F}(x)$, выступает совершенно равноправно как в выражении для плотности энергии, так и в выражении для плотности потока энергии рассматриваемой полевой системы.

В частности, уравнение (13.14) показывает, что энергия, в рассматриваемой системе, переносится не только магнитным полем, но и девиатационным джейтонным полем, что, в свою очередь, возводит девиатационное джейтонное поле в ранг физических полей.

При этом, соответствующие плотности потоков энергии определены соотношениями
$$
\vec{S}_{\mathcal{F}}(x) = \mathcal{F}(x) \cdot \vec{\mathit E}_{\mathit A}(x),                \eqno\ldots (13.15)
$$
$$
\vec{S}_{\mathcal{G}_{d}}(x) = \mathcal{G}_{d}(x) \cdot \vec{\mathit E}_{\mathit A}(x).                \eqno\ldots (13.16)
$$

Для данной полевой системы выполняются равенства
$$
 \mathcal{F}(x)\,^{2} =  \mathcal{G}_{d}(x)\,^{2},                              \eqno\ldots (13.17)
$$
$$
\mathcal{F}(x) \cdot \vec{\mathit E}_{\mathit A}(x) =   \mathcal{G}_{d}(x) \cdot \vec{\mathit E}_{\mathit A}(x),
\eqno\ldots (13.18)
$$
использование которых позволяет выражения для плотности энергии и плотности потока энергии  {\em этой} полевой системы представить в двух  {\em количественно} эквивалентных видах
$$
W(x) = \frac{\vec{E}_{ \mathit A}(x)\,^{2} + \frac{1}{2}\, \mathcal{F}_{ \mathit L}(x)\,^{2}}{2},      \eqno\ldots (13.19)
$$
$$
\vec{S}(x) = \mathcal{F}_{ \mathit L}(x) \cdot \vec{\mathit E}_{\mathit A}(x),                                     \eqno\ldots (13.20)
$$
или
$$
W(x) = \frac{\vec{E}_{ \mathit A}(x)\,^{2} + \frac{1}{2}\, \mathcal{G}_{d \mathit L}(x)\,^{2}}{2},      \eqno\ldots (13.21)
$$
$$
\vec{S}(x) = \mathcal{G}_{d \mathit L}(x) \cdot \vec{\mathit E}_{\mathit A}(x).                                      \eqno\ldots (13.22)
$$

В первом случае в качестве носителя "неэлектрической "\, части плотности энергии данной полевой системы выступает
только магнитное поле, определяемое лоренцевым тензором $\mathcal{F}_{ \mathit L}(x) $ или сопровождающим его лоренцевым псевдовектором  $ \vec{\mathcal{H}}_{\mathit L}(x)$, связанных соотношением, аналогичным (1.15).

Таким образом, определение магнитного поля как поля, представленного лоренцевым
тензором $\mathcal{F}_{ \mathit L}(x) =  2\,\mathcal{F}(x)$, с последующим использованием соотношений (13.19) и (13.20), приводит к {\em количественно} правильным выражениям для плотности энергии и плотности потока энергии рассматриваемой полевой системы, при полном игнорировании существования девиатационного джейтонного поля как {\em физического} поля, и поэтому соотношения (13.19) и (13.20) могут рассматриваться лишь как средство достижения правильного {\em количественного} результата.

Во втором случае, система уравнений (13.21) и (13.22), происходит аналогичное, то есть, определение девиатационного джейтонного поля как поля, представленного лоренцевым тензором
$  \mathcal{G}_{d \mathit L}(x) = 2\,\mathcal{G}_{d }(x)$, с последующим использованием соотношений (13.21) и (13.22), так же приводит к количественно правильным выражением для плотности энергии и плотности потока энергии рассматриваемой полевой системы, но при полном игнорировании существовании магнитного поля.

В связи с этим, в данном контексте в качестве соотношений, определяющих плотность энергии и плотность потока энергии данной полевой системы, рассматриваются выражения (13.13) и (13.14), в связи с чем,  магнитное поле и девиатационное джейтонное поле определяются как поля, представленные тензорами  $\mathcal{F}(x)$ и $\mathcal{G}_{d}(x)$, соответственно.

При этом уместно подчеркнуть, что как непосредственные силовые воздействия магнитного и джейтонного полей на движущиеся  электрически заряженные частицы, посредством магнитных и джейтонных сил \cite{Al5}, так и опосредованные силовые воздействия этих полей на системы электрически заряженных частиц, посредством порождаемых данными полями соответствующих электрических полей, не могут быть сведены, в общем случае, одно к другому ни в количественном, ни в качественном отношениях, что в свою очередь является дополнительным основанием к тому, чтобы рассматриваемые  плотность энергии и плотность потока энергии представлять соотношениями
(13.13) и (13.14), соответственно.

И так, в рассматриваемой фундаментальной классической полевой системе плотность "не электрической"\, части плотности энергии представлена  плотностями энергий магнитного и девиатационного джейтонного полей и при этом в равных количествах.

Аналогичное заключение относится и к плотности потока энергии, то есть, плотность потока энергии в данной полевой системе представлена плотностями потоков энергии электромагнитного и электроджейтонного полей и при этом в равных, по отношению к друг другу, долях, определяемых уравнениями (13.15) и (13.16).

\chapter{\!Уравнения электроджейтонного поля}

 \section*{\S \,14 Системы дифференциальных уравнений электромагнитного и электроджейтонного полей}
\addcontentsline{toc}{section}{\S \,14 Системы дифференциальных уравнений электромагнитного и электроджейтонного полей}

\qquad Рассмотрим общий  случай классической полевой системы, представленной на рис.~4.3, включающий произвольные направления излучения.

Эта система состоит из электрических полей, $ \vec{\mathit E}_{\mathit A}(x)$  и $ \vec{\mathit E}_{\mathit \varphi}(x)$, магнитного поля, определенного тензором $ \mathcal{F}(x)$, и джейтонных полей, представленных тензорами
$ \mathcal{G}_{d}(x)$ и $ \mathcal{G}_{\ast}(x)$, а 4-потенциал поля излучения удовлетворяет классическому условию Лоренца.

Для такой полевой системы, вместо (13.1) уравнение Даламбера для векторного потенциала $ \vec{\mathit A}(x)$
теперь приводит к соотношению
$$
  \vec{\bigtriangledown}\cdot\mathcal{F}(x) +  \vec{\bigtriangledown}\cdot\mathcal{G}_{d}(x)+  \vec{\bigtriangledown}\cdot\mathcal{G}_{\ast}(x) = - \partial ^{0}\vec{\mathit E}_{\mathit A}(x),                 \eqno\ldots (14.1)
$$
которое, прежде всего, говорит о том, что плотность тока смещения "порождает"\, (сопровождает), в общем
случае, не только магнитное поле, представленное в этом уравнении тензором $ \mathcal{F}(x)$, но и девиатационное  и дилатационное джейтонные  поля определяемые, соответственно, тензорами $ \mathcal{G}_{d}(x)$ и $ \mathcal{G}_{\ast}(x)$, в полном соответствии с геометрическим анализом этой полевой системы, представленным на рис.~4.3.

Уравнение (14.1), совместно с условием Лоренца, приводит не только к соответствующим уравнениям второй пары системы уравнений Максвелла  \cite{Sch}, но и к уравнениям вторых пар для электроджейтонного поля \cite{Al2}, то есть к уравнениям, в которых $ \mathcal{F}$- и $ \mathcal{G}$-полевые переменные разделены.

Действительно, переписав (14.1) в виде
 $$
 \vec{\bigtriangledown}\cdot\mathcal{F}(x) +  \vec{\bigtriangledown}\cdot\mathcal{G}(x) =
   - \partial ^{0}\vec{\mathit E}_{\mathit A}(x),                 \eqno\ldots (14.2)
$$
находим, что использование в (14.2) тождества
$$
\vec{\bigtriangledown}\cdot\mathcal{G}(x) = \vec{\bigtriangledown}\cdot\mathcal{F}(x) +
     \vec{\bigtriangledown}\cdot \partial A(x)^{ \mathit T},                 \eqno\ldots (14.3)
$$
в котором производный тензор $ \partial A(x)$ определен соотношением (1.18), приводит не только к уравнению
 $$
2\, \vec{\bigtriangledown}\cdot\mathcal{F}(x)  =  - \partial ^{0}\vec{\mathit E}_{\mathit A}(x) - \vec{\bigtriangledown}\cdot \partial A(x)^{ \mathit T},                 \eqno\ldots (14.4)
$$
но и к уравнению
$$
2\, \vec{\bigtriangledown}\cdot\mathcal{G}(x)  =  - \partial ^{0}\vec{\mathit E}_{\mathit A}(x) + \vec{\bigtriangledown}\cdot \partial A(x)^{ \mathit T}.                  \eqno\ldots (14.5)
$$

Но, в рамках условия Лоренца, имеем соотношение
$$
\vec{\bigtriangledown}\cdot \partial A(x)^{ \mathit T} = \vec{\bigtriangledown}\,div\,\vec{\mathit A}(x) = \partial^{0}\vec{\mathit E}_{\mathit \varphi}(x),                             \eqno\ldots (14.6)
$$
после использования которого система (14.4), (14.5) принимает, соответственно, вид
$$
\vec{\bigtriangledown}\cdot\mathcal{F}(x)  =  - \partial ^{0}\vec{\mathit E}_{\mathcal{F}}(x),                 \eqno\ldots (14.7)
$$
$$
\vec{\bigtriangledown}\cdot\mathcal{G}(x)  =  - \partial ^{0}\vec{\mathit E}_{\mathcal{G}}(x),                 \eqno\ldots (14.8)
$$
где
$$
\vec{\mathit E}_{\mathcal{F}}(x)  \eqdef \frac{1}{2}\,\left(\vec{\mathit E}_{\mathit A}(x) +
\vec{\mathit E}_{\mathit \varphi}(x) \right),             \eqno\ldots (14.9)
$$
$$
\vec{\mathit E}_{\mathcal{G}}(x)  \eqdef \frac{1}{2}\,\left(\vec{\mathit E}_{\mathit A}(x) -
\vec{\mathit E}_{\mathit \varphi}(x) \right).              \eqno\ldots (14.10)
$$

Система уравнений (14.7), (14.8) демонстрирует, прежде всего, что если {\em магнитное} поле в {\em электромагнитной} волне сопровождается электрическим полем $ \vec{\mathit E}_{\mathcal{F}}(x)$, то {\em джейтонное} поле в  {\em электроджейтонной} волне сопровождается электрическим полем $ \vec{\mathit E}_{\mathcal{G}}(x)$, принципиально отличающееся от $ \vec{\mathit E}_{\mathcal{F}}(x)$ например тем, что напряженность
 $ \vec{\mathit E}_{\mathcal{G}}(x)$ на продольной оси рассматриваемого поля излучения линейно вибрирующего точечного электрического заряда, рис. 4.3\,{\em б})\,, отлична от нуля и продольно ориентирована, что позволяет говорить, с учетом того, что на этой оси так же и $\mathcal{G}(x) \neq \tilde{0}$, о наличии, на данной  оси, {\em продольной электроджейтонной волны},  представленной полевым комплексом \{$\vec{\mathit E}_{\mathcal{G}}(x), \mathcal{G}(x)$\}.

По любым другим, не нормальным по отношению к рассматриваемой продольной оси, направлениям излучения напряженность $ \vec{\mathit E}_{\mathcal{G}}(x)$ содержит, наряду с поперечной составляющей, и продольную составляющую, что говорит о том, что продольная электроджейтонная волна распространяется и
 в этих направлениях, при этом максимум соответствующей диаграммы направленности  находится на рассматриваемой продольной оси данного вибратора.

 Таким образом имеем, что наряду с поперечной электромагнитоджейтонной волной, рассмотренной выше (рис.8.1),
 существует, по крайней мере в рассматриваемом математическом отношении, и продольная электроджейтонная волна с максимумом излучения на продольной оси рассматриваемой излучающей системы.

 Уравнение (14.7), представляющее собой соответствующее уравнение из второй пары системы уравнений Максвелла  \cite{La}, легко переписать и в традиционной форме.

 Действительно, воспользовавшись соотношением, идентичным (5.8),
 $$
\vec{\bigtriangledown}\cdot\mathcal{F}(x)  =  - rot\,\vec{\mathcal{H}}(x),                 \eqno\ldots (14.11)
$$
из (14.7) мгновенно находим
$$
 rot\,\vec{\mathcal{H}}_{\mathit L}(x) = \partial ^{0}\vec{\mathit E}_{\mathcal{F} L}(x),                 \eqno\ldots (14.12)
$$
где  $ \vec{\mathit E}_{\mathcal{F}L}(x)  = \vec{\mathit E}_{\mathit A}(x) +
\vec{\mathit E}_{\mathit \varphi}(x) $ -- традиционная лоренцева напряженность электрического поля, сопровождающего
лоренцево магнитное поле в традиционной электромагнитной волне.

Волновое уравнение Даламбера для скалярного потенциала, при условии Лоренца, приводит к соотношению
$$
div\,\vec{\mathit E}_{\mathit A}(x) = - div\,\vec{\mathit E}_{\varphi}(x),              \eqno\ldots (14.13)
$$
используя которое мгновенно получаем и соответствующее второе уравнение второй пары системы уравнений
Максвелла для рассматриваемой полевой системы
$$
div\,\vec{\mathit E}_{\mathcal{F}\mathit L}(x) = 0.              \eqno\ldots (14.14)
$$

Уравнение (14.8), представляющее собой уравнение из второй пары уравнений, описывающих {\em электроджейтонное} поле, в свою очередь, можно переписать в виде
$$
\vec{\bigtriangledown}\cdot\mathcal{G}_{d}(x)  =  - \partial ^{0}\vec{\mathit E}_{\mathcal{G}}(x) - \vec{\bigtriangledown}\cdot\mathcal{G}_{\ast}(x).                 \eqno\ldots (14.15)
$$

С другой стороны, используя соотношение
$$
\vec{\bigtriangledown}\cdot (\varphi (x) g) = \vec{\bigtriangledown}\,\varphi (x)       \eqno\ldots (14.16)
$$
и определение (1.8), приходим к уравнению
$$
\vec{\bigtriangledown}\cdot\mathcal{G}_{\ast}(x) =
 \frac{1}{tr g}\,\vec{\bigtriangledown}\,div\,\vec{\mathit A}(x),                         \eqno\ldots (14.17)
$$
которое, после использования условия Лоренца, принимает вид, не содержащий явно
 векторный потенциал $ \vec{\mathit A}(x)$,
 $$
\vec{\bigtriangledown}\cdot\mathcal{G}_{\ast}(x) =
 - \partial^{0}\,\vec{\mathit E}_{\mathcal{G}_{\ast}}(x),                         \eqno\ldots (14.18)
$$
где напряженность электрического поля $ \vec{\mathit E}_{\mathcal{G}_{\ast}}(x)$,
сопровождающего {\em дилатационное} джейтонное поле в рассматриваемой полевой
системе, определена соотношением
$$
\vec{\mathit E}_{\mathcal{G}_{\ast}}(x)  \eqdef  - \frac{1}{tr g}\,\vec{\mathit E}_{\varphi}(x).            \eqno\ldots (14.19)
$$

Последующее использование (14.18) в (14.15) приводит к уравнению
$$
\vec{\bigtriangledown}\cdot\mathcal{G}_{d}(x)  =  - \partial ^{0}\vec{\mathit E}_{\mathcal{G}_{d}}(x),       \eqno\ldots (14.20)
$$
в котором напряженность электрического поля, $ \vec{\mathit E}_{\mathcal{G}_{d}}(x)$, сопровождающего девиатационное джейтонное поле в данной полевой системе, определена соотношением
$$
\vec{\mathit E}_{\mathcal{G}_{d}}(x) = \frac{1}{2}\,\left(\vec{\mathit E}_{\mathit A}(x) -
\frac{1}{3}\,\vec{\mathit E}_{\varphi}(x) \right).                                                                        \eqno\ldots (14.21)
$$

Наконец, использование (14.13) и определений (14.10), (14.19) и (14.21) приводит к соотношениям
$$
div\,\vec{\mathit E}_{\mathcal{G}}(x) = \partial^{0}G^{00}(x),                                             \eqno\ldots (14.22)
$$
$$
div\,\vec{\mathit E}_{\mathcal{G}_{d}}(x) = \frac{2}{3}\,\partial^{0}G^{00}(x),                    \eqno\ldots (14.23)
$$
$$
div\,\vec{\mathit E}_{\mathcal{G}_{\ast}}(x) = \frac{1}{3}\,\partial^{0}G^{00}(x),                    \eqno\ldots (14.24)
$$
где $\mathit G^{00}(x)$ -- "временная"\, компонента 4-тензора электроджейтонного поля, $ \mathit G (x) = \partial  \stackrel{ s}{\otimes}\mathit A (x)$.

(14.22) демонстрирует, что, в отличие от $ \vec{\mathit E}_{\mathcal{F}}(x)$, напряженность электрического поля,
$ \vec{\mathit E}_{\mathcal{G}}(x)$, сопровождающего джейтонное поле, представленное  тензором  $\mathcal{G}(x)$,
имеет, вне источника поля, отличную от нуля расходимость, определяемую правой частью (14.22) \cite{Al2}.

И так, уравнения (14.7), (14.18) и (14.20), совместно с законами электродинамической индукции, (1.11), (1.12) и (1.13),
составляют следующие системы электродинамических уравнений

$$
\left\{
\begin{array}{c}
\vec{\bigtriangledown}\cdot\mathcal{F}(x)  = - \partial ^{0}\vec{\mathit E}_{\mathcal{F}}(x), \\
\partial^{0} \mathcal{F}(x)  = -  \tilde{rot}\,\vec{E}_{A}(x).
\end{array}
\right.                     \eqno\ldots (I)
$$

$$
\left\{
\begin{array}{c}
\vec{\bigtriangledown}\cdot\mathcal{G}_{d}(x)  =  - \partial ^{0}\vec{\mathit E}_{\mathcal{G}_{d}}(x), \\
\partial^{0}\, \mathcal{G}_{d}(x) = -  \tilde{def}_{d}\,\vec{E}_{A}(x).
\end{array}
\right.                     \eqno\ldots (I\!I)
$$

$$
\left\{
\begin{array}{c}
\vec{\bigtriangledown}\cdot\mathcal{G}_{\ast}(x) = - \partial^{0}\,\vec{\mathit E}_{\mathcal{G}_{\ast}}(x),  \\
\partial^{0}\, \mathcal{G}_{\ast}(x) = -  \tilde{def}_{\ast}\,\vec{E}_{A}(x).
\end{array}
\right.                     \eqno\ldots (I\!I\!I)
$$

$$
\left\{
\begin{array}{c}
\vec{\bigtriangledown}\cdot\mathcal{G}(x) = - \partial^{0}\,\vec{\mathit E}_{\mathcal{G}}(x),  \\
\partial^{0}\, \mathcal{G}(x) = -  \tilde{def}\,\vec{E}_{A}(x).
\end{array}
\right.                     \eqno\ldots (IV)
$$
устанавливающих дифференциальные связи магнитного и джейтонных полей с сопровождающими их электрическими полями в электродинамических полевых системах.

При этом важно подчеркнуть, что уравнение вида $ \vec{a}(x)\cdot T(x) = \vec{b}(x)$ -- это не просто перезапись соответствующих координатных форм, а уравнение в котором тензор второго ранга $ T(x)$ представлен инвариантным
геометрическим объектом, не зависящим от выбора того или иного базиса пространства $  \mathit V\otimes\mathit V$ \cite{Pob}.

Использование соотношений (1.15), (1.16) и (14.11) позволяет
 получить, как следствие системы~(I), систему уравнений для псевдовектора $ \vec{\mathcal{H}}(x)$, сопровождающего тензор $ \mathcal{F}(x)$ в соответствующем пространстве,
 $$
 rot\,\vec{\mathcal{H}}(x) = \partial^{0}\vec{\mathit E}_{\mathcal{F}}(x),
 $$
$$
2\,\partial^{0}\vec{\mathcal{H}}(x) = - rot\,\vec{\mathit E}_{\mathit A}(x),
$$
которая, в традиционных лоренцевых полевых переменных, принимает традиционный вид \cite{La},
$$
 rot\,\vec{\mathcal{H}}_{\mathit L}(x) = \partial^{0}\vec{\mathit E}_{\mathcal{F}L}(x),
$$
$$
\partial^{0}\vec{\mathcal{H}}_{L}(x) = - rot\,\vec{\mathit E}_{\mathit A}(x) = - rot\,\vec{\mathit E}_{\mathcal{F}L}(x).
$$

Системы (II) и (III) подобных представлений не имеют так как тензоры  $ \mathcal{G}_{d}(x)$ и
 $ \mathcal{G}_{\ast}(x)$ не имеют нетривиальных сопровождающих векторов в силу очевидных тождеств
$$
\varepsilon\cdot \cdot \, \mathcal{G}_{d}(x) \equiv \vec{0},                    \eqno\ldots (14.25)
$$
$$
\varepsilon\cdot \cdot \, \mathcal{G}_{\ast}(x) \equiv \vec{0}.                   \eqno\ldots (14.26)
$$

Однако, тензор $ \mathcal{G}_{\ast}(x)$ обладает сопровождающим
 его 3-скаляром (линейным скалярным инвариантом)
 $$
 \mathcal{G}^{\ast}(x) \eqdef g\cdot \cdot\,\mathcal{G}_{\ast}(x) = div\,\vec{ \mathit A}(x),          \eqno\ldots (14.27)
 $$
 вследствие чего легко  получить, как следствие (III), систему уравнений для 3-скалярного
  поля $ \mathcal{G}^{\ast}(x)$.

  Действительно, использование соотношений (14.17), (14.19) и (14.27) приводит к следующим выражениям для левой и правой частей первого уравнения системы  (III)
$$
 \vec{\bigtriangledown}\cdot\mathcal{G}_{\ast}(x) =\frac{1}{tr g}\,\vec{\bigtriangledown}\,\mathcal{G}^{\ast}(x)  ,              \eqno\ldots (14.28)
$$
$$
\partial^{0}\, \mathcal{G}_{\ast}(x) = - \frac{1}{tr g}\,\partial^{0}\vec{E}_{\varphi}(x),                \eqno\ldots (14.29)
$$
подстановка которых в первое уравнение системы  (III) приводит к первому уравнению для $ \mathcal{G}^{\ast}(x)$
$$
 \vec{\bigtriangledown}\,\mathcal{G}^{\ast}(x) = \partial^{0}\vec{E}_{\varphi}(x).               \eqno\ldots (14.30)
$$

В свою очередь, двойное внутреннее произведение обеих частей второго уравнения системы  (III)
на тензор $ g$  и последующее использование равенства (14.13) и соотношения
$$
g\cdot\cdot\,\tilde{def}_{\ast}\,\vec{E}_{ \mathit A}(x) = div\,\vec{E}_{ \mathit A}(x),           \eqno\ldots (14.31)
$$
дает второе уравнение для сопровождающего 3-скалярного поля $ \mathcal{G}^{\ast}(x)$
$$
\partial^{0}\, \mathcal{G}^{\ast}(x) = div\,\vec{E}_{\varphi}(x),                                   \eqno\ldots (14.32)
$$
выполняющееся тождественно, как и все первые пары электродинамических уравнений, лишь в силу
 определения полевых переменных входящих в него.

 (14.30), (14.32) представляет собой систему уравнений, сопровождающую систему (III) и представляющую собой дифференциальные связи между скалярным полем $ \mathcal{G}^{\ast}(x)$, сопровождающим дилатационное джейтонное поле представленное тензором $ \mathcal{G}_{\ast}(x)$, и электрическим полем, определяемым напряженностью $ \vec{E}_{\varphi}(x)$.

С другой стороны находим, что подобно тому как 3-тензор $ \mathcal{F}(x)$ сопровождается 3-псевдовектором
$ \vec{\mathcal{H}}(x)= -\displaystyle \frac{1}{2}\;\varepsilon\cdot\cdot\, \mathcal{F}(x)$, являющимся угловой скоростью вращения соответствующей сопровождающей области фазового пространства как абсолютно
твердого тела, так и 3-тензор  $ \mathcal{G}_{\ast}(x)$ сопровождается 3-скаляром $  \mathcal{G}^{\ast}(x)= g\cdot\cdot\,\mathcal{G}_{\ast}(x)$, так же характеризующим вид движения сопровождающей области фазового пространства, выступая в качестве скорости относительного изменения объема бесконечно малой сопровождающей области этого пространства \cite{Rash}.

И в том, и в другом случае имеем две {\em вспомогательные} характеристики динамических процессов динамических систем определяемых тензорами $ \mathcal{F}(x)$ и $ \mathcal{G}_{\ast}(x)$ в соответствии с уравнениями (2.8) и (2.12) и, тем самым, устанавливаем, одновременно, иерархию отношений данных тензоров и рассматриваемых сопровождающих их объектов (см. так же комментарий к уравнению (3.8)).

В рамках 4-мерной формулировки классической электродинамики, системы уравнений (14.7), (14.14) и (14.8), (14.22) имеют, соответственно,  более компактный лоренц-инвариантный вид
$$
\partial \! \cdot \! F(x) = \theta,                                    \eqno\ldots (14.33)
$$
$$
\partial \! \cdot \! G(x) = \theta,                                    \eqno\ldots (14.34)
$$
где 4-тензоры $ \mathit F (x)$ и $ \mathit G(x)$, через 4-потенциал $ \mathit A(x)$, определены соотношениями
$$
 F(x) \eqdef \widetilde{rot}\,A(x),                                    \eqno\ldots (14.35)
$$
$$
 G(x) \eqdef \widetilde{def}\,A(x).                                    \eqno\ldots (14.36)
$$

4-тензорные дифференциальные операторы, $ \widetilde{rot}$ и $  \widetilde{def}$ представлены соотношениями
$$
\widetilde{rot} \, \eqdef \partial  \stackrel{a}{\otimes} ,                           \eqno\ldots (14.37)
$$
$$
\widetilde{def}\, \eqdef \partial  \stackrel{ s}{\otimes} ,                                \eqno\ldots (14.38)
$$
в которых,  $ \partial \eqdef e^{\mu}\partial_{\mu}$ --  дифференциальный оператор Гамильтона в псевдоевклидовом пространстве событий.

Естественно, систему (14.33), (14.34) легко получить и с помощью релятивистски инвариантной, но более формальной, процедуры.

Действительно, переписав однородное уравнение Даламбера для 4-потенциала  $ \mathit A(x)$,
$$
\partial \cdot \partial A(x) = \theta,                                           \eqno\ldots (14.39)
$$
где символом $ \partial A(x)$ теперь обозначен 4-градиент 4-потенциала $ \mathit A(x)$,
$$
 \partial A(x)\,\eqdef \partial \otimes A(x),                                           \eqno\ldots (14.40)
$$
в виде
 $$
\partial\! \cdot\! F(x) + \partial \!\cdot\! G(x)= \theta,                                           \eqno\ldots (14.41)
$$
находим, что использование в (14.41) соотношения, идентичного (14.3),
$$
 \partial \!\cdot\! G(x)= \partial\! \cdot\! F(x) + \partial \!\cdot \! \partial A(x)^{\mathit T},                      \eqno\ldots (14.42)
$$
приводит не только к уравнению
$$
2\, \partial\! \cdot\! F(x) + \partial \!\cdot \! \partial A(x)^{\mathit T} = \theta,                      \eqno\ldots (14.43)
$$
но и к уравнению
$$
2\, \partial\! \cdot\! G(x) - \partial \!\cdot \! \partial A(x)^{\mathit T} = \theta,                      \eqno\ldots (14.44)
$$
которые, после учета соотношения
$$
 \partial \!\cdot \! \partial A(x)^{\mathit T} = \partial\, \partial\! \cdot\! A(x),                      \eqno\ldots (14.45)
$$
и условия Лоренца приводят к системе (14.33), (14.34)
 \footnote{Система уравнений (14.33), (14.34), в координатной форме, впервые представлена в дипломной работе автора, выполненной в Лаборатории Теоретической Физики Объединенного Института  Ядерных Исследований (г. Дубна) в 1970~г., в которой начато  рассмотрение классической электродинамики, {\em явно} содержащей как традиционный антисимметрический производный аффинор $  \mathit F (x)$,  так и  симметрический производный   аффинор  $\mathit G(x)$.}.

  Наконец, выполнив аналогичную релятивистски инвариантную процедуру для неоднородного
   уравнения Даламбера, представленного в виде, явно содержащим 4-тензор $ \partial A(x)$,
 $$
\partial \cdot \partial A(x) = j(x),                                           \eqno\ldots (14.46)
$$
получаем, что (14.46), в рамках условия Лоренца, приводит к двум уравнениям
$$
\partial \! \cdot \! F_{\mathit L}(x) = j(x),                                    \eqno\ldots (14.47)
$$
$$
\partial \! \cdot \! G_{\mathit L}(x) = j(x) ,                                    \eqno\ldots (14.48)
$$
первое из которых представляет собой известную вторую пару системы уравнений
Максвелла в 4-инвариантном виде, а второе представляет 4-инвариантную форму второй
пары системы уравнений для электроджейтонного поля.

В целях сопоставления уравнений (14.47) и (14.48), представим обобщение этих уравнений на случай
 наличия  в рассматриваемом пространстве событий гравитационного поля.

 В традиционной координатной форме уравнение (14.48), при соответствующем переопределении
 плотности тока  \cite{La}, в данном случае имеет вид
 $$
\frac{1}{\sqrt{-\textsl{g}(x)}}\,\partial_{m}(\sqrt{-\textsl{g}(x)}\, \tilde{ \mathit G}_{L}^{mn}(x))  + \mathit \Gamma^{n}_{\,kl}(x)\tilde{ \mathit G}_{L}^{kl}(x)  =  J^{n}(x).               \eqno\ldots (14.49)
$$
где
$$
\tilde{ \mathit G}^{L}_{mn}(x) = \mathit G^{L}_{mn}(x) - 2\, \mathit \Gamma ^{k}_{\,mn}(x)A_{k}(x),        \eqno\ldots (14.50)
$$
в то время как уравнение (14.47) в этом случае представляется известным соотношением \cite{La}
$$
\frac{1}{\sqrt{-\textsl{g}(x)}}\,\partial_{m}(\sqrt{-\textsl{g}(x)}\,  \mathit F_{L}^{mn}(x)) =  J^{n}(x),        \eqno\ldots (14.51)
$$
существенно отличающимся от (14.49).

Уравнение (14.50) одновременно демонстрирует специфическую, по отношению
 к электромагнитному полю, "реакцию"\, электроджейтонного поля на пространство событий общей теории относительности.

 Возвращаясь к 3-инвариантным формам полевых уравнений, из (14.47) и (14.48) имеем, соответственно,
$$
 \vec{\bigtriangledown}\cdot\mathcal{F}_{\mathit L}(x)  =
 - \vec{j}(x) - \partial ^{0}\vec{\mathit E}_{\mathcal{F}L}(x),   \eqno\ldots (14.52)
$$
$$
div\,\vec{\mathit E}_{\mathcal{F}L}(x) = j^{0}(x),                                   \eqno\ldots (14.53)
$$
$$
 \vec{\bigtriangledown}\cdot\mathcal{G}_{\mathit L}(x)  =
 - \vec{j}(x) - \partial ^{0}\vec{\mathit E}_{\mathcal{G}L}(x),                               \eqno\ldots (14.54)
$$
$$
div\,\vec{\mathit E}_{\mathcal{G}L}(x) = - j^{0}(x) +  \partial ^{0}G_{L}^{00}(x).                \eqno\ldots (14.55)
$$

При этом является очевидным, что, при стандартном условии Лоренца классической электродинамики,
уравнения (14.47) и (14.48), как уравнения на потенциал $ \mathit A(x)$, эквивалентны как по отношению к
друг другу, так и по отношению к уравнению Даламбера (14.46).

Однако, как уравнения на компоненты (напряженности) электромагнитного и электроджейтонного полей,
(14.47) и (14.48) описывают самостоятельные связи соответствующих полевых переменных \cite{Al2}, что и
демонстрирует сопоставление системы (14.52), (14.53) с системой (14.54), (14.55).

При этом, подобно тому как закон Ампера\,--\,Максвелла, представленный уравнением (14.52), является соответствующим обобщением теоремы Ампера, представленной уравнением (5.6), так и уравнение (14.54), в этом отношении,
выступает как соответствующее обобщение уравнения (5.7), демонстрируя, что джейтонное поле порождается не только током проводимости, но и соответствующим током смещения.

При этом уместно отметить, что (14.54), на продольной оси линейно вибрирующей точечной
электрически заряженной частицы,  не является вырожденным
 уравнением вследствие того, что как джейтонное поле, определяемое тензором $ \mathcal{G}(x)$, так и сопровождающее его электрическое поле, представленное напряженностью $ \vec{\mathit E}_{\mathcal{G}}(x)$, на этой оси отличны от нуля.

 И так, имеем хорошо известную  систему дифференциальных уравнений Максвелла, приводимую
 тут исключительно в целях сопоставления,
 $$
 \left.
\begin{array}{c}
 \vec{\bigtriangledown}\cdot\mathcal{F}_{\mathit L}(x)  =
 - \vec{j}(x) - \partial ^{0}\vec{\mathit E}_{\mathcal{F}L}(x), \\
 \partial^{0} \mathcal{F}(x)  = -  \tilde{rot}\,\vec{E}_{A}(x) = -  \tilde{rot}\,\vec{E}_{\mathcal{F}L}(x),\\
div\,\vec{\mathit E}_{\mathcal{F}L}(x) = j^{0}(x),  \\
div\,\vec{\mathcal{H}}(x) = 0,
\end{array}
\right.         \eqno\ldots (V)
$$
и систему дифференциальных уравнений описывающую электроджейтонное поле,
$$
 \left.
\begin{array}{c}
 \vec{\bigtriangledown}\cdot\mathcal{G}_{\mathit L}(x)  =
 - \vec{j}(x) - \partial ^{0}\vec{\mathit E}_{\mathcal{G}L}(x), \\
 \partial^{0} \mathcal{G}(x)  = -  \tilde{def}\,\vec{E}_{A}(x) ,\\
div\,\vec{\mathit E}_{\mathcal{G}L}(x) = - j^{0}(x) +  \partial ^{0}G_{L}^{00}(x).
\end{array}
\right.         \eqno\ldots (V\!I)
$$

В квазистационарном случае система ({\em VI}\,) имеет вид
$$
 \left.
\begin{array}{c}
 \vec{\bigtriangledown}\cdot\mathcal{G}_{\mathit L}(x)  =
 - \vec{j}(x) , \\
 \partial^{0} \mathcal{G}(x)  = -  \tilde{def}\,\vec{E}_{A}(x) ,\\
div\,\vec{\mathit E}_{\mathcal{G}L}(x) = - j^{0}(x),
\end{array}
\right.         \eqno\ldots (V\!I\,^{'})
$$
демонстрирующий, что соответствующие электрическое и джейтонное поля и в этом случае не являются
автономными по отношению к друг другу, представляя, тем самым, {\em единый} полевой
объект -- квазистационарное электроджейтонное поле.

При этом, {\em главная} полевая связь, в данной полевой системе, устанавливается дифференциальным
законом электроджейтонной индукции.

В квазистационарном электромагнитном поле соответствующая главная связь устанавливается
 соответствующим законом электромагнитной индукции.

 Вышеуказанным одновременно демонстрируется фундаментальная физическая роль дифференциальных
 законов электромагнитной и электроджейтонной индукции   в электродинамике квазистационарных полевых систем.

 Таким образом, система "основных"\, электродинамических  уравнений, в рассматриваемом квазистационарном случае,   теперь имеет вид
 $$
 \left.
\begin{array}{c}
\vec{\bigtriangledown}\cdot\mathcal{F}(x) +\vec{\bigtriangledown}\cdot\mathcal{G}(x)  =  - \vec{j}(x) , \\
 \partial^{0} \mathcal{F}(x)  = -  \tilde{rot}\,\vec{E}_{A}(x), \\
 \partial^{0} \mathcal{G}(x)  = -  \tilde{def}\,\vec{E}_{A}(x).
\end{array}
\right.         \eqno\ldots (V\!I\!I)
$$

В частности тогда имеем, что полевые системы, представленные на рис. 10.1 и рис.11.1, в областях
$\mathcal{D} _{in}$ представлены известными квазистационарными электромагнитными  полями, электрические составляющие которых, в соответствии с общепринятой теорией близкодействия, вызывают электрические
токи в проводниках, находящихся в данных областях.

В областях  $\mathcal{D} _{ex}$ полевые системы представлены квазистационарными
{\em электроджейтонными} полями, электрические составляющие которых, так же в соответствии с теорией близкодействия, вызывают электрические токи в проводниках, находящихся в этих областях.

В связи с этим тогда находим, в частности, что электрический ток вторичной обмотки
 традиционного трансформатора  порождается электрической составляющей квазистационарного
 {\em электроджейтонного} поля, "занимающего"\, внешнюю, по отношению к сердечнику трансформатора, область 3-пространства.

 Аналогично, ускорение электрически заряженных частиц в индукционных линейных ускорителях, например, так же осуществляется, и так же в соответствии с теорией близкодействия, электрической составляющей квазистационарного
 {\em электроджейтонного} поля, возникающего вне кольцевых ферритовых сердечников в процессе изменения силы тока в  обмотках индукторов.

 Интерпретация закона электродинамической индукции М. Фарадея, в рассматриваемых случаях, представлена так же
 в комментариях к уравнению (6.52).

 Система уравнений ({\em V}\,), в 4-инвариантной форме, имеет известный компактный вид
 $$
 \partial \! \cdot \! F_{\mathit L}(x) = j(x),
 $$
 $$
 \partial \! \cdot \! \widetilde{F}_{\mathit L}(x) =\theta,         \eqno\ldots (14.56)
$$
 где $ \widetilde{F}_{\mathit L}(x)$ -- тензор, дуальный по отношению к   4-тензору   Фарадея\,--\,Максвелла
 $\mathit F_{\mathit L}(x)$, определяемый соотношением
$$
  \widetilde{F}_{\mathit L}(x) = \frac{1}{2}\,\epsilon \cdot \cdot F_{\mathit L}(x)^{\mathit T},         \eqno\ldots (14.57)
$$
в котором, $ \epsilon$ -- абсолютно антисимметричный единичный тензор четвертого ранга  \cite{La}.

4-инвариантное уравнение, включающее, в частности, обсуждаемые законы электродинамической индукции,
представлено соотношением
$$
\partial_{a}G_{ \mathit L}(x) = \partial F(x),                                        \eqno\ldots (14.58)
$$
в котором, 4-градиенты соответствующих 4-тензоров определены соотношениями
$$
\partial_{a}G_{ \mathit L}(x)  \eqdef \partial \stackrel{ a}{\otimes}G_{ \mathit L}(x) =
 \partial \otimes G(x) - \partial \stackrel{ \mathit T}{\otimes} G(x),                                            \eqno\ldots (14.59)
$$
$$
 \partial F(x) \eqdef  \partial \otimes F(x).                                              \eqno\ldots (14.60)
$$

Легко видеть, что (14.58) содержит в себе, например, уравнение (14.56).

Действительно, тройное внутреннее произведение 4-тензора $\epsilon$ на обе части уравнения (14.58) и последующее использование соотношений
$$
\partial \cdot \widetilde{F}(x) = \frac{1}{2}\,\epsilon \cdot\cdot\cdot \partial F(x),     \eqno\ldots (14.61)
$$
$$
\epsilon \cdot\cdot\cdot \partial_{a}G_{ \mathit L}(x) \equiv \theta,                          \eqno\ldots (14.62)
$$
мгновенно приводит к (14.56).

Ковариантная координатная форма уравнения (14.58) имеет вид хорошо известного тождества \cite{Nov}
$$
\partial _{\nu}G_{\rho\mu}(x) - \partial _{\rho}G_{\nu\mu}(x) = \partial _{\mu}F_{\nu\rho}(x),        \eqno\ldots (14.63)
$$
которое так же мгновенно приводит к координатным формам уравнений (1.11)--(1.13).

Уместно отметить так же, что вторые пары электродинамических уравнений (14.47) и (14.48) могут
 быть получены  и в рамках использования традиционного лагранжева формализма  стартующего из лагранжианов
$$
\mathfrak{L}_{ \mathit F}(x) = - F(x)^{2} - j(x)\!\cdot\! A(x),                                        \eqno\ldots (14.64)
$$
$$
\mathfrak{L}_{ \mathit G}(x) = - G(x)^{2} - j(x)\!\cdot\! A(x),                                        \eqno\ldots (14.65)
$$
соответствующие уравнения Лагранжа\,--\,Эйлера для которых мгновенно приводят к
 соотношениям (14.47), (14.48), соответственно.

 Аналогично, и традиционное уравнение (14.46) получается  в рамках лагранжева формализма  стартующего из общеизвестного классического лагранжиана
 $$
\mathfrak{L}(x) = - \frac{1}{2}\, \partial A(x)^{2} - j(x)\!\cdot\! A(x).                                        \eqno\ldots (14.66)
$$

Как и в части (I), $ F(x)^{2}$, $ G(x)^{2}$ и $ \partial A(x)^{2}$ в (14.64)--(14.66) определены традиционно --
как скалярные квадраты соответствующих 4-тензоров, то есть, для произвольного 4-тензора
 второго ранга, $ T(x)$, имеем
 $$
 T(x)^{2}\,\eqdef T(x)\cdot\cdot \,T(x)^{\mathit T}\, \eqdef \,\left(T(x), T(x) \right).
 $$

 Сопоставление лагранжианов $ \mathfrak{L}_{0 \mathit F}(x) = - F(x)^{2}$ и
$ \mathfrak{L}_{0 }(x) = -\displaystyle \frac{1}{2}\, \partial A(x)^{2}$ рассмотрено,    например, в \cite{BSh} и
  тут не обсуждаются.

  Следует однако отметить, что, в силу соотношения
 $$
 \partial A(x)^{2} = F(x)^{2}+ G(x)^{2},                        \eqno\ldots (14.67)
 $$
 лагранжиан (14.66) принимает вид
$$
\mathfrak{L}(x) = - \frac{1}{2}\, F(x)^{2} - \frac{1}{2}\, G(x)^{2} - j(x)\!\cdot\! A(x),                    \eqno\ldots (14.68)
$$
{\em явно} демонстрирующий, что традиционный лагранжиан (14.66) является, в действительности,
лагранжианом электромагнитного и электроджейтонного полей, в связи с чем, соответствующее уравнение
Лагранжа\,--\,Эйлера для лагранжиана (14.66) и приводит  одновременно как к уравнению (14.47), так и к
уравнению (14.48).

С другой стороны, стартуя с лагранжиана (14.68), уравнение Лагранжа\,--\,Эйлера, соответствующее варьированию
соответствующего действия по 4-потенциалу $ A(x)$, непосредственно приводит к уравнению
$$
\partial \! \cdot\!  F(x) + \partial\! \cdot\!  G(x) =  j(x),                          \eqno\ldots (14.69)
$$
явно демонстрирующему, подобно уравнению (5.4), что 4-ток $ j(x)$   "порождает"\, как электромагнитное, так и
электроджейтонное поля.

Как и в (14.41), последующее использование (14.42) в (14.69) приводит, на этом пути, так же к системе {\em двух}
 вторых пар электродинамических уравнений,
$$
\partial \!\cdot \! F_{\mathit L}(x) + \partial\! \cdot \! \partial A(x)^{\mathit T} =  j(x),                          \eqno\ldots (14.70)
$$
$$
\partial \!\cdot \! G_{\mathit L}(x) - \partial\! \cdot \! \partial A(x)^{\mathit T} =  j(x),                          \eqno\ldots (14.71)
$$
последующее использование в которых соотношения (14.45) и классического условия Лоренца дает систему
(14.47), (14.48), из которой следуют уравнения (14.52)--(14.55).

Альтернативно, по-прежнему сохраняя явное наличие $ \mathcal{F}$-- и
$ \mathcal{G}$-- полей, из (14.69) имеем, в частности, уравнение
$$
\partial^{0}\vec{E}_{\mathit A}(x) + \vec{\bigtriangledown}\cdot \mathcal{F}(x) +
  \vec{\bigtriangledown}\cdot \mathcal{G}(x)= - \vec{j}(x),                                                   \eqno\ldots (14.72)
$$
последующее использование в котором соотношения (14.3) так же приводит к системе {\em двух} уравнений,
 $$
\partial^{0}\vec{E}_{\mathit A}(x) + \vec{\bigtriangledown}\cdot \mathcal{F}_{ \mathit L}(x) +
  \vec{\bigtriangledown}\cdot \partial A(x)^{\mathit T} = - \vec{j}(x),                                                   \eqno\ldots (14.73)
$$
 $$
\partial^{0}\vec{E}_{\mathit A}(x) + \vec{\bigtriangledown}\cdot \mathcal{G}_{ \mathit L}(x) -
  \vec{\bigtriangledown}\cdot \partial A(x)^{\mathit T} = - \vec{j}(x),                                                   \eqno\ldots (14.74)
$$
в которых тензор $  \partial A(x)$ определен уже соотношением (1.18).

Последующее использование (14.6) в (14.73) и (14.74) так же приводит к системе (14.52), (14.54) полученной выше как
непосредственное следствие уравнений (14.47) и  (14.48), то есть,
$$
\partial^{0}\!\left(\vec{E}_{\mathit A}(x) + \vec{E}_{\varphi}(x)\right) +
\vec{\bigtriangledown}\cdot \mathcal{F}_{ \mathit L}(x)  = - \vec{j}(x),        \eqno\ldots (14.75)                                                                                                                         $$
$$
\partial^{0}\!\left(\vec{E}_{\mathit A}(x) - \vec{E}_{\varphi}(x) \right)+
 \vec{\bigtriangledown}\cdot \mathcal{G}_{ \mathit L}(x)  = - \vec{j}(x).                   \eqno\ldots (14.76)
$$

Рассматриваемая альтернатива явно демонстрирует тот "механизм"\,, по которому каждое
 из полей, $ \mathcal{F}_{\mathit L}(x)$ и $ \mathcal{G}_{\mathit L}(x)$,  "выбирает"\, для себя  соответствующее электрическое поле.

 Вне полевого источника, (14.75), (14.76) принимает вид системы уравнений
 $$
\partial^{0}\vec{E}_{\mathcal{F}}(x)  + \vec{\bigtriangledown}\cdot \mathcal{F}(x)  = \vec{0},    \eqno\ldots (14.77)
$$
$$
\partial^{0}\vec{E}_{\mathcal{G}}(x)  + \vec{\bigtriangledown}\cdot \mathcal{G}(x)  = \vec{0},               \eqno\ldots (14.78)
$$
выступающей, в частности, в качестве системы   {\em уравнений связи} соответствующих полевых переменных
электромагнитного и электроджейтонного полей.

С другой стороны, уравнение (14.69) мгновенно приводит к традиционному неоднородному
 уравнению Даламбера (14.46).

 Таким образом, уравнение (14.69) рассматриваемой теории, стартующей с лагранжиана (14.68),
 в отличие от соответствующего уравнения Максвелла (14.47)  \cite{Sch}, является эквивалентным, как
 уравнение на 4-потенциал $ A(x)$, традиционному уравнению Даламбера без необходимости введения,
 на данном этапе, каких-либо дополнительных условий на 4-потенциал  $ A(x)$.

 Именно это обстоятельство позволило, при анализе выше рассмотренных полевых систем, стартовать
 с рассматриваемого уравнения Даламбера.

 Уместно подчеркнуть так же, что {\em стартовой} аксиомой при построении обсуждаемой теории (обсуждаемой классической  электродинамики) является  утверждение о  {\em физическом} существовании электроджейтонного поля как  самостоятельного и равноправного, по отношению к электромагнитному полю, материального полевого объекта,
 представленного 4-тензором $  \mathit G(x)$ .

 Именно эта стартовая позиция, математическое обоснование которой представлено в \cite{Al1, Al2}, и приводит,
 в рамках изложения теории на основе лагранжева формализма, к последующему использованию, в качестве стартового
 лагранжиана обсуждаемой электродинамики, плотности функции Лагранжа (14.68), {\em явно} демонстрирующей
 равноправный учет как электромагнитного, так и электроджейтонного полей.

 При этом, фундаментальное полевое уравнение традиционной классической электродинамики, неоднородное
 уравнение Даламбера для 4-потенциала  $\mathit A(x)$, обсуждаемой электродинамикой {\em не модифицируется},
 однако его решения теперь рассматриваются в качестве носителей как электромагнитного, так и электроджейтонного полей, то есть,  $\mathit A(x)$ теперь является 4-потенциалом как электромагнитного, так и электроджейтонного полей
 (электромагнитоджейтонным потенциалом), в связи с чем, лагранжиан взаимодействия, содержащийся в правых частях
равенств (14.66) и (14.68), теперь выступает в качестве классического лагранжиана "минимального"\,  взаимодействия
электрон-позитронного поля с электромагнитным и электроджейтонным полями.

В количественном отношении, электромагнитному полю теперь отводится более скромная роль определяемая
4-тензором $\mathit F(x)$  представленным соотношением (14.35), то есть, в обсуждаемой электродинамике
электромагнитное поле ассоциируется {\em с данным} тензором, что так же является принципиальным отличительным признаком данной теории.

Уместно вновь подчеркнуть, что вышеукзанная стартовая аксиома обсуждаемой электродинамики одновременно значительно ограничивает "калибровочный произвол"\, \cite{IZ} 4-потенциала  $\mathit A(x)$,  допуская лишь такие преобразования $ \mathit A(x)\mapsto\mathit A(x)+\partial \Lambda(x)$, при которых $ \Lambda(x) \in \{ C_{\mu}x^{\mu}+ C \}$, то есть, калибровочный произвол ограничен рамками группы "ультраспециализированных"\, градиентных преобразований второго рода, в которой подгруппа  $R_{c}:\{\Lambda(x) = C_{\mu}x^{\mu}\}$  представляет собой известную группу $R$-калибровочных преобразований полевых функций и связанную с ней  $R$-инвариантность  \cite{Br}.

 \section*{\S \,15 Волновые уравнения для полевых переменных электромагнитного и электроджейтонного полей}
\addcontentsline{toc}{section}{\S \,15 Волновые уравнения для полевых переменных электромагнитного и электроджейтонного полей}

\qquad Переходя к определению волновых свойств рассматриваемых полевых переменных, применим последовательно
дифференциальные операторы $ \tilde{rot}$, $ \tilde{def}_{d}$ и $ \tilde{def}_{\ast}$ к обеим частям стандартной формы
неоднородного уравнения Даламбера для векторного потенциала,
$$
\Box\,\vec{A}(x) = - \vec{j}(x),                        \eqno\ldots (15.1)
$$
в результате чего, после последующего использования определений (1.5)--(1.8), получаем следующую систему
локально линейно независимых, по отношению к друг другу, неоднородных волновых уравнений
$$
\Box\,\mathcal{F}(x) = - \tilde{rot}\,\vec{j}(x),                        \eqno\ldots (15.2)
$$
$$
\Box\,\,\mathcal{G}_{d}(x) = - \tilde{def}_{d}\,\,\vec{j}(x),                        \eqno\ldots (15.3)
$$
$$
\Box\,\,\mathcal{G_{\ast}}(x) = - \tilde{def}_{\ast}\,\,\vec{j}(x).                        \eqno\ldots (15.4)
$$

Уравнения (15.2)--(15.4) прежде всего демонстрируют, что как магнитное, так и джейтонные поля
вне точек источника удовлетворяют волновым уравнениям, то есть, каждое из этих полей, в случае их нестационарности, обладает волновым характером распространения в 3-пространстве.

С другой стороны, эти уравнения демонстрируют так же, что источниками этих полей могут выступать
электрические токи, конгруэнции   векторных линий плотностей тока которых определены производными
тензорами правых частей этих уравнений.

 Например, уравнение (15.2) показывает, что переменный электрический ток, конгруэнция векторных линий плотности
 тока которого представлена  {\em центром}, может выступать как источник волнового
 тензорного поля $ \mathcal{F}(x)$, геометрически представленного {\em центром} векторных линий
 векторного потенциала $ \vec{\mathit A}_{\mathcal{F}}(x)$ (соответствующим фазовым   портретом
  тензора $ \mathcal{F}(x)$).

  Аналогично, уравнение (15.3) демонстрирует, что переменный электрический ток,  конгруэнция векторных линий плотности  тока которого представлена  {\em седлом}, может выступать как источник волнового  тензорного поля
  $ \mathcal{G}_{d}(x)$ (волнового девиатационного джейтонного поля), геометрически представленного {\em седлом} векторных линий  векторного потенциала $ \vec{\mathit A}_{\mathcal{G}_{d}}(x)$ (соответствующим фазовым
  портретом  тензора $ \mathcal{G}_{d}(x)$).

  В свою очередь, и уравнение (15.4) так же демонстрирует, что переменный электрический ток,  конгруэнция векторных линий плотности  тока которого представлена  {\em дикритическим узлом}, может выступать как источник волнового  тензорного поля $ \mathcal{G}_{\ast}(x)$ (волнового дилатационного джейтонного поля), геометрически представленного {\em дикритическим узлом} векторных линий  векторного потенциала $ \vec{\mathit A}_{\mathcal{G}_{\ast}}(x)$ (соответствующим фазовым   портретом  тензора $ \mathcal{G}_{\ast}(x)$).

  Подобные процессы генерации магнитного и джейтонных полей, представленные уравнениями (15.2)--(15.4),
  наблюдаются, например, при возбуждении $ \mathit H_{01}$-,   $ \mathit H_{11}$-  и $ \mathit E_{01}$-волн
  в волноводах, представленных на рис. 12.1--12.3.

  Таким образом, в соответствии с данными уравнениями, на рассматриваемых продольных осях данных
  волноводов имеем, соответственно, магнитное, девиатационное джейтонное и дилатационное джейтонное
  поля, волновым образом распространяющиеся вдоль этих осей, о чем было констатировано выше в комментариях
   к  данным рисункам.

   С другой стороны, в соответствии с законами электродинамической индукции (1.11)--(1.13), данные поля,
   в процессе своего распространения, порождают {\em наблюдаемые} в этих волноводах электрические поля,
    конгруэнции векторных линий которых, в поперечных плоскостях данных волноводов, представлены, соответственно, центром,  каноническим седлом и дикритическим (звездным) узлом.

В свою очередь, применяя оператор $ \partial^{0}$ к обеим частям каждого из уравнений (15.2)--(15.4), с последующим использованием рассматриваемых законов электродинамической индукции, получаем систему локально линейно независимых, по отношению к друг другу, уравнений
$$
\Box\, \tilde{rot}\,\vec{E}_{\mathit A}(x) =  \tilde{rot}\,\partial^{0}\,\vec{j}(x),                        \eqno\ldots (15.5)
$$
$$
\Box\,\,  \tilde{def}_{d}\,\vec{E}_{\mathit A}(x)=  \tilde{def}_{d}\,\,\partial^{0}\,\vec{j}(x),                        \eqno\ldots (15.6)
$$
$$
\Box\,\, \tilde{def}_{\ast}\,\vec{E}_{\mathit A}(x) =  \tilde{def}_{\ast}\,\,\partial^{0}\,\vec{j}(x),                        \eqno\ldots (15.7)
$$
которые, как и уравнения (15.2)--(15.4), прежде всего демонстрируют, что конгруэнции векторных линий
электрических полей, определяемых тензорами левых частей этих уравнений, то есть, электрических полей
сопровождающих магнитное и джейтонные поля, так же волновым образом распространяются внутри рассматриваемых направляющих систем, в полном соответствии с наблюдаемыми в этих системах процессами.

С другой стороны, данные уравнения демонстрируют, что источниками рассматриваемых электрических полей
могут выступать переменные во времени электрические токи, конгруэнции векторных линий плотности тока которых
определены соответствующими тензорами правых частей этих уравнений.

В частности, уравнение (15.5) демонстрирует, что переменный во времени электрический ток,  конгруэнция векторных линий плотности тока которого представлена  {\em центром}, может выступать  и как непосредственный  источник волнового  тензорного поля $ \tilde{rot}\,\vec{E}_{\mathit A}(x)$, геометрически представленного {\em центром} векторных линий  электрического поля   $\vec{E}_{\mathit A\mathcal{F}} (x)$ (фазовым   портретом данного тензора).

Именно это и используется, например, при возбуждении $  \mathit H_{01}$-волны в круглом волноводе,
то есть, уравнение (15.5) имеет технологическое применение, что относит данное уравнение, а вместе с ним и уравнение (15.2), к классу {\em физических}  уравнений имеющих практическое использование.

Аналогичная интерпретация проводится и для уравнений (15.6), (15.7) и (15.3), (15.4), с последующим заключением
как о физическом ранге этих уравнений, так и физическом статусе полевых переменных входящих в них.

Еще раз следует подчеркнуть, что в данном контексте, когда говорится о центре, седле или узле, то имеются в виду
соответствующие фазовые портреты соответствующих динамических систем  {\em на плоскости}, например,
на плоскостях поперечных сечений соответствующих волноводов, а поля  этих направляющих систем
 представляются в рамках концепции Л.~Бриллюэна.

Уместно привести и известное \cite{Nik} неоднородное волновое уравнение для псевдовектора
$ \vec{\mathcal{H}}_{ \mathit L}(x)$, традиционно представляющего традиционное магнитное поле,
$$
\Box\,\vec{\mathcal{H}}_{\mathit L}(x) = - rot \,\vec{j}(x),                        \eqno\ldots (15.8)
$$
которое мгновенно следует или из (15.1), после применения оператора $ rot$ к обеим частям этого
уравнения и последующего использования определения $ \vec{\mathcal{H}}_{ \mathit L}(x)\eqdef rot\,\vec{\mathit A}(x)$,
или из (15.2), после двойного внутреннего умножения обеих частей этого уравнения на тензор третьего ранга
$ \varepsilon$ и последующего учета (1.15) и общего соотношения дифференциальных операторов
$ \tilde{rot}$  и $ rot$,
$$
\varepsilon\cdot\cdot \,\tilde{rot}\,\vec{a}(x)= - rot\,\vec{a}(x),                    \eqno\ldots (15.9)
$$
использованного для получения (1.16).

Соответствующие неоднородные волновые уравнения для полевых переменных  $ \vec{\mathit E}_{\mathcal{F}L}(x)$   и
$ \vec{\mathit E}_{\mathcal{G}L}(x)$ так же легко получаются из стандартного неоднородного волнового уравнения для
4-потенциала, $ \Box\,A(x) = - j(x)$, путем соответствующих дифференцирований левых и правых частей
этого уравнения с последующим использованием определений этих полевых переменных.

Таким образом, имеем соотношения
$$
\square\,\vec{\mathit E}_{\mathcal{F}L}(x) = \partial^{0}\,\vec{j}(x) + \vec{\bigtriangledown}\,j^{0}(x),         \eqno\ldots (15.10)
$$
$$
\square\,\vec{\mathit E}_{\mathcal{G}L}(x) = \partial^{0}\,\vec{j}(x) - \vec{\bigtriangledown}\,j^{0}(x),         \eqno\ldots (15.11)
$$
в свою очередь демонстрирующие, в частности, соответствующий волновой характер распространения переменных
во времени электрических полей, сопровождающих переменные во времени магнитное и джейтонное поля.

 \chapter{\,Электроджейтонные волны и продольное электроджейтонное излучение}

 \section*{\S \,16 Плотность энергии и плотность потока энергии электромагнитного и электроджейтонного полей в общем случае}
\addcontentsline{toc}{section}{\S \,16 Плотность энергии и плотность потока энергии электромагнитного и электроджейтонного полей в общем случае}

\qquad  Рассмотрим основные динамические характеристики, плотность энергии и плотность потока
энергии, а так же соответствующие законы сохранения, для полевых систем, представленных электромагнитным и электроджейтонным полями.

Как и в выше рассмотренном случае плоской электродинамической волны Ландау, выражения для этих
величин, через решения уравнений движения данных полевых систем, можно легко получить непосредственно из
самих уравнений движения этих систем.

Системы уравнений (V) и (VI) позволяют получить рассматриваемые динамические характеристики
для каждой из полевых систем описывающихся этими системами уравнений.

По-прежнему, в целях сопоставления, проведем соответствующие процедуры для каждой из этих полевых систем.

Рассматривая систему представленную электромагнитным полем, вновь выпишем необходимые для данной цели
полевые уравнения системы (V)
$$
\vec{\bigtriangledown}\!\cdot \!\mathcal{F}(x)  =  - \partial ^{0}\vec{\mathit E}_{\mathcal{F}}(x) - \frac{1}{2}\, \vec{j}(x), \eqno\ldots (16.1)
$$
$$
\partial^{0} \mathcal{F}(x)  = -  \tilde{rot}\,\vec{E}_{A}(x).                                          \eqno\ldots (16.2)
$$

Следуя стандартной процедуре  \cite{La}, после скалярного умножения обеих частей
 уравнения (16.1) на $ \vec{\mathit E}_{\mathcal{F}}(x)$, а обеих частей уравнения (16.2) на $  \mathcal{F}(x)$, получаем
 систему скалярных уравнений
 $$
 \frac{1}{2}\,\partial^{0}\,\vec{E}_{\mathcal{F}}(x)^{2} = - \left(\vec{\bigtriangledown} \!\cdot \!\mathcal{F}(x), \vec{E}_{\mathcal{F}}(x) \right) - \frac{1}{2}\,\vec{j}(x) \!\cdot\!  \vec{E}_{\mathcal{F}}(x),                           \eqno\ldots (16.3)
 $$
 $$
\frac{1}{2}\,\partial^{0}\,\mathcal{F}(x)^{2} = - \left(\tilde{rot}\,\vec{E}_{\mathit A}(x),  \mathcal{F}(x)\right),   \eqno\ldots (16.4)
 $$
последующее сложение которых, с использованием соотношения, идентичного (13.8),
$$
\left(\vec{\bigtriangledown}\!\cdot \!\mathcal{F}(x), \vec{E}_{\mathcal{F}}(x) \right)= \vec{\bigtriangledown}\!\cdot\! \left(\mathcal{F}(x) \!\cdot\! \vec{E}_{\mathcal{F}}(x) \right) -  \left(\mathcal{F}(x), \tilde{rot}\,\vec{E}_{\mathcal{F}}(x) \right),              \eqno\ldots (16.5)
$$
приводит к дифференциальному закону сохранения энергии для электромагнитного поля
\begin{multline*}
 \partial^{0}\,\frac{E_{\mathcal{F}}(x)^{2}+\mathcal{F}(x)^{2}}{2}= -  \vec{\bigtriangledown}\cdot \left(\mathcal{F}(x)\cdot \vec{E}_{\mathcal{F}}(x) \right) -  \\
 \left(\mathcal{F}(x), \tilde{rot}\,\vec{E}_{\mathcal{G}}(x) \right) - \frac{1}{2}\,\vec{j}(x)\cdot  \vec{E}_{\mathcal{F}}(x).                                                               \qquad\ldots(16.6)
\end{multline*}

Использование тождества $ \tilde{rot}\,\vec{E}_{\varphi}(x)\equiv\tilde{0}$ позволяет скалярное произведение
во втором члене правой части (16.6) представить в виде
$$
\left(\mathcal{F}(x), \tilde{rot}\,\vec{E}_{G}(x) \right) =
 \frac{1}{2}\,\left(\mathcal{F}(x), \tilde{rot}\,\vec{E}_{\mathit A}(x) \right) =
  - \frac{1}{4}\,\partial^{0}\,\mathcal{F}(x)^{2},                                                  \eqno\ldots (16.7)
$$
вследствие чего уравнение (16.6) принимает вид
$$
\partial^{0}\,\frac{E_{\mathcal{F}}(x)^{2}+\mathcal{F}(x)^{2}/ 2}{2}= -  \vec{\bigtriangledown}\cdot \left(\mathcal{F}(x) \!\cdot\! \vec{E}_{\mathcal{F}}(x) \right)  - \frac{1}{2}\,\vec{j}(x)\!\cdot \! \vec{E}_{\mathcal{F}}(x),                      \eqno\ldots (16.8)
$$
представляющий собой дифференциальный закон сохранения энергии для системы, представленной полевыми переменными $ \vec{E}_{\mathcal{F}}(x)$ и $ \mathcal{F}(x)$.

Для системы, представленной традиционными лоренцевыми полевыми переменными, $ \vec{E}_{\mathcal{F}L}(x)$ и $ \mathcal{F}_{\mathit L}(x)$, (16.8) приводит к известной теореме Пойнтинга
$$
\partial^{0}\,\frac{E_{\mathcal{F}L}(x)^{2}+\mathcal{F}_{L}(x)^{2}/ 2}{2}= -  \vec{\bigtriangledown}\cdot \left(\mathcal{F}_{L}(x) \!\cdot\! \vec{E}_{\mathcal{F}L}(x) \right)  - \vec{j}(x)\cdot  \vec{E}_{\mathcal{F}L}(x),                      \eqno\ldots (16.9)
$$
из которой следуют хорошо известные выражения для плотности энергии  и плотности потока энергии данной
полевой системы
$$
W_{\mathcal{F}L}(x) = \frac{E_{\mathcal{F}L}(x)^{2}+\mathcal{F}_{L}(x)^{2}/ 2}{2},            \eqno\ldots (16.10)
$$
$$
\vec{S}_{\mathcal{F}L}(x) = \mathcal{F}_{L}(x) \!\cdot\! \vec{E}_{\mathcal{F}L}(x).             \eqno\ldots (16.11)
$$

В частном случае плоской волны Ландау, рассмотренной выше, $ \vec{E}_{\mathcal{F}L}(x) = \vec{E}_{\mathit A}(x)$,
в результате чего соотношения (16.10) и (16.11) приводят, соответственно, к (13.19) и (13.20), совместно с выше приведенным анализом этих выражений.

Наконец, использование соотношения (3.6) и равенства $\mathcal{F}(x)^{2}= 2\,\vec{\mathcal{H}}(x)^{2}$ позволяет
выразить уравнения (16.9)--(16.11) через псевдовектор $ \vec{\mathcal{H}}_{ \mathit L}(x)$, сопровождающий
лоренцев тензор $ \mathcal{F}_{ \mathit L}(x)$, получив традиционный вид этих уравнений  \cite{La}.

Теперь рассмотрим полевую систему  представленную электроджейтонным полем, так же вновь выписав
необходимые для нашей цели полевые уравнения системы (VI)
$$
\vec{\bigtriangledown}\!\cdot \mathcal{G}(x)  =  - \partial ^{0}\vec{\mathit E}_{\mathcal{G}}(x) - \frac{1}{2}\, \vec{j}(x), \eqno\ldots (16.12)
$$
$$
\partial^{0}\, \mathcal{G}(x)  = -  \tilde{def}\,\vec{E}_{A}(x).                           \eqno\ldots (16.13)
$$

Вновь следуя прежней процедуре, обе части уравнения (16.12) скалярно умножаем на $ \vec{\mathit E}_{\mathcal{G}}(x)$,
а обе части уравнения (16.13) на $ \mathcal{G}(x)$, в результате чего получаем систему соответствующих скалярных уравнений
$$
 \frac{1}{2}\,\partial^{0}\,\vec{E}_{\mathcal{G}}(x)^{2} = - \left(\vec{\bigtriangledown}\!\cdot \mathcal{G}(x), \vec{E}_{\mathcal{G}}(x) \right) - \frac{1}{2}\,\vec{j}(x) \! \cdot \!  \vec{E}_{\mathcal{G}}(x),                           \eqno\ldots (16.14)
 $$
 $$
\frac{1}{2}\,\partial^{0}\,\mathcal{G}(x)^{2} = - \left(\tilde{def}\,\vec{E}_{\mathit A}(x),  \mathcal{G}(x)\right),   \eqno\ldots (16.15)
 $$
 последующее сложение которых, с использованием соотношения, идентичного (13.9),
 $$
\left(\vec{\bigtriangledown}\! \cdot  \mathcal{G}(x), \vec{\mathit E}_{\mathcal{G}}(x) \right) =
 \vec{\bigtriangledown}\!\cdot \! \left(\mathcal{G}(x)\! \cdot\! \vec{\mathit E}_{\mathcal{G}}(x) \right) - \left(\mathcal{G}(x), \tilde{def} \,\vec{\mathit E}_{\mathcal{G}}(x) \right),                                                                                   \eqno\ldots (16.16)
$$
приводит к дифференциальному закону сохранения энергии для электроджейтонного поля
\begin{multline*}
 \partial^{0}\,\frac{E_{\mathcal{G}}(x)^{2}+\mathcal{G}(x)^{2}}{2}= -  \vec{\bigtriangledown}\cdot \left(\mathcal{G}(x) \!\cdot\! \vec{E}_{\mathcal{G}}(x) \right) -  \\
  \left(\mathcal{G}(x), \tilde{def}\,\vec{E}_{\mathcal{F}}(x) \right) - \frac{1}{2}\,\vec{j}(x)\cdot  \vec{E}_{\mathcal{G}}(x).                                                                         \qquad\ldots(16.17)
\end{multline*}

Скалярное произведение во втором члене правой части (16.17) теперь представлено соотношением
$$
 \left(\mathcal{G}(x), \tilde{def}\,\vec{E}_{\mathcal{F}}(x) \right) = - \frac{1}{4}\,\partial^{0}\,\mathcal{G}(x)^{2}+ \frac{1}{2}\,
  \left(\mathcal{G}(x), \tilde{def}\,\vec{E}_{\varphi}(x) \right),                    \eqno\ldots (16.18)
$$
вследствие чего уравнение (16.17) принимает вид
\begin{multline*}
 \partial^{0}\,\frac{E_{\mathcal{G}}(x)^{2}+\mathcal{G}(x)^{2}/2}{2}= -  \vec{\bigtriangledown}\cdot \left(\mathcal{G}(x) \!\cdot\! \vec{E}_{\mathcal{G}}(x) \right) -   \\
   \frac{1}{2}\,\left(\mathcal{G}(x), \tilde{def}\,\vec{E}_{\varphi}(x) \right) - \frac{1}{2}\,\vec{j}(x)\!\cdot\!  \vec{E}_{\mathcal{G}}(x),                               \qquad\ldots(16.19)
\end{multline*}
качественно отличающееся от (16.8) наличием $ \vec{E}_{\varphi}(x)$-члена в правой части данного уравнения.

Для  системы, представленной лоренцевыми полевыми переменными,
$ \vec{E}_{\mathcal{G}L}(x)$ и $ \mathcal{G}_{\mathit L}(x)$, (16.19) приводит к соотношению
\begin{multline*}
 \partial^{0}\,\frac{E_{\mathcal{G}L}(x)^{2}+\mathcal{G}_{L}(x)^{2}/2}{2}= -  \vec{\bigtriangledown}\cdot \left(\mathcal{G}_{L}(x) \!\cdot\! \vec{E}_{\mathcal{G}L}(x) \right) -   \\
   \frac{1}{2}\,\left(\mathcal{G}_{L}(x), \tilde{def}\,\vec{E}_{\varphi}(x) \right) - \vec{j}(x)\!\cdot \! \vec{E}_{\mathcal{G}L}(x).                                                \qquad\ldots(16.20)
\end{multline*}

В случаях когда  $ \vec{E}_{\varphi}(x)= \vec{0}$, (16.20) принимает стандартную форму  закона сохранения
$$
 \partial^{0}\,\frac{E_{\mathcal{G}L}(x)^{2}+\mathcal{G}_{L}(x)^{2}/2}{2}= -  \vec{\bigtriangledown}\cdot \left(\mathcal{G}_{L}(x) \!\cdot\! \vec{E}_{\mathcal{G}L}(x) \right) - \vec{j}(x)\!\cdot \! \vec{E}_{\mathcal{G}L}(x),                         \eqno\ldots (16.21)
$$
из которой следуют выражения для плотности энергии и плотности потока энергии рассматриваемой полевой
системы в данном случае
$$
W_{\mathcal{G}L}(x) = \frac{E_{\mathcal{G}L}(x)^{2}+\mathcal{G}_{L}(x)^{2}/ 2}{2},            \eqno\ldots (16.22)
$$
$$
\vec{S}_{\mathcal{G}L}(x) = \mathcal{G}_{L}(x) \!\cdot\! \vec{E}_{\mathcal{G}L}(x).             \eqno\ldots (16.23)
$$

Для плоской волны Ландау "автоматически"\, выполняется соотношение $ \vec{E}_{\varphi}(x)= \vec{0}$, а
выражения (16.22) и (16.23) приводят к (13.21) и (13.22), соответственно, совместно  с соответствующими
 комментариями к ним.

  \section*{\S \,17 Продольные и поперечные электроджейтонные волны, генерируемые линейно вибрирующей
  точечной электрически заряженной частицей}
\addcontentsline{toc}{section}{\S \,17 Продольные и поперечные электроджейтонные волны, генерируемые линейно вибрирующей
  точечной электрически заряженной частицей}

\qquad  Второй важный случай, когда $ \vec{\mathit E}_{\varphi}$-член не доставляет "хлопот"\,, представлен
системой рассматриваемых полевых переменных волновой зоны излучения на продольной оси линейно
вибрирующей точечной электрически заряженной частицы.

В этом случае $ \vec{\mathit E}_{\varphi}(x) = - \vec{\mathit E}_{\mathit A}(x)$, поэтому скалярное произведение
второго члена правой части (16.19) принимает вид
$$
\left(\mathcal{G}(x), \tilde{def}\,\vec{E}_{\varphi}(x) \right) = \frac{1}{2}\,\partial^{0}\,\mathcal{G}(x)^{2},      \eqno\ldots (17.1)
$$
в результате чего уравнение (16.19), в рассматриваемой области данного вибратора, так же принимает стандартную
дифференциальную форму закона сохранения энергии
$$
\partial^{0}\,\frac{E_{\mathcal{G}}(x)^{2}+\mathcal{G}(x)^{2}}{2}= -  \vec{\bigtriangledown}\!\cdot\! \left(\mathcal{G}(x) \!\cdot\! \vec{E}_{\mathcal{G}}(x) \right),                                          \eqno\ldots (17.2)
$$
из которого следуют выражения для плотности энергии и плотности потока энергии {\em электроджейтонного} поля
на рассматриваемой {\em продольной} оси рассматриваемого источника излучения
$$
W_{\mathcal{G}}(x) = \frac{E_{\mathcal{G}}(x)^{2}+\mathcal{G}(x)^{2}}{2},            \eqno\ldots (17.3)
$$
$$
\vec{S}_{\mathcal{G}}(x) = \mathcal{G}(x) \!\cdot\! \vec{E}_{\mathcal{G}}(x).             \eqno\ldots (17.4)
$$

Система уравнений (16.12), (16.13), в данном случае, принимает симметричный, по отношению к входящим в неё
полевым переменным, вид
$$
\vec{\bigtriangledown}\!\cdot \mathcal{G}(x)  =  - \partial ^{0}\vec{\mathit E}_{\mathcal{G}}(x) , \eqno\ldots (17.5)
$$
$$
\partial^{0}\, \mathcal{G}(x)  = -  \tilde{def}\,\vec{E}_{ \mathcal{G}}(x).                           \eqno\ldots (17.6)
$$

Стартуя с этих уравнений так же легко получить соотношения (17.2)--(17.4).

4-потенциал Лиенара\,--\,Вихерта в зоне излучения удовлетворяет однородному волновому уравнению
Даламбера, из которого следует соответствующий волновой характер распространения полевых переменных,
входящих в систему уравнений (17.5), (17.6), определяющих дифференциальные связи этих переменных,
вследствие чего эти полевые переменные, в рассматриваемой области, представляют собой единый полевой объект
(электроджейтонное поле), волновым образом распространяющийся  вдоль продольной оси
рассматриваемого излучателя и представляющий собой электроджейтонную волну.

При этом, в силу того, что электрическая составляющая данной волны,
 $ \vec{ \mathit E}_{\mathcal{G}}(x)= \vec{ \mathit E}_{\mathit A}(x)$, ориентирована вдоль направления распространения
 этой волны, естественно классифицировать данную электроджейтонную волну как {\em продольную} электроджейтонную волну.

 Рассмотрение случая произвольного направления излучения продольной электроджейтонной волны данного источника
 электродинамического излучения сводится, подобно тому как это было при анализе поперечной составляющей данного  излучения представленного на  рис.8.1, к простой замене: $ \vec{\mathit A}(x)\mapsto \vec{\mathit A}^{l}(x)$,
 $ \vec{\mathit E_{\mathcal{G}}}(x)\mapsto \vec{\mathit E}_{\mathcal{G}}^{l}(x)= \vec{\mathit E}_{\mathit A}^{l}(x)$.

 И так, классическое электродинамическое излучение линейно вибрирующей точечной электрически заряженной частицы представлено не только поперечной электромагнитоджейтонной волной, представленной на рис. 4.3\,{\em a})\, и  рис. 8.1, но и продольной  электроджейтонной волной, рассмотренной выше и представленной на рис. 4.3 {\em б})\, и  рис. 17.1.

 Рис. 4.3\,{\em б})\, {\em геометрически} демонстрирует, что "неэлектрическая"\, часть данной волны является
 суперпозицией девиатационного джейтонного и дилатационного джейтонного полей, при тождественном отсутствии,
 на рассматриваемой продольной оси, магнитного поля.

\setcounter{section}{17}
\setcounter{figure}{0}
\begin{figure}[!h]
\begin{center}
\vspace{-6pt}
\includegraphics[width=125mm]{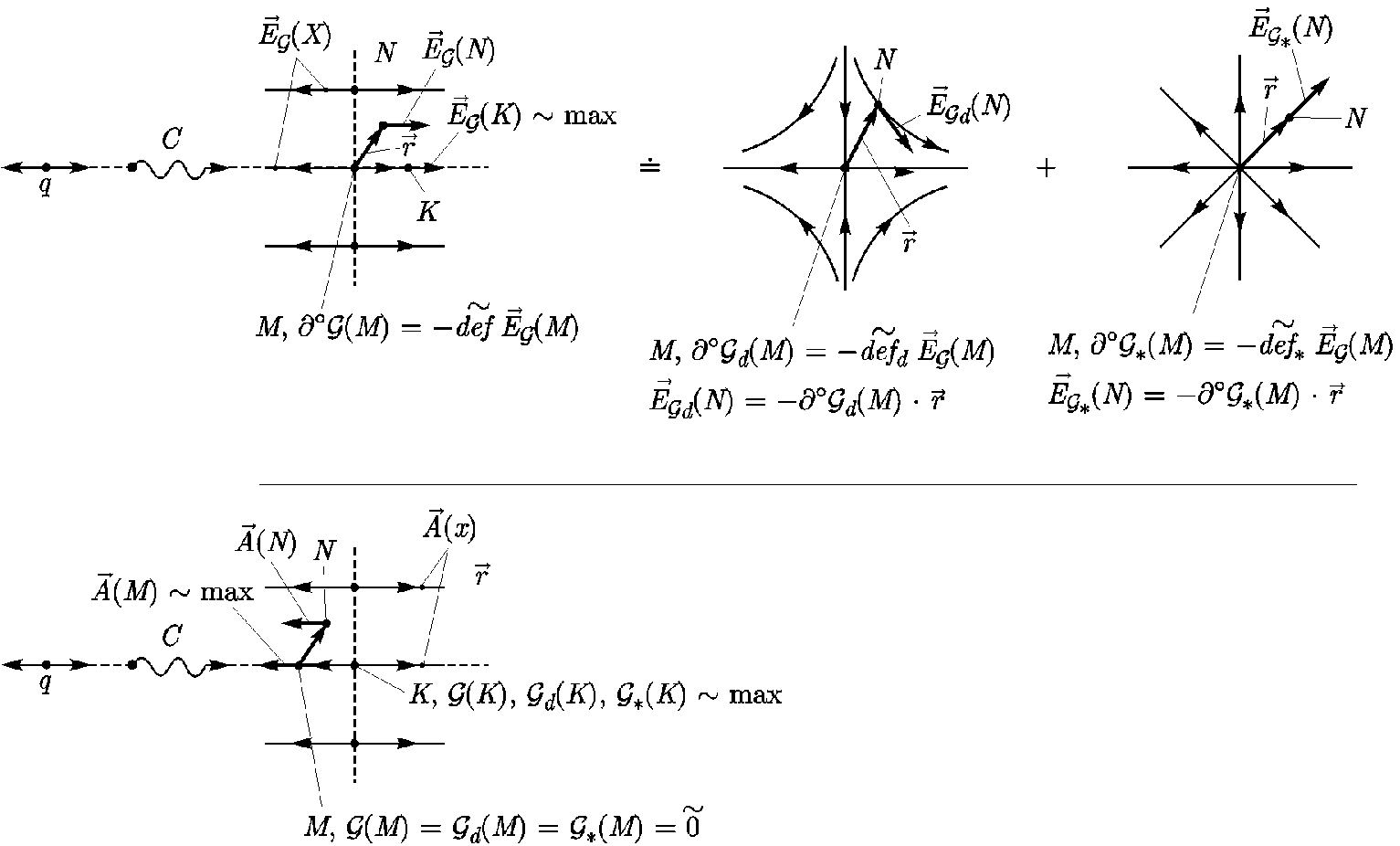}
\vspace{1pt}
\caption{Конгруэнтное разложение векторного поля напряженности электрического поля продольной электроджейтонной волны в меридиональной плоскости  линейно вибрирующей точечной электрически заряженной частицы (продольное направление волновой зоны), законы электроджейтонной индукции, действующие в этой волне, и их геометрическая интерпретация. Случай, когда точка $M$ лежит вне собой линии плоскости электрического поля ${\vec E}_{\mathcal G}(x)$ сводится лишь к дополнительному учету соответствующего однородного электрического поля, что содержательной стороны рассматриваемых процессов не затрагивает.}
\end{center}
\end{figure}

 Рис. 17.1, с одной стороны, выступает в качестве наглядной демонстрации участия законов электроджейтонной индукции, (1.12) и (1.13), в формировании рассматриваемого вида излучения, с другой стороны, является демонстрацией
 очередной конкретной роли джейтонных полей в этом процессе.

 И так, в соответствии с выше рассмотренным  разложением (1.21), электрическое поле, определяемое   напряженностью $ \vec{\mathit E}_{\mathit A}(x)$, является, в общем случае, топологической суперпозицией электрических полей, конгруэнции   векторных линий которых представлены не только классическим центром, но и классическим (каноническим) седлом   и дикритическим узлом.

Вместе с тем, уравнение (1.11) описывает "рождение"\, изменяющимся во времени магнитным полем лишь только
 {\em вихревого} электрического поля, определяемого тензором $ \tilde{rot}\,\vec{E}_{\mathit A}(x)$.

 Таким образом, в данном отношении, данное уравнение, а с ним и уравнение Фарадея\,--\,Максвелла
 $$
 \partial^{0}\,\vec{\mathcal{H}}_{\mathit L}(x) = - rot\,\vec{E}_{\mathcal{F}L}(x),
 $$
 не является полным описанием реальной действительности.

 Уравнения (1.12) и (1.13) восстанавливают полноту описания рассматриваемых процессов {\em полевого}
 "рождения"\, электрических полей.

 С другой стороны находим, что теперь класс электрических полей классической электродинамики
 пополняется  {\em безвихревыми} электрическими полями, порождаемыми изменяющимся во времени соответствующими джейтонными полями, то есть, как и общеизвестное вихревое электрическое поле, имеющими
   {\em полевое}  происхождение, однако обладающими принципиально иной конфигурацией  силовых линий.

  В качестве основного итога настоящего контекста, вновь констатируем, что подобно тому как магнитная составляющая электромагнитного поля (магнитное поле)    оказывает  {\em опосредованное} силовое воздействие на системы электрически заряженных частиц посредством порождаемого ей (им), в соответствии с законом электромагнитной индукции (1.11),  {\em вихревого} электрического поля и это является одним из фактов, относящих магнитное поле к классу  {\em физических}  полей, так и джейтонная составляющая электроджейтонного поля ({\em джейтонное поле}) оказывает {\em опосредованное} силовое воздействие на системы электрически заряженных частиц посредством порождаемого ей (им), в соответствии с законами электроджейтонной индукции (1.12) и (1.13),  {\em деформирующих} электрических полей, что так же следует рассматривать как факт, относящий джейтонное поле к классу {\em самостоятельных физических} полей.

В заключение данной части, автор считает своим приятным долгом выразить глубокую благодарность профессору РАН, доктору физико-математических наук Арбузову А. Б. (Лаборатория Теоретической Физики  Объединенного Института Ядерных Исследований (г. Дубна)),  доктору физико-математических  Залиханову Б. Ж. и кандидату  физико-математических наук Баранову В. А.(Лаборатория Ядерных Проблем Объединенного Института Ядерных Исследований (г. Дубна)) за проявленное внимание  к  обсуждаемым работам.



\renewcommand{\bibname}{\em Литература к части II}

\renewcommand{\contentsname}{Оглавление}
\tableofcontents
\end{document}